\documentclass[rmp,aps,twocolumn,nofootinbib]{revtex4}

\usepackage{graphicx}
\usepackage{amssymb}
\usepackage{amsbsy}
\usepackage{mathbbol}
\usepackage{psfrag}
\usepackage[english]{babel}

\def\bGbar{\bar{\bf G}}
\def\bG{{\bf G}}
\def\bg{{\bf g}}
\def\bGscript{\pmb{\cal G}}

\def\bSigma{{\bf\Sigma}}
\def\bchi{{\pmb \chi}}

\def\bGamma{{\bf\Gamma}}

\def\bU{{\bf U}}
\def\bUbar{\bar{\bf U}}

\def\chibar{\bar{\chi}}
\def\chic{{\chi_c}}

\def\Dbar{{\bar{D}}}

\def\epsbar{\bar{\epsilon}}

\def\Gammac{{\Gamma_c}}
\def\Gbar{\bar{G}}
\def\G{{\bf G}}
\def\Gscript{{\cal{G}}}

\def\Gc{{G_c}}

\def\k{{\bf{k}}}
\def\K{{\bf{K}}}
\def\kt{{\tilde{\k}}}
\def\tk{{\tilde{k}}}

\def\q{{\bf{q}}}
\def\Q{{\bf{Q}}}

\def\tq{{\tilde{q}}}

\def\si{\sigma}

\def\t{{\bf t}}
\def\J{{\bf J}}

\def\Ubar{\bar{U}}

\def\Ubar{\bar{U}}

\def\X{{\bf X}}
\def\x{{\bf x}}
\def\xt{{\tilde{\x}}}
\def\tx{{\tilde{x}}}

\def\vereq#1#2{\lower3pt\vbox{\baselineskip1.5pt \lineskip1.5pt
\ialign{$#1\hfill##\hfil$\crcr#2\crcr\sim\crcr}}}

\begin{document}

\title{Quantum Cluster Theories}

\author{Thomas Maier}
\email{maierta@ornl.gov}
\affiliation{Computational Science and Math Division, 
Oak Ridge National Laboratory, Oak Ridge, TN 37831-6114}
\author{Mark Jarrell}
\affiliation{Department of Physics, University of Cincinnati, 
Cincinnati, Ohio 45221-0011}
\author{Thomas Pruschke}
\affiliation{Theoretical Physics, University of G\"ottingen, 
Tammannstr.\ 1, 37077 G\"ottingen, Germany}
\author{Matthias H. Hettler}
\affiliation{Forschungszentrum Karlsruhe, Institut f\"ur Nanotechnologie, 
Postfach 3640, 76021 Karlsruhe, Germany}

\date{\today}

\begin{abstract}
  Quantum cluster approaches offer new perspectives to study the
  complexities of macroscopic correlated fermion systems. 
  These approaches can be understood as generalized mean-field theories. 
  Quantum cluster approaches are non-perturbative and are always in the thermodynamic limit.
Their quality can be systematically improved, and they provide
  complementary information to finite size simulations.  They have
  been studied intensively in recent years and are now well established.
After a brief historical review, this article comparatively discusses 
the nature and advantages of these cluster techniques. 
Applications to common models of correlated electron systems are reviewed. \footnote{This article has been submitted to Reviews of Modern Physics.}
\end{abstract}

\maketitle

\tableofcontents 

\section{Introduction} 
\label{sec:intro}

\subsection{Brief history} 
\label{subsec:motiv} 


%
%
%
%
%
%


The theoretical description of interacting many-particle systems
remains one of the grand challenges in condensed matter physics.
Especially the field of strongly correlated electron systems has
regained theoretical and experimental interest with the discovery of
heavy Fermion compounds and high-temperature superconductors. In this
class of systems the strength of the interactions between particles is
comparable to or larger than their kinetic energy, i.e.\ any theory
based on a perturbative expansion around the non-interacting limit is
at least questionable.  Theoretical tools to describe these systems
are therefore faced with extreme difficulties, due to the
non-perturbative nature of the problem.  A large body of work has been
devoted to a direct (numerically) exact solution of finite size
systems using exact diagonalization or Quantum Monte Carlo methods.
Exact diagonalization however is severely limited by the exponential
growth of computational effort with system size, while Quantum Monte
Carlo methods suffer from the sign problem at low temperatures.
Another difficulty of these methods arises from their strong finite
size effects, often ruling out the reliable extraction of low energy
scales that are important to capture the competition between different
ground states often present in strongly correlated systems.

Mean-field theories are defined in the thermodynamic limit and
therefore do not face the finite size problems. With applications to a
wide variety of extended systems from spin models to models of
correlated electrons and/or bosons, mean-field theories are extremely
popular and ubiquitous throughout science.
The first mean-field theory which gained wide acceptance was developed
by P. Weiss for spin systems (\onlinecite{weiss:1907}).
The Curie-Weiss mean-field theory reduces the complexity of the
thermodynamic lattice spin problem by mapping it onto that of a
magnetic impurity embedded in a self-consistently determined mean
magnetic field.

Generally, mean-field theories divide the infinite number of degrees
of freedom into two sets. A small set of degrees of freedom is treated
explicitly, while the effects of the remaining degrees of freedom are
summarized as a mean-field acting on the first set. Here, by
mean-field theory, we refer to the class of approximations which
account for the correlations between spatially localized degrees of
freedom explicitly, while treating those at longer length scales with
an effective medium.  Such local approximations become exact in the
limit of infinite coordination number or equivalently infinite
dimensions $D$ (\onlinecite{itzykson:89}); however non-local
corrections become important in finite dimensions.  The purpose of
this review is to discuss methods for incorporating non-local
corrections to local approximations.

Many different local approximations have been developed for systems
with itinerant degrees of freedom.  Early attempts focused on
disordered systems, and included the virtual crystal approximation
(\onlinecite{VCA:nordheim1,VCA:nordheim2,VCA:parmenter,VCA:schoen}) 
and the average-T matrix approximation (\onlinecite{ATA:beeby,ATA:ehrenreich}).  However, the most successful local approximations
for disordered systems is the Coherent Potential Approximation (CPA)
developed by \textcite{soven:CPA} and others
(\onlinecite{taylor:CPA,shiba:RCPA}).  This method is distinguished
from the others in that it becomes exact in both the limit of dilute
and concentrated disordered impurity systems, as well as the limit of
infinite dimensions.

There have been many attempts to extend the CPA formalism to
correlated systems, starting with the Dynamical CPA (DCPA) of
\textcite{Sumi:DCPA,Kakehashi:DCPA}, the XNCA of
\textcite{Kuramoto:NCA2,Kim:NCA} and the LNCA of
\textcite{Grewe:LNCA,Grewe:LNCA2}.  A great breakthrough was achieved
with the formulation of the Dynamical Mean-Field Theory (DMFT) (for a
review see \onlinecite{georges:dmftrev,pruschke:dmftrev}) in the limit
of infinite dimensions by \textcite{metzner:dmft} and
\textcite{hartmann:dmft}.  The DCPA and the DMFT have been the most
successful approaches and employ the same mapping between the cluster
and the lattice problems.  They differ mostly in their starting
philosophy.  The DCPA employs the CPA equations to relate the impurity
solution to the lattice whereas in the DMFT the irreducible quantities
calculated on the impurity are used to construct the lattice
quantities.

Despite the success of these mean-field approaches, they share the
critical flaw of neglecting the effects of non-local fluctuations.
Thus they are unable to capture the physics of, e.g. spin waves in
spin systems, localization in disordered systems, or spin-liquid
physics in correlated electronic systems.  Non-local corrections are
required to treat even the initial effects of these phenomena and to
describe phase transitions to states described by a non-local order
parameter.

The first attempt to add non-local corrections to mean-field theories
was due to \textcite{bethe:cluster} by adding corrections to the
Curie-Weiss mean-field theory. This was achieved by mapping the
lattice problem onto a self-consistently embedded finite-size spin
cluster composed of a central site and $z$ nearest neighbors embedded
in a mean-field.  For small $z$, the resulting theory provides a
remarkably large and accurate correction to the transition temperature
(\onlinecite{Kikuchi:BCising,Suzuki:CAM}).

Many attempts have been made to apply similar ideas to disordered
electronic systems (\onlinecite{gonis:gf}).  Most approaches were
hampered by the difficulty of constructing a fully causal theory, with
positive spectral functions.  Several causal theories were developed
including the embedded cluster method (\onlinecite{gonis:gf}) and the
molecular CPA (MCPA) by \textcite{tsukada:MCPA} (for a review see
\onlinecite{duc:mcpa}).  These methods generally are obtained from the
local approximation by replacing the impurity by a finite size cluster
in real space. As a result these approaches suffer from the lack of
translational invariance, since the cluster has open boundary
conditions and only the surface sites couple to the mean-field.

Similar effort has been expended to find cluster extensions to the
DMFT, including most notably the Dynamical Cluster Approximation (DCA)
(\onlinecite{hettler:dca1,hettler:dca2}) and the Cellular Dynamical
Mean-Field Theory (CDMFT) (\onlinecite{kotliar:cdmft}). Both cluster
approaches reduce the complexity of the lattice problem by mapping it
to a finite size cluster self-consistently embedded in a mean-field.
As in the classical case, the self-consistency condition reflects the
translationally invariant nature of the original lattice problem. The
main difference with their classical counterparts arises from the
presence of quantum fluctuations. Mean-field theories for quantum
systems with itinerant degrees of freedom cut off spatial fluctuations
but take full account of temporal fluctuations.  As a result the
mean-field is a time- or respectively frequency dependent quantity.
Even an effective cluster problem consisting of only a single site
(DMFT) is hence a highly non-trivial many-body problem.  CDMFT and DCA
mainly differ in the nature of the effective cluster problem. The
CDMFT shares an identical mapping of the lattice to the cluster
problem with the MCPA, and hence also violates translational
symmetries on the cluster.  The DCA maps the lattice to a periodic and
therefore translationally invariant cluster.

A numerically more tractable cluster approximation to the
thermodynamic limit was developed by \textcite{gross:cluster}. In this
formalism the self-consistent coupling to a mean-field is neglected.
This leads to a theory in which the self-energy of an isolated finite
size cluster is used to approximate the lattice propagator. As shown
by \textcite{senechal:cluster2}, this cluster extension of the
Hubbard-I approximation is obtained as the leading order approximation
in a strong-coupling expansion in the hopping amplitude and hence this
method was named Cluster Perturbation Theory (CPT).

Generally, cluster formalisms share the basic idea to approximate the
effects of correlations in the infinite lattice problem with those on
a finite size quantum cluster. We refer to this class of techniques as
quantum cluster theories. In contrast to Finite System Simulations
(FSS), these techniques are built for the thermodynamic limit. In this
review we focus on the three most established quantum cluster
approaches, the DCA, the CDMFT and the CPT formalisms.  The CDMFT
approach was originally formulated for general, possibly
non-orthogonal basis sets. In this review we restrict the discussion
to the usual, completely localized orthogonal basis set and refer the
reader to \textcite{kotliar:cdmft} for the generalization to arbitrary
basis sets.

The organization of this article is as follows: To familiarize the
reader with the concept of cluster approaches, we develop in section
\ref{subsec:ccw} a cluster generalization of the Curie-Weiss
mean-field theory for spin systems. Section \ref{sec:qct} sets up the
theoretical framework of the CDMFT, DCA and CPT formalisms by
presenting two derivations based on different starting philosophies.
The derivation based on the locator expansion in Sec.~\ref{subsec:es}
is analogous to the cluster generalization of the Curie-Weiss
mean-field method and thus is physically very intuitive.  The
derivation based on the cluster approximation to diagrams defining the
grand potential in Sec.~\ref{subsec:freeeg} is closely related to the
reciprocal space derivation of the DMFT by \textcite{hartmann:dmft}.
The nature of the different quantum cluster approaches together with
their advantages and weaknesses are assessed in
Sec.~\ref{subsec:disc}.  Discussions of the effective cluster problem,
generalizations to symmetry broken states and the calculation of
response functions are presented in Secs.~\ref{subsec:ecm},
\ref{subsec:brokensym} and \ref{subsec:suscept}. The remainder of this
section is devoted to describe the application of the DCA formalism to
disordered systems in Sec.~\ref{subsec:dis} and to a brief discussion
of alternative methods proposed to introduce non-local corrections to
the DMFT method in Sec.~\ref{subsec:alt}. In Sec.~\ref{sec:qcs} we
review the various perturbative and non-perturbative techniques
available to solve the effective self-consistent cluster problem of
quantum cluster approaches. We include a detailed assessment of their
advantages and limitations. Although numerous applications of quantum
cluster approaches to models of many-particle systems are found in the
literature, this field is still in its footsteps and currently very
active. A large body of work has been concentrated on the Hubbard
model. We review the progress made on this model in Section
\ref{sec:results} together with applications to several other strongly
correlated models. Finally, Sec.~\ref{sec:conc} concludes the review
by stressing the limitations of quantum cluster approaches and
proposing possible directions for future research in this field.


\subsection{Corrections to Curie-Weiss theory}
\label{subsec:ccw}

As an intuitive example of the formalism developed in the next
sections we consider a systematic cluster extension of the Curie-Weiss
mean-field theory for a lattice of classical interacting spins. This
discussion is especially helpful to illustrate many new aspects of
cluster approaches as compared to finite size simulations. The quality
of this approach, and its convergence and critical properties will be
demonstrated with a simple example, the one-dimensional Ising model
\begin{equation}
  \label{eq:cw1}
  H=-J\sum_{i} \sigma_i \sigma_{i+1} - h \sum_{i} \sigma_i
\end{equation}
where $\sigma_i=\pm 1$ are classical spins, $h$ is an external
magnetic field and the exchange integral $J>0$ acts between nearest
neighbors only, favoring ferromagnetism. The generalization of this
approach to higher dimensions and quantum spin systems is straightforward.
 
We start by dividing the infinite lattice into $N/N_c$ clusters of
size $N_c$ (see Fig.  \ref{fig:Nc4_clusters}) with origin $\tx$ and
the exchange integral $J_{ij}$ into intra- ($\J_c$) and inter-cluster
($\delta\J$) parts
\begin{equation}
  \label{eq:cw2}
  \J(\tx_i-\tx_j) = \J_c\delta_{\tx_i,\tx_j} +\delta\J(\tx_i-\tx_j) 
\end{equation}
where each of the terms is a matrix in the $N_c$ cluster sites.  The
central approximation of cluster theories is to retain correlation
effects within the cluster and neglect them between the clusters.  A
natural formalism to implement this approximation is the locator
expansion.  The spin-susceptibility $\chi_{ij}=\beta(\langle
\sigma_i\sigma_j\rangle - \langle
\sigma_i\rangle\langle\sigma_j\rangle)$, where $\beta=1/T$ is the
inverse temperature, can be written as a locator expansion in the
inter-cluster part $\delta \J$ of the exchange interaction, around the
cluster limit $\bchi^o=\bchi(\delta \J=0)$ as
\begin{equation}
  \label{eq:cw3}
  \bchi(\tx_i-\tx_j)=\bchi^o\delta_{\tx_i,\tx_j}+\bchi^o\sum_l\delta\J(\tx_i-\tx_l)
  \bchi(\tx_l-\tx_j)
\end{equation}
where we used again a matrix notation in the $N_c$ cluster sites.  By
using the translational invariance of quantities in the superlattice
$\tx$, this expression can be simplified in the reciprocal space $\tq$
of $\tx$ to
\begin{equation}
  \label{eq:cw4}
  \bchi(\tq)=\bchi^o+\bchi^o\delta\J(\tq)\bchi(\tq)\,.
\end{equation} 
This locator expansion has two well-defined limits. For an infinite
size cluster it recovers the exact result since the surface to volume
ratio vanishes making $\delta \J$ irrelevant, and thus
$\bchi=\bchi^o$. For a single site cluster, $N_c=1$, it recovers the
Curie-Weiss mean-field theory. This is intuitively clear since for
$N_c=1$ fluctuations between all sites are neglected. With the
susceptibility of a single isolated site $\chi^o=1/T$ and $\delta
J(\tq=0)=J(q=0)=J$, we obtain for the uniform susceptibility
\begin{eqnarray}
  \label{eq:cw5}
  \chi(q=0) = \frac{1}{1/\chi^o-J(q=0)}=\frac{1}{T-T_c}
\end{eqnarray}
the mean-field result with critical temperature $T_c=J$.
 
For cluster sizes larger than one, translational symmetry within the
cluster is violated since the clusters have open boundary conditions
and $\delta \J$ only couples sites on the surface of the clusters. As
detailed in the next section, this shortcoming can be formally
overcome and translational invariance restored by considering an
analogous expression to the locator expansion (\ref{eq:cw4}) in the
Fourier space $Q$ of the cluster
\begin{eqnarray}
  \label{eq:cw6}
  \chi(Q,\tq)&=&\chi^o(Q)+\chi^o(Q)\delta J(Q,\tq) \chi(Q,\tq)\nonumber\\
             &=&\frac{1}{1/{\chi^o}(Q)-\delta J(Q,\tq)}\,,
\end{eqnarray}
with analogous relations for the intra- and inter-cluster parts of $J$
\begin{eqnarray}
  \label{eq:cw7}
  \delta J(Q,\tq) &=& J(Q+\tq) - {\bar J}(Q) \\\label{eq:cw7_2}
  {\bar J}(Q) &=& \frac{N_c}{N} \sum_\tq J(Q+\tq)\,.  
\end{eqnarray}
Here, $\tq$ is a vector in the reciprocal space of $\tx$, and $Q$ is a
vector in the reciprocal space of the cluster sites. The
Fourier transform of the exchange integral is given by $J(Q+\tq)=J
\cos (Q+\tq)$, the intra-cluster exchange is $\bar{J}(Q)$, while the
inter-cluster exchange is $\delta J(Q,\tq)$. As we will see in the
next section, the resulting formalism is analogous to the dynamical
cluster approximation for itinerant fermion systems.

In analogy to the Curie-Weiss theory, the lattice system can now be
mapped onto an effective cluster model embedded in a mean-field since
correlations between the clusters are neglected. The susceptibility
restricted to cluster sites is obtained by averaging or
coarse-graining over the superlattice wave-vectors $\tq$
\begin{eqnarray}
  \label{eq:cw8}
  {\bar \chi}(Q)=\frac{N_c}{N}\sum_{\tq} \chi(Q,\tq) =
  \frac{1}{1/\chi^o(Q) - \Gamma(Q)}\,.
\end{eqnarray}
with the hybridization function
\begin{equation}
  \label{eq:20}
  \Gamma(Q)=\frac{\frac{N_c}{N}\sum_\tq \delta J^2(Q,\tq)\chi(Q,\tq)}{1+\frac{N_c}{N}\sum_\tq \delta J(Q,\tq) \chi(Q,\tq)}\,.
\end{equation}
This follows from the fact that the isolated cluster susceptibility
$\chi^o(Q)$ does not depend on the integration variable $\tq$ in
Eq.~(\ref{eq:cw8}).

This expression defines the effective cluster model
\begin{eqnarray}
  \label{eq:8}
  {\cal H}_c & = & -\sum_Q {\bar J}(Q) \sigma(Q) \sigma(-Q) -h \sigma(Q=0)\nonumber\\
   & & - \sum_{Q,\tq}  \delta
  J(Q,\tq) \sigma(Q) \langle \sigma(-Q-\tq)  \rangle\,,
\end{eqnarray}
where $\sigma(Q)$ ($\sigma(q)$) denotes the cluster (lattice)
Fourier transform of $\sigma_i$ and $\langle \dots \rangle$ the
expectation value calculated with respect to the cluster Hamiltonian
${\cal H}_c$. As in the Curie-Weiss theory, the cluster model is used
to self-consistently determine the order parameter $\langle
\sigma(Q+\tq) \rangle=\langle \sigma(Q) \rangle \delta(\tq)$ in the
ferromagnetic state. In the paramagnetic state, the susceptibility
calculated in the cluster model takes the same form as the
coarse-grained result Eq.~(\ref{eq:cw8}) obtained from the locator
expansion.

The uniform susceptibility $\chi(Q=0,\tq=0)$ contains information
about the nature of this cluster approach, its critical properties and
its convergence with cluster size. The sum in Eq.~(\ref{eq:cw7_2}) may
be solved analytically
\begin{equation}
{\bar J}(Q=0)= J(N_c/\pi) \sin(\pi/N_c)\,.
\end{equation}
The isolated cluster susceptibility $\chi^o(Q)$ can also be calculated
analytically by using the transfer matrix method to give
\cite{tmatrix:goldenfeld}
\begin{equation}
\label{eq:transfer}
\chi^o(Q=0)= \beta \exp(2K)
\frac{ 1- \left( \tanh(K)\right)^{N_c}}
{ 1+ \left( \tanh(K)\right)^{N_c}}\,,
\end{equation}
where $K=\beta{\bar J}(Q=0)=\beta J(N_c/\pi) \sin(\pi/N_c)$.  With
these expressions the uniform lattice susceptibility
Eq.~(\ref{eq:cw6}) becomes
\begin{eqnarray}
\label{eq:isingdcachi}
\chi(T) &=& \frac{1}{1/\chi^o(Q=0)-\delta J(Q=0,\tq=0)}\\ 
&=& \frac{1}{1/\chi^o(Q=0)-J\left(1- (N_c/\pi) \sin(\pi/N_c) \right)}\nonumber\,.
\end{eqnarray}
The cluster estimate of the lattice susceptibility interpolates
between the Curie-Weiss result and the exact lattice result as $N_c$
increases. It may be used to reveal some of the properties of cluster
approximations and to compare the cluster results to both the
finite-size calculation and the exact result in the thermodynamic
limit.


\begin{figure}[htb]
  \includegraphics*[width=3.0in]{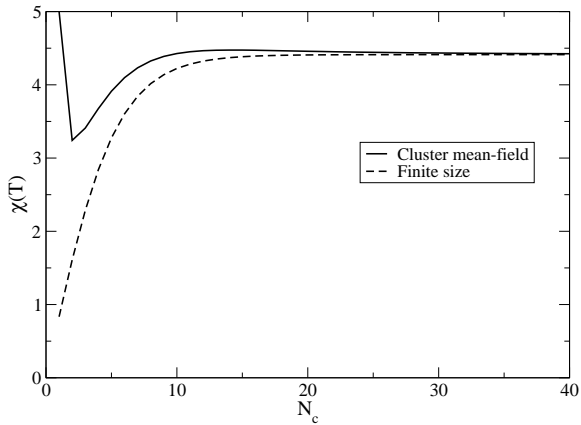}
\caption{The cluster and finite-size estimates of the uniform lattice 
  susceptibility versus cluster size when $J=1$ and $T=0.7$.}
\label{fig:1dising_comp}
\end{figure}

First, both the cluster mean-field result Eq.~(\ref{eq:isingdcachi}) and 
the finite-size result Eq.~(\ref{eq:transfer}) with $K=\beta J$ may be 
regarded as an approximation to the thermodynamic result. However, as 
illustrated in Fig.~\ref{fig:1dising_comp}, the cluster mean-field result
converges more quickly as a function of cluster size $N_c$ than the
finite size result.  This reflects the superior starting point of the
cluster approximation compared to the finite-size calculation.  The
cluster approximation is an expansion about the mean-field result,
whereas the finite-size calculation is an expansion about the atomic
limit.

It is instructive to explore the convergence of the cluster result
analytically.  For large $N_c$, the character of the susceptibility
Eq.~(\ref{eq:isingdcachi}) can be split into three regimes.  At very
high temperatures
\begin{equation}
\chi(T) \approx \frac{1}{T-\Theta} \;\;\;{\mbox{for}}\;\;\; T\gg J
\end{equation}
where $\Theta\approx 2J + \frac{J}{6} \left( \frac{\pi}{N_c} \right)^2$.
At intermediate temperatures, 
\begin{equation}
\chi(T) \approx \beta e^{2\beta J}\left(1-\frac{\beta J}{3}\left(\frac{\pi}{N_c} \right)^2 \right) \;\;\;{\mbox{for}}\;\;\; J\gg T\gg T_c\,.
\end{equation}
The true critical behavior of the system can be resolved by studying
the properties of this intermediate temperature regime.  At both high
and intermediate temperatures, the susceptibility differs from the
exact result by corrections of order ${\cal{O}}(1/N_c^2)$.  In
general, cluster methods with periodic boundary conditions have
corrections of order ${\cal{O}}(1/L_c^2)$, where $L_c=N_c^{1/D}$ is
the linear size of the cluster.

At low temperatures, very close to the transition to the ferromagnetic
state, deviations from the exact result are far larger.  Here, for
large clusters
\begin{equation}
\chi(T) \sim \frac{N_c}{T-T_c}\,,
\end{equation}
with the critical temperature $T_c>0$, whereas the exact
susceptibility in this regime 
$\chi(T) \approx \beta \exp{\left( 2\beta J\right) }$
does not diverge until zero temperature.  This discrepancy is expected
in cluster approximations, since they treat long length scales which
drive the transition in a mean-field way and therefore neglect long
wave-length modes which eventually suppress the transition. Hence,
cluster approximations generally predict finite transition
temperatures independent of dimensionality due to their residual
mean-field character. With increasing cluster size however, the
transitions are expected to be systematically suppressed by the
inclusion of longer-ranged fluctuations.

\begin{figure}[htb]
\includegraphics*[width=3.0in]{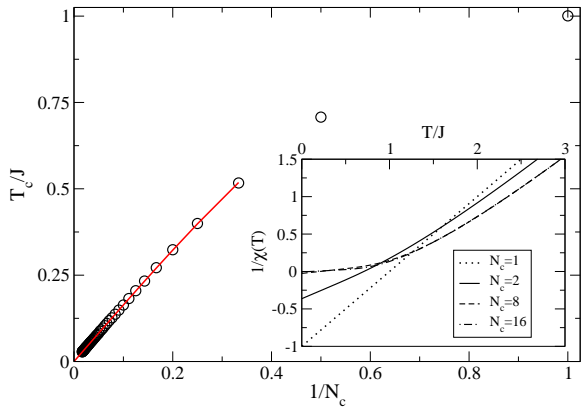}
\caption{Transition temperature for the 1D Ising model versus the inverse 
  cluster size when $J=1$ obtained with the cluster mean-field
  approach.  For large clusters, $T_c \sim 1/N_c$, shown as a solid
  line.  Inset: The inverse susceptibility versus temperature.  The
  Curie-Weiss behavior at low $T$ where $1/\chi(T)$ is linear in $T$,
  illustrates that the transition is always mean-field like.  The true
  critical behavior of the transition is seen in the crossover region
  near $T=J$.}
\label{fig:1dising_Tc}
\end{figure}

For cluster sizes larger than one, all three regions are evident in the 
plot of the cluster mean-field estimate of the inverse susceptibility, 
shown in the inset to Fig.~\ref{fig:1dising_Tc}.  For $N_c=8$ and $N_c=16$,
the high and low temperature parts are seen as straight lines on the 
plot in the inset, with the crossover region in between.  In numerical
simulations with significant sources of numerical noise especially
close to the transition, it is extremely difficult to resolve the 
true low-temperature mean-field behavior.  Exponents extracted from 
fits to the susceptibility in these simulations will more likely reflect 
the properties of the intermediate temperature regime.

Despite the large deviations of the cluster result from the exact
result low temperatures, we may still extract the correct physics
through finite size extrapolation. In general, for a system where the
correlations build like $\xi \sim
\left|\frac{T-T_c}{T_c}\right|^{-\nu}$, we expect $T_c \sim T_c^* -a
L_c^{-1/\nu}$, where $T_c^*$ is the exact transition temperature,
$L_c$ is the linear cluster size, and $a$ is a positive real
constant \cite{Suzuki:CAM}. However, for the 1D Ising system, $\xi
\sim \frac12 \exp{\left(\beta J\right)}$, so more care must be taken.
Fortunately, an analytic expression for the transition temperature may
be extracted from Eq.~(\ref{eq:isingdcachi}).  For large clusters,
$T_c \approx J N_c \left( \frac16\left(\frac{\pi}{N_c}\right)^2 -
  \frac1{120}\left(\frac{\pi}{N_c}\right)^4\right)$.  This behavior is
shown in the main frame of Fig.~\ref{fig:1dising_Tc} with the circles
depicting the numerical values for $T_c$ and the solid line their
asymptotic behavior.


\section{Quantum cluster theories} 
\label{sec:qct} 

In this section we provide two derivations of quantum cluster
approaches for systems with itinerant quantum degrees of freedom. The
locator expansion in Sec.~\ref{subsec:es} is analogous to the cluster
extension of the Curie-Weiss mean-field theory developed in the
preceding section.
Sec.~\ref{subsec:freeeg} provides a microscopic derivation based on
cluster approximations to the thermodynamic grand potential. A
detailed discussion of the nature of quantum cluster approaches and
the effective cluster model is presented in Secs.~\ref{subsec:disc}
and \ref{subsec:ecm}. Generalizations for symmetry broken phases, the
calculation of susceptibilities and the application to disordered
systems is explained in Secs.~\ref{subsec:brokensym},
\ref{subsec:suscept} and \ref{subsec:dis} and a brief discussion of
alternative cluster methods is presented in Sec.~\ref{subsec:alt}.

\subsection{Cluster approximation to the locator expansion}
\label{subsec:es}

In this section, we derive a number of cluster formalisms for
itinerant many body systems using an analogous approach to that
discussed in Sec.~\ref{subsec:ccw} for classical spin systems.  For
simplicity we assume in this section that no symmetry breaking occurs;
the treatment of symmetry broken phases is discussed in
Sec.~\ref{subsec:brokensym}. The basic idea is to write down a locator
expansion, i.e. an expansion in space around a finite-size cluster.
This approach is not only intuitive but also allows us to assess the
nature of quantum cluster approximations.  As with their classical
counterparts, quantum cluster theories approximate the lattice problem
with many degrees of freedom by an effective cluster problem with few
degrees of freedom embedded in an external bath or mean-field created
from the remaining degrees of freedom.  By neglecting correlations
that extend beyond the cluster size, one can then formulate a theory
in which the lattice system is replaced by an effective cluster
embedded in a mean-field host. While the formalism derived here is
analogous to the formalism discussed in Sec.~\ref{subsec:ccw} for spin
systems, there are significant differences. Since we are dealing with
itinerant fermions, the theory is built upon the single-particle Green
function instead of the two-particle spin correlation function, and
the mean-field is dynamical due to the itinerant nature of the
particles.

This derivation is illustrated on the example of the extended Hubbard model
\begin{eqnarray}
  \label{eq:loc1}
    H = \sum_{ij,\sigma}
  t_{ij}
  c^\dagger_{i\sigma}c^{}_{j\sigma} 
  + \frac{1}{2}\sum_{ij,\sigma\sigma'} U_{ij} n_{i\sigma} n_{j\sigma'}\,.
\end{eqnarray}
Here $i$ and $j$ are lattice site indices, the operators
$c_{i\sigma}^\dagger$ ($c_{i\sigma}^{}$) create (destroy) an electron
with spin $\sigma$ on site $i$,
$n_{i\sigma}=c^\dagger_{i\sigma}c^{}_{i\sigma}$ is their corresponding
number density, and $U_{ij}$ denotes the Coulomb repulsion between
electrons on sites $i$ and $j$. The hopping amplitude between sites
$i$ and $j$ is denoted by $t_{ij}$, its local contribution
$t_{ii}=\epsilon_o$ and its Fourier transform to reciprocal space is
the dispersion $\epsilon_\k$. In this section we limit the discussion
to the regular Hubbard model with a purely local interaction
$U_{ij}=U(1-\delta_{\sigma\sigma'})\delta_{ij}$. The more general case
of finite non-local interactions $U_{ij}$ for $i\neq j$ is
discussed in Sec.~\ref{subsec:freeeg}.

The central quantity upon which we build the locator expansion is the
single-particle thermodynamic Green function ($\tau$ is the imaginary
time, $T_\tau$ the corresponding time ordering operator, $\beta=1/T$
the inverse temperature and $\omega_n$ are the fermionic Matsubara
frequencies)
\begin{equation}
  G_{ij,\sigma}(\tau)=-\langle T_\tau c^{}_{i\sigma}(\tau) c^\dagger_{j\sigma}\rangle
\end{equation}
\begin{equation}
G_{ij,\sigma}(i\omega_n)=\int\limits_0^\beta d\tau
e^{i\omega_n\tau}G_{ij,\sigma}(\tau)\,;\,\, \omega_n=\frac{(2n+1)\pi}{\beta}
\end{equation}
or respectively its analytical continuation
$G_{ij,\sigma}(z)=\langle\langle c^{}_{i\sigma}, c^\dagger_{j\sigma}
\rangle\rangle_z$ to complex frequencies $z$.

\begin{figure}[htb]
\centerline{
\includegraphics*[width=3.5in]{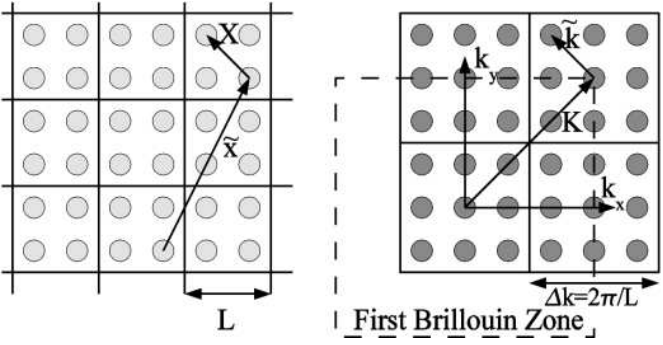}}
\caption{Definition of the coordinates in real (left) and reciprocal (right) space illustrated for $N_c=4$ site clusters. The origin of a cluster is labeled by $\xt$, the sites within a cluster by $\X$. The reciprocal space to $\X$ is labeled by $\K$, the wave-vectors of the superlattice, i.e. within a cell by $\kt$.}
\label{fig:Nc4_clusters}
\end{figure}

To set up a suitable notation for cluster schemes, we divide the
$D$-dimensional lattice of $N$ sites into a set of finite-size
clusters each with $N_c$ sites of linear size $L_c$ such that
$N_c=L_c^D$, and resolve the first Brillouin zone into a corresponding
set of reduced zones which we call cells. This notation is illustrated
in Fig.~\ref{fig:Nc4_clusters} for $N_c=4$ site clusters. For larger
cluster sizes and more complex cluster geometries we refer the reader
to \textcite{jarrell:dca3}. Care should be taken so that the point
group symmetry of the clusters does not differ too greatly from that
of the original lattice \cite{betts:cubicclusters}.  We use the
coordinate $\xt$ to label the origin of the clusters and $\X$ to label
the $N_c$ sites within a cluster, so that the site indices of the
original lattice $\x=\X+\xt$.  The points $\xt$ form a superlattice
with a reciprocal space labeled by $\kt$.  The reciprocal space
corresponding to the sites $\X$ within a cluster shall be labeled
$\K$, with $K_\alpha=n_\alpha \cdot 2\pi/L_c$ and integer $n_\alpha$.
Then the wave-vectors in the full Brillouin zone are given by
$\k=\K+\kt$.

With these conventions, the Fourier transforms of a given function
$f(\X,\xt)$ for intra- and inter-cluster coordinates are defined as
\begin{eqnarray}
  f(\X,\xt) &=& \frac{N_c}{N}\sum_\kt e^{i\kt\cdot\xt} f(\X,\kt) \label{eq:FT1}\\
  f(\X,\kt) &=& \sum_\xt e^{-i\kt\cdot\xt} f(\X,\xt)\label{eq:FT2}\\
  f(\X,\kt) &=& \frac{1}{N_c}\sum_\K e^{i(\K+\kt)\cdot\X} f(\K,\kt)\label{eq:FT3}\\
  f(\K,\kt) &=& \sum_\X e^{-i(\K+\kt)\cdot\X} f(\X,\kt)\label{eq:FT4}\,.
\end{eqnarray}

To separate out the cluster
degrees of freedom, the hopping amplitude $t$ and the self-energy
$\Sigma$ (defined from the Green function via the Dyson equation
$G^{-1}=G_0^{-1}-\Sigma$ with the non-interacting Green function
$G_0$) is split into intra- and inter-cluster parts
\begin{eqnarray}
  \label{eq:loc2}
  \t(\xt_i-\xt_j)&=&\t_c\delta_{\xt_i,\xt_j}+\delta
  \t(\xt_i-\xt_j)\\
  \bSigma(\xt_i-\xt_j,z)&=&\bSigma_c(z)\delta_{\xt_i,\xt_j}
  + \delta\bSigma(\xt_i-\xt_j,z)\,.
\end{eqnarray}
All the quantities are $N_c\times N_c$ matrices in the cluster sites,
$\t_c=\t(\xt=0)$ and $\bSigma_c(z)=\bSigma(\xt=0,z)$ are the
intra-cluster hopping and self-energy, while $\delta\t(\xt)$ and
$\delta\bSigma(\xt,z)$ are the corresponding inter-cluster quantities
only finite for $\xt\neq 0$.

With these definitions we write the Green function using a locator
expansion, an expansion in $\delta\t$ and $\delta\bSigma$ around the
cluster limit. In matrix notation in the $N_c$ cluster sites it reads
\begin{eqnarray}
  \label{eq:loc3}
  \bG(\xt_i-\xt_j,z)&=&\bg(z)\delta_{\xt_i\xt_j}+\bg(z)\sum_l[\delta{\bf t}(\xt_i-\xt_l)\nonumber\\
                  &+&\delta{\bf \Sigma}(\xt_i-\xt_l,z)]\bG(\xt_l-\xt_j,z)
\end{eqnarray}
where the $N_c\times N_c$ matrix
\begin{equation}
  \label{eq:loc4}
  \bg(z)=[(z+\mu)\mathbb{1}-\t_c-\bSigma_c(z)]^{-1}
\end{equation}
is the Green function of the cluster decoupled from the remainder of
the system ($\mu$ is the chemical potential). Since translational
invariance in the superlattice $\xt$ is preserved, this expression may
be simplified by Fourier transforming the inter-cluster coordinates to
give
\begin{eqnarray}
     \label{eq:loc5}
\bG(\kt,z)=\bg(z)+\bg(z)[\delta\t(\kt)+\delta\bSigma(\kt,z)]\bG(\kt,z)\,.     
\end{eqnarray}

The central approximation that unites all cluster formalisms is to
truncate the self-energy to the cluster, by neglecting $\delta\bSigma$ 
to arrive at
\begin{eqnarray}
  \label{eq:loc6}
  \bG(\kt,z)&=&\bg(z)+\bg(z)\delta\t(\kt)\bG(\kt,z)\nonumber\\
          &=& [\bg^{-1}(z)-\delta \t(\kt)]^{-1}\,.
\end{eqnarray}
As we discuss below, this approximation corresponds to truncating the
potential energy to the cluster while keeping all the contributions
to the kinetic energy.  Therefore quantum cluster approaches are good
approximations to systems with significant screening, where non-local
correlations are expected to be short-ranged.
Quantum cluster approaches are particularly powerful because the
remaining self-energy term $\bSigma_c(z)$ implicitly contained in the
propagator $\bg(z)$ in Eq.~(\ref{eq:loc6}) is restricted to the
cluster degrees of freedom.  Hence it can be calculated
non-perturbatively in an effective cluster model
as a functional $\bSigma_c(z)={\cal F}[\bGbar(z)]$, where
\begin{equation}
  \label{eq:loc7}
  \bGbar(z)=\frac{N_c}{N}\sum_{\kt} \bG(\kt,z)
\end{equation}
is the $\kt$-averaged or coarse-grained $\bG(\kt,z)$, i.e. the Green
function restricted to the cluster.
As discussed in the next section this approximation is consistent with
neglecting inter-cluster momentum conservation, i.e.  neglecting the
phase factors $e^{i\kt\cdot\xt}$ on the vertices of the self-energy
diagrams.

Using the expression (\ref{eq:loc6}) for the lattice Green function and the 
fact that $\bg(z)$ does not depend on $\kt$, the coarse-grained Green 
function $\bGbar$ can be written as
\begin{equation}
  \label{eq:loc8}
  \bGbar(z)=[\bg^{-1}(z)-\bGamma(z)]^{-1}\;\;,
\end{equation}
with a hybridization function $\bGamma$ defined by
\begin{eqnarray}
  \label{eq:loc9}
    \bGamma(z)&=&
  \left[\mathbb{1}+\frac{N_c}{N}\sum_\kt \delta\t(\kt)\G(\kt,z)\right]^{-1}\nonumber\\
 &\times& \left[\frac{N_c}{N}\sum_{\kt}\delta\t(\kt)\G(\kt,z)\delta\t(\kt)\right]\,.
\end{eqnarray}
Its physical content is that of an effective amplitude for fermionic
hopping processes from the cluster into the host and back again into
the cluster. The denominator in Eq.~(\ref{eq:loc9}) is a correction
that excludes the cluster from the effective medium.  $\bGamma(z)$ 
thus plays an analogous role to that of the internal magnetic field 
in mean-field approximations of spin systems. However, due to the 
itinerant character of the fermionic degrees of freedom, it is a 
dynamical quantity.

Both the CPT and the CDMFT formalisms may be defined at this point.  A
self-consistent set of equations is formed from $\bG$ as a functional
of $\bSigma_c$ using Eq.~(\ref{eq:loc6}) together with
Eq.~(\ref{eq:loc4}), and with an appropriate choice of a cluster
solver (see Sec.~\ref{sec:qcs}), $\bSigma_c$ as a functional of
$\bGbar$. In the CDMFT approximation the hybridization $\bGamma$ is
determined self-consistently with Eq.~(\ref{eq:loc9}), i.e. from the
translational invariance of the super-lattice. The resulting
self-consistency cycle is discussed in Sec.~\ref{subsub:scs}.  The CPT
formalism is obtained when $\bGamma$ is neglected. The Green function
$\bg(z)$ then becomes the Green function of an isolated cluster and
the CPT result for the lattice Green function is obtained immediately
via Eq.~(\ref{eq:loc6}) without self-consistency. Thus the
renormalization of the cluster degrees of freedom due to the coupling
to the host described by $\bGamma$ is neglected in the CPT but
included in the CDMFT formalism.

The DCA formalism may be motivated by the demand to restore translational 
invariance within the cluster.  Since the inter-cluster hopping $\delta\t(\kt)$ 
is finite for sites on the surface of the cluster and zero for bulk sites, 
only surface sites hybridize with the host. Hence translational invariance 
with respect to the cluster sites $\X$ is violated.  The cause of this 
violation can be seen by representing the hopping integral $\t(\kt)$ as 
the intra-cluster Fourier transform of the dispersion $\epsilon_{\K+\kt}$
using Eq.~(\ref{eq:FT3}),
\begin{equation}
  \label{eq:loc10}
  [\t(\kt)]_{\X_i\X_j}=\frac{1}{N_c}\sum_\K e^{i(\K+\kt)\cdot(\X_i-\X_j)}\epsilon_{\K+\kt}\,.
\end{equation}
The violation of translational
symmetry is caused by the phase factors $e^{i\kt\cdot(\X_i-\X_j)}$
associated with the superlattice wave-vectors $\kt$. Thus
translational symmetry can be restored by neglecting these
phase-factors, or equivalently, by multiplying $[\t(\kt)]_{\X_i\X_j}$
with the $\kt$-dependent phase $e^{-i\kt\cdot(\X_i-\X_j)}$,
\begin{eqnarray}
  \label{eq:loc11}
  [\t_{DCA}(\kt)]_{\X_i\X_j} &=& [\t(\kt)]_{\X_i\X_j}
  e^{-i\kt\cdot(\X_i-\X_j)} \nonumber\\
&=& \frac{1}{N_c}\sum_{\K} e^{i\K\cdot(\X_i-\X_j)} \epsilon_{\K+\kt}\,.
\end{eqnarray}
Since $\t_{DCA}$ is fully cyclic in the cluster sites, the DCA intra-
and inter-cluster hopping integrals can be written as cluster
Fourier transforms
\begin{eqnarray}
[\t_{c,DCA}]_{\X_i\X_j} & = &\frac{1}{N_c}\sum_\K e^{i\K\cdot(\X_i-\X_j)} \epsbar_\K\\
\left[\delta\t_{DCA}(\kt)\right]_{\X_i\X_j} & = &\frac{1}{N_c}\sum_\K e^{i\K\cdot(\X_i-\X_j)} \delta t(\K+\kt) 
\end{eqnarray}
with 
\begin{eqnarray}
  \label{eq:loc13}
  \epsbar_\K &=& \frac{N_c}{N}\sum_\kt \epsilon_{\K+\kt}\\
  \delta t(\K+\kt) &=&  \epsilon_{\K+\kt} - \epsbar_\K\,.
\end{eqnarray}

Since the DCA intra- and inter-cluster hopping integrals retain
translational invariance within the cluster, the DCA cluster
self-energy $\bSigma_c$ and hybridization function $\bGamma$ are
translationally invariant. The lattice Green function,
Eq.~(\ref{eq:loc6}) hence becomes diagonal in cluster
Fourier space 
\begin{eqnarray}
  G(\K+\kt,z) &= & g(\K,z) + g(\K,z)\delta t(\K+\kt) G(\K+\kt,z)\nonumber\\\label{eq:loc14}
              &=&  \frac{1}{g^{-1}(\K,z)-\delta t(\K+\kt)}
\end{eqnarray}
with the Green function decoupled from the host
\begin{equation}
  \label{eq:loc15}
  g(\K,z)=[z-\epsbar_\K+\mu-\Sigma_c(\K,z)]^{-1}\,.
\end{equation}
Along the lines presented above, the DCA cluster self-energy
$\Sigma_c(\K,z)$ is calculated as a functional of the coarse-grained
Green function
\begin{eqnarray}
  \label{eq:loc16}
  \Gbar(\K,z)&=&\frac{N_c}{N}\sum_\kt G(\K+\kt,z)\nonumber\\
           &=& \frac{1}{g^{-1}(\K,z)-\Gamma(\K,z)}
\end{eqnarray}
which defines the DCA hybridization function
\begin{equation}
  \label{eq:loc17}
  \Gamma(\K,z) = \frac{\frac{N_c}{N}\sum_\kt
  \delta t^2(\K+\kt)G(\K+\kt,z)}{1+\frac{N_c}{N}\sum_\kt
  \delta t(\K+\kt)G(\K+\kt,z)}\,.
\end{equation}
The self-consistent procedure to determine the DCA cluster self-energy
$\Sigma_c(\K,z)$ is analogous to CDMFT and discussed in detail in
Sec.~\ref{subsub:scs}.

This locator expansion yields a very natural physical interpretation
of cluster approximations. We note that the potential energy may be
written as ${\rm{Tr}} \left(\bSigma \G \right)$ \cite{QTMPS:fetter},
where the trace runs over cluster sites, superlattice wave-vectors,
frequency and spin. As detailed above, the central approximation of
cluster expansions is the neglect of the term
$\delta\bSigma(\kt,z)\bG(\kt,z)$ in Eq.~(\ref{eq:loc5}).  Thus the
approximation $\delta\bSigma(z)=0$ essentially neglects the
inter-cluster corrections to the potential energy in all calculated
lattice quantities.  On the other hand, the kinetic energy is
identified as ${\rm{Tr}} \left(\t \G \right)$.  Since its
inter-cluster contribution is {\em{not}} neglected, the kinetic and
potential energy contributions are {\em{not}} treated on equal
footing.  Indeed this is the essential difference between cluster
mean-field approximations and finite size calculations.  In the former
the potential energy of the lattice is truncated to that of the
cluster whereas the kinetic energy is not.  This leads to a
self-consistent theory, generally (but not always) with a
single-particle coupling between the cluster and the host.  In the
latter both the kinetic and potential energies of the lattice are
truncated to their cluster counterparts.
Therefore we might expect cluster methods to
converge more quickly as a function of cluster size,
compared to finite size techniques, for metallic systems with extended
states and significant screening. That this is indeed the case was
illustrated in the previous section for classical spin systems (see
Fig.~\ref{fig:1dising_comp}).

For completeness we note that non-local interaction terms, e.g.\ a
nearest neighbor Coulomb repulsion $U_{ij}$ can be treated in a
similar manner by splitting it into intra- ($U_c$) and inter-cluster
($\delta U$) parts. A similar locator expansion to Eq.~(\ref{eq:loc6})
or respectively Eq.~(\ref{eq:loc14}) is then written down with respect
to the inter-cluster part $\delta U$ for the corresponding
susceptibility (in this case this would be the charge susceptibility).
As a result, an additional coarse-grained interaction $U_c$ acts
within the cluster (see also Sec.~\ref{subsec:freeeg}), and the
locator expansion in $\delta U$ leads to an additional
self-consistency on the two-particle level. For a cluster size of
$N_c=1$ this formalism corresponds to the extended DMFT
\cite{edmft:si}. A first application of this extended cluster
algorithm to the 2D t-J model for $N_c>1$ was discussed in
\textcite{maier:03}.




\subsection{Cluster approximation to the grand potential}
\label{subsec:freeeg}
\label{subsubsec:dca}

In this section we provide a microscopic derivation of the CPT, the
CDMFT and the DCA formalisms based on different cluster approximations
to the diagrammatic expression for the grand potential. The advantage
of this approach is that it allows us to employ almost all of the
diagrammatic technology which has been developed in the past several
decades to a new set of cluster formalisms. Furthermore we are able to
asses the quality of cluster approximations regarding their
thermodynamic properties. The most significant disadvantage is that
the formalism developed here only applies to systems which are
amenable to a diagrammatic expansion.

The following ideas will be illustrated on the extended Hubbard model
Eq.~(\ref{eq:loc1}).  We use the notation introduced in
Sec.~\ref{subsec:es}, Fig.~\ref{fig:Nc4_clusters}, i.e. the cluster
centers are denoted by $\tilde{\bf{x}}$ and sites within the cluster
by ${\bf{X}}$.  The wave-vectors $\tilde{\bf{k}}$ and ${\bf{K}}$ are
their respective conjugates.

\textcite{baym:61} (see also \onlinecite{baym}) showed that
thermodynamically consistent approximations may be constructed by
requiring that the single-particle self-energy $\bSigma$ fulfills
\begin{equation}
\label{eq:conserv}
\bG_0^{-1}-\bG^{-1}=\bSigma=\frac{\delta \Phi[\bG]}{\delta \bG}\,,  
\end{equation}
i.e. is obtained as a functional derivative of the Baym-Kadanoff
$\Phi$-functional with respect to the Green function $\bG$ and that
the approximation is self-consistent (via the left hand identity).
The Baym-Kadanoff generating functional $\Phi[\bG,\bU]$ is
diagrammatically defined as a skeletal graph sum over all distinct
compact closed connected diagrams constructed from the Green function
$\bG$ and the interaction $\bU$. Thus, the diagrammatic form of the
approximate generating functional together with an appropriate set of
Dyson and Bethe-Salpeter equations, completely defines the
diagrammatic formalism.

As described in standard textbooks \cite{AGD} the relation between 
the grand potential functional $\Omega$ and the $\Phi$-functional 
is expressed in terms of the linked cluster expansion as
\begin{equation}
\Omega[\G,\bU] = -k_BT\left\{\Phi[\bG,\bU] - {\rm{Tr}}\ln (-\bG)
              - {\rm{Tr}} (\bSigma \bG )\right\}\,,
\label{eq:omegatophi}
\end{equation}
where the trace indicates summation over cluster sites $\X$,
superlattice wave-vectors $\kt$, frequency and spin.  With the
condition~(\ref{eq:conserv}), the grand potential is stationary with
respect to $\bG$, i.e. $\delta \Omega/\delta \bG = 0$. Such approximations 
are thermodynamically consistent, i.e. observables calculated from the 
Green function $\bG$ agree with those calculated as derivatives of the 
grand potential $\Omega$.  As shown by \textcite{baym} the requirement 
(\ref{eq:conserv}) together with momentum and energy conservation at the 
vertices also assures that the approximation preserves Ward identities, 
i.e.\  satisfies conservation laws.

Prominent examples of conserving approximations include the
Hartree-Fock theory and the fluctuation exchange approximation
\cite{bickers:89}. As exemplified by these theories, the typical
approach to construct a conserving approximation is to restrict the
diagrams in $\Phi$ to a certain sub-class, usually the lowest-order
(in the interaction $\bU$) diagrams. The resulting weak-coupling
approximations however usually fail for systems where the interaction
$\bU$ is of the same order or larger than the bandwidth.

Quantum cluster approaches go a different route: Instead of neglecting
classes of diagrams in $\Phi$, quantum cluster approaches reduce the
infinite number of degrees of freedom over which $\Phi$ is evaluated
to those of a finite size cluster. In contrast to perturbative
approaches however, all classes of diagrams are kept.

\subsubsection{Cluster perturbation theory}

The simplest way to reduce the degrees of freedom in $\Phi[\bG]$ is to
replace the full lattice Green function $\bG(\kt,z)$ by the Green
function $\bg(z)=[(z+\mu)\mathbb{1}-\t_c-\bSigma_c(z)]^{-1}$ of an
isolated cluster of size $N_c$. Consequently, the self-energy
$\bSigma_c=\delta\Phi[\bg]/\delta \bg$ obtained from $\Phi$ is the
self-energy of an isolated finite size cluster. This however leads to
a theory which lacks self-consistency.
Moreover one has to make the ad-hoc assumption that the lattice
self-energy is identical to the one obtained from the cluster,
$\bSigma_c$. The left hand side of Eq.~(\ref{eq:conserv}) then yields
the form for the CPT lattice Green function
\begin{equation}
  \bG(\kt,z) = [\bG_0^{-1}(\kt,z))-\bSigma_c(z)]^{-1}\,,
\label{eq:cptgros}
\end{equation}
where all the quantities are $N_c\times N_c$ matrices in the cluster
sites.  Since the bare lattice Green function is given by
$\bG_0(\kt,z)=[(z+\mu)\mathbb{1}-\t(\kt)]^{-1}$ and the hopping can be
split into intra- and inter-cluster parts (see Eq.~(\ref{eq:loc2})),
$\t(\kt)=\t_c+\delta \t(\kt)$, we obtain
\begin{equation}
\label{eq:cpt1}
\bG(\kt,z) = [{\bg}^{-1}(z)-\delta \t(\kt)]^{-1}
\end{equation} 
with $\bg^{-1}(z)=(z+\mu)\mathbb{1}-\t_c-\bSigma_c(z)$. This form was
derived in Sec.~\ref{subsec:es} from the locator expansion
Eq.~(\ref{eq:loc6}) by ignoring the hybridization $\bGamma$ between
cluster and host. According to Eq.~(\ref{eq:cptgros}), the CPT can be
viewed as the approximation that is obtained by replacing the
self-energy in the Dyson equation of the lattice Green function $\bG$
by the self-energy of an isolated cluster $\bSigma_c$.  This idea was
first developed by \textcite{gross:cluster} and applied to the 3-band
Hubbard model. A different approach to derive the CPT was taken by
\textcite{pairault:strc2} (see also
\onlinecite{pairault:strc,senechal:cluster}). They showed that
Eq.~(\ref{eq:cpt1}) is obtained as the leading order term in a strong
coupling expansion in the hopping $\delta\t$ between sites on
different clusters. This derivation of the CPT provided a fundamental
theoretical basis to assess the nature of the approximation
as well as to systematically improve the quality of the approach by
including higher order terms in the perturbative expansion.

The CPT becomes exact in the weak coupling limit $U/t=0$ and the
strong-coupling limit $t/U=0$ as well as in the infinite cluster size
limit $N_c\rightarrow\infty$ \cite{senechal:cluster}. The limit
$t/U=0$ is reproduced exactly since the CPT is the perturbative result
in the hopping. In this limit, all the sites in the lattice are
decoupled, and the system is solved exactly by the single-site Green
function $g_c$. In the opposite limit $U/t=0$, the cluster self-energy
$\bSigma_c$ in Eq.~(\ref{eq:cptgros}) vanishes and $\bG(\kt)=\bG_0(\kt)$
is the exact solution. In the limit $N_c\rightarrow \infty$, the
cluster Green function $\bg_c$ becomes the exact Green function of the
full system. At finite $t/U$ and cluster size $N_c=1$, the CPT
recovers the Hubbard-I approximation \cite{hubbard:63} where the
self-energy is approximated by the self-energy $U^2/4\omega$ (at
half-filling) of an isolated atom
\cite{gross:cluster,senechal:cluster,senechal:cluster2}.

According to the derivation of the CPT, the cluster Green function
$\bg_c$ is to be calculated on a cluster with open boundary
conditions. Since the hopping $\t_c$ between sites inside the cluster
is treated exactly whereas the inter-cluster hopping $\delta\t$
between surface sites on different clusters is treated perturbatively,
translational invariance for sites $\X$ in the cluster is violated
while it is preserved for sites $\xt$ in the superlattice. As a
result, the cluster wave-vector $\K$ is not a good quantum number and
we have as a generalization of the Fourier transform
Eq.~(\ref{eq:FT4}) (we omit the frequency dependence for convenience)
\begin{eqnarray}
  G(\k,\k') &=& \frac{1}{N_c}\sum_\Q \sum_{\X_i,\X_j} e^{-i\k\cdot\X_i}G(\X_i,\X_j,\k) e^{i\k'\cdot \X_j}\nonumber\\
  & & \times\,\, \delta(\k-\k'-\Q)\,.   \label{eq:cpt5}
\end{eqnarray}
where $\k$ and $\k'$ are wave-vectors in the full Brillouin zone and
$\Q$ is a wave-vector in the cluster reciprocal space. Here we used
the relation $\bG(\kt)=\bG(\k)$ which follows from Eq.~(\ref{eq:FT2})
by replacing $\kt$ by $\k=\kt+\K$.  To restore translational
invariance in the full lattice Green function, the CPT approximates
$G(\k,\k')$ by the $\Q=0$ contribution to obtain
\begin{equation}
  \label{eq:2}
  G_{CPT}(\k)=\frac{1}{N_c}\sum_{\X_i,\X_j} e^{-i\k\cdot(\X_i-\X_j)} G(\X_i,\X_j,\k)\,,
\end{equation}
as the translational invariant propagator used to calculate spectra.
With this approximation, the CPT provides a very economical method to
calculate the lattice Green function of an infinite size
($N\rightarrow\infty$) Hubbard-like model from the Green function (or
equivalently self-energy) of an isolated cluster of finite size $N_c
\ll N$. From $G_{CPT}(\k)$ one can calculate single-particle
quantities such as photoemission spectra, kinetic and potential
energies, double occupancy, etc.


To reduce the numerical cost, it was suggested to use periodic
boundary conditions on the CPT cluster by adding the appropriate
hopping terms to the intra-cluster hopping $\t_c$ and subtracting them
from the inter-cluster hopping $\delta\t$ \cite{hanke:cluster}.
However, as discussed by \textcite{senechal:cluster}, periodic
boundary conditions lead to less accurate spectra for the 1D Hubbard
model than open boundary conditions. This a-posteriori argument for
open boundary conditions is substantiated by calculations within
Potthoff's self-energy functional approach (see Sec.~\ref{subsec:alt})
which show that the grand potential of the system is only stationary
in the limit of open boundary conditions \cite{potthoff:cluster1}.

\subsubsection{Cellular dynamical mean-field theory}

A superior approximation may be obtained if, instead of the isolated
cluster Green function $\bg$, the full lattice Green function $\bG$
restricted to cluster sites
is used to evaluate the functional $\Phi$. This approximation can be
motivated microscopically by approximating the momentum conservation
on internal vertices in the diagrams defining $\Phi$. Momentum
conservation at each vertex is described by the Laue function
\begin{equation}
\label{eq:Laue}
\Delta=\sum_\x e^{i\x\cdot(\k_1+\k_2+\cdots,-\k_1'-\k_2'-\cdots)}=
N\delta_{\k_1+\k_2+\cdots,\k_1'+\k_2'+\cdots}\,,
\end{equation}
where $\k_1$, $\k_2$ ($\k_1'$, $\k_2'$) are the momenta entering
(leaving) the vertex. \textcite{hartmann:dmft} showed that the DMFT
may be derived by completely ignoring momentum conservation at each
internal vertex by setting $\Delta=1$.  Then one may freely sum over
all of the internal momentum labels, and the Green functions in the
diagrams are replaced by the local Green function $G_{ii}=1/N\sum_\k
G(\k)$.

The CDMFT and DCA (see below) techniques may also be defined by their
respective approximations to the Laue function. In the CDMFT the Laue
function is approximated by
\begin{equation}
\label{eq:LMCA}
\Delta_{CDMFT}=\sum_{\X}e^{i\X\cdot (\K_1+{\tilde{\bf{k}}}_1+
\K_2+{\tilde{\bf{k}}}_2+\cdots -\K_1'-{\tilde{\bf{k}}}_1'-\K_2'-{\tilde{\bf{k}}}_2'-\cdots)}\,.
\end{equation}
Thus the CDMFT omits the phase factors $e^{i{\tilde{\bf{k}}}\cdot\xt}$
resulting from the position of the cluster in the original lattice,
but keeps the phase factors $e^{i{\tilde{\bf{k}}}\cdot\X}$. The
latter are directly responsible for the violation of translational
invariance. Consequently, all quantities in the CDMFT are functions
of two cluster momenta $\K_1$, $\K_2$ or two sites $\X_1$, $\X_2$
respectively.

If the CDMFT Laue function Eq.~(\ref{eq:LMCA}) is applied to diagrams
in $\Phi$, each Green function leg is replaced by the CDMFT
coarse-grained Green function (the frequency dependence is dropped for
notational convenience)
\begin{eqnarray}
\label{eq:cgMCA}
\bar{G}(\X_1,\X_2)=G(\X_1,\X_2;\xt=0)&=&\nonumber\\
&&\hspace*{-5.3cm}\frac{1}{N^2}\!\!\!\sum_{\stackrel{\K_1,\K_2}{{\tilde{\bf{k}}}_1,{\tilde{\bf{k}}}_2}}
e^{i(\K_1+{\tilde{\bf{k}}}_1)\cdot\X_1}G(\K_1,\K_2;{\tilde{\bf{k}}}_1,{\tilde{\bf{k}}}_2)e^{-i(\K_2+{\tilde{\bf{k}}}_2)\cdot\X_2}
=\nonumber\\
&&\hspace*{-5.3cm}\frac{N_c^2}{N^2}\sum_{{\tilde{\bf{k}}}_1,{\tilde{\bf{k}}}_2}G(\X_1,\X_2,{\tilde{\bf{k}}}_1,{\tilde{\bf{k}}}_2)\,,
\end{eqnarray}
or in matrix notation for the cluster sites $\X_1$ and $\X_2$
\begin{equation}
\label{eq:cgGMCA}
\bGbar(z)=\frac{N_c}{N}\sum_{{\tilde{\bf{k}}}} \bG({\tilde{\bf{k}}},z)\,,
\end{equation}   
since $\bG$ is diagonal in ${\tilde{\bf{k}}}_1, {\tilde{\bf{k}}}_2$
due to the translational invariance of the superlattice. Similarly
each interaction line is replaced by its coarse-grained result
\begin{equation}
\label{eq:cgVMCA}
\bUbar=\frac{N_c}{N}\sum_{{\tilde{\bf{k}}}}\bU({\tilde{\bf{k}}})\,.
\end{equation}   
The summations over the cluster sites $\X$ within each diagram remain
to be performed. As a consequence of coarse-graining the propagators in
$\Phi$, the CDMFT self-energy
\begin{equation}
  \label{eq:6}
  \bSigma_c(z)=\frac{\delta\Phi[\bGbar(z),\bUbar]}{\delta \bGbar(z)}
\end{equation}
is restricted to cluster sites and consequently independent of $\kt$.
Note that by definition, $\bGbar$ and $\bUbar$ are truncated outside
the cluster, i.e. if the interaction $\bU$ is non-local, $\bUbar$
includes only interactions within, but not between clusters.

The CDMFT estimate of the lattice grand potential is obtained by
substituting the CDMFT approximate generating functional
$\Phi[\bGbar,\bUbar]$ into Eq.~(\ref{eq:omegatophi}).  From the
condition that the grand potential is stationary with respect to the
lattice Green function, $\delta \Omega/\delta \bG =0$, one obtains a
relation between the lattice self-energy and the cluster self-energy
\begin{eqnarray}
\label{eq:MCAS}
\Sigma(\K_1,\K_2;\kt_1,\kt_2)&=&\nonumber\\
&&\hspace*{-3cm}\sum_{\X_1,\X_2} e^{-i(\K_1+\kt_1)\cdot\X_1}
\Sigma_c(\X_1,\X_2) e^{i(\K_2+\kt_2)\cdot\X_2}\,.
\end{eqnarray}
With Eq.~(\ref{eq:cgMCA}) the left hand side of Eq.~(\ref{eq:conserv})
then becomes the coarse-graining relation
\begin{equation}
  \label{eq:17}
  \bGbar(z)=\frac{N_c}{N} \sum_\kt [\bG_0^{-1}(\kt,z)-\bSigma_c(z)]^{-1}
\end{equation}
with the bare Green function $\bG_0(z)=[(z+\mu)\mathbb{1}-\t(\kt)]^{-1}$.

\subsubsection{Dynamical cluster approximation}

\begin{figure}[htb]
  \includegraphics*[width=3.3in]{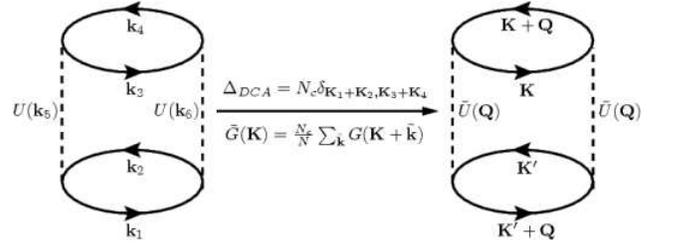} 
  \caption{A second-order term in the generating functional $\Phi$ of the
    Hubbard model. The dashed line represents the interaction $U$, and
    the solid line on the left hand side (right hand side) the lattice
    (coarse-grained) single-particle Green function $G$ ($\Gbar$).
    When the DCA Laue function is used to describe momentum
    conservation at the internal vertices, the wave-vectors collapse
    onto those of the cluster and each lattice Green function is
    replaced by its coarse-grained average.}
  \label{fig:phicol} 
\end{figure}

In the DCA the phase factors $e^{i{\tilde{\bf{k}}}\cdot\X}$ are
omitted too, so that the DCA approximation to the Laue function
becomes
\begin{equation}
\label{eq:LDCA}
\Delta_{DCA}=N_c\delta_{\K_1+\K_2+\cdots,\K_1'+\K_2'+\cdots}
\end{equation}
and Green function legs in $\Phi$ are
replaced by the DCA coarse grained Green function
\begin{equation}
\label{eq:cgGDCA}
\Gbar(\K,z)=\frac{N_c}{N}\sum_{{\tilde{\bf{k}}}}G(\K+{\tilde{\bf{k}}},z)\,,
\end{equation} 
since Green functions can be freely summed over the $N/N_c$
wave-vectors ${\tilde{\bf{k}}}$ of the superlattice.  Similarly, the
interactions are replaced by the DCA coarse grained interaction
\begin{equation}
\label{eq:cgVDCA}
{\bar{U}}(\K)=\frac{N_c}{N}\sum_{{\tilde{\bf{k}}}}U(\K+{\tilde{\bf{k}}})\, .
\end{equation} 
As with the CDMFT, the effect of coarse-graining the interaction is to
reduce the effect of non-local interactions to within the cluster.
This collapse of the diagrams in the $\Phi$ functional onto those of
an effective cluster problem is illustrated in Fig.~\ref{fig:phicol}
for a second order contribution.

The resulting compact graphs are functionals of the coarse grained
Green function $\bar{G}(\K)$ and interaction ${\bar{U}}(\K)$, and thus
depend on the cluster momenta $\K$ only.  For example, when $N_c=1$,
only the local part of the interaction survives the coarse graining.
As with the CDMFT, within the DCA it is important that {\em{both}} the
interaction and the Green function are coarse-grained
\cite{hettler:dca2}. As a consequence of the collapse of the
$\Phi$-diagrams, the DCA self-energy
\begin{equation}
  \label{eq:18}
  \Sigma_c(\K,z)=\frac{\delta \Phi[\Gbar(\K,z),\Ubar]}{\delta \Gbar(\K,z)}
\end{equation}
only depends on the cluster momenta $\K$.

To obtain the DCA estimate of the lattice grand potential, we
substitute the DCA approximate generating functional
$\Phi[\Gbar(\K),\Ubar(\K)]$ into Eq.~(\ref{eq:omegatophi}).  The grand
potential is stationary with respect to ${\bf{G}}$ when
\begin{equation}
\frac{\delta \Omega[\Gbar(\K),\Ubar(\K)]}{\delta G(\k)}=
\Sigma_c(\K)-\Sigma(\k)=0\,,
\end{equation}
which means that $\Sigma(\k)=\Sigma_c(\K)$ is the proper approximation
for the lattice self-energy corresponding to $\Phi[\Gbar(\K),\Ubar]$.
The self-consistency condition on the left hand side of
Eq.~(\ref{eq:conserv}) then becomes the coarse-graining relation
\begin{equation}
\Gbar(\K,z) =\frac{N_c}{N}\sum_\kt [G_0^{-1}(\K+\kt)-\Sigma_c(\K,z)]^{-1} \, .
\label{G_DCA}
\end{equation}
with the bare Green function $G_0(\K+\kt,z)=[z-\epsilon_{\K+\kt}+\mu]^{-1}$.

Both the CDMFT and the DCA have well defined limits. In
the infinite size cluster limit $N_c\to\infty$, the CDMFT and DCA
approximations to the Laue function recover the exact Laue function,
$\Phi$ is evaluated with the full lattice Green function and
interactions and thus the exact result is recovered.  When $N_c=1$,
both the Laue functions reduce to $\Delta=1$, $\Phi$ is evaluated with
the local Green function and local contributions of the interactions,
and the DMFT result is recovered.

\subsubsection{Self-consistency scheme}
\label{subsub:scs}
\begin{figure}[htb]
  \includegraphics*[width=3.0in]{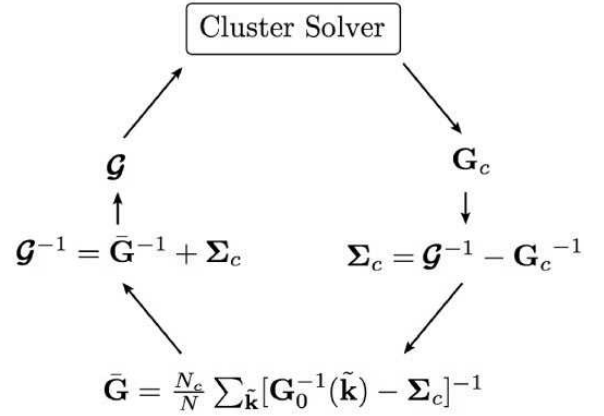} \caption{Sketch of the
    CDMFT and DCA embedded cluster algorithms. The iteration starts
    with computing the coarse grained Green function $\bGbar$ using an
    initial guess for the cluster self-energy $\bSigma_c.$ The cluster
    excluded Green function $\bGscript$ is then used to define the
    effective cluster problem which yields a new estimate of
    $\bSigma_c$.}
  \label{fig:clusteralg} 
\end{figure}

The two equations (\ref{eq:conserv})
form a non-linear set of equations which have to be solved
self-consistently to determine the cluster self-energy
$\bSigma_c[\bGbar]$ with the use of a suitable cluster solver. For
cluster solvers that sum up all diagrams of $\bSigma_c$, i.e. in
contrast to a skeletal expansion of $\bSigma_c$, an additional step is
necessary in the self-consistent cycle. In order to not overcount
self-energy diagrams, $\bSigma_c$ is to be calculated as a functional
of the corresponding bare propagator to $\bGbar$, the cluster excluded
Green function
\begin{equation}
  \label{eq:gscdmft}
  {\bGscript}(z) = [\bGbar^{-1}(z)+\bSigma_c(z)]^{-1} 
\end{equation}
This equation\footnote{A unifying matrix notation is used. In the
  CDMFT, the quantities are matrices in the $N_c$ cluster sites and in
  particular
  $[\bG_0^{-1}(\kt,z)]_{\X_i\X_j}=(z+\mu)\delta_{\X_i\X_j}-[\t(\kt)]_{\X_i\X_j}$.
  For the DCA, the matrices are diagonal in the cluster momenta $\K$
  and $[\bG_0^{-1}(\kt,z)]_{\K\K} =z+\mu-\epsilon_{\K+\kt}$.}
unambiguously defines the self-consistent iteration procedure
illustrated in Fig.~\ref{fig:clusteralg}:
\begin{enumerate}
\item The iteration is started by guessing an initial cluster
  self-energy $\bSigma_c(z)$, usually zero or the result from second
  order perturbation theory, to
  
\item Calculate the coarse grained quantities
$$
 \bGbar(z) = \frac{N_c}{N}
  \sum_\kt [\bG_0^{-1}(\kt,z)-\bSigma_c(z)]^{-1}\,\,,\,\bar{\bf U}=\frac{N_c}{N}\sum_\kt {\bf U}(\kt)
$$

\item The effective cluster problem is then set up with the cluster
  excluded Green function $\bGscript(z)$ and $\bar{\bf U}$.
  
\item The self-energy $\bSigma_c(z)$ or respectively the cluster Green
  function $\bG_c(z)$ is calculated in the effective cluster model (see
  Sec.~\ref{subsec:ecm}) by using any of the quantum cluster solvers
  discussed in Sec.~\ref{sec:qcs}.
  
\item For techniques that produce the cluster Green function $\bG_c$
  rather than the self-energy, the new estimate of the cluster
  self-energy is calculated as
  $\bSigma_c(z)=\bGscript^{-1}(z)-{\bG_c}^{-1}(z)$.

\end{enumerate}
The iteration closes by re-calculating the coarse-grained Green
function $\bGbar(z)$ in step (2) with the new estimate of the cluster
self-energy. This procedure is repeated until the cluster Green
function $\bG_c(z)$ equals the coarse-grained Green function
$\bGbar(z)$ to within the desired accuracy.

\subsection{Discussion}
\label{subsec:disc}
In contrast to the DMFT, a unique setup for {\em the} embedded cluster
theory does not exist. Depending on the treatment of e.g.  boundary
conditions (see Sec.~\ref{subsec:es} and \onlinecite{kotliar:cdmft2})
differences in the coupling of the cluster to its environment arise
(see comparison in Sec.~\ref{subsubsec:comp} below).  In fact, there
exist infinitely many realizations of embedded cluster theories for
any given model Hamiltonian
\cite{kotliar:cdmft2,okamoto:cluster,potthoff:cluster1,potthoff:cluster2},
two of which we focus on in this review.

The fundamental approximation common to all approaches is that they
try to go beyond conventional mean-field approximations and introduce
non-local physics by replacing the unsolvable lattice Hamiltonian by
some manageable finite portion -- possibly with effective model
parameters -- and reintroduce the thermodynamic limit by a mean-field
type treatment of the remaining system.  Thus, the influence of truly
long-ranged correlations, i.e.\ those beyond the cluster size, is
still not incorporated, but short-ranged correlations and in
particular the local dynamics with respect to the cluster can ideally
be treated exactly. That this can already lead to substantial
renormalizations has been demonstrated in Sec.~\ref{subsec:ccw} and is
also well-known from the DMFT. The combination of short- to
medium-ranged correlations with mean-field treatment of long-ranged
physics enables the investigation of a system's tendency to certain
types of order not accessible by conventional mean-field theory. One
can thus more clearly identify the possible existence of long-ranged
correlations which are normally hard to establish in conventional
finite-system calculations.  On the other hand, it is this mean-field
likeness that practically disables a proof for the appearance of a
phase transition in the real model, even though the behavior of
transition temperatures etc.\ with cluster size can give valuable
hints about the system (see Sec.~\ref{subsec:ccw}).

\subsubsection{Conservation and thermodynamic consistency}

Both the DCA and CDMFT approximations are self-consistent and are
$\Phi$-derivable since they satisfy Eq.~(\ref{eq:conserv}).  Thus,
they are thermodynamically consistent in the Baym-Kadanoff sense.
Observables calculated from the lattice Green function $\bG$ agree
with those calculated as derivatives of the lattice grand potential
$\Omega$.  Since the CPT is not self-consistent, it is not
thermodynamically consistent.  However, none of these approaches is
conserving in the Baym-Kadanoff sense since they all violate
momentum conservation.  Thus, each of these approaches is likely to
violate some set of the Ward identities \cite{hettler:dca2}.

\subsubsection{Causality}

One problem in any formation of an embedded cluster theory is the
manifestation of causality, i.e.\ a physical Green function cannot
have poles anywhere except on the real axis. In particular, for the
fundamental quantity of the theory, the single particle Green
function, this means that the proper self-energy in momentum space has
to obey $\Im m\Sigma(\k,\omega+i0^+)<0$.  Early attempts to formulate
cluster extensions to DMFT \cite{schiller:cluster} ran into exactly
this problem, which e.g.\ manifests itself in negative single-particle
spectral functions.  Explicit proofs for causality can be given for
the DCA \cite{hettler:dca2} and the CDMFT (\onlinecite{kotliar:cdmft};
see also \onlinecite{biroli:cdmft}). More precisely, any embedded
cluster theory consistent with the locator expansion (or a suitably
defined cavity construction) obeys, due to Eqs.\ (\ref{eq:loc4}) and
(\ref{eq:loc8}), causality.  A closer look at Eqs.\ (\ref{eq:loc4})
and (\ref{eq:loc8}) also reveals how problems can arise. It is, for
example, tempting to replace the well defined cluster quantity ${\bf
  \Sigma}_c(z)$ in (\ref{eq:loc4}) by some approximation to ${\bf
  \Sigma}(z)$, i.e.\ the full self-energy. As has been discussed in
some detail by \textcite{okamoto:cluster} (see also
Sec.~\ref{subsubsec:oka}), such a procedure will in general introduce
ringing phenomena and lead to acausal behavior. How strongly such a
violation of causality will eventually influence the interesting
low-energy results is quite likely a question of the model parameters
under investigation.  In any case, it must be taken as serious reason
to at least doubt the quantitative accuracy of such results.

\subsubsection{Reducible and irreducible quantities}

Fundamental quantities like the one-particle self-energy
${\bf\Sigma}(z)$, or its many-particle counterparts entering e.g.\ 
susceptibilities, carry the whole fragile information about the
many-body physics of a given model. In the language of diagrammatic
perturbation theory they are built of so-called irreducible diagrams.
Hence they are also frequently called irreducible quantities; in
contrast, the single-particle Green function or a susceptibility is a
reducible quantity.  It is an important aspect, that the cluster
theories discussed use approximants for these irreducible quantities
only, i.e. the quantities obtained as derivatives of the Baym-Kadanoff
$\Phi$ functional. In fact, in the formulation in
Sec.~\ref{subsec:freeeg} any attempt to replace reducible quantities
like the one-particle Green function directly by approximants is at
least dangerous.

To see this, consider the grand potential functional
Eq.~(\ref{eq:omegatophi}). It is expressed as a sum over all closed
connected distinct graphs constructed from the Green function $\G$ and
interaction $\bU$.  The subset of compact graphs comprise the
Baym-Kadanoff generating functional $\Phi[\bG,\bU]$ which is expressed
as a skeletal graph sum over all distinct compact closed connected
graphs. Compact diagrams have no internal parts representing
corrections to the Green function $\bG$.

\begin{figure}[htb]
\includegraphics*[width=1.5in]{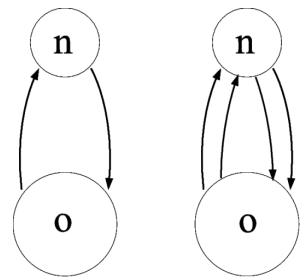}
\caption{Non-compact (left) and compact (right) non-local 
  corrections to the grand potential.  Here the upper (lower) circle
  is meant to represent a set of graphs which are closed except for
  the external lines shown, and restricted to site $n$ (the origin).}
\label{fig:phi_corrections} 
\end{figure} 

        
In quantum cluster theories the graphs for $\Phi$ are approximated by
their cluster counterparts.  As an example, consider the limit of
infinite dimensions, used by Metzner and Vollhardt to derive the DMFT
\cite{metzner:dmft}. In this limit most closed connected graphs are
local since the Green function falls off quickly with distance $r$,
$G(r)\sim D^{-r/2}$.  In fact, only a small set of graphs,
corresponding to non-compact corrections, remain finite.  To see this,
consider the simplest non-local corrections to non-compact and compact
parts of the grand potential of a Hubbard-like model, illustrated in
Fig.~\ref{fig:phi_corrections}.  Here the upper (lower) circle is a
set of graphs composed of intra-site propagators restricted to site $n$
(the origin).  Consider all such non-local corrections on the shell of
sites which are $n$ mutually orthogonal unit translations from the
origin.  In the limit of high dimensions, there are $2^n D!/((D-n)!
n!)\sim {\cal{O}} (D^n)$ such sites.  Since as $D\to\infty$, $G(r)\sim
D^{-r/2}$\cite{metzner:dmft}, the legs on the compact correction
contribute a factor ${\cal{O}}(D^{-2n})$ whereas those on the
non-compact correction contribute ${\cal{O}}(D^{-n})$.  Therefore the
compact non-local correction falls as $D^{-n}$ and vanishes as
$D\to\infty$; whereas, the non-compact correction remains of order
one, regardless of how far site $n$ is from the origin. As a result,
the generating functional, which is composed of only compact graphs,
is a functional of the local Green function and interaction in this
limit
\begin{equation}
\Phi[\bG,\bU] = \Phi[G_{ii},U_{ii}] + {\cal{O}}(1/D)\,.
\end{equation}
A similar analysis was done for the DCA cluster problem by
\textcite{karan:dca3}. They find that the corrections from sites outside 
the cluster associated with compact diagrams are quite small (i.e.\
high order in the linear cluster size $1/L_c$) justifying the approximation
\begin{equation}
\Phi[\bG,\bU]\approx \Phi[\Gbar(\K),\Ubar(\K)]\,,
\end{equation} 
while those associated with non-compact diagrams are large and cannot 
be neglected. The same analysis may be done for the CDMFT, simply 
by replacing the DCA graphs by those for the CDMFT.

Thus, the essential approximation for the DMFT, the DCA and the CDMFT
is to approximate the lattice generating functional $\Phi$ by its
cluster counterpart in the estimate of the lattice grand potential,
Eq.~(\ref{eq:omegatophi}).  Concommittantly the derivatives of $\Phi$,
i.e. the lattice self-energy and vertex functions are approximated by
their respective cluster counterparts.  This once more underlines why
in embedded cluster theories it is necessary to always work with the
cluster irreducible quantities; they are the only quantities
compatible with the cluster and any attempt to include more
information by e.g.\ Fourier transformation will lead to consistency
problems, which eventually express themselves as causality violation.
Irreducible quantities are also those important to discuss the
convergence behavior of a cluster method with increasing cluster size
as discussed in the next section.


\subsubsection{Comparison}

\label{subsubsec:comp}
Detailed comparisons of the DCA and CDMFT algorithms were presented in
\textcite{maier:dca3}, \textcite{maier:dca4},
\textcite{kotliar:cdmft2} and
\textcite{biroli:cdmft}. 
Both approximations share the underlying idea and general algorithm,
and differ only in the form used for the hopping matrix $\t$ (see
Eq.~(\ref{eq:loc11})). The purpose of this section is to convey the
consequences of this, at first sight, small difference on the
effective cluster problem, convergence properties and the calculation
of the lattice self-energy.

\paragraph{Nature of effective cluster problem.}

For notational convenience, we use a 1D model with nearest neighbor
hopping $t$ only, set the on-site energy $\epsilon_o=0$ and denote the
cluster size by $L_c$. The generalization to higher dimension or
longer-ranged hopping terms is straightforward.

The CDMFT uses the original form for the hopping matrix $\t(\tk)$
which is obtained e.g. as an inter-cluster Fourier transform (see
Eq.~(\ref{eq:FT2})) of $\t(\tx)$.
Only entries between neighboring sites inside the cluster
$[\t(\tk)]_{X_i, X_i\pm 1}=-t$ and between neighboring sites on the
surface of the cluster $[\t(\tk)]_{X_i, X_i\pm (L_c-1)}=-te^{\mp i\tk
  L_c}$ are finite.  The former entries form the intra-cluster hopping
matrix $\t_c=N_c/N\sum_\tk \t(\tk)$ while the latter entries form the
inter-cluster hopping matrix $\delta \t(\tk)=\t(\tk)-\t_c$. Both
amplitudes are given by the original hopping $t$. For the effective
cluster problem this translates to the fact that only sites on the
surface of the cluster couple to the effective medium, while sites
inside the cluster only couple to their neighboring sites in the
cluster. Hence the cluster problem has open boundary conditions and
translational invariance is violated within the cluster. The lattice
Green function $\bG(\tk)$ (see Eq.~(\ref{eq:loc6})) is a matrix in the
cluster sites and cannot be diagonalized by going over to cluster
$K$-space. Therefore the coarse-graining step Eq.~(\ref{eq:loc7}) is
done in real space.

The DCA restores translational invariance by setting
$[\t_{DCA}(\tk)]_{X_i X_j}=[\t(\tk)]_{X_i X_j}e^{-i\tk\cdot(X_i-X_j)}$
(see Eq.~(\ref{eq:loc11})). As a consequence, its matrix elements
become identical and are given by $-te^{\mp i\tk}$ between sites
$X_j=X_i\mp 1$ and $X_j=X_i\pm(L_c-1)$.  Hence the DCA hopping matrix
$\t_{DCA}(\tk)$ is fully cyclic with respect to the cluster sites and
the lattice Green function $\bG(\tk)$ is diagonalized by going over to
cluster $K$-space. The DCA intra-cluster hopping matrix
$\t_{c,DCA}=N_c/N\sum_\tk \t_{DCA}(\tk)$ is also cyclic with finite
matrix-elements
\begin{equation}
  \label{eq:comp1}
   [\t_{c,DCA}]_{X_i, X_j} = -t\frac{L_c}{\pi}\sin\frac{\pi}{L_c} 
\end{equation}
between sites $X_j=X_i\pm 1$ and $X_j=X_i\pm(L_c-1)$ and the DCA
cluster problem therefore has periodic boundary conditions.  At finite
cluster size $L_c$, the intra-cluster hopping is reduced by the factor
$1/6(\pi/L_c)^2+{\cal O}((\pi/L_c)^4)$ compared to its lattice
counterpart $t$. In the infinite cluster size limit it becomes $t$.
This reduction in the intra-cluster coupling is compensated by the
inter-cluster hopping which is of long-ranged nature,
\begin{equation}
  \label{eq:comp2}
  [\delta \t_{DCA}(\tx)]_{X_iX_j} = -t[\frac{\sin[(\tx\mp 1)\pi/L_c]}{(\tx\mp 1)\pi/L_c}-\frac{\sin (\pi/L_c)}{\pi/L_c}\delta_{\tx,0}]
\end{equation}
between sites $X_j=X_i\mp 1$ and $X_j=X_i\pm(L_c-1)$. It is important
to note that $\delta \t_{DCA}$ couples {\em all} the sites in the
cluster to sites in the effective medium. It vanishes for $\tx=0$ and
decreases as $1/\tx$ with increasing $\tx$. We also notice that
\begin{equation}
\label{eq:dcascal}
\delta \t_{DCA}\sim 1/L_c
\end{equation}
for large linear cluster sizes $L_c$ and emphasize that this result
holds generally in any dimension $D$.

The restoration of translational invariance in the DCA is achieved by
mapping the lattice onto a cluster with periodic boundary conditions with 
reduced hopping integrals $t_{c,DCA}$ and coupling every site in the cluster
to a neighboring site in the effective medium through long-ranged
hopping integrals $\delta t_{DCA}(\tx)$. The sum of all finite intra-
and inter-cluster couplings for a given site is again given by the
original value $t$.  Similar conclusions about the nature of the
effective CDMFT and DCA cluster problems were reached in a study of
the large $U$ limit of the Falicov-Kimball model (FKM, see
Eq.~(\ref{eq:2dfkm})), i.e. the classical Ising model
\cite{biroli:cdmft}.

We stress that clusters with linear size $L_c=2$ are special. Here both
terms in the intra-cluster hopping Eq.~(\ref{eq:comp1}) give a
contribution to the same matrix-element. Hence the nearest-neighbor
hopping is given by $-2t \left(L_c/\pi \right) \sin \left(\pi/L_c\right)$, 
i.e.\ with the prefactor
$-2t$ instead of $-t$ for larger clusters. This reflects the fact that
every site sees its nearest neighbor twice due to the periodic
boundary conditions.  Non-local fluctuations are thus enhanced in
clusters with linear size $L_c=2$ as seen e.g. in an over-proportional
suppression of transition temperatures (see Secs.~\ref{subsec:FKM} and
\ref{subsubsec:af}).

\paragraph{Convergence with cluster size.}

The differences in boundary conditions translate directly to different
asymptotic behaviors for large cluster sizes $N_c$. In leading order
the hybridization with the mean-field, $\bGamma$ vanishes like
$\delta \t^2$ as the cluster size increases (see Eqs.~(\ref{eq:loc9})
and (\ref{eq:loc17})). In the CDMFT the magnitude of $\delta\t$ is of
order one for the sites on the surface of the cluster and zero
otherwise. The average hybridization per cluster site in the CDMFT
thus scales like
\begin{equation}
  \label{eq:3}
  \bar{\Gamma}_{CDMFT}=\frac{1}{N_c}{\rm Tr}\,[\bGamma_{CDMFT}]\sim {\cal O}(\frac{1}{L_c})\,,
\end{equation}
where the trace runs over cluster sites and frequency. This behavior
is evident since only the $2D(L_c^{D-1})$ sites on the surface
contribute to the sum and $N_c=L_c^D$.  In the DCA, $\delta \t={\cal
  O}(1/L_c)$ (see Eq.~(\ref{eq:dcascal})). The average
hybridization of the DCA cluster to the effective medium hence scales
faster to zero as
\begin{equation}
  \label{eq:7}
  \bar{\Gamma}_{DCA}=\frac{1}{N_c}{\rm Tr}\, [\bGamma_{DCA}]\sim{\cal O}(\frac{1}{L_c^2})
\end{equation}
since each of the $N_c$ terms contributes a term of the order ${\cal
  O}(1/L_c^2)$.  

It can further be shown that $\bGamma$ acts as the small parameter in
these theories: The approximation performed by the DCA and the CDMFT
is to replace the lattice Green function $G$ by its coarse-grained
quantity $\Gbar$ in diagrams for the generating functional $\Phi$ (see
Sec.~\ref{subsec:freeeg}).  Both Green functions differ by the
inter-cluster hopping $\delta \t(\kt)$, the self-energy and the
hybridization function $\bGamma$.  Since the diagrams in $\Phi$ are
summed over $\kt$, the terms having the same order as $\delta \t(\kt)$
vanish since $N_c/N\sum_\kt \delta\t(\kt)=0$. If we assume that the
self-energy has corrections of the same or higher order in $1/L_c$ as
$\bGamma$, the convergence of $\Phi$ is entirely determined by
$\bGamma$. With the scaling relations (\ref{eq:3}) and (\ref{eq:7}) we
find for the CDMFT that $\Phi_{CDMFT}\approx\Phi+{\cal O}(1/L_c)$
while the DCA converges like $\Phi_{DCA}\approx\Phi+{\cal
  O}(1/L_c^2)$.  Since $\bSigma=\delta \Phi/\delta \bG$, it converges
with $L_c$ as the corresponding $\Phi$ confirming the assumption. The
generating functional $\Phi$ and hence the grand potential thus
converges faster in the DCA than in the CDMFT with increasing cluster
size.

\paragraph{Numerical comparison.}

The scaling behavior Eqs.~(\ref{eq:3}) and (\ref{eq:7}) of the CDMFT
and DCA average hybridization strengths was verified numerically in
the 1D FKM in \textcite{maier:dca3} (see also
\onlinecite{maier:dca4}). Here we review the effects of the different
scaling behaviors of the average hybridization on the phase transition
in this model. The Hamiltonian of the FKM is discussed in
Sec.~\ref{subsec:FKM}, Eq.~(\ref{eq:2dfkm}).
It can be considered as a simplified Hubbard model with only one
spin-species being allowed to hop. However it still shows a complex
phase diagram including a Mott gap for large $U$ at half filling, an
Ising-like charge ordering with the corresponding transition
temperature $T_c$ being zero in 1D, and phase separation in all
dimensions. Since the 1D FKM is in the 1D Ising universality class we
expect similar scaling behavior as observed in the results for the 1D
Ising model in Sec.~\ref{subsec:ccw}. In particular, we expect finite
transition temperatures within both cluster approximations due to
their residual mean-field character.  The CDMFT and DCA effective
cluster models were solved with a QMC approach described in
\textcite{hettler:dca2}.

\begin{figure}[htb]
  \centerline{ \includegraphics*[width=3.0in]{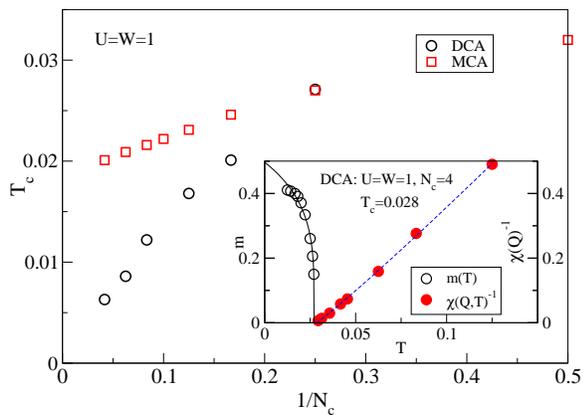}}
  \caption{Transition temperature versus inverse cluster size
    calculated with DCA/QMC (circles) and CDMFT/QMC (squares) when
    $U=W=4t=1$. Inset: DCA order parameter $m(T)$ and inverse charge
    susceptibility $\chi(Q)^{-1}$ versus temperature. Taken from
    \textcite{maier:dca4}.}
\label{fig:comp1} 
\end{figure}

The DCA transition temperature $T_c$ was determined from the divergence
of the lattice charge susceptibility $\chi(Q)$ calculated from the
particle-hole correlation function as detailed in
Sec.~\ref{subsec:suscept}. In the CDMFT formalism the calculation of
lattice susceptibilities is difficult if not impossible due to the
lack of translational invariance. Here $T_c$ is determined from the
calculation of the charge order parameter $m$ as detailed in
Sec.~\ref{subsec:brokensym}. For the DCA both techniques are
illustrated in the inset to Fig.~\ref{fig:comp1}.

As for the 1D Ising model (see Fig.~\ref{fig:1dising_Tc} in
Sec.~\ref{subsec:ccw}), the DCA result for $T_c$ scales to zero almost
linearly in $1/N_c$ for large $N_c$.  Moreover, $T_c$ obtained from
DCA is smaller and thus closer to the exact result than the $T_c$
obtained from CDMFT. The CDMFT does not show any scaling behavior and
in fact seems to tend to a finite value for $T_c$ as
$N_c\rightarrow\infty$.  As explained above, this striking difference
can be attributed to the different boundary conditions. The open
boundary conditions of the CDMFT cluster result in a large surface
contribution so that $\bar{\Gamma}_{CDMFT}>\bar{\Gamma}_{DCA}$. This
engenders pronounced mean-field behavior that stabilizes the finite
temperature transition for the cluster sizes treated in
Fig.~\ref{fig:comp1}. For larger cluster sizes the bulk contribution
to the CDMFT grand potential should dominate so that $T_c$ is expected
to fall to zero.

Complementary results are found in simulations of finite size
systems. In general, systems with open boundary conditions are expected
to have a surface contribution in the grand potential of order ${\cal
  O}(1/L_c)$ \cite{fisher:72}. This term is absent in systems with
periodic boundary conditions. As a result, simulations of finite size
systems with periodic boundary conditions converge much more quickly
than those with open boundary conditions \cite{landau:76}.

The DCA converges faster than the CDMFT for critical properties as
well as extended cluster quantities due to the different boundary
conditions and the coupling to the mean-field.  As detailed above,
each site in the DCA cluster experiences the same coupling to the
effective medium, while in the CDMFT only the sites on the boundary of
the cluster couple to the mean-field host.  Provided that the system
is far from a transition, the sites in the center of the CDMFT cluster
couple to the mean-field only through propagators which fall
exponentially with distance. Local results such as the single-particle
density of states thus converge more quickly in the CDMFT when
measured on a central site (see \onlinecite{maier:dca4}).


\paragraph{Calculation of the lattice self-energy.} 

Another significant difference between the two cluster techniques
appears in the calculation on the lattice self-energy. The DCA
approximates the lattice self-energy by a constant within a DCA cell
in momentum space, $\Sigma(\K+\kt,z)=\Sigma_c(\K,z)$.  Therefore the
self-energy is a step function in $\k$-space. In order to obtain
smooth non-local quantities such as the Fermi surface or the band
structure, an interpolated $\Sigma(\k,z)$ may be used. Bilinear
interpolation in 2D is guaranteed to preserve the sign of the function
however leads to kinks in $\Sigma(\k,z)$. Yielding the smoothest
possible interpolation of $\Sigma(\K,z)$, the use of an Akima spline
which does not overshoot is consistent with the DCA assumption that
the self-energy is a smoothly varying function in $\k$-space. However
it is important to note that this interpolated self-energy should not
be used in the self-consistent loop as this can lead to violations of
causality as discussed above.

In the CDMFT, the lattice self-energy is given by the
Fourier transform of the cluster self-energy (see Eq.~(\ref{eq:MCAS}))
\begin{eqnarray}
  \label{eq:11}
  \Sigma(\k,\k',z)&=&\frac{1}{N_c}\sum_\Q\sum_{\X_i,\X_j}e^{-i\k\cdot\X_i} \Sigma_c(\X_i,\X_j,z) e^{i\k'\cdot\X_j}\nonumber\\
 & & \times\,\, \delta(\k-\k'-\Q)\,.
\end{eqnarray}
Since the CDMFT cluster violates translational invariance, the lattice
self-energy depends on two momenta $\k$ and $\k'$ which can differ by
a wave-vector $\Q$ of the cluster reciprocal space. To restore
translational invariance, the CDMFT approximates the lattice
self-energy by the $\Q=0$ contribution to give \cite{kotliar:cdmft}
\begin{equation}
  \label{eq:12}
  \Sigma(\k,z)=\frac{1}{N_c}\sum_{\X_i,\X_j} e^{-i\k\cdot(\X_i-\X_j)} \Sigma_c(\X_i,\X_j,z)\,.
\end{equation}
In real space, the lattice self-energy
\begin{equation}
  \label{eq:13}
  \Sigma(\x_i,\x_j,z)=\frac{1}{N_c}\sum_{\X_i,\X_j}\Sigma_c(\X_i,\X_j)\delta_{\X_i-\X_j,\x_i-\x_j}
\end{equation}
is thus obtained by averaging over those cluster self-energy elements
$\Sigma(\X_i,\X_j)$ where the distance $\X_i-\X_j$ equals the distance
$\x_i-\x_j$.  As explained in \textcite{kotliar:cdmft2}, the factor
$1/N_c$ leads to an underestimation of non-local self-energy
contributions at small cluster sizes, since the number of
contributions for fixed $\x_i-\x_j>0$ in the sum Eq.~(\ref{eq:13}) is
always smaller than $N_c$. As a possible solution to this problem,
\textcite{kotliar:cdmft2} suggested to replace the form (\ref{eq:13})
for the lattice self-energy by a weighted sum which preserves
causality.  One could e.g. weight the terms in the sum by their number
instead of $N_c$ to achieve better results.

It is important to note that as in the DCA, the lattice self-energy
Eq.~(\ref{eq:12}) or (\ref{eq:13}) does not enter the self-consistent
loop. \textcite{biroli:cdmft} however realized that a translational
invariant formulation of the CDMFT algorithm can be obtained by
replacing the cluster self-energy $\bSigma_c(z)$ by the
translationally invariant lattice self-energy $\bSigma(\kt)$,
Eq.~(\ref{eq:12}) in the coarse-graining step Eq.~(\ref{eq:cgGMCA}).
Despite the dependence on $\kt$, this form of $\bSigma(\kt)$ can be
shown to preserve causality \cite{biroli:cdmft}.



%
%
%
%
%









\subsection{Effective cluster model}
\label{subsec:ecm}



Quantum cluster approaches reduce the complexity of the lattice
problem with infinite degrees of freedom to a (numerically) solvable
system with $N_c$ degrees of freedom.  As detailed in
Sec.~\ref{subsubsec:dca} this is achieved through the approximation of
$\Phi[\bG,\bU]$, the exact Baym-Kadanoff functional of the exact Green
function $\bG$ and interaction $\bU$, by a spatially localized
quantity $\Phi[\bGbar,\bUbar]$ which is a functional by the
corresponding (coarse-grained) quantities restricted to the cluster
sites, $\bGbar=N_c/N\sum_\kt \bG(\kt)$ and $\bUbar=N_c/N\sum_\kt
\bU(\kt)$.

$\Phi[\bGbar,\bUbar]$ may be calculated non-perturbatively as the
solution of a quantum cluster model
\begin{equation}\label{eq:Hc}
{\cal H}_c=H_{c,0}+H_{c,I}\,.
\end{equation}
${\cal H}_c$ consists of a non-interacting term $H_{c,0}$ describing the
bare cluster degrees of freedom and their coupling to a host.
The interacting term $H_{c,I}$ 
is related to the corresponding term in the original lattice model
through the coarse-grained interaction $\bUbar$. This ensures that the
functional dependencies of the cluster functional $\Phi_c$ and its
lattice counterpart $\Phi$ are identical.


The non-interacting term $H_{c,0}$ is fixed by the requirement that the
Green function ${\bf G}_c$ of the cluster model equals the
coarse-grained Green function $\bGbar$ of the original model
\begin{equation}\label{eq:cgDyson}
{\bf G}_c \equiv \bGbar=[\bGscript^{-1}-\bSigma_c]^{-1}\,,
\end{equation}
and hence is specified by the cluster-excluded Green function
$\bGscript$ (see Eq.~(\ref{eq:gscdmft})). 

For Green function or respectively action based cluster solvers, like
e.g. perturbation theory or QMC, $H_{c,0}$ can hence be encoded in the
cluster-excluded Green function $\bGscript$. The corresponding cluster
action $S_c$ for the fermionic cluster degrees of freedom represented
by the Grassman variables $\gamma$, $\gamma^*$ reads in imaginary time
and cluster real space
\begin{eqnarray}
  \label{eq:19}
  &&S_c[\gamma,\gamma^*]=-\int\limits_0^\beta d\tau\int\limits_0^\beta d\tau' \sum_{ij,\sigma} 
  \gamma^*_{i\sigma}(\tau) \Gscript_{ij,\sigma}(\tau-\tau') \gamma_{j\sigma}(\tau')\nonumber\\
 &&+\int\limits_0^\beta d\tau
  \sum_{ij,\sigma\sigma'} \frac{\Ubar_{ij}}{2}\gamma^*_{i\sigma}(\tau) \gamma^*_{j\sigma'}(\tau)
  \gamma^{}_{j\sigma'}(\tau) \gamma^{}_{i\sigma}(\tau)\,,
\end{eqnarray}
where we used the short hand notation $i,j$ for the cluster sites
$\X_i,\X_j$. Note that for the CDMFT the quantities $\Gscript_{ij}$
and $\Ubar_{ij}$ are given by Eqs.~(\ref{eq:gscdmft}) and
(\ref{eq:cgVMCA}) respectively, while for the DCA they are given by
the cluster Fourier transforms of $\Gscript(\K)=[\Gbar^{-1}(\K,z)+\Sigma_c(\K,z)]^{-1}$
and respectively of $\Ubar(\K)$
(see Eq.~(\ref{eq:cgVDCA})).

For Hamiltonian-based techniques, like e.g. the non-crossing
approximation, exact diagonalization or numerical renormalization
group, the need for an explicit formulation of $H_{c,0}$ is
inevitable. To setup the bare part $H_{c,0}$, it is convenient to use
Eq.~(\ref{eq:loc8}) for the CDMFT or respectively (\ref{eq:loc16}) for
the DCA together with Eq.~(\ref{eq:cgDyson}) to represent the cluster
excluded Green function $\bGscript$.  In the CDMFT, we have with
Eq.~(\ref{eq:loc4})
\begin{equation}
  \label{eq:41}
  \bGscript(z)=[(z+\mu)\mathbb{1}-\t_c-\bGamma(z)]^{-1}\,,
\end{equation}
and the matrix-elements of the intra-cluster hopping $\t_c$ are given
by the hopping amplitudes of the original lattice, $t_{ij}$ as
detailed in Sec.~\ref{subsubsec:comp}. The non-interacting problem is
thus split into two parts, cluster degrees of freedom with hopping
integrals $t_{ij}$ and their coupling to a dynamic host described by
the hybridization function $\bGamma(z)$. The CDMFT cluster model can
hence be written as (see also \onlinecite{bolech:cluster})
\begin{eqnarray}
  \label{eq:cmcdmft}
  {\cal H}_{c}=H_{c,0}+H_{c,I}&=&\sum_{ij,\sigma} (t_{ij}-\mu\delta_{ij}) c^\dagger_{i\sigma} c^{}_{j\sigma}\nonumber\\ &+& 
 \sum_{ij,\kt,\sigma} \lambda^{}_{ij\kt} [a^\dagger_{i\kt\sigma} a^{}_{j\kt\sigma}+h.c.]\nonumber\\
& + &  
 \sum_{ij,\kt,\sigma}
 [V^{}_{ij}(\kt) c_{i\sigma}^\dagger a_{j\kt\sigma}^{} + h.c.]\nonumber\\ 
&+& \sum_{ij,\sigma\sigma'} \frac{\Ubar_{ij}}{2}c^\dagger_{i\sigma}c^{\dagger}_{j\sigma'} c^{}_{j\sigma'}c^{}_{i\sigma}\,.
\end{eqnarray}
The first part describes the isolated cluster degrees of freedom with
fermionic creation (destruction) operators $c^\dagger_{i\sigma}$
($c^{}_{i\sigma}$). The second term simulates the host degrees of
freedom as a non-interacting conduction band with the help of
auxiliary operators $a^\dagger_{i\kt\sigma}$ ($a^{}_{i\kt\sigma}$) and
energy levels $\lambda_{ij\kt}$. The coupling between the cluster
states and the bath with an amplitude $V_{ij}(\kt)$ is described by
the third term and the interacting term is given by the last term. The
sums over $\kt$ run over the $N/N_c$ wave-vectors of the superlattice.
Integrating out the auxiliary degrees of freedom yields an action of
the form (\ref{eq:19}) with
\begin{eqnarray}
  \label{eq:1}
  \Gscript_{ij}(z)&=&[(z+\mu)\mathbb{1}-\t_c-\bGamma_c(z)]_{ij}^{-1}\\
  \Gamma_{c,ij}(z)&=&\sum_{lm,\kt}V_{il}^*(\kt)[z\mathbb{1}-{\bf \lambda}(\kt)]^{-1}_{lm}V_{mj}(\kt)\,.
\end{eqnarray}
Self-consistency then requires that the auxiliary parameters
$\lambda_{ij\k}$ and $V_{ij}(\kt)$ are chosen in a way such that the
cluster hybridization function $\bGamma_c(z)$ is identical to its
lattice counterpart $\bGamma(z)$ defined in Eq.~(\ref{eq:41}). It is
important to note that, while the isolated cluster parameter $\t_{c}$
can be deduced directly from the non-interacting part of the lattice
system, the energy levels $\lambda_\k$ and coupling constants
$V_{ij}(\k)$ of the auxiliary particles are not known a-priori, but
determined by the self-consistency condition $\bGamma_c=\bGamma$.
Since $\Gamma_{ij}$ is only finite on the surface of the cluster (see
discussion in Sec.~\ref{subsubsec:comp}), the coupling $V_{ij}(\kt)$
between the cluster and the host is only finite for sites $i$ on the
surface of the cluster which effectively reduces the number of baths.
This was numerically verified in CDMFT exact diagonalization studies
by \textcite{bolech:cluster}.

For the DCA we have with Eqs.~(\ref{eq:loc15}) and (\ref{eq:loc16})
\begin{equation}
\label{eq:Gscdca}
\Gscript(\K,z)=[z-\epsbar_\K+\mu-\Gamma(\K,z)]^{-1}\,,
\end{equation}
and hence the DCA effective cluster model is best represented in
cluster $\K$-space as
\begin{eqnarray}
  \label{eq:cmdca}
  {\cal H}_{c}&=&H_{c,0}+H_{c,I}\nonumber\\&=&\sum_{\K,\sigma} (\epsbar^{}_\K-\mu) 
  c^\dagger_{\K\sigma} c^{}_{\K\sigma} + 
 \sum_{\k,\sigma} \lambda^{}_\k a^\dagger_{\k\sigma} a^{}_{\k\sigma}\nonumber\\
& + &  
 \sum_{\K,\kt,\sigma}[V^{}_{\K}(\kt) 
  c_{\K\sigma}^\dagger a_{\K+\kt\sigma}^{} + h.c.]\nonumber\\ 
&+& \sum_{\stackrel{\K,\K'}{\Q}}\sum_{\sigma\sigma'} \frac{\Ubar(\Q)}{2N_c}
    c^\dagger_{\K+\Q\sigma}c^\dagger_{\K'-\Q\sigma'}
    c^{}_{\K'\sigma'} c^{}_{\K\sigma} \,.
\end{eqnarray}
The operators $c^\dagger_{\K\sigma}$ ($c^{}_{\K\sigma}$) create
(destroy) an electron with momentum $\K$ and spin $\sigma$ on the
cluster. $\Ubar(\Q)$ is the Coulomb repulsion between electrons on the
cluster defined in Eq.~(\ref{eq:cgVDCA}) and the sum over $\kt$ in the
coupling term again is restricted to the $N/N_c$ wave-vectors of the
superlattice. Analogous to the CDMFT case it is easy to see that the
DCA cluster model yields an action of the form (\ref{eq:19}) (in
cluster Fourier space) with a $\Gscript$ of the form (\ref{eq:Gscdca})
and the cluster hybridization function
\begin{equation}
  \label{eq:4}
  \Gamma_c(\K,z)=\frac{N_c}{N} \sum_{\kt} \frac{|V_\K(\kt)|^2}{z-\lambda_{\K+\kt}}\,.
\end{equation}
The auxiliary parameters of the DCA cluster model are then determined
by the condition $\Gamma_c(\K)=\Gamma(\K)$.


For $N_c=1$ both the CDMFT and the DCA cluster models reduce to the
single-impurity Anderson model. If self-consistency is also
established on the two-particle level (see discussion at the end of
Sec.~\ref{subsec:es}) via a susceptibility, an additional coupling to
a bosonic field representing the corresponding fluctuations has to be
considered in the cluster model (for details see
\onlinecite{maier:03}). For $N_c=1$ the cluster model then reduces to
the effective impurity model used in the EDMFT approach
\cite{edmft:si}.

As detailed in Sec.~\ref{subsec:es}, the CPT formalism sets the
hybridization function $\bGamma$ to zero, i.e considers an isolated
cluster without the coupling to a host.  Thus the CPT cluster model is
identical to the original lattice model restricted to cluster sites,
i.e. given by the first and last terms of the CDMFT cluster model
Eq.~(\ref{eq:cmcdmft}).


\subsection{Phases with broken symmetry}
\label{subsec:brokensym}

For simplicity, in the preceding sections the self-consistent
equations of quantum cluster theories have been derived assuming the
absence of symmetry-breaking.
In Sec.~\ref{subsec:suscept} we review how
instabilities to ordered phases can be identified by the computation
of response functions.  However, to explore the nature as well as
possible coexistence of symmetry-broken states, generalizations of the
cluster algorithms that explicitly account for the possibility of
symmetry-breaking on the single-particle level are necessary.

The applicability and modifications required to treat symmetry broken
phases depend on the cluster approach.  The CPT formalism is not
amenable to the description of ordered phases because its self-energy
is that of a finite isolated cluster in which spontaneous symmetry
breaking cannot occur. However, Dahnken {\em et al.} developed a
variational extension of the CPT \cite{potthoff:cluster3} based on the
self-energy functional approach by \textcite{potthoff:cluster2} which
yields a self-consistent scheme to study ordered phases (for details
see Sec.~\ref{subsec:alt}).  The CDMFT formalism can describe ordered
phases which are identifiable by a broken translational symmetry (such
as antiferromagnetism) by construction, since the translational
symmetry of the CDMFT cluster is already broken (see
\onlinecite{maier:dca3,maier:dca4,biroli:cdmft}).  Hence translational
invariant solutions are often found to be unstable against the ordered
one \cite{biroli:cdmft}.  The DCA formalism is translationally
invariant by construction, and therefore generalizations of the
algorithm are necessary to treat ordered phases. To keep this section
concise, we exemplify the necessary generalizations of the DCA
formalism to a selection of relevant types of symmetry-broken phases
along with the mapping onto the corresponding cluster models. The
adoption of the presented concepts to the CDMFT approach is
straightforward.

Once the equations are generalized to account for symmetry breaking,
the requisite algorithmic changes are relatively simple. A finite 
conjugate external field is used to initialize the calculation
and break the symmetry. The field is then switched off after a few
iterations and the system relaxes to its equilibrium state in the
absence of external fields.  On the other hand, if the field remains
small and finite, the dependence of the order parameter on the
field can be determined and used as an alternate way to
calculate the susceptibility (by extrapolation to zero field).  
This approach is especially useful 
for cluster solvers such as the non-crossing approximation or the 
fluctuation-exchange approximation where the computation of two-particle 
correlation functions is 
numerically too expensive.


\subsubsection{Uniform magnetic field -- Ferromagnetism}

We first consider the formalism necessary to treat ferromagnetism.
A finite homogeneous external magnetic field $h$ is introduced which 
acts on the spin $\sigma$ of the fermions according to the Zeeman term
\begin{equation}
  \label{eq:field}
  -h\sum_{i,\sigma}\sigma c^\dagger_{i\sigma} c^{}_{i\sigma}\,.
\end{equation}
The effect of $h$ on the motion of the spatial degree of freedom of
the electrons, i.e. the diamagnetic term, is neglected for
simplicity\footnote{In 2D systems the magnetic field can be applied
parallel to the plane to avoid orbital effects.}. 

In the presence of finite $h$ or a uniform magnetization, the single-particle 
Green functions for up- and down-electrons are not equivalent. To account 
for this $SU(2)$ symmetry-breaking, the spin-index of the Green function
$G_\sigma$, self-energy $\Sigma_\sigma$ and effective medium
$\Gscript_\sigma$ (and hence $\Gamma_\sigma$) in the derivation of the
DCA-equations has to be retained. For a finite uniform magnetic field
$h$ the DCA lattice Green function reads
\begin{equation}
  \label{eq:Gh}
  G_\sigma(\k,z)=\frac{1}{z-\epsilon_{\K+\kt}+\mu+h\sigma - \Sigma_{c,\sigma}(\K,z)}\,
\end{equation}
and the coarse grained and corresponding cluster-excluded Green function 
\begin{eqnarray}
  \label{eq:29}
  \Gbar_\sigma(\K,z)&=&\frac{N_c}{N}\sum_\kt G_\sigma(\K+\kt,z)\,,\\
\Gscript^{-1}_\sigma(\K,z)&=&G^{-1}_\sigma(\K,z)+\Sigma_{c,\sigma}(\K,z)
\end{eqnarray}
become spin-dependent. 

The action of the effective cluster model is
identical to the action in the paramagnetic state, Eq.~(\ref{eq:19}),
but the spin indices have to be explicitly retained.
It then describes electrons in an external magnetic field $h$ coupled
to a spin-dependent host and self-consistency is established by
equating the Green function of the cluster model with the
coarse-grained Green function (\ref{eq:29}). 

In analogy, for Hamiltonian based cluster solvers, an additional term
\begin{equation}
  \label{eq:24}
  -h\sum_{\K,\sigma}\sigma c^\dagger_{\K\sigma}c^{}_{\K\sigma}
\end{equation}
is added to the cluster Hamiltonian, Eq.~(\ref{eq:cmdca}), in the
presence of a finite external magnetic field $h$. The coarse-grained
Green function $\Gbar$, Eq.~(\ref{eq:Gh}), is then used to calculate
the magnetization $m:=1/N\sum_{i\sigma}\sigma \langle
n_{i\sigma}\rangle$ according to
\begin{equation}
  \label{eq:25}
  m=\frac{1}{N_c}\sum_{\K,\sigma}\sigma \Gbar_\sigma(\K,\tau=0^-)
\end{equation}
and after analytical continuation
\begin{equation}
  \label{eq:23}
  m=-\frac{1}{\pi}\frac{1}{N_c}\sum_{\K\sigma}\sigma\int\limits_{-\infty}^{+\infty} d\omega f(\omega)\Im m\Gbar_\sigma(\K,\omega+i\delta)\,.
\end{equation}

\subsubsection{Superconductivity}

In this and the next section we 
generalize the DCA formalism to treat phases with superconducting and
antiferromagnetic order, respectively. For better readability, we
refrain from discussing the description of phases with coexisting
superconducting and antiferromagnetic order. The extension to an
integrated formalism is straightforward and has been discussed in
\textcite{kats:dca}.

We consider superconducting phases where the electrons are paired in
spin singlet or triplet states with $S_z=0$ indicated by finite values
of the order parameter $\Delta_\k:=\langle
c_{\k\uparrow}c_{-\k\downarrow}\rangle$ for some $\k$. In addition to
the normal Green function $G(\k,\tau)$ it is therefore necessary to
introduce the anomalous Green function $F(\k,\tau)=-\langle T_\tau
c_{\k\uparrow}(\tau) c_{-\k\downarrow} \rangle$.  The spatial and
temporal symmetry of the pairing state can then be inferred from the
symmetries of $F$. Since $F$ describes the pairing of fermions, it
necessarily is antisymmetric under the exchange of two particles. The
spatial symmetry of the pairing state is determined by the
$\k$-dependence of the anomalous Green function $F(\k,\tau)$.  If we
assume conventional even-frequency pairing $F(\k,-\tau)=F(\k,\tau)$,
in the case of spin-singlet pairing, $F$ has to be symmetric in $\k$,
i.e. $F(-\k,\tau)=F(\k,\tau)$ as is the case for even parity order
parameters such as e.g. $s$-wave and $d$-wave.  In the case of
spin-triplet pairing $F$ is antisymmetric in $\k$. i.e.\ 
$F(-\k,\tau)=-F(\k,\tau)$ as e.g. in a $p$-wave state.

The allowed symmetry of the pairing state is restricted by the
cluster geometry.  It depends upon the $\k$-dependence of the
dispersion $\epsilon_\k$ and the $\K$-dependence of the cluster
self-energy ${\bf \Sigma}(\K,\tau)$.  When $N_c=1$, ${\bf \Sigma}$ is
local and the $\k$-dependence of $F(\k,\tau)$ is entirely through
$\epsilon_\k$. Hence only pairing states with the symmetry of the
lattice such as $s$-wave and extended $s$-wave can be described by
this formalism \cite{jarrell:92,jarrell:93}.  Larger cluster sizes are
necessary to describe order-parameters with a symmetry less than the
lattice symmetry, e.g. simulations with $N_c=4$ are necessary to
describe phases with a $d_{x^2-y^2}$-wave order parameter which
transforms according to $\cos k_x - \cos k_y$.

By utilizing the concept of Nambu-spinors
\begin{equation}
\label{eq:BS1}
 \Psi^\dagger_\k:=(c^\dagger_{\k\uparrow}, c^{}_{-\k\downarrow})\,;\,
  \Psi_\k=(\Psi^\dagger_\k)^\dagger
\end{equation}
the self-consistent equations can be written in a more compact form,
since the corresponding Green function matrix in Nambu space
\begin{equation}
  \label{eq:27}
  \bG(\k,z):=\langle\langle \Psi^{}_\k; \Psi^\dagger_\k
  \rangle\rangle_z= \left( \begin{array}{cc}
      G(\k,z) & F(\k,z)\\
      F^*(\k,z^*) & -G^*(-\k,-z^*)
      \end{array}\right)
\end{equation}
contains information about both the normal and anomalous Green
functions. In the presence of an external pairing field
$\eta(\k)=\eta'(\k)+i\eta''(\k)$ which couples to
$c_{-\k\downarrow}c_{\k\uparrow}$, the non-interacting part of the
Hubbard Hamiltonian can be written as $H_0=\sum_{\k} \Psi^\dagger_\k
[\epsilon_\k\sigma_3-\eta'(\k)\sigma_1+\eta''(\k)\sigma_2 ]
\Psi^{}_\k$ so the the DCA lattice Green function in the
superconducting state becomes
\begin{eqnarray}
  \bG(\k,z)&=&[z\sigma_o-(\epsilon_{\k}-\mu)\sigma_3\nonumber\\
               && -\eta'(\k)\sigma_1-\eta''(\k)\sigma_2-\bSigma_c(\K,z)]^{-1}\,,
  \label{eq:28}
\end{eqnarray}
where $\k=\K+\kt$ and $\sigma_i$ are the Pauli-spin matrices.  The
diagonal parts of the Nambu-matrix $\bSigma_c(\K,z)$ describe
quasiparticle renormalizations and the off-diagonal parts contain
information about the $\K$- and frequency dependence of the pairing
state. Again, the coarse-grained Green function
\begin{eqnarray}
  \label{eq:30}
  \bGbar(\K,z)&=&\frac{N_c}{N}\sum_\kt \bG(\K+\k,z) \nonumber\\
 &=& \left(
    \begin{array}{cc}
      \Gbar(\K,z) & \bar{F}(\K,z)\\
      \bar{F}^*(\K,z^*) & -\Gbar^*(-\K,-z^*)
    \end{array}\right)
\end{eqnarray}
is used to calculate the cluster-excluded Green function
\begin{equation}
  \label{eq:31}
  \bGscript(\K,z)=[\bGbar^{-1}(\K,z)+\bSigma_c(\K,z)]^{-1}
\end{equation}
which together with the coarse-grained interaction $\Ubar(\K)$ defines
the action of the corresponding effective cluster model
\begin{eqnarray}
  \label{eq:32}
  &&S_c=-\int\limits_0^\beta d\tau \int\limits_0^\beta d\tau' \sum_{ij}
  {\pmb \psi}_{i}^\dagger(\tau) \bGscript_{ij}(\tau-\tau')
  {\pmb\psi}_j(\tau')\nonumber\\
 &&+ \int\limits_0^\beta d\tau \sum_{ij}
  \frac{\Ubar_{ij}}{2}\left[{\pmb\psi}_i^\dagger(\tau)\sigma_3{\pmb\psi}_i^{}(\tau)\right]\left[({\pmb\psi}_j^\dagger(\tau)\sigma_3{\pmb\psi}_j^{}(\tau)\right]\,.
\end{eqnarray}
As in the normal state case (see Eq.~(\ref{eq:19})), the cluster
action is represented in cluster real space and all the quantities are
cluster Fourier transforms of the corresponding quantities in cluster
$\K$-space.  The spinors of Grassmann-variables
${\pmb\psi}_i^\dagger=(\gamma_{i\uparrow}^*,\gamma^{}_{i\downarrow})$
and ${\pmb\psi}_i=({\pmb\psi}_i^\dagger)^\dagger$ generate coherent
states corresponding to the cluster Fourier transform of the
Nambu-spinors Eq.~(\ref{eq:BS1}). In analogy, the corresponding
cluster Hamiltonian in the superconducting state is obtained from the
normal state cluster Hamiltonian Eq.~(\ref{eq:cmdca}) by representing
it with the Nambu-spinors (\ref{eq:BS1}) and adding a U(1) symmetry
breaking term
\begin{equation}
  \label{eq:21}
  -\sum_{\K} \Psi^\dagger_\K [\bar{\eta}'(\K)\sigma_1-\bar{\eta}''(\K)\sigma_2] \Psi^{}_\K\,, 
\end{equation}
where $\bar{\eta}(\K)=N_c/N\sum_\kt \eta(\K+\kt)$ is the
coarse-grained pair-field.

After self-consistency is established by requiring that the Green
function of the effective cluster model calculated with the action
Eq.~(\ref{eq:32}) 
equals the coarse-grained Green function Eq.~(\ref{eq:30}), the order
parameter $\Delta_\k$ can be calculated. Within the DCA the resolution
in $\k$-space is restricted to the cluster $\K$-points and the order
parameter is coarse-grained
\begin{equation}
  \label{eq:26}
  \bar{\Delta}_\K=\frac{N_c}{N}\sum_\kt \langle c_{\K+\kt\uparrow} c_{-(\K+\kt)\downarrow} \rangle = \bar{F}(\K,\tau=0^+)
\end{equation}
and given by the equal-time coarse-grained anomalous Green function
$\bar{F}$ as 
(after analytical continuation)
\begin{eqnarray}
  \label{eq:34}
  && \bar{F}(\K,\tau=0^+)=\frac{1}{\beta}\sum_n \bar{F}(\K,i\omega_n)
  \nonumber\\
 &&= \frac{1}{\pi}\int\limits_0^\infty d\omega \tanh \left(\frac{\beta\omega}{2}\right) \Im m \bar{F}(\K,\omega+i\delta)\,.
\end{eqnarray}


\subsubsection{Antiferromagnetic order}

In this section we derive the DCA cluster formalism for
antiferromagnetism on a bipartite lattice.  This formalism is
appropriate when $N_c>1$.  A formalism suitable for the case when
$N_c=1$ is discussed in detail in \textcite{georges:dmftrev} in the
context of DMFT.

The antiferromagnetically ordered state is characterized by a spatial
variation of the magnetization according to
\begin{equation}
  \label{eq:35}
  {\bf m}(\x)={\bf m}e^{i\Q\cdot\x}\,,
\end{equation}
where $\x$ denotes the sites in the lattice and $\Q$ is the antiferromagnetic 
wave-vector ($\Q=(\pi,\pi)$ in 2D). Hence bipartite lattices can be divided 
into two inequivalent sublattices according to
\begin{equation}
  \label{eq:36}
  e^{i\Q\cdot \x} = \left \{ \begin{array}{rcl}
      1 & : & \x \in \mbox{A sub-lattice}\\
     -1 & : & \x \in \mbox{B sub-lattice}\end{array}
\right.\,.
\end{equation}
The magnetic ordering thus reduces the translational symmetry of the
original lattice. The volume of the magnetic unit cell is doubled
compared to the structural unit cell. Accordingly, the volume of the
first Brillouin zone in the antiferromagnetic state is reduced to half
the original volume and $\Q=(\pi,\pi)$ becomes a reciprocal lattice
vector. Hence in neutron scattering experiments Bragg peaks are found
at the wave-vector $\Q$ corresponding to a period of two lattice
spacings. 

As a consequence of the translational symmetry breaking, the
correlation function
\begin{equation}
  \label{eq:37}
  G_\sigma(\k,\k+\Q;\tau)=-\langle T_\tau c^{}_{\k\sigma}(\tau) c^\dagger_{\k+\Q\sigma} \rangle
\end{equation}
becomes finite. Along the lines of the formalism for the
superconducting state it is thus convenient to introduce spinors
\begin{equation}
  \label{eq:38}
  \Psi^\dagger_{\k\sigma}=(c^\dagger_{\k\sigma},
  c^\dagger_{\k+\Q\sigma})\,; \,\, \Psi_{\k\sigma}=(\Psi_{\k\sigma}^\dagger)^\dagger
\end{equation}
for the antiferromagnetic state.  In the presence of a staggered
external magnetic field $h(\x) = h \exp(i\Q\cdot\x)$ the
non-interacting part of the Hamiltonian for bipartite lattices with
$\epsilon_{\k+\Q}=-\epsilon_\k$ then becomes $H_0=\sum_{\k\sigma}'
\Psi^\dagger_{\k\sigma}(\epsilon_\k\sigma_3 -\sigma \frac{h}{2}\sigma_1)
\Psi^{}_{\k\sigma}$ where the prime on the sum indicates summation
over the reduced Brillouin zone only. The corresponding Green function
\begin{eqnarray}
  \label{eq:42}
  \bG_\sigma(\k,z)&=&\langle\langle \Psi^{}_{\k\sigma}; \Psi^\dagger_{\k\sigma} \rangle\rangle_z\\ 
 & = & [z\sigma_o-(\epsilon_\k-\mu)\sigma_3-\sigma\frac{h}{2}\sigma_1-\bSigma_c(\K,z)]^{-1}\nonumber\,,
\end{eqnarray}
with $\k=\K+\kt$, is coarse-grained over the DCA cells
\begin{eqnarray}
  \label{eq:43}
  \bGbar_\sigma(\K)&=&\frac{N_c}{N}\sum_\kt \bG_\sigma(\K+\kt)\\
       &=& \left(\begin{array}{cc}
     \Gbar_\sigma(\K,\K) & \Gbar_\sigma(\K,\K+\Q)\\
     \Gbar_\sigma(\K+\Q,\K) & \Gbar_\sigma(\K+\Q,\K+\Q)
\end{array}\right)\nonumber\,,    
\end{eqnarray}
where we dropped the frequency argument for convenience.  The cluster
excluded Green function
\begin{eqnarray}
  \label{eq:44}
  \bGscript_\sigma(\K,z)&=&[\bGbar^{-1}_\sigma(\K,z)+\bSigma_{c,\sigma}(\K,z)]^{-1}\,.
\end{eqnarray}
then has two elements, $\Gscript_\sigma(\K,\K)$ and
$\Gscript_\sigma(\K,\K+\Q)$. As a result, its Fourier transform to
real space
\begin{eqnarray}
  \label{eq:45}
  \Gscript_{ij,\sigma}&=&\frac{1}{N_c}\sum_\K e^{i\K\cdot(\X_i-\X_j)}\Gscript_\sigma(\K)\\
 &+& \frac{1}{N_c}\sum_\K e^{i\K\cdot(\X_i-\X_j)}e^{-i\Q\cdot\x_j}\Gscript_\sigma(\K,\K+\Q)\nonumber
\end{eqnarray}
breaks translational symmetry. The action of the corresponding cluster
model in the antiferromagnetic state
is then formally identical to the action in the paramagnetic state,
Eq.~(\ref{eq:19}). As in the superconducting state the corresponding
cluster model is obtained from the paramagnetic cluster model
Eq.~(\ref{eq:cmdca}) by changing the representation to the Nambu
spinors Eq.~(\ref{eq:38}) and adding a symmetry breaking term
\begin{equation}
  \label{eq:40}
  -\sum_{\K\sigma} \sigma\frac{h}{2}\Psi^\dagger_{\K\sigma} \sigma_1 \Psi^{}_{\K\sigma}\,,
\end{equation}
to account for the external staggered magnetic field $h$.

After convergence, the sub-lattice magnetization
$m=1/N\sum_{i\sigma}e^{i\Q\cdot \x_i} \sigma \langle
n_{i\sigma}\rangle$ may be calculated from the off-diagonal component
of the Green function matrix Eq.~(\ref{eq:43}) according to
\begin{eqnarray}
  \label{eq:49}
  m&=&\frac{1}{N_c}\sum_{\K\sigma}\sigma \Gbar_\sigma(\K,\K+\Q;\tau=0^-)\\
   &=&-\frac{1}{N_c\pi}\sum_{\K\sigma}\int\limits_{-\infty}^{+\infty} d\omega
   f(\omega)\sigma \Im m \Gbar_\sigma(\K,\K+\Q;\omega+i\delta)\nonumber\,.
\end{eqnarray}

\subsection{Calculation of susceptibilities}
\label{subsec:suscept}

A convenient way to identify continuous phase transitions is to search
for divergences of susceptibilities. One particular advantage of
quantum cluster theories is that they allow us to consistently
calculate these susceptibilities from the corresponding cluster
susceptibility\cite{hettler:dca2}.  Unfortunately, the calculation of
two-particle correlation functions in the CDMFT formalism is strongly
hampered by the violation of translational invariance on the cluster.
So, in this section, we will restrict our attention to the calculation
of two-particle quantities in th DCA following \textcite{hettler:dca2}
and \textcite{jarrell:dca3}.

As a specific example, we describe here the calculation of the
two-particle Green function
\begin{eqnarray}
\label{chiph}
\nonumber
\chi_{\si\si'}(q,k,k')&=&
\int_{0}^{\beta}\int_{0}^{\beta}\int_{0}^{\beta}\int_{0}^{\beta} 
d\tau_1 d\tau_2 d\tau_3 d\tau_4   \\ 
& \times & e^{i\left( (\omega_{n}+\nu) \tau_1 - \omega_{n} \tau_2 
                +\omega_{n'} \tau_3 - (\omega_{n'}+\nu) \tau_4 \right)} \nonumber \\ 
& \times & 
\langle T_\tau
c_{\k+\q \si}^{\dag}(\tau_1)
c_{\k \si}(\tau_2)
c_{\k' \si'}^{\dag}(\tau_3)
c_{\k'+\q \si'}(\tau_4) 
\rangle\,,
\nonumber
\end{eqnarray}             
where we adopt the conventional notation \cite{AGD}
$k=(\k,i\omega_n)$, $k'=(\k,i\omega_{n'})$, $q=(\q,i\nu_n)$ and
$T_\tau$ is the time ordering operator.

$\chi_{\si\si'}(q,k,k')$ and the irreducible two-particle vertex
function $\Gamma_{\si\si'}(q,k,k')$ (not to be confused with the
single-particle hybridization function) are related to each other
through the Bethe-Salpeter equation
\begin{eqnarray} 
\label{betheph}
\nonumber 
&&\chi_{\si\si'}(q,k,k')= 
\chi^0_{\si\si'}(q,k,k') + \chi^0_{\si\si''}(q,k,k'') \\  
&&\times\,\, \Gamma_{\si''\si'''}(q,k'',k''') \chi_{\si'''\si'}(q,k''',k')
\end{eqnarray} 
where $\chi^0_{\si\si'}(q,k,k'')$ is the non-interacting
susceptibility constructed from a pair of fully-dressed
single-particle Green functions. As usual, a summation is to be made
for repeated indices.

We now make the DCA substitution $\Gamma_{\si\si'}(\q,\K+\kt,\K'+\kt')
\to \Gammac_{\si\si'}\left(\q,\K,\K'\right)$ in Eq.~(\ref{betheph})
where $\Gamma_c$ is the irreducible two-particle vertex calculated on
the cluster (frequency labels have been suppressed).  Note that only
the bare and dressed two-particle Green functions $\chi$ depend upon
the superlattice wave-vectors $\kt$. Since $\chi$ and $\chi^0$ in the
product on the RHS of Eq.~(\ref{betheph}) share no common momentum
labels, we may freely sum over the wave-vectors $\kt$, yielding
\begin{eqnarray} 
\label{betheph_CG}
\nonumber 
\lefteqn{ \chibar_{\si\si'}(q,K,K')= 
\chibar^0_{\si\si'}(q,K,K') + \chibar^0_{\si\si''}(q,K,K'')} \nonumber\\  
&\times & \Gammac_{\si''\si'''}(q,K'',K''') 
\chibar_{\si'''\si'}(q,K''',K')\,.
\end{eqnarray} 
By coarse-graining the Bethe-Salpeter equation, we have greatly
reduced its complexity; each of the matrices above is sufficiently
small that they may be easily manipulated using standard techniques.

In contrast with the single-particle case where the coarse-grained
quantities are identical to those of the cluster, the cluster quantity
$\chic_{\si\si'}(q,K,K')$ is not equal to $\chibar_{\si\si'}(q,K,K')$.
This is because the self-consistency is established only at the
single-particle level.  Unlike the single-particle case where both
$\Sigma_c(K)$ and ${\bar G}(K)$ are directly calculated, neither
$\Gammac_{\si\si'}(q,K,K')$ nor the coarse-grained susceptibility
$\chibar_{\si\si'}(q,K,K')$ are calculated during the
self-consistency.  Instead, the coarse-grained non-interacting
susceptibility $ \chibar^0_{\si\si'}(q,K,K')$ is calculated in a
separate program after the DCA converges using the relation
\begin{eqnarray}
\label{chinot}
\nonumber
\chibar^0_{\si\si'}[(\q,i\nu_n);(\K,i\omega_n);(\K',i\omega_{n'})] =
\delta_{\si\si'}\delta_{ \K\K'} \delta_{\omega_n\omega_{n'}} \\
\times \frac{N_c}{N} 
\sum_{\tk}G_{\si}( \K+{\bf \tk},i\omega_n)
G_{\si}({ \K+{\bf \tk}+\q}, i\omega_n+\nu_n)
  \quad\mbox{.}
\end{eqnarray}
The corresponding cluster susceptibility is calculated by the cluster
solver, e.g. the QMC process, as discussed in Sec.~\ref{subsubsec:qmc}
and the vertex function is extracted by inverting the cluster
two-particle Bethe-Salpeter equation
\begin{eqnarray} 
\label{betheph_C}
\nonumber 
\chic_{\si\si'}(q,K,K')= \chic^0_{\si\si'}(q,K,K') + 
\chic^0_{\si\si''}(q,K,K'') \\  
          \times \Gammac_{\si''\si'''}(q,K'',K''') 
                     \chic_{\si'''\si'}(q,K''',K')\,.
\end{eqnarray} 
If we combine Eqs.~(\ref{betheph_C}) and (\ref{betheph_CG}), then the
coarse-grained susceptibility may be obtained after elimination of
$\Gammac_{\si\si'}(q,K,K')$ between the two equations.  It reads
\begin{equation}
\label{chicoars2}
\nonumber
{\bar \bchi}^{-1} = {\pmb \chi}_c^{-1} -\bchi_c^{0^{-1}} +  {\bar \bchi}^{0^{-1}}\,,
\end{equation}      
where, for example, ${\bar \bchi}$ is the matrix formed from
$\chibar_{\si,\si'}(q,K,K')$ for fixed $q$.  The charge $(ch)$ and
spin $(sp)$ susceptibilities ${\chi}_{ch,sp}(q,T)$ are deduced from
$\chibar$ according to
\begin{eqnarray}
\label{chichsp}
{\chi}_{ch,sp}(q,T)=\frac {(k_BT)^2}{N_c^2}
\sum_{ KK'\sigma\sigma'}
\lambda_{\sigma \sigma'} 
\chibar_{\si\si'}(q,K,K')\,, 
\end{eqnarray}
where $\lambda_{\sigma\sigma'}=1$ for the charge channel and
$\lambda_{\sigma\sigma'}=\sigma\sigma'$ for the spin channel.  The
calculation of particle-particle (i.e.\ pairing) susceptibilities
follows from a straightforward generalization of this formalism. The
reader is referred to prior articles on cluster QMC for more details
on these topics \cite{jarrell:dca3}.


\subsection{Disordered systems}
\label{subsec:dis}
In this section, we describe cluster approximations for disordered
systems.  The mapping between the lattice and the cluster in the CDMFT
and the molecular CPA (MCPA) is identical, and when the CDMFT is
applied to disordered systems, it becomes identical to the MCPA
\cite{tsukada:MCPA,duc:mcpa}. The MCPA and related techniques for
disordered systems are extensively reviewed in \textcite{gonis:gf} and
will not be discussed here.

In the remainder of this section we describe a DCA for disordered
systems that recovers the CPA for $N_c=1$ and becomes exact when
$N_c\to\infty$.  This is an extension of the formalism introduced by
\textcite{jarrell:dca1}, modified to include the effects of correlated
disorder \cite{jarrell:ddca}.  We consider an Anderson model with
diagonal disorder, described by the Hamiltonian
\begin{equation}
H =
\displaystyle - t  \sum_{<ij>,\sigma} 
\left( 
c^{\dagger}_{i\sigma} c_{j\sigma} + 
c^{\dagger}_{j\sigma} c_{i\sigma}
\right)  \\[5mm]
\displaystyle+
 \sum_{i \sigma}(V_i- \mu) n_{i\sigma}
\label{eqn:Hquenched}
\end{equation}
where $c^{\dagger}_{i\sigma}$ creates a quasiparticle on site $i$ with
spin $\sigma$ and $n_{i\sigma}=c^{\dagger}_{i\sigma} c_{i\sigma}$.
The disorder occurs in the site occupancies labeled by $\xi_i$ and in
the local orbital energies $V_i$ which describe the electrostatic
potential at site $i$.  We assume that $V_i$ are quenched random
variables distributed according to some specified probability
distribution which includes the effects of intersite correlations.

The effect of the disorder potential $\sum_{i \sigma}V_i n_{i,\sigma}$
may be described using standard diagrammatic perturbation theory
(although we eventually sum to {\em{all}} orders).  We perturb around
the ordered state, described by the first term in
Eq.~(\ref{eqn:Hquenched}).
\begin{figure}[htb]
\centerline{
\includegraphics*[width=2.5in]{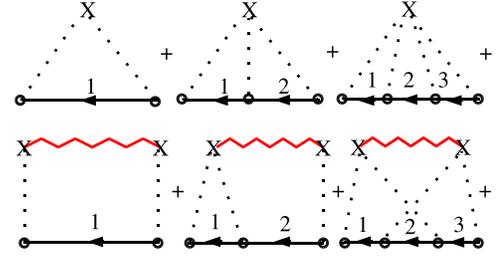}}
\caption{A few low-order diagrams in the irreducible self-energy of a
  quenched diagonally disordered system.  Each circle represents the
  scattering of a state $\k$ from sites denoted by X and a dotted
  line.  The correlations between the electrostatic potentials on
  different sites are denoted by the wavy line.}
\label{fig:DCA_dis_fig1}
\end{figure}

Translational invariance and momentum conservation are restored by
averaging over all allowed values of the site occupancies $\{\xi_i\}$
and the corresponding disorder potentials $V_i$.  We study this
effect, e.g., on the sixth graph shown in Fig.~\ref{fig:DCA_dis_fig1}
which makes a contribution to the self-energy
\begin{eqnarray}
&&\frac{1}{N^4}\sum_{i\neq j,\k_1,\k_2,\k_3} \langle V_i^2 V_j^2 \rangle
G(\k_1)G(\k_2)G(\k_3)\nonumber \\
&&\hspace*{1cm}\times\,\, e^{i\x_i\cdot(\k_1-\k+\k_3-\k_2)}e^{i\x_j\cdot(\k_2-\k_1+\k'-\k_3)}
\end{eqnarray}  
After averaging over the disorder configurations, $\langle V_i^2 V_j^2
\rangle$ becomes a function of $\x_i-\x_j$.  We identify this average
as $D_{ij}^{22}$.  With translational invariance restored, we may
complete the Fourier transform, and obtain
\begin{eqnarray}
\frac{\delta_{\k\k'}}{N^3}\sum_{\k_1,\k_2,\k_3,\q} 
D^{22}(\q)G(\k_1)G(\k_2)G(\k_3) && \nonumber \\
\left( N\delta_{\q+\k_2+\k,\k_1+\k_3} -1 \right)\,.
\end{eqnarray}
It is easy to extend this argument to all orders in perturbation theory.
All graphs are composed of sums of products of Laue functions (e.g.\
$\Delta=N\delta_{\q+\k_2+\k,\k_1+\k_3}$), and Green functions $G(\k)$ 
and $D^{nm}(\q)$, where $D^{nm}(\q)$ is the Fourier transform of 
$D_{ij}^{nm}=\langle V_i^n V_j^m \rangle$.

A hierarchy of approximations may then be constructed by approximating
the Laue functions within the graphs.  These include the CPA where
conservation of the internal momentum labels is completely neglected.
Here, all of the Laue functions involving the internal momentum labels
are set to one, $\Delta=N\delta_{\k_1+\k_2\cdots} \to 1$.  In this
case we may freely sum over all internal momentum labels, and all
terms describing non-local correlations, such as those on the bottom
of Fig.~\ref{fig:DCA_dis_fig1} vanish, whereas the CPA graphs shown on
top remain.  Different cluster approximations, including the MCPA and
the DCA may be constructed by systematically restoring momentum
conservation through the appropriate choice of Laue function, as
discussed in Sec.~\ref{subsec:freeeg}
        
For the DCA, we divide the lattice into clusters as described in
Fig.~\ref{fig:Nc4_clusters}, write wave-vectors as $\k=\K+\kt$ and
employ the DCA Laue function Eq.~(\ref{eq:LDCA}),
\begin{equation}
\Delta_{DCA} = 
N_c \delta_{\K_1+\K_2,\K_3+\K_4  ...}\,, 
\label{Laue_DCA}
\end{equation}
such that momentum conservation is preserved only within the cluster
reciprocal space $\K$. With this choice for the Laue function, we may
freely sum over superlattice wave-vectors $\kt$ within each DCA
coarse-graining cell (see Fig.~\ref{fig:Nc4_clusters}).  This leads to
the replacement of the lattice propagators $G(\k)$ and $D^{nm}(\k)$
with coarse-grained propagators $\Gbar(\K)$ and $\Dbar^{nm}(\K)$,
respectively,
\begin{equation}
\bar{G}(\K,z) = \frac{N_c}{N}\sum_{\kt}G(\K+\kt,z),
\label{eq:gbar}
\end{equation}
\begin{equation}
\bar{D}^{nm}(\K) = \frac{N_c}{N}\sum_{\kt}D^{nm}(\K+\kt),
\label{eq:dbar}
\end{equation}
The first of these sums is straightforward; however, the second
requires some investigation due to the powers of the potential.

Here, we calculate $\bar{D}^{nm}(\K)$ for a binary $A$,$B$ alloy where
the concentration of $A$ atoms is $c$ and that of $B$ atoms is $1-c$.
This calculation can be generalized to multicomponent alloys and may
easily be extended for more complex alloys.  For the binary alloy, we
employ an idempotent formalism where the idempotent $\xi_i=1$
indicating that the site is occupied by an $A$ atom, or $\xi_i=0$
indicating a $B$ atom site.  If we associate $V_i=V_A$ or $V_B$ for an
$A$ or $B$ atom, respectively, then $V_i=\xi_i V_A + (1-\xi_i)V_B$.
Then, since $\xi_i^n=\xi_i$ and $(1-\xi_i)^n=1-\xi_i$,
\begin{eqnarray}
\langle V_i^n V_j^m \rangle &=& \langle (\xi_i V_A + (1-\xi_i)V_B)^n 
                                        (\xi_j V_A + (1-\xi_j)V_B)^m\rangle
                                        \nonumber\\
                            &=& \langle (\xi_i V_A^n + (1-\xi_i)V_B^n)
                                        (\xi_j V_A^m + (1-\xi_j)V_B^m)\rangle
                                        \nonumber\\
                            &=& V_\alpha^n V_\beta^m g_{ij}^{\beta\gamma}
\end{eqnarray}
where repeated indices in the last line are summed over and
\begin{eqnarray}
g_{ij}^{AA} &=& \langle \xi_i\xi_j\rangle\\
g_{ij}^{AB} &=& \langle \xi_i(1-\xi_j)\rangle\\
g_{ij}^{BB} &=& \langle (1-\xi_i)(1-\xi_j)\rangle
\end{eqnarray}
are the joint probabilities for occupation of sites $i$ and $j$ by
atoms of the designated types.

This formalism may be generalized to a multicomponent alloy $\alpha =
A, B,C,\dots$. Since a site may only be occupied by an atom of one
type only, in general
\begin{equation}
\langle V_i^n V_j^m \rangle = V_\alpha^n V_\beta^m g_{ij}^{\beta\gamma}\,.
\end{equation}
$\langle V_i^n V_j^m \rangle$ is a {\em{linear}} function of the 
probabilities $g_{ij}^{\beta\gamma}$ for all $n$ and $m$.  Thus, the 
effect of coarse-graining $D^{nm}(\k)$ is equivalent to coarse-graining 
$g^{\beta\gamma}(\k)$ for all $n$ and $m$.

As an example, consider a binary alloy with only near-neighbor
configurational correlations.  Here we may write
\begin{equation}
g_{ij}^{\beta\gamma}=g^{0,\beta\gamma}+
\alpha\left(2\delta_{\beta\gamma}-1\right) \delta_{i+\epsilon,j}
\end{equation}
where $\epsilon$ indexes the near-neighbors to site $i$, and
$g^{0,AA}=c^2$, $g^{0,BB}=(1-c)^2$ and $g^{0,AB}=c(1-c)$ are the joint
probabilities for the occupation of different sites for a system
without configurational correlations.  On a hypercubic lattice of
dimension $D$
\begin{equation}
g^{\beta\gamma}(\k)=g^{0,\beta\gamma}\delta_{\k,0}+
\alpha\left(2\delta_{\beta\gamma}-1\right)  \sum_{l=1}^D \cos(k_l)\,.
\end{equation}
The corresponding coarse-grained result is 
\begin{equation}
\bar{g}^{\beta\gamma}(\K) = g^{0,\beta\gamma}\delta_{\K,0}+
\alpha\left(2\delta_{\beta\gamma}-1\right)  R \sum_{l=1}^D \cos(K_l)
\end{equation}
where $R=(L_c/\pi)\sin(\pi/L_c)$ is a coarse-graining factor
($L_c=N_c^{1/D}$ is the linear cluster size).
If we transform back to the cluster coordinates, then
\begin{equation}
\bar{g}_{ij}^{\beta\gamma}=g^{0,\beta\gamma}+
R \alpha\left(2\delta_{\beta\gamma}-1\right)   \delta_{i+\epsilon,j}
\end{equation}
are the configurational probabilities for the cluster, where
$\epsilon$ labels the sites adjacent to $i$. In the CPA limit, $L_c=1$
and thus $R$ vanishes indicating the lack of any configurational
correlations.  In the limit as $N_c=L_c^D\to\infty$, $R=1$, so
correlations are systematically restored.

The cluster problem generated by the substitution $\Delta \to
\Delta_{DCA}$ may be solved numerically.  Each of the diagrams in
Fig.~\ref{fig:DCA_dis_fig1} representing $N_c$ independent scatters,
or less, remain finite; however, scattering diagrams for greater than
$N_c$ scatters vanish.  The complexity of the problem is further
reduced since the nontrivial sums involve only the cluster momenta
$\K$ ( numbering $N_c$ instead of $N$).  Furthermore, since these
diagrams are the same as those from a finite-sized periodic cluster of
$N_c$ sites, we can easily sum this series to all orders by
numerically solving the corresponding cluster problem.  The resulting
algorithm is identical to that presented in Sec.~\ref{subsub:scs},
except that: 1.  In the coarse-graining step we must calculate both
the coarse-grained correlation function
\begin{equation}
{\bar{g}^{\beta\gamma}}(\K) = \frac{N_c}{N} \sum_\kt g^{\beta\gamma}(\K+\kt).
\end{equation}
and the coarse-grained cluster Green function
\begin{equation}
{\bar{G}}(\K,z) = \frac{N_c}{N} \sum_\kt 
\frac{1}{z +\mu-\epsilon_{\K+\kt} - \Sigma_c(\K,z)}
\end{equation}
2. We then solve the cluster problem by performing a weighted average
of the cluster Green function (in matrix notation in the cluster
sites)
\begin{equation}
{\bf{G}} = \left < 
\left(\bGscript^{-1} - {\bf{V}}\right)^{-1} \right>
\end{equation}
over all disorder configurations. The weighting of each configuration is
determined by the Fourier transform of ${\bar{g}^{\beta\gamma}}(\K)$ to 
obtain the cluster configurational probabilities ${\bar{g}^{\beta\gamma}_{ij}}$.  
After convergence is reached, the irreducible cluster quantities may be 
used to calculate the properties of the lattice.  


\subsection{Alternative cluster methods}
\label{subsec:alt}

In this review we decided to focus on the three, in our view most
established cluster methods. This section reviews several other ideas
proposed to introduce non-local correlations, some of which are
complementary to the approaches discussed.

\setcounter{paragraph}{0}
\subsubsection{Self-energy functional theory} 

The self-energy-functional approach developed by
\textcite{potthoff:cluster2} (see also \onlinecite{potthoff:cluster1})
is a very general unifying concept for existing cluster approaches and
in addition provides the power to construct novel cluster algorithms.
Similar to the formalism presented in Sec.~\ref{subsec:freeeg}, this
approach views the grand potential $\Omega$ as the central quantity.
Here, the self-energy $\bSigma$ is considered the basic dynamic
variable and a self-energy functional $\Omega[\bSigma]={\rm
  Tr}\,ln(-(\bG_0^{-1}-\bSigma)^{-1})+F[\bSigma]$ is constructed from
the Legendre-transform $F[\bSigma]=\Phi[\bG[\bSigma]]-{\rm
  Tr}(\bSigma\bG[\bSigma])$ of the $\Phi[\bG]$ functional. This
approach proceeds with setting up a general variational scheme to use
dynamical information from a (numerically) solvable reference system
$H'=H_0(\t')+H_1({\bf U})$ to approximate the physics of the original
system $H=H_0(\t)+H_1({\bf U})$. While the single-particle parts $H_0$
are generally different, the interaction part $H_1({\bf U})$ is kept
fixed, to ensure that the functionals $F[\bSigma]$ of the reference
system and the original system share the same functional dependence.
It can then be shown that the grand potential of the original system
$\Omega_\t[\bSigma]$ can be evaluated exactly according to
\begin{eqnarray}
  \label{eq:alt1}
  \Omega_\t[\bSigma(\t')]=\Omega'&+&{\rm Tr\,ln}(-(\bG_0^{-1}-\bSigma(\t'))^{-1})\nonumber\\
 &-& {\rm Tr\,ln}(-\bG') 
\end{eqnarray}
from the grand potential $\Omega'$, the trial self-energy
$\bSigma(\t')$ and Green function $\bG'$ of the reference system.
Variation is then performed with respect to the single-particle
parameters $\t'$ of the reference system and the stationary point is
determined by $\partial \Omega_\t[\bSigma(\t')]/\partial \t'=0$.

\begin{figure}
\centerline{
\includegraphics*[width=3.5in]{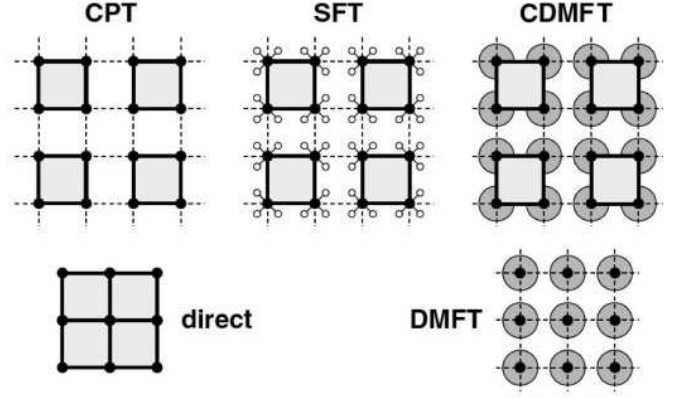}}
\caption{Taken from \textcite{potthoff:cluster1}. Various reference systems $H'$ for the 2D Hubbard model on a square lattice. The solid and dashed lines represent the intra- and inter-cluster hopping respectively, the filled circles the on-site interaction $U$. Additional $n_b$ bath sites are depicted by open circles, while the shaded circles represent $n_b=\infty$ bath sites.}
\label{fig:sft}
\end{figure}

The choice of a suitable and solvable reference system $H'$ given the
original Hamiltonian $H$ is in principle arbitrary. Within the SFT,
cluster approximations are constructed from a reference system $H'$
that represents decoupled clusters of size $N_c$ which are optionally
coupled to additional $n_b$ uncorrelated bath sites. A schematic
illustration for the Hubbard model is shown in Fig.~\ref{fig:sft}. A
set of decoupled sites as a reference system $H'$ (not shown) yields
the Hubbard-I approximation.  Non-local correlations can be included
by considering a system of decoupled clusters of size $N_c>1$.  This
approach is identical to the CPT when the intra-cluster parameters
$\t'$ are fixed to the original values $\t$ : The self-energy of the
reference system, $\bSigma(\t')={\bG'_0}^{-1}-{\bG'}^{-1}$ is
calculated once and used to approximate the Green function of the
original lattice model, $\bG^{-1}=\bG_0^{-1}-\bSigma(\t')$. This
yields the expression~(\ref{eq:cpt1}) of the CPT,
$\bG^{-1}={\bG'}^{-1}-\delta\t$, since the bare Green-functions of the
original and reference systems only differ by the inter-cluster
hopping $\delta\t$.

Alternatively within the SFT the intra-cluster hopping $\t'$ can be
treated variationally. Calculations for the 1D Hubbard model
\cite{potthoff:cluster1} however show that the SFT functional
Eq.~(\ref{eq:alt1}) is stationary for $\t'$ slightly larger but very
close to the original value $\t$, with the difference becoming smaller
as $N_c$ increases. The SFT can also be used to find the correct
boundary conditions of the CPT cluster. Calculations with fixed
intra-cluster hopping $\t'=\t$ and additional variational hopping
$t_r$ between the surface sites to simulate periodic boundary
conditions show that the self-energy functional is only stationary at
$t_r=0$ (open boundary conditions), not at $t_r=t$ (periodic boundary
conditions) \cite{potthoff:cluster1}.

One shortcoming of the CPT is its inability to describe phases with
broken symmetry due to the finite size of the clusters. The SFT can be
used to construct a variational CPT, an extension of the CPT to
account for long-range order \cite{potthoff:cluster3}. This is done by
adding a fictitious symmetry breaking single-particle term to the
reference system and subtracting it in the inter-cluster part so that
it has no effect on the original lattice Hamiltonian. The magnitude of
this term is then optimized in the SFT variational treatment.
Calculations for the 2D Hubbard model show very good agreement with
variational Monte Carlo and auxiliary field Quantum Monte Carlo
results for the ground-state energy and the staggered magnetization
\cite{potthoff:cluster3}.

Embedded cluster approximations are constructed if additional $n_b$
uncorrelated bath sites are introduced. The DMFT is obtained by
choosing as a reference system a set of decoupled sites coupled to
$n_b=\infty$ uncorrelated bath sites. The CDMFT is identical to a
system of decoupled clusters coupled to $n_b=\infty$ bath sites with
the intra-cluster parameters being fixed at the original values $\t$
and the bath parameters being treated as variational. It can then be
shown \cite{potthoff:cluster1} that the SFT self-energy functional is
stationary at the bath parameters that fulfill the CDMFT
self-consistency. A corresponding analysis of the DCA formalism within
this approach is still lacking.

Intermediate approximations can be constructed by considering a finite
number $0<n_b<\infty$ of bath sites. This approach is superior to the
CPT and inferior to the CDMFT.  
For a reference system with $N_c=1$ and $n_b=1$ this approach can be
used to study the Mott transition in the Hubbard model
\cite{potthoff:cluster4} analytically and it was shown to yield the
same qualitative picture as the full DMFT ($N_c=1$,
$n_b=\infty$) analysis (see \onlinecite{georges:dmftrev}).


\subsubsection{Fictive impurity models}
\label{subsubsec:oka}
Similar to the SFT approach discussed above, this very general
approach by \textcite{okamoto:cluster} is centered on the self-energy
$\bSigma$ as the basic dynamic variable. It is based on the idea that
the cluster model is merely an algorithm to calculate coefficients in
an orthogonal function expansion of the momentum dependence of the
electronic self-energy
\begin{equation}
  \label{eq:alt2}
  \Sigma_{\rm approx}(\k,z) = \sum_{i=0\dots n} \Phi_i(\k)\Sigma_i(z)\,.
\end{equation}
The coefficients $\Sigma_i(z)$ can then be obtained from the solution
of a $n+1$-site fictive impurity model involving $n+1$ mean-fields
which are fixed by the requirement that the impurity model Green
functions $G_i$ equal the corresponding integrals over the lattice
Green function
\begin{equation}
  \label{eq:alt3}
  G_i(z)=\sum_\k \Phi_i(\k)[G^{-1}_o(\k,z)-\Sigma_{\rm approx}(\k,z)]^{-1}\,.
\end{equation}
To include local and nearest neighbor correlations, the momentum
dependence of the self-energy may be expanded up to second order using
the orthogonal functions $\Phi_0(\k)=1$ and $\Phi_1(\k)=e^{ika}$.
Since in general the orthogonal functions $\Phi_i(\k)$ change sign
over the Brillouin zone except for the local term $i=0$, causality is
not guaranteed when the expansion is truncated at low order. However
it is shown that simple filtering of the higher-order terms may be
used to circumvent these problems \cite{okamoto:cluster}.

The DCA may be viewed as a specific example of this approach, where
the indices $i$ correspond to the centers $\K$ of the DCA cells and
the functions $\Phi_\K(\k)$ are set to 1 if $\k$ is contained in the
cell represented by $\K$ and 0 otherwise (see
Fig.~\ref{fig:Nc4_clusters}). Causality problems are thus avoided by
using orthogonal functions which are non-negative everywhere. The
resulting approximants however have discontinuities in momentum space.

\subsubsection{Non-local effects via spectral density approximation}

\textcite{laad:cluster} proposed to include non-local
$1/D$-corrections by combining DMFT with the Spectral Density
Approximation (SDA) \cite{roth:sda} and applied this approach to the
FKM (see Eq.~(\ref{eq:2dfkm})) for which it becomes particularly
simple. In the SDA, the moments of the spectral function are
determined (via repeated evaluation of commutators with the
Hamiltonian) by complicated but static correlators. For the FKM, to
order $1/D$ the SDA self-energy of the $d$ electrons,
$\Sigma_{0,d}(\k,z)$ can be expressed in terms of the static
susceptibility of the $f$-electrons.
This self-energy is purely real, but momentum dependent
and is used to approximate the bath self-energy in the hybridization
function
\begin{equation}
\Gamma(z) = \sum_{\k} \frac{t^2}{z -\epsilon_{\k} + \mu 
-\Sigma_{0,d}(\k,z)}
\end{equation}
for the effective impurity problem in $D=\infty$. The impurity self
energy, $\Sigma_{imp,d}(z)$ (see \onlinecite{brandt:fkm3})
combines with the bath self-energy $\Sigma_{0,d}(\k,z)$ to 
a dynamical, non-local self-energy of the form
\begin{equation}
 \Sigma_{d}(\k,z)= \Sigma_{imp,d}(z) + 
\Sigma_{0,d}(\k,z)- \sum_{\k} \Sigma_{0,d}(\k,z)\,.
\label{eq:laadself} 
\end{equation}
This finally determines the Green function of the mobile $d$-
electrons of the usual form, $ G_{d}(\k,z) = [z -\epsilon_{\k} + \mu
-\Sigma_{d}(\k,z)]^{-1}$,
which is used together with $\Sigma_d(\k,z)$ to estimate the
susceptibility of the $f$-electrons (see \onlinecite{laad:cluster} for
details). $\Sigma_{0,d}(\k,z)$ can then be recalculated to close the
self-consistency loop.

With this method, Laad and van den Bossche studied the DOS and the
spectral function $A(\k,\omega)$ of the FKM on a 2D square lattice.
Their results agree well with known results and other studies like
\textcite{hettler:dca2}.  On the other hand, the FKM is a particularly
gentle test bed for their method, due to its effective impurity nature
(mobile electrons in a static background).  For a model with true
dynamics like the Hubbard model, it is unclear whether the method is
even feasible (due to the additional spin-flip and pair-hopping
correlators), and the limitation of a purely real bath self-energy is
likely to be too restrictive.

\subsubsection{Non-local corrections via projection technique}

Using the projection technique Tien has developed a cluster extension
of the DMFT by taking into account both local and non-local
contributions to the dynamics within a relevant subspace of Liouville
or operator space
\cite{tran:cluster1,tran:cluster2,tran:cluster3,tran:cluster4,tran:cluster5}.
The information of a given subspace is stored in static susceptibility
and frequency functions while the effects of the remaining subspace
are collected in a dynamic memory function.
The idea of Tien's approach is
to approximate this memory function by a local quantity.
Hence, non-local
correlations are taken into account through static quantities
(susceptibilities), while dynamical correlations are approximated by a
local memory function. If the relevant subspace is spanned by
operators acting on the same or nearest-neighbor sites only, these
quantities can be calculated in an effective impurity model
\cite{tran:cluster1,tran:cluster2} or, in an improved version, in a
two-site cluster model
\cite{tran:cluster3,tran:cluster4,tran:cluster5}. 

This approach has not been rigorously proven to be causal, however its
application to the Falicov-Kimball \cite{tran:cluster1,tran:cluster3}
and Hubbard model \cite{tran:cluster4} shows that the spectral
function is positive definite and the sum-rules of the first few
moments of the spectral densities are preserved. The neglect of
non-local dynamical correlations however leads to spurious behavior at
low temperatures: A Kondo resonance emerges in the in 2D half-filled
Hubbard model at low temperatures similar to the behavior observed in
DMFT simulations but inconsistent with other cluster calculations
which show a pseudogap down to the lowest temperatures (see
Sec.~\ref{subsubsec:af}).  This shortcoming is due to the fact that
the nearest-neighbor static correlations are proportional to the
band-dispersion $\epsilon_\k$. At the Fermi energy, $\epsilon_\k=0$,
hence non-local correlations vanish and a Kondo-resonance is generated
due to the local nature of the dynamical correlations.

\subsubsection{Two-site correlations with composite operators}

A dynamical non-perturbative two-site approximation for the Hubbard
model based on the composite operator method was developed by
\textcite{mancini:cluster} and later adopted and improved by Stanescu
and Phillips \cite{phillips:cluster1,phillips:cluster2}. By using
Hubbard operators as a local basis which exactly diagonalize the
interaction part of the Hamiltonian, this approach recovers both the
weak coupling $U\ll t$ and strong coupling $U\gg t$ limits of the Hubbard
model. The memory function $\delta m(\k,\omega)$ which collects the
effects of dynamical correlations is expanded in a two-site
approximation as $\delta m(\k,\omega)=\delta m_0(\omega) + \alpha_\k
\delta m_1(\omega)$ where $\alpha_\k=(\cos k_x+\cos k_y)/2$ and
$\delta m_0$ ($\delta m_1$) are the local (nearest-neighbor)
contributions. Hence this method contains on-site and nearest-neighbor
dynamical correlations. Unknown quantities are expressed in terms of
resolvents for the eigenstates of a two-site impurity system. The
renormalization of these resolvents due to their coupling to the
surrounding of the two-site system is treated within the non-crossing
approximation.

The application of this technique to the Hubbard model shows
qualitative agreement of the single-particle spectra
\cite{mancini:cluster} in 2D with finite size QMC results and high
accuracy of specific heat results \cite{phillips:cluster1} as compared
to the Bethe ansatz solution in 1D. Although this technique includes
only on-site and nearest-neighbor correlations, it already captures
important signatures of correlations consistent with DCA/QMC
results for larger clusters such as the existence of a Mott-Hubbard gap
in the 2D Hubbard model for all values of $U>0$ at low $T$
\cite{phillips:cluster1} and the emergence of a pseudogap in the
density of states due to antiferromagnetic correlations
(\onlinecite{phillips:cluster2}; see also Sec.~\ref{subsec:2dhm}).


\section{Quantum cluster solvers}
\label{sec:qcs}

Cluster techniques map the lattice system onto a self-consistently
embedded quantum cluster model. This chapter discusses the most
promising numerical approaches used to solve this cluster problem.
After stressing some general difficulties faced by potential cluster
solvers in Sec.~\ref{subsec:gr}, we present several perturbative
techniques, including second order perturbation theory in
Sec.~\ref{subsubsec:2pt}, the fluctuation-exchange approximation in
Sec.~\ref{subsubsec:flex}, and the non-crossing approximation in
Sec.~\ref{subsubsec:nca}. Section~\ref{subsec:npt} reviews the
application of non-perturbative, (numerically) exact techniques
including Quantum Monte Carlo in Sec.~\ref{subsubsec:qmc}, Exact
Diagonalization (ED) in Sec.~\ref{subsubsec:ed} and the numerical
renormalization group in Sec.~\ref{subsubsec:rg}.

\subsection{General remarks}
\label{subsec:gr}


The fundamental difference between a finite size cluster and the
effective cluster problem of quantum cluster theories is the existence
of additional quantum mechanical bath degrees of freedoms in the
latter. The simplest realization of such a system is of course the
well-known Anderson impurity model \cite{pwa:siam}.  In general, its
ground state is a nontrivial many-body state, which is not
perturbatively connected to states of simpler starting points like a
non-interacting or a free impurity. Furthermore, the excitations also
are of many-body character and typically involve dynamically generated
low-energy scales which depend non-analytically on system parameters.
Consequently, any perturbation theory is faced with severe limitations
concerning its region of applicability and the most successful
techniques used to solve this fundamental problem of solid state
theory are non-perturbative \cite{ah:siam}.

Nevertheless, a variety of tools to approximately or numerically solve
this model have been developed over the last 25 years \cite{ah:siam}.
Since the physics of the Anderson impurity model is very well
understood \cite{ah:siam}, this knowledge can be employed to judge the
quality of results and region of applicability of these various
analytical or computational techniques at hand, a priori as well as a
posteriori. This statement does also apply to a large extent to the
DMFT, where an effective Anderson impurity model plays the central
role \cite{pruschke:dmftrev,georges:dmftrev}.

The situation becomes much more involved for quantum cluster
problems. First, from a purely technical point of view, the complexity
of the system can limit the applicability of a method on principal
grounds or even rule it out as a potential cluster solver
altogether. This aspect will be discussed for every technique in
detail in the following sections.

Similarly important is a novel complexity of the physics entering
already on the level of the bare cluster and largely complicating the
task to judge the reliability and quality of results obtained with a
certain method.  With increasing cluster size not only high-energy
states (charging energies)
-- which are the only one present in the impurity case -- will couple to the
quantum bath but also medium- and to some extent low-energy states
(e.g.\ non-local spin fluctuations). Moreover, the latter already can
and in general will contain interesting many-body effects.  The
coupling to the bath can then lead to a subtle interplay or even
competition between such intrinsic many-body effects and those
introduced by quantum fluctuations in the external bath. A
well-established example is the occurrence of a quantum
phase-transition in the two-impurity Kondo model
\cite{jones:tiam,jones:tiam2} driven by the competition between the
RKKY magnetic interaction favoring a non-local singlet and the Kondo
effect with its local singlet formation.
 
In contrast to the single impurity case where the qualitative results
to be expected from a calculation can in principle be read off the
input parameters, such a vital possibility of a plausibility check for
a given method does not exist in case of quantum cluster
models. Experience tells that even at first sight physically plausible
results need to be checked carefully. Again, the two-impurity Kondo
model can serve as a pedagogical example. The original, quite
intuitive result by \textcite{jones:tiam2} has later been reexamined
by others and found to be valid only under very special circumstances
\cite{sakai:tiam,sakai:tiam2}.  A fully satisfactory understanding of
the physics of this simplest quantum cluster model is in fact lacking
until today.  Things are further complicated for the case of cluster
mean-field theories due to the ``backflow'' of the complex local
physics via the effective bath.  An a priori understanding of the
behavior of the system is, at least given the current level of
knowledge, virtually impossible, but also an a posteriori plausibility
check is rather based on subjective physical intuition than on solid
understanding of the basic physics.

Thus, in order to obtain a reliable and consistent understanding of
the physics of correlated electron systems in the framework of cluster
mean-field theories, it is vital to employ a variety of complementary
and possibly non-perturbative tools to solve the effective quantum
cluster model. We therefore discuss in detail different computational
and analytical tools regarding their applicability and reliability.


\subsection{Perturbative techniques}
\label{subsec:pt}

The numerical effort to compute the cluster self-energy increases
rapidly with the cluster size, in principle exponentially for exact
methods. This calls for simpler methods for which the complexity of
the problem can be reduced. This can be achieved by perturbation
theory in its many variations. One example is the standard
weak-coupling perturbation theory in the electronic interaction $U$.
Second Order Perturbation Theory (SOPT) and the more elaborate
Fluctuation Exchange Approximation (FLEX) are discussed in the
following sections.  A complementary strong-coupling approach handles
the interaction exactly, but treats the coupling to the host $\Gamma$
in a perturbative expansion.  An example for this approach is the
Non-Crossing Approximation (NCA), discussed in
Sec.~\ref{subsubsec:nca}.

\subsubsection{Second order perturbation theory}
\label{subsubsec:2pt}
Second order perturbation theory has proven very useful to solve the
effective impurity problem of the DMFT. This iterated perturbation
theory (IPT) becomes exact in the weak coupling limit $U/t \ll 1$ and
coincidently at half-filling even in the strong coupling limit $U/t
\gg 1$ \cite{georges:dmftrev}. The IPT was shown to give results in
good qualitative agreement with non-perturbative DMFT results, e.g. it
captures the Mott-transition at half-filling.

The application of second order perturbation theory in the context of
quantum cluster theories follows immediately from the formulation in
Sec.~\ref{subsec:freeeg} by considering the first and second order
diagrams in the cluster functional $\Phi[\Gbar]$ and neglecting
higher-order diagrams. Here we illustrate the formalism for the
Hubbard model with local interaction $U$. For the DCA, an expression
for the cluster self-energy $\Sigma_c=\delta\Phi[\Gbar]/\delta \Gbar$
may be obtained by introducing the bare particle-hole susceptibility,
i.e. the "bubble". With the notation $K=(\K,i\omega_m),Q=(\Q,i\nu_n)$
we have
\begin{eqnarray}
\label{eq:chiph}
\chi_{ph}(Q)= - \frac{U T}{N_c}
\sum_{K} \Gbar(K+Q)\Gbar(K)\,,
\end{eqnarray}
where $T$ is the temperature and $\Gbar$ is the coarse-grained Green
function defined in Eq.~(\ref{eq:cgGDCA}). Note that $\omega_{m}$ is a
fermionic Matsubara frequency, while $\nu_{n}$ is of bosonic type. The
cluster self-energy to second order in $U$ is then given by
\begin{eqnarray}
\Sigma_c(K)= \frac{UT}{N_c}\sum_{Q}
\chi_{ph}(Q) \Gbar(K-Q) \,.
\label{eq:slf_sopt}
\end{eqnarray}
We ignored the constant (first order) Hartree term which just shifts
the chemical potential.
Physically, the self-energy accounts for the interaction of electrons
with virtually excited electron-hole fluctuations described by
$\chi_{ph}$.  Eq.~(\ref{eq:slf_sopt}) is a convolution of the
susceptibility and the Green function in both momentum and Matsubara
frequencies, and thus can be readily computed via Fast Fourier
transform (FFT).  FFTs scale only logarithmically with size, so a much
larger range of cluster sizes is available for perturbative studies of
this kind. For real frequencies, the use of FFT is less appealing,
since FFT also requires an equidistant grid of frequencies.
Compared to the usual expression for a finite size system, in the DCA
we use the coarse-grained Green function $\Gbar$ for evaluating the
susceptibility and the self-energy, so the momentum sums run over the
$N_c$ cluster momenta $\K, \Q$.
For the CDMFT, one can formulate similar expressions, but, as
discussed in Sec.~\ref{sec:qct}, the self-energy and Green function
are matrices in cluster real space.  The matrix inversions needed
during the iteration become costly as the cluster size is increased.

It must be emphasized that simple minded application of FFT to Green
functions (in particular with fermionic Matsubara frequencies) is
likely to incur large errors, due the discontinuity at zero imaginary
time that leads to the slow $1/i\omega_n$ decay of the Green functions
and self-energies.  The way to deal with this is to manually subtract
the slowly decaying parts before application of the FFT. This is
possible as the high frequency behavior can be analytically
determined. The subtraction leads to functions that decay sufficiently
fast to have a numerically well defined transform, even though one has
to cut off the Matsubara sums in
Eqs.~(\ref{eq:chiph},\ref{eq:slf_sopt}) at a finite value.  Details on
this subtraction technique can be found in \textcite{flex:dhs}.



\subsubsection{Fluctuation exchange approximation}
\label{subsubsec:flex}
FLEX \cite{bickers:89,flex:bickers2} extends the second order
perturbation theory through higher order diagrams. It rests on the
assumption that the interacting electron system
can be considered as a problem of electrons exchanging
self-consistently determined fluctuations of various kinds, i.e.
density, spin and pair fluctuations. 
In the context of quantum cluster theories, the FLEX was first applied
within the DCA by \textcite{karan:dca1}. Here we briefly review the
corresponding formalism. The generating functional $\Phi$ of the FLEX
is a sum of the three fluctuation contributions \cite{bickers:89}
\begin{equation}
\label{eq:FLEX_phi}
\Phi=\Phi_{ph}^{df}+\Phi_{ph}^{sf}+\Phi_{pp}\,,
\end{equation} 
where
\begin{eqnarray}
\label{FLEX_phi_a}
\hspace{-0.4cm}\Phi_{ph}^{df} =  -\frac{1}{2}\hspace{0.1cm}\rm{Tr}\hspace
{0.1cm}[\chi_{ph}]^2+\frac{1}{2}\hspace{0.1cm}\rm{Tr}\hspace{0.1cm}[\rm{ln}
(1+\chi_{ph})-\nonumber\\
&&\hspace{-3.cm}\chi_{ph}+\frac{1}{2}
{\chi^2}_{ph}]\,,
\end{eqnarray}
\begin{equation}
\label{eq:FLEX_phi_b}
\Phi_{ph}^{sf} = \frac{3}{2}\hspace{0.1cm}\rm{Tr}\hspace{0.1cm}[\rm{ln}
(1-\chi_{ph})+\chi_{ph}+\frac{1}{2}{\chi^2}_{ph}]\,,
\end{equation} 
\begin{equation}
\label{eq:FLEX_phi_c}
\hspace{-0.4cm}\Phi_{pp} = \rm{Tr}\hspace{0.1cm}[\rm{ln}(1+\chi_{pp})-
\chi_{pp}+\frac{1}{2}{\chi^2}_{pp}]\,,
\end{equation}
and the trace $Tr$ denotes $(T/N_c)\sum_{\Q}\sum_{n}$.  In addition to
the particle-hole bubble Eq.~(\ref{eq:chiph}), the particle-particle
susceptibility $\chi_{pp}$ also appears and is defined by
\begin{eqnarray}
\hspace*{-0.3cm}\chi_{pp}(Q)=
\frac{U T}{N_c} \sum_{K} \Gbar(Q-K)\Gbar(K)\,.
\label{eq:chipp}
\end{eqnarray}
With the generating functional of Eq.~(\ref{eq:FLEX_phi}) the cluster
self-energy can be written as
\begin{eqnarray}
\label{eq:slfpot}
\Sigma_c(K)=\frac{U T}{N_c}\sum_{Q}
\big[V^{(ph)}(Q) \Gbar(K-Q) \nonumber\\
&& \hspace*{-4cm}-  V^{(pp)}(Q) \Gbar(Q-K)\big]\,,
\end{eqnarray}  
where the so called FLEX potentials are given by
\begin{eqnarray}
\label{eq:flex-pots1}
V^{(ph)} = -\chi_{ph} + \frac{1}{2}\, \frac{\chi_{ph}}{1 + \chi_{ph}} 
+\frac{3}{2}\, \frac{\chi_{ph}}{1 - \chi_{ph}}\,,
\end{eqnarray}
and 
\begin{equation}
\label{eq:flex-pots2}
V^{(pp)} = -\chi_{pp}+ \frac{\chi_{pp}}{1 + \chi_{pp}} \,.
\end{equation}
The FLEX potentials constitute an infinite sum of fluctuation diagrams
similar to the series of density fluctuations also known as the Random
Phase Approximation, which is in fact a subset of the FLEX.  However,
the FLEX also includes significant spin and pair fluctuations.  Note
that second order perturbation theory is reproduced by expanding
Eq.~(\ref{eq:flex-pots1}) to first order in $\chi_{ph}$.

Naturally, the drawback of using these weak-coupling methods lies in
the fact that some strong coupling phenomena such as the opening of a
Mott gap at half-filling simply cannot be described by e.g. the FLEX
approach.
Also, quite often the self-consistent evaluation simply does not
converge, particularly at low temperatures. This is due to divergences
showing up in perturbatively evaluated response functions.  As a
result, the variety of systems that can be studied is limited.

To obtain real time (or frequency) data, an
analytic continuation of the Green function or self-energy is
necessary.  For the FLEX, this is achieved by means of Pad\'e
approximation \cite{vidberg}.  Real frequency results and details on
the implementation of the Pad\'e approximation can be found in
\textcite{karan:dca2}.



\subsubsection{Non-crossing approximation}
\label{subsubsec:nca}

The Non-Crossing Approximation (NCA)
\cite{Grewe:NCA,Keiter:NCA2,Keiter:NCA3,Kuramoto:NCA} was originally
developed for the single-impurity Anderson model and is based on a
diagrammatic perturbation theory \cite{Kuramoto:NCA,Keiter:NCA1}
around the atomic limit of this model.  A comprehensive and detailed
description of the NCA and its limitations can be found in the review
by \textcite{Bickers:NCA}.  The NCA has been extensively applied as an
approximate solution of the effective impurity model in the DMFT
context to study the single-band Hubbard model
\cite{Pruschke:NCA1,Pruschke:NCA2}, the three-band Hubbard model
(\onlinecite{Schmalian:NCA1,Schmalian:NCA2};
\onlinecite{Maier:NCA1,Maier:NCA2,Zoelfl:NCA}) and the Anderson
lattice model \cite{Kuramoto:NCA2,Kim:NCA}.  Within quantum cluster
theories, the NCA has been extended to solve the effective cluster
problem of the DCA by \textcite{maier:dca1}.  The aim of this section
is to convey the general concepts of the NCA method and discuss the
extensions necessary to solve the effective cluster model of quantum
cluster theories.

The perturbation theory used to construct the NCA is based on a
resolvent technique and is in general applicable to problems where a
discrete level system not representable by standard fermionic or
bosonic operators is coupled to continuous degrees of freedom
\cite{Kuramoto:NCA,keller:nca}. The discrete level system in the
context of the DCA is given by the eigenstates of the local part of
the cluster Hamiltonian, Eq.~(\ref{eq:cmdca}),
\begin{eqnarray}
  \label{eq:NCA1}
 H_{c,loc}&=&\sum_{\K\sigma}(\epsbar_\K-\mu)
c^\dagger_{\K\sigma}c^{}_{\K\sigma}
\\
 &+&
\sum_{\stackrel{\K,\K'}{\Q}}\sum_{\sigma\sigma'} \frac{\Ubar(\Q)}{2N_c}
    c^\dagger_{\K+\Q\sigma}c^\dagger_{\K'-\Q\sigma'}
    c^{}_{\K'\sigma'} c^{}_{\K\sigma}\nonumber \,.
\end{eqnarray} 
The perturbative expansion is performed with respect to the coupling
\begin{equation}
  \label{eq:NCA2}
  H_{hyb}=\sum_{\K,\kt,\sigma} [V_\K(\kt)
  c^\dagger_{\K\sigma} a^{}_{\K+\kt\sigma} + h.c.]
\end{equation}
to the continuous degrees of freedom, i.e. the auxiliary
non-interacting fermions $a^{}_{\k\sigma}$ ($a^\dagger_{\k\sigma}$).
Hence the NCA should be an especially useful approximation where the
local energy scales (in this case $\Ubar$) exceeds the magnitude of
the coupling $V$ to the host.

To apply the resolvent technique, the fermionic cluster operators
$c^{}_{\K\sigma}$, ($c^\dagger_{\K\sigma}$) are expanded in terms of
the Hubbard-operators $X_{mn}= |m\rangle \langle n|$
\begin{equation}
   \label{eq:NCA3}
   c^{}_{\K\sigma}=\sum_{m,n}C_{\K\sigma}^{mn}X_{mn}\quad,\quad c^\dagger_{\K\sigma}=\sum_{m,n} {C_{\K\sigma}^{mn}}^* X_{nm}
\end{equation}
where $\{ |m\rangle\}$ are the eigenstates of the local Hamiltonian,
Eq.~(\ref{eq:NCA1}),
\begin{equation}
  \label{eq:NCA4}
  H_{c,loc}=\sum_m E_m X_{mm}
\end{equation}
with eigenenergies $E_m$ and $C_{\K\sigma}^{mn}=\langle
m|c_{\K\sigma}|n\rangle$. In this representation, the hybridization
term Eq.~(\ref{eq:NCA2}) becomes
\begin{equation}
  \label{eq:NCA5}
  H_{hyb}=\sum_{\K,\kt,\sigma}\sum_{m,n} [V_\K(\kt)
  {C_{\K\sigma}^{mn}}^* X_{nm} a_{\K+\kt\sigma}+h.c.]\,.
\end{equation}
Since the Hubbard operators $X_{mn}$ do not obey standard fermionic
commutation relations, the conventional Feynman diagram technique
cannot be used for a perturbational expansion and the concept of
resolvents must be introduced instead \cite{Kuramoto:NCA}. The
matrix-elements of these resolvents in the space of the local
eigenstates have the form
\begin{equation}
  \label{eq:NCA6}
  R_{mn}^{-1}(z)=(z-E_m)\delta_{mn}-\Sigma^R_{mn}(z)\,, 
\end{equation}
where the resolvent self-energy $\Sigma^R(z)$ collects renormalization
effects of the individual molecular states $\{|m\rangle\}$ due to the
hybridization Eq.~(\ref{eq:NCA5}) to the host. We note that
$\Sigma^R(z)$ and hence $R(z)$ reduce to diagonal matrices if the
hybridization term $H_{hyb}$ commutes with the local part $H_{c,loc}$
of the cluster Hamiltonian. While this is the case in the paramagnetic
state we stress that the non-diagonal elements become essential in the
symmetry-broken phases \cite{maier:dca2,maier:dca0}.

\begin{figure}[htb]
\centerline{
\includegraphics*[width=2.5in]{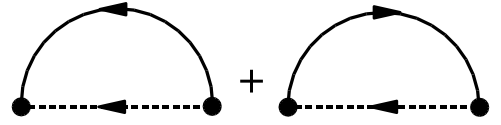}}
\caption{Non-crossing diagrams for the resolvent self-energy $\Sigma^R$. The
dashed line stands for the dressed resolvent $R$ defined in
Eq.~(\ref{eq:NCA6}), the full line for the host electrons.}
\label{fig:NCA}
\end{figure}

Keeping only the lowest order, self-consistent diagrams for the
resolvent self-energy $\Sigma^R$ illustrated in Fig.~\ref{fig:NCA}
defines the NCA. Since the hybridization with the host degrees of
freedom enters the expressions only in the form of a density of states
\begin{eqnarray}
  \label{eq:NCA7}
  \frac{1}{N}\sum_{\kt}|V_\K(\kt)|^2 
  \delta(\omega-\lambda_{\K+\kt})
 &=&-\frac{1}{\pi}\Im
  m \Gamma_c(\K,\omega)\,,
\end{eqnarray}
there is no need to explicitly calculate the auxiliary host parameters
$V_\K(\kt)$ and $\lambda_\k$ of the cluster Hamiltonian,
Eq.~(\ref{eq:cmdca}). To ensure that the solution of the cluster
problem is also the solution of the coarse-grained lattice problem,
$\Gamma_c(\K,z)$ is replaced with the hybridization function
$\Gamma(\K,z)$ (see Eq.~(\ref{eq:Gscdca})) determined from
coarse-graining the lattice Green function.  After observing energy,
momentum and spin conservation, the analytical expression for the
resolvent self-energy within the NCA reads
\begin{eqnarray}
  \label{eq:NCA8}
  &&\Sigma^R_{mn}(z)=-\frac{1}{\pi}\sum_{\K,\sigma}\sum_{l,l'}\times\\
 &&\left[ 
  \int\limits_{-\infty}^{+\infty} d\varepsilon f(\varepsilon) \Im m
  \Gamma(\K,\varepsilon )C_{\K\sigma}^{ml} R_{ll'}(z+\varepsilon)
  {C_{\K\sigma}^{nl'}}^* \right.\nonumber\\
  &&+ \left. \int\limits_{-\infty}^{+\infty}d\varepsilon f(-\varepsilon) 
  \Im m \Gamma(\K,\varepsilon) {C_{\K\sigma}^{lm}}^* R_{ll'}(z-\varepsilon) C_{\K\sigma}^{l'n}\right]\nonumber\,.
\end{eqnarray}
The coupled equations (\ref{eq:NCA6}) and (\ref{eq:NCA8}) are solved
self-consistently to compute the matrix-elements of the resolvent
${\bf R}(z)$. By using the expansion Eq.~(\ref{eq:NCA3}), the cluster Green
function $G_c(\K,z)=\langle\langle c^{}_{\K\sigma};
c^\dagger_{\K\sigma} \rangle\rangle_z$ is written in terms of Hubbard
operators
\begin{equation}
  \label{eq:NCA9}
  G_c(\K,z)=\sum_{m,n,m',n'}C_{\K\sigma}^{mm'}{C_{\K\sigma}^{nn'}}^*
  \langle\langle X_{mm'}; X_{n'n}  \rangle\rangle_z\,.
\end{equation}
Within the NCA, the correlation function on the right hand side
reduces to a convolution between resolvents
\begin{eqnarray}
  \label{eq:NCA10}
  \langle\langle X_{mm'}; X_{n'n}  \rangle\rangle_z &=& \frac{1}{Z_c}
  \int\limits_{-\infty}^{+\infty} d\varepsilon e^{-\beta\varepsilon}
  [\rho_{nm}(\varepsilon)
  R_{m'n'}(\varepsilon+\omega)\nonumber\\
  &&-\rho_{m'n'}(\varepsilon) R^*_{nm}(\varepsilon-\omega)]\,,
\end{eqnarray}
where $\rho_{nm}(\omega)=-1/\pi\Im m R_{nm}(\omega+i\delta)$ is the
spectral density of the resolvents,
$Z_c=\sum_n\int_{-\infty}^{+\infty}d\varepsilon e^{-\beta\varepsilon}
\rho_{nn}(\varepsilon)$ the cluster partition function and $\beta=1/T$
the inverse temperature.

Although the application of the NCA was illustrated for the DCA
cluster model, the NCA can in
principle also be applied to solve the CDMFT cluster problem,
Eq.~(\ref{eq:cmcdmft}).  The fact that in the CDMFT the hybridization
function $\Gamma$ is a matrix in the $N_c$ cluster sites however
complicates this task considerably.
 
One of the great advantages of this resolvent perturbation theory and
the NCA as approximation is that it can be solved for real
frequencies, i.e.\ a cumbersome and problematic analytic continuation
as in the case of QMC simulations is not necessary.  Moreover, one can
show analytically that it can produce a low-energy scale depending
non-analytically on the bare model parameters \cite{MuHa:NCA} and that
certain general features like universality in this energy scale are
present, too \cite{Fischer:NCA}.  Surprisingly, the NCA is capable of
recovering a variety of complex features in thermodynamic and dynamic
quantities of the single impurity Anderson model at $U=\infty$
\cite{Bickers:NCA}. However it was realized rather quickly from the
numerical solution of the NCA equations \cite{Grewe:NCA,Kuramoto:NCA}
as well as an approximate analytical treatment at $T=0$
\cite{MuHa:NCA} that the NCA shows serious deficiencies in its low
temperature/low energy behavior.  More precisely, it generates
unphysical power-law structures in the low-energy behavior of physical
quantities \cite{MuHa:NCA}.  Nevertheless there exists a finite window
from high down to temperatures of the order of an effective Fermi
liquid scale where the NCA produces reliable and physically sensible
results \cite{Bickers:NCA,CoxGrewe,Fischer:NCA,Pruschke:NCA2}.
 
A rather different question is whether these results are only
qualitatively or even quantitatively correct. Here, the answer
critically depends on the system under consideration. Going beyond a
single Anderson impurity at $U=\infty$ introduces the necessity to
include diagrams beyond the lowest order NCA
\cite{BCW:NCA,Pruschke:NCA} to at least ensure a correct reproduction
of energy scales or exactly known limits (like for instance uncoupled
sites in a cluster calculation). An inclusion of these so-called
vertex corrections to the bare NCA defined by Eq.~(\ref{eq:NCA8}) has
up to now been possible only for the single impurity model with finite
$U$ \cite{Pruschke:NCA}, because the complexity and number of the
additional diagrams rather fast exceeds all practical limits
\cite{Heindl:NCA}.
 
Neglecting vertex corrections in practice means that the magnitude of
a possible Fermi liquid scale will be grossly underestimated.
However, in the case of strong non-local fluctuations, as one would
expect in 2D problems, such a scale will in general be extremely small
or even zero anyway.  Here, the influence of vertex corrections can be
expected to be less severe and the NCA even quantitatively reliable.
 
Even with these restrictions, the NCA still becomes formidable as the
cluster size $N_c$ increases. The number of eigenstates and hence
number of coupled equations (\ref{eq:NCA6}, \ref{eq:NCA8}) that have
to be solved self-consistently grows like $n_{imp}^{N_c}$, where
$n_{imp}$ is the number of Fock states of the isolated impurity (e.g.
$n_{imp}=4$ in the Hubbard model, $n_{imp}=3$ in the t-J model).
Thus, although the study of larger clusters is in principle possible,
the NCA technique is very limited in cluster size and so far has only
been applied to $N_c=4$ size clusters.

\subsection{Non-perturbative techniques}
\label{subsec:npt}

Non-perturbative techniques solve the effective cluster problem
(numerically) exact. This advantage comes at the expense that only
small cluster sizes can be treated. The size restriction on the
cluster QMC technique described in the following section is basically
the same as in the case of FSS QMC. For the ED and NRG approaches
discussed in the next sections, however, a Hamiltonian for both the
cluster and host degrees of freedom needs to be explicitly simulated,
further limiting the usefulness of these methods.

\subsubsection{Quantum Monte Carlo}
\label{subsubsec:qmc}

\paragraph{Introduction.}
Quantum Monte Carlo (QMC) is a powerful and general technique for
quantum cluster problems.  QMC has several advantages including the
ability to treat relatively large clusters, the simplicity of the
required code, and the fact that only the cluster excluded Green
function $\Gscript$ and the coarse-grained interaction $\Ubar$ are
required as inputs.  QMC is also numerically exact with small and
controllable sources of systematic and statistical error.  Its
disadvantages include the fact that there is a minus sign problem
which is unpredictable, difficulties in calculating real-frequency
results, and the numerical expense of the approach.

The QMC algorithm for clusters is based on the Hirsch and Fye
algorithm which was developed to simulate the Anderson impurity
problem \cite{hirsch:QMC}.  It was later generalized to solve the DMFT
and DCA effective impurity problems by \textcite{jarrell:92} and
\textcite{jarrell:dca3}, respectively.  Although the algorithm is
formulated using a path integral in imaginary time, the Maximum
Entropy Method may be used to analytically continue the QMC data to
obtain real-frequency spectra \cite{jarrell:MEM}.

The QMC algorithm uses the action of the effective cluster model,
Eq.~(\ref{eq:19}) as a starting point and therefore can be equally
applied within the CDMFT and the DCA. It thus requires as inputs the
initial bare Green function $\Gscript$, and the form of the
coarse-grained interaction\footnote{It is also possible to treat the
  interaction in a self-consistent manner, as was described in
  \cite{hettler:dca2}} $\Ubar$.  This is an advantage, since due to
the required self-consistency of embedded cluster techniques, we
generally do not know the auxiliary parameters of the cluster
Hamiltonian.

Several steps are required to evaluate the path integrals of this
action using QMC.  We first introduce Hubbard-Stratonovich (HS) fields
required to decouple the interaction, transforming a problem of
interacting electrons and bosons to one of non-interacting particles
coupled to time-dependent HS fields.  The fermionic and bosonic fields
are then integrated out, and the integrals over the decoupling
fields are performed with a Monte Carlo algorithm.  Measurements of
any diagrammatic quantity are accomplished by decomposing the
associated operators using Wick's theorem (both connected and
disconnected contractions must be included) and then averaging the
result over the Monte Carlo generated field configurations.

\paragraph{QMC for the simple Hubbard Model.}
As a specific example, we consider the simple Hubbard model with a
local interaction of strength $U$.  Since this interaction is local, it
is unaffected by coarse-graining, so $\Ubar=U$.  To approximate the
time-integrals in the action, we introduce a discrete time grid of
length $L$ and time step $\Delta\tau = \beta/L$.  The interacting part
of the action is then decoupled by mapping it to an auxiliary Ising
field via a discrete Hirsch-Hubbard-Stratonovich (HHS) transformation
\cite{hirsch:HHS} ,
\begin{eqnarray}
\nonumber
&&e^{-\Delta\tau U\sum_{i}(n_{i \uparrow}-1/2)(n_{i \downarrow}-1/2) }\\
&&= \frac{1}{2} e^{-\Delta\tau U/4} 
\prod_{i} \sum_{s_{\bf i}=\pm 1} e^{\alpha s_{\bf i} (n_{i \uparrow} - n_{i \downarrow})}\,,
\end{eqnarray}
where $\cosh(\alpha)=e^{\Delta\tau U/2}$ and the index $i$ denotes the
cluster sites.  With this change, the cluster action takes the form
\begin{eqnarray}
S_c[\gamma,\gamma^*] &=&  \sum_{i,l;i',l',\sigma}
\gamma_{i,l,\sigma}^* \Gscript^{-1}(i,l;i',l')\gamma_{i',l',\sigma} \\
            &-& \sum_{{\bf{i}}=1}^{N_c}  \sum_{l=1}^{N_l} \sum_\sigma 
\alpha  \gamma_{{\bf{i}},l,\sigma}^{*} \sigma s_{{\bf{i}},l} \gamma_{{\bf{i}},l-1,\sigma}
\label{SC}
\end{eqnarray}
where $l$ denotes the time slice $\tau_l$ and
$\Gscript(i,l;i',l')\equiv\Gscript_{ii'}(\tau_l-\tau_{l'})$ the
cluster excluded Green function defined in Eq.~(\ref{eq:cgDyson}).
Now we integrate out the remaining cluster Grassmann variables.  The
partition function then becomes
\begin{eqnarray}
Z&\propto&\int {\cal D}[\gamma]{\cal D}[\gamma^*]e^{-S_c[\gamma,\gamma^*]}\nonumber\\ 
&=&Tr_{\{s_{i,l}\}} \prod_{\sigma} 
\det \left(\Gc_{\sigma; s_{i,l}} \right)^{-1} 
\label{Z3}
\end{eqnarray}
where factors which are fixed during the QMC process have been
ignored.  $\left( \Gc_{\sigma; s_{i l}} \right)^{-1}$ is the inverse
cluster Green function matrix with elements
\begin{equation}
\left(\Gc_{\sigma; s_{i l}}\right)^{-1}_{i,j,l,l'}
= \Gscript^{-1}_{i,j,l,l'}
-\delta_{i,j} \delta_{l',l-1}
\alpha \sigma s_{i,l}\,.
\label{ithG}
\end{equation}

A Monte Carlo algorithm is used to perform the remaining integral over
the HHS fields.  The Markov process in this algorithm proceeds by
suggesting local changes of the HHS fields at one point in space-time.
These changes are accepted or rejected according to the change in
their Boltzmann weight, the argument of the trace in Eq.~(\ref{Z3}).
If the change is accepted, the Green function must be updated
accordingly.  Several approximations, i.e. changes to these equations
that are beyond linear order in $\Delta\tau$, are necessary to obtain
an efficient algorithm.  First, we re-exponentiate the first term on
the right-hand side of Eq.~(\ref{ithG}), obtaining in a simple matrix
notation in space-time
\begin{eqnarray}
\Gc_\sigma^{-1} = \Gscript^{-1}- T \left( e^{{V}_{\sigma}}-1\right) \;,
\label{Grexp}
\end{eqnarray}
where $T$ is $\delta_{i,j}\delta_{l-1,l'}$ and ${V}_\sigma(i,l) \equiv \alpha s_{i,l}\sigma$.
Note that the term in the parenthesis is beyond zeroth order in $\Delta\tau$.  Therefore, to first 
order in $\Delta\tau$, we may write
\begin{eqnarray}
\Gc_\sigma^{-1} = \Gscript^{-1}+  
\left( \Gscript^{-1} - 1 \right) \left( e^{{V}_{\sigma}}-1\right) \;,
\label{Grexp1}
\end{eqnarray}
since $\Gscript^{-1} - 1 = -T +{\cal{O}}(\Delta\tau)$.  Therefore, the
inverse Green functions for two different field configurations,
$\{s_{{\bf{i}} l}\}$ and $\{s_{{\bf{i}} l}'\}$, are related by
\begin{equation}
{\Gc_\sigma'}^{-1} e^{-V'_\sigma}=\Gc_\sigma^{-1} e^{-{V}_{\sigma}}
- e^{-V_{\sigma}}
+ e^{-V'_{\sigma}} \,.
\end{equation}
Or, after multiplying by $e^{V'_\sigma}$, and collecting terms
\begin{eqnarray}
         {\Gc_\sigma'}^{-1}-\Gc_\sigma^{-1} =
         (\Gc_\sigma^{-1}-1)e^{-{V}_{\sigma}}(e^{V'_{\sigma}}-e^{V_\sigma}) \,.
\end{eqnarray}
Multiplying from the left by $\Gc$ and from the right by $\Gc'$, we find
\begin{equation}
\Gc_\sigma'=\Gc_\sigma+
(\Gc_\sigma-1)(e^{V'_{\sigma} - {V}_{\sigma}}-1)\Gc_\sigma'
\label{GGprime1}
\end{equation}
or
\begin{equation}
\Gc_\sigma{\Gc_\sigma'}^{-1}=
1+(1-\Gc_\sigma)(e^{V'_{\sigma} - V_{\sigma} }-1) \,.
\label{GGprime2}
\end{equation}
The QMC algorithm proposes changes in the Hubbard-Stratonovich field
configuration ${\{s_{i,l}\}} \to {\{s'_{i,l}\}}$, and
accepts these changes with the transition probability $P_{s\to s'}$.
Thus, to implement the algorithm, we need $P_{s\to s'}$ and a relation
between the cluster Green functions $\Gc$ and $\Gc'$ for the two
different auxiliary field configurations.  To simplify the notation,
we introduce a combined space-time index $i=(i,l)$, and
consider only local changes in the fields $s_m\to s'_m=-s_m$.  As
can be inferred from Eq.~(\ref{Z3}), the probability of a
configuration $\{s_i\}$ is $P_{s} \propto
\det(\Gc_{\uparrow\{s_i\}}^{-1})\det(\Gc_{\downarrow\{s_i\}}^{-1})$;
\footnote{If $P_{s}$ is not positive definite, then $\left|
    P_{s}\right| $ is used as the sampling weight, and its sign is
  appended to the measurement.  I.e.\ for a measurement $m$ and sign
  S, $\left<m\right>_P= \left<mS\right>_{|P|}/\left<S\right>_{|P|}$.}
on the other hand detailed balance requires $P_{s'}P_{s'\to
  s}=P_{s}P_{s\to s'}$ for all $s'$.  We may satisfy this requirement
either by defining the transition probability $P_{s' \to s}=R/(1+R)$,
where
\begin{equation}
R\equiv \frac{P_{s}}{P_{s'}}=
\frac{ \det(\Gc_{\uparrow}')\det(\Gc_{\downarrow}') }
{ \det(\Gc_{\uparrow})\det(\Gc_{\downarrow}) } 
\end{equation}
is the relative weight of two configurations, or by setting $P_{s' \to
  s}= \mbox{minimum}(R,1)$.  The first choice is called the ``heat
bath'' algorithm, and the second the ``Metropolis'' algorithm which is
generally, but not always, superior.  If the difference between two
configuration is due to a flip of a single Hubbard Stratonovich field
at the $m$th location in the cluster space-time \cite{hirsch:QMC}, we
obtain from Eq.~(\ref{GGprime2})
\begin{eqnarray}
R = \prod_{\sigma}[1 + (1-\Gc_{\sigma, mm})(e^{-\alpha\sigma (s_m-s_m')} -1)] ^{-1}\,.
\end{eqnarray}
For either the Metropolis or the heat bath algorithm, if the change is
accepted, we must update the Green function accordingly. The
relationship between $\Gc$ and $\Gc'$ is given by
Eq.~(\ref{GGprime1})
\begin{eqnarray}
\Gc_{\sigma, ij}'&=&\Gc_{\sigma, ij}\nonumber \nonumber \\
                  &+&\frac{(\Gc_{\sigma, i m}-\delta_{im}) 
         (e^{-\alpha\sigma (s_m-s'_m)}-1)}{1+(1-\Gc_{\sigma, mm})
                (e^{-\alpha\sigma (s_m-s'_m)}-1)}\Gc_{\sigma, mj}\,.
\label{updateG}
\end{eqnarray}

The QMC procedure is initialized by setting $\Gc_{\sigma, i j }={\cal
  G}_{i j}$, and choosing the corresponding field configuration with
all $s_i=0$.  Then we use Eq.~(\ref{updateG}) to create a Green
function corresponding to a meaningful field configuration (i.e.\ 
$s_i=\pm 1$, for each $i=(i,l)$ or the $\{s_i\}$ from a
previous run or iteration) and
proceed by stepping through the space-time of the cluster, proposing
local changes $s_i\to-s_i$.  We accept the change if $P_{s'\to s}$ is
greater than a random number between zero and one and update the Green
function according to Eq.~(\ref{updateG}).  After roughly one hundred
warm-up sweeps through the space-time lattice of the cluster, the
system generally comes into equilibrium and we begin to make
measurements.

\paragraph{Measurements.}  Several points must be considered when making
measurements in the QMC procedure.  First, for a given configuration
of the HHS fields, the problem is non-interacting.  Thus, the
estimators of any correlation function may be constructed by taking
all allowed Wick's contractions (both connected and disconnected).
Therefore, any quantity which may be represented in terms of the Green
functions (and perhaps the HHS fields themselves), may be measured.
Second, great care must be taken when constructing the estimators of
the measurements.  Different estimators may yield different results
due the systematic and statistical error in the QMC procedure.  It is
important to choose the optimal form of the estimator of each
measurement, and then use all the prior knowledge (exact limits,
symmetries, etc.) that we have to reduce the error.

For example, one difficulty encountered with the QMC algorithm is that
a reliable transform from imaginary-time quantities to Matsubara
frequencies is required.  A careful treatment of the frequency
summation or the imaginary-time integration is crucial in order to
ensure the accuracy and the stability of the algorithm and to maintain
the correct high-frequency behavior of the Green function. We need to
evaluate the following integral
\begin{eqnarray}
G_c({\bf K},i\omega_n) = 
\int_{0}^{\beta} d\tau e^{i\omega_n \tau}G_c({\bf K},\tau)  \quad\mbox{.} 
\label{FTG}
\end{eqnarray}
But from the QMC, we know the function $G_c({\bf K}, \tau)$ only at a
discrete subset of the interval $[0,\beta]$. As may be readily seen by
discretizing the above equation, the estimation of $G_c({\bf K},
i\omega_n)$ becomes inaccurate at high-frequencies. This is formalized
by Nyquist's theorem which tells us that above the frequency
$\omega_c= \frac{\pi}{\Delta \tau}$ unpredictable results are produced
by conventional quadrature techniques.  For example, a rectangular
approximation to the integral in Eq.~(\ref{FTG}) yields a
$G_c(\K,i\omega_n)$ that is periodic in $\omega_n$. This presents a
difficulty since causality requires that
\begin{eqnarray}
\lim_{\omega_n \rightarrow \infty} G_c(\K,i\omega_n) = \frac {1}{i\omega_n}\,.
\label{limit}
\end{eqnarray}
 
This problem may be avoided by using the high frequency information
from other sources.  For example, the second-order perturbation theory
result, Eq.~(\ref{eq:slf_sopt}) has the correct asymptotic behavior.
Alternatively, we can use the Green function from the previous
iteration which has the correct high frequency behavior.  We then
compute the Matsubara-frequency Green function from the imaginary-time
QMC Green function as \cite{jarrell:qmc}
\begin{eqnarray}
G_c({\bf K},i\omega_n) &=&  G_{c,pt}({\bf K},i\omega_n) + \\
                       & & \int_{0}^{\beta} d\tau e^{i\omega_n \tau}
(G_c({\bf K},\tau)-G_{c,pt}({\bf K},\tau)) \quad\mbox{.} \nonumber 
\end{eqnarray}
where $G_{c,pt}$ is any 
Green function 
with the correct high-frequency behavior. The integral is computed by
first interpolating the difference $G_c({\bf K},\tau)-G_{c,pt}({\bf
  K},\tau)$ using an Akima spline, and then integrating the spline (a
technique often called oversampling). The smooth Akima spline
suppresses spurious high frequency behavior so that $G_c({\bf
  K},i\omega_n) = G_{c,pt}({\bf K},i\omega_n)$ when $\omega_n \gg
\omega_c$.  The resulting self-energy, extracted from
$\Sigma_c(\K,i\omega_n) = 1/\Gscript({\bf K},i\omega_n)-1/G_c({\bf
  K},i\omega_n)$ is still not accurate at high frequencies since it
reflects either the perturbation theory or the previous iteration.
However, the exact high frequency form is known.  Thus, we can fit the
high frequency self energy, at frequencies at or below the Nyquist
cutoff $\pi/(\Delta\tau)$, to the form
\begin{equation}
\Sigma_c(\K,i\omega_n) \approx a(\K)/(i\omega_n) + b(\K)/\omega_n^2\,.
\end{equation}
In some cases $a$ and $b$ are known exactly.  We may then append this
form onto the self-energy at frequencies higher than
$\pi/(\Delta\tau)$.




\subsubsection{Exact diagonalization}
\label{subsubsec:ed}
The Lanczos ED method \cite{haydock} provides a numerical way to
diagonalize the Hamiltonian of a finite size system. Since the cluster
model in the CPT formalism is identical to the model of a finite size
system with open boundary conditions, the regular Lanczos ED method
can be applied in this case without modification
\cite{senechal:cluster2}.  For the cluster model of embedded cluster
theories however, it needs to be generalized to account for the
self-consistent coupling to the host. ED has been used to solve the
impurity problem of the DMFT \cite{caffarel,si:ed}.  It is natural
that it can be extended to be used as a cluster solver in quantum
cluster theories.  This was demonstrated by \textcite{bolech:cluster}
in an application of the CDMFT to the 1D Hubbard model.  The general
method of exact diagonalization \cite{haydock} need not be reviewed
here. We only outline the specific implementation necessary for a
cluster theory.

ED is a wave function based method, i.e. it is applied to diagonalize
the effective cluster Hamiltonian, e.g. the Hamiltonian in
Eq.~(\ref{eq:cmcdmft}). To this end, the hybridization function
$\bGamma(z)$ (see Eq.~(\ref{eq:41})) is fitted to the form
\begin{equation}
 \Gamma_{ij}(z) = \sum_{lm,\kt}V_{il}(\kt)[z\mathbb{1}-{\bf \lambda}(\kt)]^{-1}_{lm}V^*_{mj}(\kt)\,,
\label{eq:eddelta}
\end{equation}
to obtain estimates for the auxiliary parameters of the host,
$\lambda_{ij}(\kt)$ and $V_{ij}(\kt)$. In order to apply ED to the
cluster Hamiltonian, the auxiliary host degrees of freedom, i.e. the
sum over $\kt$, must be discretized to a finite set of $N_h$ orbitals.
Applying ED to the resulting Hamiltonian, one can then compute a
cluster self-energy $\bSigma^c$ \cite{caffarel}, and from the cluster
self-energy a new estimate of the coarse grained Green function
$\bGbar$ to close the self-consistency loop.

The Hilbert space of the resulting Hamiltonian increases exponentially
with the cluster size $N_c$ \it and \rm the number of wave-vectors
$N_h$ representing the host. For ED to be feasible, the total number
of "sites" $N=N_c + N_h$ must be of the order of $N \sim 20$.
Furthermore, for the distinction of the lattice into cluster and host
to make sense, $N_h \ge N_c$.  This size limitation is possibly
prohibitive for anything but 1D systems. Moreover, by applying ED to
solve the cluster model one abandons the thermodynamic limit, i.e.
one of the advantages of cluster mean-field theories. Still, ED is an
exact method and can deal with more complicated interactions than the
simple on-site Hubbard repulsion, as exemplified in
\textcite{bolech:cluster}, where nearest neighbor interactions were
considered.


\label{subsec:perspectives}
%
%
%
\subsubsection{Wilson's numerical renormalization group}
\label{subsubsec:rg}
%
Over the past ten years, Wilson's numerical renormalization group
(NRG) \cite{nrg:wilson,nrg:krish1,nrg:krish2} has become a major
computational tool to study quantum impurity problems.  Its advantages
are that it is (i) non-perturbative, (ii) can handle exponentially
small energy scales with unprecedented accuracy and (iii) does not
suffer from any principle limitations regarding the parameter space of
the model.  In addition, through the inspection of the flow of the
spectrum of the Hamiltonian under the renormalization group
transformation, the method provides direct very detailed information
about the structure of the low-energy spectrum.

Originally, its application was limited to static quantities
\cite{nrg:wilson,nrg:krish1,nrg:krish2}, but the calculation of
dynamical properties is also possible (for a comprehensive overview of
the early work see e.g.~\onlinecite{ah:siam}) and thus its application
to the DMFT impurity problem \cite{dmftnrg:sakai,dmftnrg:shimizu}. 
The direct calculation of the one-particle self-energy from the NRG
\cite{nrg:bulla1}
triggered a variety of applications within the  DMFT framework
\cite{dmftnrg:bulla1,dmftnrg:pruschke1,dmftnrg:bulla2,dmftnrg:zitzler1,dmftnrg:zitzler2,dmftnrg:zitzler3}
at both $T=0$ and finite temperatures.

\begin{figure}[htb]
\centerline{%
\includegraphics*[width=2.5in,clip]{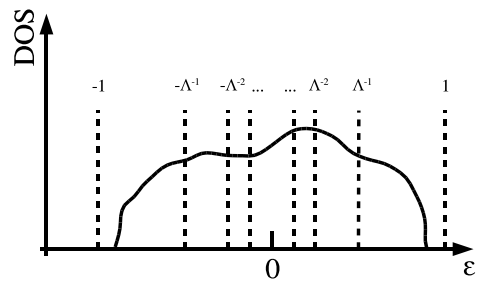}}
\caption{Schematic setup of the NRG. The energy axis is partitioned into
exponentially decreasing intervals.}
\label{fig:NRGsetup}
\end{figure}
Given this tremendous success, the obvious question is if one can
extend the NRG to become useful for quantum cluster problems. Let us
therefore briefly repeat the basic steps in the NRG scheme. The
procedure starts by picking a number $\Lambda>1$ and an interval on
the energy axis that contains the support of the density of states of
the bath degrees of freedom. As depicted in Fig.~\ref{fig:NRGsetup}
this interval is then partitioned into exponentially decreasing
intervals $]\Lambda^{-(n+1)},\Lambda^{-n}]$ for $\epsilon>0$ with
$n=0,1,\ldots$; likewise for $\epsilon<0$. Within each interval, a
Fourier decomposition of the bath operators is performed; the
fundamental approximation of the NRG consists of neglecting all
Fourier components except for the homogeneous one. 
After tri-diagonalization, the resulting Hamiltonian reads
\begin{eqnarray}
H_{\rm QI}&\approx&
H_{\rm Imp}+\label{equ:NRGtrid}\\ 
&\displaystyle+&\sum\limits_{\sigma}\sum\limits_{n=1}^\infty
\left(\varepsilon_n a^\dagger_{n\sigma}a^{\phantom{\dagger}}_{n\sigma}
+t_{n-1} a^\dagger_{n-1\sigma}a^{\phantom{\dagger}}_{n\sigma}+\mbox{h.c.}\right)\nonumber
\end{eqnarray}
where $H_{\rm Imp}$ denotes the Hamiltonian of the ``impurity'' and
the operators $a^{(\dagger)}_{n\sigma}$ represent the bath degrees of
freedom.
Note that the Hamiltonian
(\ref{equ:NRGtrid}) represents a semi-infinite chain with the impurity
at its left end. 
\textcite{nrg:wilson} showed that the quantities $t_n\propto\Lambda^{-n/2}$. If
$\bar t_n=\sqrt{\Lambda^n}t_n$ and
$\bar\varepsilon_n=\sqrt{\Lambda^n}\varepsilon_n$, one can cast the
Hamiltonian into a recurrency form
\begin{equation}\label{equ:NRGrec}
\bar H_{N+1}=\begin{array}[t]{l}\displaystyle
\sqrt{\Lambda}\bar H_N+
\sum\limits_\sigma\bar\varepsilon_{N+1} a^\dagger_{N+1\sigma}a^{\phantom{\dagger}}_{N+1\sigma}\\[5mm]
\displaystyle
+\sum\limits_\sigma\left(\bar t_{N} a^\dagger_{N\sigma}a^{\phantom{\dagger}}_{N+1\sigma}+\mbox{h.c.}\right)\,.
\end{array}
\end{equation}
Given the eigenvalues and eigenvectors of $\bar H_N$, $\bar H_{N+1}$
can be constructed straightforwardly. Since at each of these steps the
dimension of the Hilbert space increases by a factor of four, the
practical use would be limited even more severely as conventional
exact diagonalization, because the construction of $\bar H_{N+1}$
requires the knowledge of the {\em whole} set of eigenvectors and
eigenvalues of $\bar H_{N}$.  However, the multiplication of $\bar
H_{N}$ with $\sqrt{\Lambda}>1$ expands the bandwidth of the spectrum
and, because $\bar t_{N}={\rm O}(1)$, the low energy properties of
$\bar H_{N+1}$ are determined by a restricted set of low-lying states
of $\bar H_{N}$ only. This observation is put into a practical
computational scheme by the following algorithm:
(i) At step $N\ge0$ ($N=0$ corresponds to
$H_{\rm Imp}$), diagonalize $\bar H_{N}$ and calculate all interesting
local properties for that particular chain length. (ii) Use a suitable
number $N_{\rm NRG}$ of the lowest eigenstates of $\bar H_{N}$ to
construct the next Hamiltonian $\bar H_{N+1}$ according to
(\ref{equ:NRGrec}). (iii) Continue with (i) until the desired accuracy
for the ground state is reached.  Note that step (ii) ensures that, no
matter how long the chain becomes, the dimension of the Hamilton
matrix to diagonalize can be fixed to a manageable number.

As was discussed extensively in the literature
\cite{nrg:wilson,nrg:krish1,nrg:krish2,ah:siam}, the knowledge of the
local properties at chain length $N$ provides a means to calculate
physical interesting quantities (thermodynamics and dynamics) at
energy respectively temperature scales $\propto\sqrt{\Lambda^{-N}}$.
This exponential decrease of successive energy scales explains why the
NRG is suitable for studies of problems with extremely small dynamical
energy scales.
This however comes at the price of a loss of accuracy in high energy 
features \cite{ah:siam}.

Up to now the discussion was restricted to spin degeneracy only.
However, it is obvious that additional degrees of freedom do not
influence the general lines of argument. In fact, for 
a problem with $L\ge1$ internal degrees of freedom in addition
to the spin (like cluster sites or several orbitals per site), 
the result (\ref{equ:NRGtrid}) acquires the form
\begin{eqnarray}
H_{\rm QI}&\approx&
H_{\rm Imp}+\\\label{equ:NRGtridmb}
&\displaystyle+&\sum\limits_{l=1}^L\sum\limits_{\sigma}\sum\limits_{n=1}^\infty
\left(\varepsilon_n^l a^\dagger_{nl\sigma}a^{\phantom{\dagger}}_{nl\sigma}
+t_{n-1}^l a^\dagger_{n-1l\sigma}a^{\phantom{\dagger}}_{nl\sigma}+\mbox{h.c.}\right)\nonumber
\end{eqnarray}
where $H_{\rm Imp}$ is again of arbitrary structure. Likewise, the
recurrency relation (\ref{equ:NRGrec}) can be set up and the algorithm
extended. However, a simple example demonstrates that for $L>1$ the
technique can easily become useless. For a typical application to the
single impurity Anderson model (SIAM) one chooses a $\Lambda=2$ and
$N_{\rm NRG}=1000$.  This is sufficient to obtain very accurate
results for all relevant physical quantities. Let us now consider the
next step, i.e.\ an orbitally degenerate problem with $L=2$. Without
coupling between the orbitals, this corresponds to two independent
SIAM we try to solve in a single calculation. Obviously, to obtain the
same accuracy as in the true single impurity case, one needs at least
$N_{\rm NRG}=1000^2$ or $\Lambda=2^2$ (for a more detailed discussion
of the issue of the accuracy of the NRG see \textcite{NRG:paula} and
references therein).

While $N_{\rm NRG}=1000^2$ is beyond all numerical possibilities,
a strongly increasing $\Lambda$ introduces a huge loss in accuracy,
both at high and low energies \cite{NRG:oliveira}.  As long as $L=2$
the problem can be partially compensated by respecting the additional
symmetries in the system \cite{NRG:sakai} or other numerical tricks
\cite{NRG:oliveira}.  However, as a generic tool to solve the
effective quantum cluster problem arising in embedded cluster
techniques much larger values of $L$ must be accessible. Thus, while
for a study of qualitative effects of non-local correlations with
$N_c=2$ the NRG is still applicable (see results in section
\ref{subsubsec:mi}), the preceding discussion makes it clear that
already for $N_c=4$ the method is currently not feasible.


%
%

\section{Applications to strongly correlated models}
\label{sec:results}

In this section we review the application of various quantum cluster
approaches to a selection of standard models of strongly correlated
electron systems. We
put special emphasis on the
capabilities and advantages of these techniques
over both finite system simulations 
and DMFT. 
In Sec.~\ref{subsec:compl} we show that quantum cluster approaches are
complementary to FSS, i.e.  that taken together the information
obtained from both techniques can yield conclusive results. 
The effects of non-local correlations on single- and two-particle
spectra as well as phase diagrams are emphasized throughout the
discussions of the Falicov-Kimball model in Sec.~\ref{subsec:FKM}, the
1D Hubbard model in Sec.~\ref{subsec:1DHM} and the 2D Hubbard model in
Sec.~\ref{subsec:2dhm}. Due to space restrictions we have to omit
recent applications of the CPT and DCA to electron-phonon systems and
refer the reader to the articles by \textcite{hohenadler:03} and
\textcite{hague:03b}, respectively.


\subsection{Complementarity of finite size and quantum cluster simulations}
\label{subsec:compl}

FSS and quantum cluster approaches yield exact solutions in the
infinite cluster size limit. At finite cluster size $N_c$ quantum
cluster approaches differ from FSS by the coupling to a
self-consistent dynamic host. At cluster size $N_c=1$ this difference
is most pronounced: While FSS reduce to the atomic limit, quantum
cluster approaches reduce to the DMFT, i.e. a highly non-trivial
approximation to the infinite size lattice. Thus it is instructive to
compare quantum cluster results with those obtained from FSS
systematically as a function of cluster size.  This has been done in
the half-filled 2D Hubbard model using the DCA/QMC algorithm, i.e. DCA
combined with QMC as a cluster solver, by \textcite{huscroft:dca1} and
\textcite{moukouri:dca1} and with DCA/FLEX by \textcite{karan:dca1}.

In the 2D Hubbard model at half-filling, the antiferromagnetic
correlation length $\zeta$ increases with decreasing temperature and
diverges at $T=0$. In FSS, i.e. in a finite size lattice with periodic
boundary conditions, the system freezes when the correlation length
exceeds the system size and a gap to excitations opens. As the system
size is increased this tendency is reduced. Correlation induced gaps
are thus generally overestimated in FSS for smaller clusters since the
system is artificially closer to criticality.  In contrast, in quantum
cluster approaches the system is in the thermodynamic limit with
correlations restricted to the cluster size. Hence the system never
freezes. As the cluster size is increased, longer ranged correlations
are progressively included. The effects of correlations therefore
increase with cluster size.

\begin{figure}[htb]
\centerline{
\includegraphics*[width=3.0in]{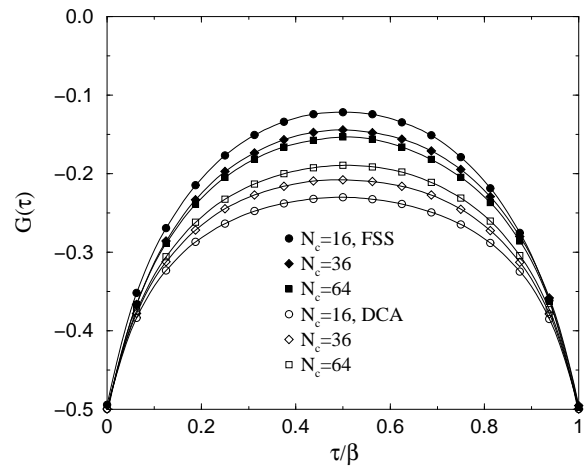}}
\caption{The imaginary time
  Green function $G(\k,\tau)$ at $\k=(\pi,0)$ in the half-filled 2D
  Hubbard model, calculated with finite size QMC (filled symbols) and
  DCA (empty symbols) when $U=4.4t$ and $\beta =4/t$. Taken from
  \textcite{moukouri:dca1}. See also \textcite{karan:dca1}.}
\label{fig:rescomp}
\end{figure} 

\begin{figure}
  \includegraphics*[width=2.5in]{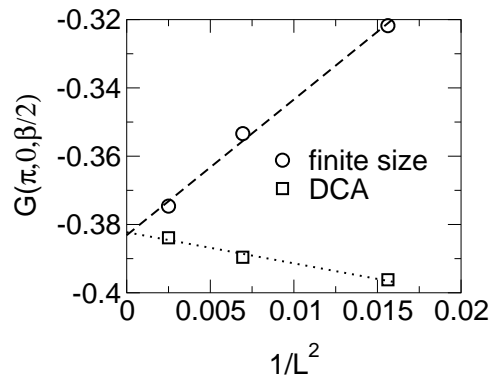}
\caption{The imaginary time
  Green function $G(\k,\tau)$ at $\k=(\pi,0),\tau=\beta/2$ in the
  half-filled 2D Hubbard model, calculated with FSS/FLEX (circles) and
  DCA/FLEX (squares) when $U=1.57t$ for various cluster sizes
  $N_c=L\times L$). Both methods scale as $1/L^2$ and converge to a
  single value as $L\rightarrow \infty$. Taken from
  \textcite{karan:dca1}.}
\label{fig:extrpl1}
\end{figure}

This behavior is illustrated in Fig.~\ref{fig:rescomp}, where we
reproduce the results obtained from DCA/QMC and FSS/QMC simulations
for the imaginary time Green function $G(\k,\tau)$ at the Fermi
wave-vector $\k=(\pi,0)$ for different cluster sizes $N_c$.  This
quantity has a more rapid decay from its maximum at $\tau=\beta/2$
when correlation effects are stronger so the gap is more pronounced.
With increasing cluster size, the DCA and FSS results for $G(\k,\tau)$
converge from opposite directions. Consistent with the expectation
that correlation effects are overestimated in FSS but underestimated
in the DCA, the decay is stronger for smaller system sizes in FSS,
while the DCA results show the opposite behavior.  Since the two
techniques become identical in the infinite cluster size limit, the
exact $G(\k,\tau)$ curve is bracketed by the FSS and DCA curves. For
methods like the FLEX (see Sec.~\ref{subsubsec:flex}) for which
calculations with much larger clusters are feasible, scaling of the
results to the infinite system is possible (see
Fig.~\ref{fig:extrpl1}). The extrapolations of FSS/FLEX and DCA/FLEX
to the infinite system ($1/L^2\rightarrow 0$) coincide within
numerical uncertainties, thus allowing the determination of the
infinite lattice result with unprecedented accuracy.

This complementarity is also seen in results for the spectral gap in
the 2D half-filled Hubbard model (see \onlinecite{huscroft:dca1} and
\onlinecite{karan:dca1}). In the DCA the gaps converge from small to
large as the cluster size increases, while the converse occurs in FSS.
Although we have only shown results of the DCA, we expect the CDMFT to
show similar size dependence,
since DCA and CDMFT share the same nature of the approximation (see
discussion in Sec.~\ref{subsec:disc}). Results obtained with the CPT
algorithm however can be viewed to some extent as a periodic
continuation of FSS. Thus it is an open question if the CPT method
shows similar complementarity.



\subsection{2D Falicov-Kimball model}
\label{subsec:FKM}
The usefulness of the discussed cluster theories was first
demonstrated by the application of the DCA to the 2D FKM by
\textcite{hettler:dca1,hettler:dca2}.  While the FKM is a particularly
gentle test bed for novel approaches, it allows the study of the
effects of non-local fluctuations.

The FKM can be considered as a simplified Hubbard model in which one
spin species has zero hopping amplitude.  
The Hamiltonian reads
\begin{equation}
\label{eq:2dfkm}
H = - t \sum_{<i,j>} d^{\dagger}_i d_j  + 
U  \sum_i (n^d_i-\frac{1}{2})(n^f_i-\frac{1}{2}) \ , 
\end{equation} 
with $n^d_i = d^{\dagger}_i d^{\phantom\dagger}_i$ and $n^f_i =
f^{\dagger}_i f^{\phantom\dagger}_i$.  For a 2D square lattice the
bandwidth of the noninteracting system is $W=8t$.  At half filling and
$D \ge 2$ the system has a second order phase transition from a
homogeneous high temperature phase
to a charge density wave with ordering vector ${\bf{Q}}=(\pi,\pi)$
for any nonzero $U$ \cite{brandt:fkm1,brandt:fkm2}. The universality
class is that of the 2D Ising model, the strong coupling limit $U/t
\gg 1$ of the FKM.
\begin{figure}[htb]
\includegraphics*[width=2.8 in]{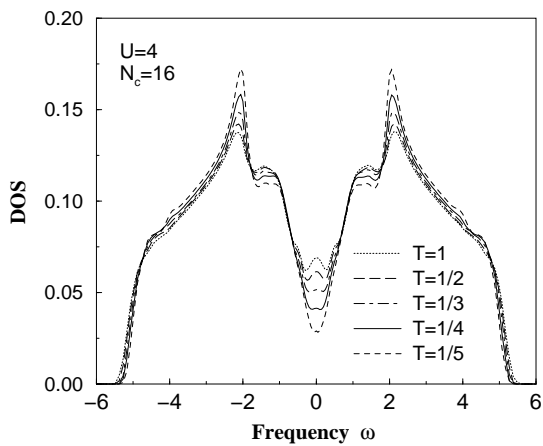}
\caption{DOS of the 2D FKM for a $4\times4$--cluster
  at various temperatures calculated with DCA. The DOS develops a
  pseudogap as the temperature approaches $T_c \approx 0.189t$
  ($U=4t$) due to the non-local CDW fluctuations present in the DCA
  $(N_c >1)$. In the DMFT ($N_c =1$), there is no $T$--dependence of
  the DOS above $T_c$.  Figure from \textcite{hettler:dca2}.  }
\label{fig:fkmdosvst}
\end{figure}
In \textcite{hettler:dca2} the FKM was evaluated within the DCA by a
combination of QMC methods and exact enumeration for small clusters.
Since the DMFT is a single-site theory ($N_c=1$), it yields an
unphysical temperature-independent DOS of the mobile $d$-electrons
\cite{brandt:fkm3} due to the constraint of half filling (one electron
per site of either the $d$ or $f$ variety).  In a cluster theory with
$N_c > 1$ this artifact is absent, since a redistribution of Boltzmann
weight with temperature is possible among the various configurations
of $d$ or $f$ electrons on the cluster sites, while maintaining the
condition of half filling on average.  This temperature dependence is
demonstrated in Fig.~\ref{fig:fkmdosvst}, where a pseudogap develops
in the local DOS with decreasing temperature.
This pseudogap can be interpreted as a precursor of the eventual
transition to a CDW phase, which features a full gap at the Fermi
level.  In addition to the gap, there is fine structure in the DOS,
related to an exchange energy $J_{eff}$. This is better observed in
the momentum resolved spectral function (see
\onlinecite{hettler:dca2,laad:cluster}).

\begin{figure}[htb]
\includegraphics*[width=2.8 in]{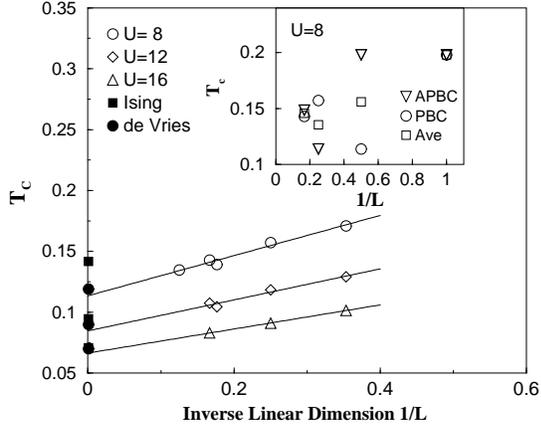}
\caption{$T_c$ in the 2D FKM as a function of the inverse linear cluster size for  $U/t=8, 12, 16$ calculated with DCA.  The Ising limit, and FSS estimates
  of $T_c$ are shown for comparison.  The inset shows that the
  influence of the cluster boundary conditions on $T_c$ disappears rapidly
  with increasing cluster size.  Figure from \textcite{hettler:dca2}.}
\label{fig:fkmtcscale}
\end{figure}

As stated above, the FKM has an instability to a phase with CDW order.
As discussed in Sec.~\ref{subsec:ccw}, embedded cluster theories
exhibit phase transitions at some temperature that, due to their
residual mean-field character, lies above the exact $T_c$ of the
infinite system.  As the cluster size increases, one expects the
effect of the mean-field to decrease, leading to a
decreasing $T_c$ with increasing cluster size. 
The $N_c$ dependence of the transition temperature $T_c$ is shown in Fig.~\ref{fig:fkmtcscale}, together with a comparison with the $T_c$ obtained from FSS
methods \cite{deVries:fkm1,deVries:fkm2,deVries:fkm3} and  $T_c$ of 
the 2D Ising model with exchange coupling $J=1/(2U)$.
The extrapolated cluster results agree with the FSS estimates and, for
large values of $U$, also with the results obtained from the 2D Ising
model.  For smaller $U$ however, charge fluctuations begin to play a
larger role, suppressing the $T_c$ compared to that of the Ising model
which lacks charge fluctuations.

The effect of different boundary conditions in the DCA cluster is
illustrated in the inset\footnote{According to the derivation of the
  DCA formalism (see Secs.~\ref{subsec:es}, \ref{subsec:freeeg} and
  Fig.~\ref{fig:Nc4_clusters}) the DCA cluster has periodic boundary
  conditions. By shifting the set of cluster $\K$-points however,
  different boundary conditions can be simulated.}.  In small
clusters the effect is strong, but already in a $6\times 6$- cluster
the bulk of the cluster dominates and the boundaries play a minor
role.

\begin{figure}
\includegraphics*[width=3 in]{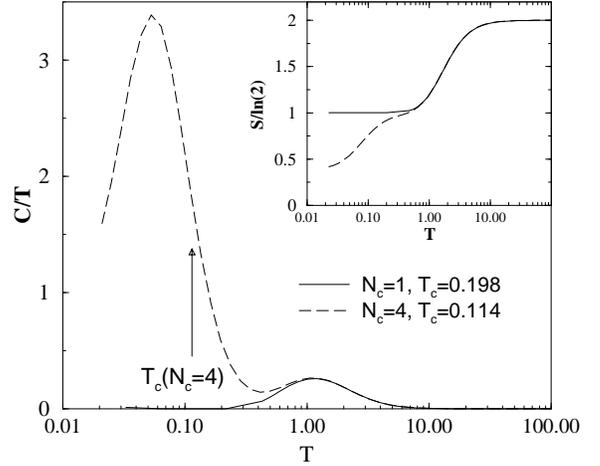}
\caption{Specific heat versus temperature for single site  and 
  $2\times 2$- clusters calculated with DCA and exact enumeration when
  $U=8t$.  For $N_c=1$, there is a single peak with integrated weight
  $\ln(2)$ associated with the suppression of local charge
  fluctuations. For $N_c=4$, there is an additional peak at lower
  temperatures associated with critical fluctuations near the charge
  ordering transition temperature. $T_c$ for $N_c=4$ is indicated by
  an arrow.  In the inset the entropy $S(T)= \int_0^{T} dT'
  \frac{C(T')}{T'}$ is shown in units of $\ln(2)$.  Figure from
  \textcite{hettler:dca2}.  }
\label{fig:fkmCoverT}
\end{figure}

When $N_c=1$ only charge fluctuations with an energy scale $U$ are
present. In a cluster theory the non-local ``spin'' fluctuations with
an effective energy scale $J_{eff}$ ($\propto 1/2U$ for large $U$)
must also be observable in thermodynamic quantities like the entropy
and the specific heat. In \textcite{hettler:dca2} the entropy and
specific heat was computed from the total energy of the cluster via a
maximum-entropy method \cite{huscroft:thermo}. For better comparison,
the calculations were performed in the uniform phase, even at
temperatures below the CDW ordering $T_c$ by not allowing for the
symmetry breaking.  The results for a $2\times 2$- cluster are shown
in Fig.~\ref{fig:fkmCoverT}, where the ratio of specific heat $C$ over
the temperature $T$ is plotted for the DMFT and the DCA on a $2\times
2$- cluster. The appearance of the a second peak at lower temperature
is a clear indication for the additional non-local fluctuations
present on the cluster. The effect on the entropy is also strong, as
shown in the inset.

\subsection{1D Hubbard model}
\label{subsec:1DHM}

In this section we discuss the application of quantum cluster
approaches to the 1D Hubbard model (in usual notation)
\begin{equation}
\label{eq:1dhm1}
{\cal H}=-t\sum_{i,\sigma} (c^\dagger_{i+1\sigma}c^{}_{i\sigma}+h.c.)+U\sum_i n_{i\uparrow} n_{i\downarrow}\,,
\end{equation}
which provides a non-trivial test ground for these techniques. In 1D
quantum fluctuations are stronger than in higher dimensions.  Hence
quantum cluster approaches which cut off correlations beyond the
length-scale set by the cluster size are expected to be less efficient
than in higher dimensions. Therefore if quantum cluster approaches
accurately describe the physics in 1D, they are highly likely to
capture the physics in 2D and 3D. In addition, since the exact
ground-state of the 1D Hubbard model is known from the Bethe ansatz
solution \cite{lieb:1dhm}, a quantitative comparison of certain static
quantities is possible. Fairly reliable results for dynamical
quantities can be obtained from the density matrix renormalization
method.

\begin{figure}[htb]
\centerline{
  \includegraphics*[width=3.0in]{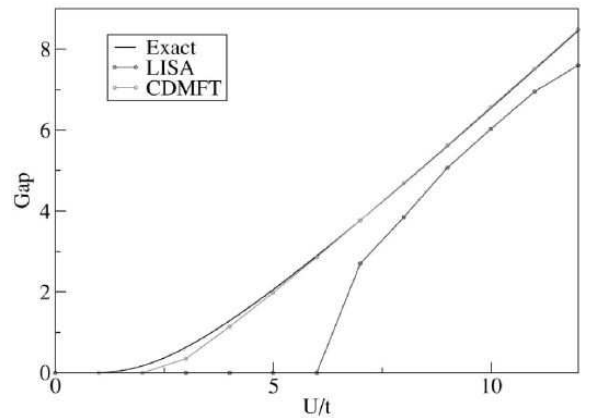}}
\caption{Spectral gap in the half-filled 1D Hubbard model as a function of $U/t$ calculated with DMFT (LISA) and CDMFT for a 2-site cluster, $N_c=2$ compared to the exact result. Taken from \textcite{bolech:cluster}.}
\label{fig:res1dhm0}
\end{figure}

\textcite{bolech:cluster} applied the CDMFT/ED method to the 1D
Hubbard model and systematically compared the results with those
obtained from the DMFT and DMRG approaches. As an example of this
study we show in Fig.~\ref{fig:res1dhm0} a comparison of the exact
result with that obtained from DMFT (referred to as LISA for ``Local
Impurity Self-consistent Approximation'') and CDMFT for the
single-particle spectral gap $\Delta(U)$ as a function of the on-site
Coulomb repulsion $U$ in the half-filled case. The total number of
sites in the effective cluster model including cluster and bath sites
was fixed to six in the ED approach (see Sec.~\ref{subsubsec:ed}).

The behavior of the cluster results (CDMFT for $N_c=2$) is
fundamentally different from the single-impurity result (LISA,
$N_c=1$): When $N_c=1$, the spectral gap is reduced to values much
smaller than the exact value and even vanishes at finite $U$. The
existence of this Mott-transition in 1D for $N_c=1$ is a well known
artifact of the DMFT method, reflecting the physics of infinite
dimensions where the Mott-transition is indeed present
\cite{georges:dmftrev}.  In contrast, the $N_c=2$ result follows the
exact result quite accurately yielding an insulating solution for all
finite values of $U>0$ in agreement with the well known physics of the
1D model.  This excellent reproduction of the exact gap for $N_c=2$ is
quite encouraging since larger cluster sizes should produce even
better results. And indeed, a comparison of the nearest neighbor Green
function with DMRG results shows systematic improvements with
increasing cluster size \cite{bolech:cluster}.

\begin{figure}[htb]
\centerline{
  \includegraphics*[width=3.0in]{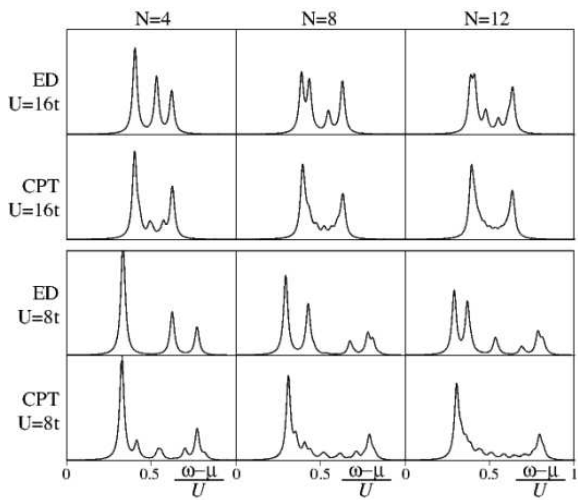}}
\caption{The spectral function $A(k,\omega)$ at $k=\pi/2$ in the half-filled 1D Hubbard model when $U=16t$ (top) and $U=8t$ (bottom), calculated with ordinary ED and with CPT for cluster sizes $N=4,8,12$. Taken from \textcite{senechal:cluster2}.}
\label{fig:res1dhm1}
\end{figure}

A related fundamental feature in 1D correlated systems is the
breakdown of the Fermi-liquid picture because of spin-charge
separation as described in the concept of Luttinger-liquids (see e.g.
\onlinecite{voit:1d}). In a Fermi-liquid, the spectral weight
$A(\k,\omega)$ is centered around a single quasiparticle peak at
$\omega=\epsilon_\k$, while in a Luttinger-liquid, $A(\k,\omega)$ is
distributed between two singularities associated respectively with
spin- and charge-excitations (spinons and holons).
\textcite{senechal:cluster2} have calculated $A(\k,\omega)$ in the 1D
Hubbard model using the CPT formalism.  Fig.~\ref{fig:res1dhm1} shows
a comparison of this quantity at the Fermi wave-vector $k=\pi/2$ as
calculated by ordinary ED and its infinite lattice extension within
the CPT method for various cluster sizes $N$ when $U=8t$ (bottom) and
$U=16t$ (top).

While no sign of spin-charge separation is seen in the pure ED
results, the CPT method reveals the two branches of the spectral
weight indicative of spin-charge separation. The two-peak structure
resolves more clearly as the cluster size increases. Since propagation
between clusters requires the spinon and holon to recombine,
spin-charge separation can only exist on length- and time-scales
limited by the cluster size. Consequently it takes fairly large
clusters to clearly resolve this feature.

\begin{figure}[htb]
  \centerline{
    \includegraphics*[width=3.5in]{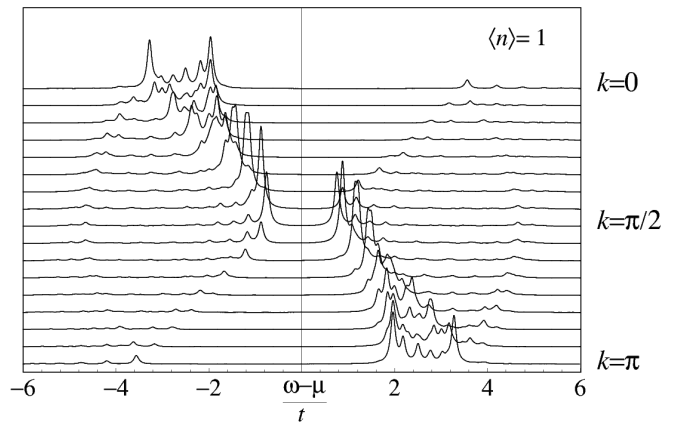}}
  \caption{The spectral function $A(k,\omega)$ in the half-filled 1D
    Hubbard model when $U=4t$, calculated with CPT for $N_c=12$.
    Taken from \cite{senechal:cluster2}.}
\label{fig:res1dhm2} 
\end{figure}

The analysis of the $k$-dependent spectral function in
Fig.~\ref{fig:res1dhm2} reveals the dispersion of the two branches.
The spinon- and holon-branches show different dispersions and can be
clearly identified together with the gap at the chemical potential at
$k=\pi/2$.  These $T=0$ simulations were extended to finite low
temperatures $T>0$ by \textcite{aichhorn:03} who implemented a novel
low-temperature Lanzcos algorithm which connects the exact
ground-state Lanczos method with the established finite temperature
Lanzcos method \cite{prelovsek:00}.  They showed that the spin-charge
separation observed in the $T=0$ results persists at finite, low
temperatures.

As exemplified by these studies, quantum cluster approaches have been
proven very useful to explore the complex behavior of the 1D Hubbard
model. Even studies with small cluster sizes are consistent with
well-known results such as the opening of a Mott gap at half-filling
for all $U>0$ and the existence of spin-charge separation reflected in
the two branches in the electronic dispersion. This success in the 1D
case shows great promise for applications of cluster methods in 2D or
3D where cluster methods are expected to be even more efficient since
correlations are less pronounced.



\subsection{2D Hubbard model}
\label{subsec:2dhm}

The interest in the 2D Hubbard model (in usual notation) 
\begin{equation}
  \label{eq:51}
  {\cal H}=\sum_{ij,\sigma} t_{ij}c^\dagger_{i\sigma} c^{}_{j\sigma} +U\sum_i n_{i\uparrow} n_{i\downarrow}
\end{equation}
has been revived recently in particular since it is believed to
capture the physics of the superconducting planes in high-temperature
superconductors (HTSC) (\onlinecite{anderson:htsc,zhang:88,zhang:90}).
This section reviews various applications of quantum cluster
approaches to the 2D Hubbard model at half-filling and at finite
doping, including results for a possible Mott-Hubbard transition,
antiferromagnetism and its precursors, pseudogap phenomena and
superconductivity.

\subsubsection{Metal-insulator transition}
\label{subsubsec:mi}

The study of a possible metal-insulator transition (MIT) in the 2D
Hubbard model at half-filling ($\epsilon_o=t_{ii}=-U/2$) is under
active research.  This problem was studied in the unfrustrated model,
i.e. with only nearest neighbor hopping
$t_{ij}=\epsilon_o\delta_{ij}-t\delta_{\langle ij\rangle}$ within
DCA/QMC by \textcite{moukouri:dca1} and with the two-site composite
operator method (see Sec.~\ref{subsec:alt}) by
\textcite{phillips:cluster1}. The frustrated case (additional
next-nearest neighbor hopping $t'\delta_{\langle\langle ij
  \rangle\rangle}$) was investigated by \textcite{parcollet:cdmft}.

Numerical calculations have shown that the ground state of the
unfrustrated model is an antiferromagnetic insulator with the N\'{e}el
temperature $T_{\rm N}=0$ constrained by the Mermin-Wagner theorem.  A
spectral gap hence exists at $T=0$. However the central question about
the origin of the gap and its relation to antiferromagnetic ordering
is less understood. Is the gap a direct consequence of the
antiferromagnetic ordering at $T=0$ or does it arise from strong
correlations at higher temperatures?

To appreciate the significance of this issue it is important to
understand the fundamental difference between antiferromagnetic
insulators, i.e. insulators due to magnetic ordering and Mott
insulators, i.e. insulators due to electronic correlations.
Antiferromagnetic insulators result from the doubling of the unit cell
in the ordered state and are therefore adiabatically connected to band
insulators which have an even number of electrons per unit cell. In
contrast, paramagnetic Mott insulators have an odd number of electrons
per unit cell and are therefore fundamentally different from band
insulators.

At strong coupling ($U\gg W$) the situation is well understood: A
charge gap of order $U$ develops in the spectrum below temperatures
$T\approx U$ due entirely to strong electronic correlations. The spins
are coupled by the exchange interaction $J=4t^2/U$ and govern the
low-energy physics.  As a result spin and charge are separated.
Systems in this regime are hence Mott insulators and the
antiferromagnetic ordering at $T=0$ is merely the result of the
Mott-transition at higher temperatures.

Different scenarios however exist for the weak-coupling regime ($U \ll
W$): In the weak coupling point of view, a spin density wave forms at
$T=0$ due to the nesting of the Fermi surface and leads to the
doubling of the unit cell. Hence the gap in the spectrum is a direct
consequence of the antiferromagnetism at $T=0$. This perturbative
point of view is referred to as Slater mechanism. It is in contrast to
the second opinion due to \textcite{anderson:1dhm,anderson:htsc} who
has argued that the 2D half-filled Hubbard model is always in the
strong-coupling regime, so that a Mott gap is present for all $U>0$ as
in 1D (see Sec.~\ref{subsec:1DHM}). As the temperature decreases,
local moments develop because of the opening of the Mott gap which
then order at $T=0$. Thus the antiferromagnet at $T=0$ is a
consequence of the Mott transition.

\begin{figure}[htb]
  \centerline{
    \includegraphics*[width=3.0in]{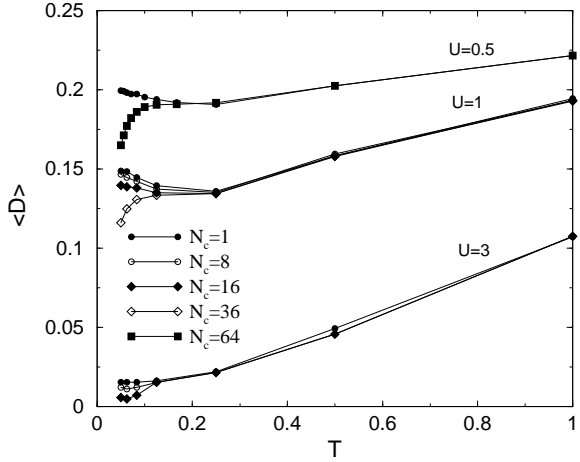}}
  \caption{Double occupancy as a function of temperature for different values of $U$  in the half-filled 2D Hubbard model calculated with DCA/QMC for various cluster sizes $N_c$. Here, $t=0.25$, so the lines for $U=0.5, 1, 2$ correspond to $U/t=2, 4, 12$. Taken from \textcite{moukouri:dca1}.}
\label{fig:res2dhmmi2} 
\end{figure}

The MIT in the half-filled Hubbard model has been extensively studied
within the DMFT (for a review see \onlinecite{georges:dmftrev}). In
the DMFT one can easily disentangle the effects leading to
antiferromagnetic and Mott gaps.  The DMFT equations for the
paramagnetic state of the bipartite Bethe lattice are identical with
the equations of the fully frustrated infinite dimensional model
\cite{georges:dmftrev}. This justifies the study of the paramagnetic
solution within the antiferromagnetic phase of the unfrustrated model
which shows a first order Mott MIT ending at a finite temperature
critical point \cite{georges:dmftrev}.  Although this justification
does not hold for $N_c>1$, one can still study the paramagnetic
solution by enforcing the spin symmetry and hence avoiding the opening
of a full spectral gap due entirely to magnetic ordering. Following
this approach, \textcite{moukouri:dca1} studied the MIT in the
unfrustrated 2D half-filled Hubbard model using DCA/QMC systematically
as a function of cluster size $N_c$. The MIT can be identified by
analyzing the behavior of the double occupancy $\langle D\rangle =
\langle n_\uparrow n_\downarrow \rangle$. This quantity is shown in
Fig.~\ref{fig:res2dhmmi2} for different values of the Coulomb
repulsion $U$ and cluster size $N_c$.

When $N_c=1$ the double occupancy displays evidence for a MIT when $U$
is of the order of the bandwidth $W=8t=2$: $\langle D \rangle$ is
monotonically increasing with temperature when $U=3$, but displays a
minimum for $U=0.5$ and $1$ indicating the emergence of quasiparticle
states at the chemical potential at low temperatures.  When $N_c>1$
the situation is radically different: In the strong coupling regime
($U=3$), local fluctuations dominate and $\langle D \rangle$ is
essentially independent of $N_c$ except at very low temperatures. In
contrast, in the weak coupling regime, the minimum found for $N_c=1$
flattens progressively as $N_c$ increases from 8 to 16. When $N_c\geq
36$ a downturn in $\langle D \rangle$ appears at low temperatures. By
opening a gap and hence localizing the moments, the system can gain
free energy by taking advantage of the short-ranged magnetic
correlations.

\begin{figure}[htb]
  \centerline{
  \includegraphics*[width=3.0in]{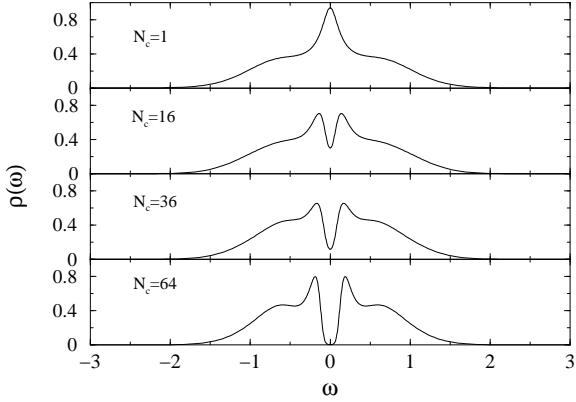}}
  \caption{DOS when $U/t=4$ for different temperatures in the half-filled 2D Hubbard model calculated with DCA/QMC for various cluster sizes $N_c$. Taken from \textcite{moukouri:dca1}.}
\label{fig:res2dhmmi2b} 
\end{figure}

The results for the DOS shown in Fig.~\ref{fig:res2dhmmi2b} support
the evidence from the double occupancy. The quasiparticle Kondo-like
resonance at the chemical potential for $N_c=1$ is destroyed by
non-local correlations when $N_c>1$. As $N_c$ increases, a gap opens
at the chemical potential and the Hubbard sidebands become more
pronounced consistent with the suppression of $\langle D \rangle$.
Given the fact that DCA always underestimates correlation induced
spectral gaps (see Sec.~\ref{subsec:compl}), these simulations
indicate the absence of a weak-coupling regime in the unfrustrated 2D
Hubbard model at half-filling consistent with Anderson's point of
view.
\begin{figure}[htb]
  \centerline{
    \includegraphics*[width=1.8in]{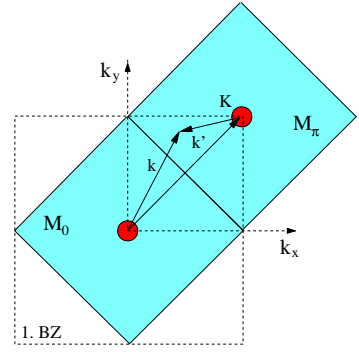}}
  \caption{Tiling of the first Brillouin zone for $N_c=2$.}
\label{fig:Nctwo} 
\end{figure} 

\begin{figure}[htb]
  \centerline{
    \includegraphics*[width=3.5in]{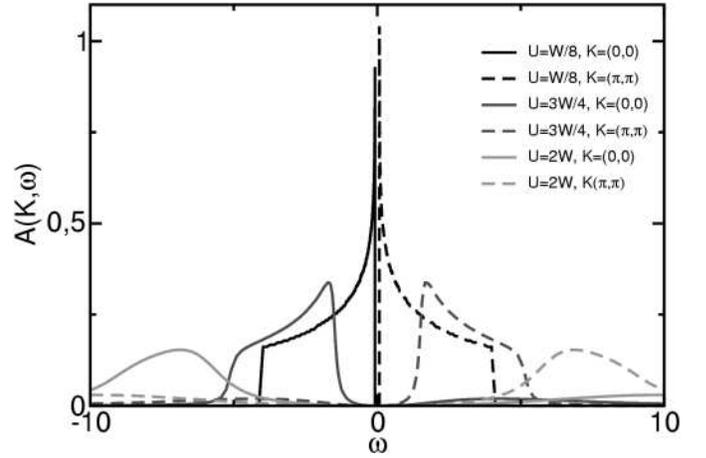}}
  \caption{DCA coarse-grained spectral functions for different values of $U$ obtained from
    $N_c=2$ DCA/NRG calculations at $T=0$. Apparently, no finite
    critical value of $U$ for a conventional Mott-Hubbard transition
    seems to exist. In contrast, for all $U>0$ a gap exists at the
    Fermi energy.}
\label{fig:NctwoNRG} 
\end{figure}

Another interesting aspect is whether this result changes at zero
temperature, $T=0$ and the MIT predicted by DMFT returns. As evidence
that even at $T=0$ one must expect a gap in the spectra for any $U>0$,
we show here first DCA results at $T=0$ obtained with Wilson's NRG for
a cluster size of $N_c=2$. The appropriate tiling of the Brillouin
zone is shown in Fig.~\ref{fig:Nctwo} and the resulting coarse-grained
spectral functions $\bar{A}(\K,w)=-\frac{1}{\pi}\Im m \Gbar(\K,w)$ for
the two cluster $\K$-points are shown in Fig.~\ref{fig:NctwoNRG}.
Even for small values of $U$ a well defined gap exists in the spectrum
at the Fermi energy. Note that all calculations were done in the
paramagnetic phase, i.e.\ the concept of a Slater insulator does not
apply here. The gap quickly increases with increasing $U$ and at the
same time the system gains more spectral weight in the incoherent
parts of the spectrum, i.e.\ begins to resemble what one expects from
Mott localized states.

\begin{figure}[htb]
  \centerline{
    \includegraphics*[width=2.5in]{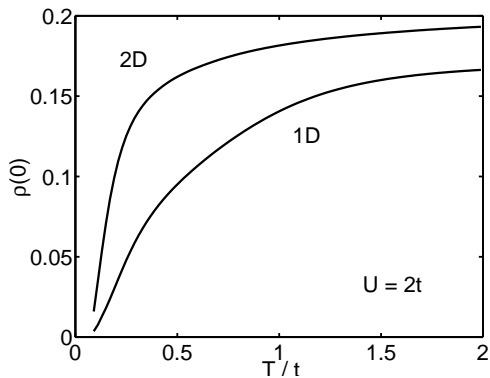}}
  \caption{DOS as a function of temperature at the chemical potential in the 1D and 2D half-filled Hubbard model when $U=2t$ calculated with a two-site composite operator approximation. Taken from \textcite{phillips:cluster1}.}
\label{fig:res2dhmmi1} 
\end{figure}

The DCA results were confirmed in a similar cluster study by
\textcite{phillips:cluster1} using the two-site composite operator
method discussed in Sec.~\ref{subsec:alt}. Figure \ref{fig:res2dhmmi1}
shows the results of this study for the temperature dependence of the
value of the DOS at the chemical potential, $\rho(0)$ in the 1D and 2D
models for small $U=2t$. As a consequence of the shape of the
non-interacting DOS, the 2D result for $\rho(0)$ is enhanced over the
1D result.  However, in both 1D and 2D, $\rho(0)$ falls to zero as the
temperature decreases indicating the absence of a metallic state at
half-filling even for small $U$.

\begin{figure}[htb]
  \centerline{
    \includegraphics*[width=3.in]{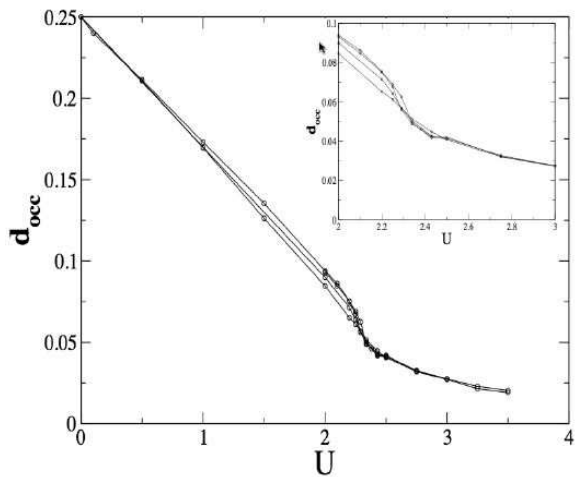}}
  \caption{Double occupancy as a function of $U$ in the frustrated 2D
    Hubbard model with $t'=t$ for different temperatures (from top to
    bottom) $T/t =1/5, 4/30, 1/10, 1/11$, calculated with CDMFT/QMC. Taken
    from \textcite{parcollet:cdmft}.}
  \label{fig:res2dhmmi4} 
\end{figure}

To elucidate the role of antiferromagnetic correlations in the opening
of the Mott gap, the frustrated Hubbard model may be studied. In the
$t$-$t'$ Hubbard model, a next-nearest neighbor hopping $t'$ between
sub-lattices strongly frustrates antiferromagnetic correlations. This
model was studied by \textcite{parcollet:cdmft} for $t'=t$ on a
$2\times 2$ cluster using CDMFT/QMC. We reproduce their results for
the $U$ dependence of the double occupancy $d_{occ}=1/4\sum_{i=1}^4
\langle n_{i\uparrow} n_{i\downarrow} \rangle$ for different
temperatures in Fig.~\ref{fig:res2dhmmi4}. Similar to the behavior
found in DMFT, $d_{occ}$ displays a downturn at a critical value
$U_c$, indicating a transition from a metallic to an insulating state.
An inspection of the spectral weight $A(\k,0)$ at the chemical
potential reveals that the gap opens first in the region around
$\k=(\pi,0)$ \cite{parcollet:cdmft}.
These results were substantiated by a CPT study of the frustrated 2D
Hubbard model \cite{senechal:cluster3} for $t'=-0.4t$. Although not
the focus of this study, the results show further evidence of a Mott
transition at a finite value of $U$ in the filling dependence of the
chemical potential.

The existence of the Mott transition in the frustrated Hubbard model
and its absence in the unfrustrated model seems to indicate that
antiferromagnetic correlations play a key role in the opening of a
Mott gap at weak coupling. Since the opening of the gap occurs in the
paramagnetic solution, it cannot be attributed to the existence of
antiferromagnetic ordering. Thus the conclusion reached from these
cluster studies is that a symbiosis of local moment formation and
short-ranged antiferromagnetic correlations cause the gap to open at
finite temperatures \cite{moukouri:dca1}.

\subsubsection{Antiferromagnetism and precursors}
\label{subsubsec:af}

If the simulations are performed without enforcing the spin symmetry
or frustrating the lattice, the system is given the possibility to
transform to a state with antiferromagnetic order. Since the system is
two-dimensional, we know from the Mermin-Wagner theorem that the true
N\'{e}el temperature $T_{\rm N}$ is necessarily zero. As found in
infinite dimensions \cite{jarrell:qmc}, the DMFT predicts a finite
temperature transition even in 2D due to its mean-field character. As
discussed in Sec.~\ref{subsec:ccw}, cluster approaches restore
non-local fluctuations and thus are expected to drive the N\'{e}el
temperature systematically to zero as the cluster size increases.

\begin{figure}[htb]
  \centerline{
    \includegraphics*[width=2.5in,angle=0]{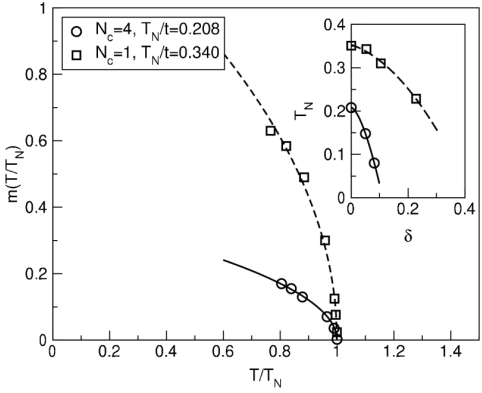}}
  \caption{Sublattice magnetization as a function of temperature in
    the half-filled 2D Hubbard model calculated with DCA/NCA for
    cluster sizes $N_c=1, 4$ when $U=4t$. Inset: Ne\'el-temperature
    versus doping. Taken from \textcite{maier:dca0}.} 
\label{fig:res2dhmaf1}
\end{figure}

As discussed in Secs.~~\ref{subsec:suscept} and
\ref{subsec:brokensym}, phase transitions can be identified from the
disordered (here: paramagnetic) state by calculating the corresponding
susceptibility, or from the ordered state by computing the order
parameter.  The calculation of order parameters is exemplified in
Fig.~\ref{fig:res2dhmaf1} where we plot the DCA/NCA result for the
sub-lattice magnetization $m=1/N\sum_{i,\sigma} e^{i\Q\cdot\x_i}\sigma
n_{i\sigma}$ (see Eq.~(\ref{eq:49})) as a function of the reduced
temperature $t=T/T_{\rm N}$ in the 2D half-filled Hubbard model for
the cluster sizes $N_c=1$ and $N_c=4$ when $U=4t$.  The $N_c=4$ N\'eel
temperature $T_{\rm N}=0.208t$ is reduced from the $N_c=1$ result
$T_{\rm N}=0.304t$ and the order parameter is strongly suppressed. As
expected, non-local spin fluctuations suppress antiferromagnetism.

\begin{figure}[htb]
  \centerline{
    \includegraphics*[width=3.0in]{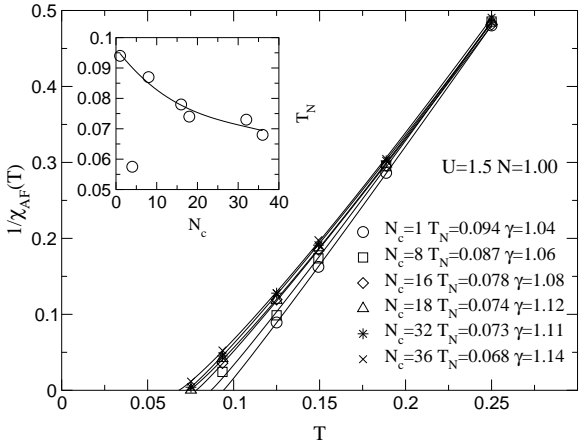}}
  \caption{Inverse antiferromagnetic susceptibility versus temperature
    in the half-filled 2D Hubbard model calculated with DCA/QMC for
    various cluster sizes $N_c$ when $U=6t$. The lines are fits to the
    function $(T-T_{\rm N})^\gamma$. Inset: Corresponding
    Ne\'el-temperatures as a function of the cluster size. Taken from
    \textcite{jarrell:dca3}.}
 \label{fig:res2dhmaf2} 
\end{figure}

Figure \ref{fig:res2dhmaf2} shows the DCA/QMC result for the
temperature dependence of the inverse antiferromagnetic susceptibility
$1/\chi_{AF}$ at $U=6t$ for various cluster sizes $N_c$ in the
paramagnetic state. At high temperatures the susceptibility is
independent of $N_c$ due to the lack of non-local fluctuations. In
contrast to FSS, the low temperature susceptibility diverges at
$T=T_{\rm N}$ indicating the instability to the antiferromagnetic
state. When $N_c=1$ the susceptibility diverges with a critical
exponent $\gamma\approx 1$ as expected for a mean-field theory.
Consistent with the NCA results the susceptibility diverges at lower
temperatures when $N_c>1$ with larger exponents indicative of
fluctuation effects. However as discussed in Sec.~\ref{subsec:ccw}
these critical exponents reflect the behavior at intermediate
temperatures.  Very close to the transition, there must a region of
mean-field behavior.  However, this region is very difficult to
resolve with DCA/QMC, due to numerical noise, which is especially
large near the transition.

As shown in the inset, the transition temperatures fall very slowly
with the cluster size $N_c$, but the $N_c=4$ falls off the curve.  As
detailed in Sec.~\ref{subsubsec:comp}, fluctuation effects in clusters
with linear size $L=2$ are over-proportionally enhanced since its
coordination number is reduced compared to the original system.  Hence
the $N_c=4$ result does not fall on the curve, similar to the behavior
seen in DCA studies of the FKM \cite{hettler:dca2}.

The question arises of whether the same non-local fluctuations which
are responsible for suppressing the antiferromagnetism result in
precursers of the antiferromagnetic phase transition. The onset of
antiferromagnetic correlations on short time- and length-scales may be
signaled by a pseudogap in the DOS as a precursor to the
antiferromagnetic gap. This was predicted by \textcite{kampf:prec}
using a phenomenological ansatz for the weak coupling Hubbard model
based on the presence of strong antiferromagnetic spin-fluctuations.
On a microscopic level, this question has been addressed by FSS/QMC in
the 2D Hubbard model in \textcite{vekic:93} and
\textcite{creffield:95} and by approximate many-body techniques in
\textcite{deisz:96} and \textcite{moukouri:99}.  But the results have
been inconclusive as to the existence of a pseudogap at low
temperatures, due to the limitations of these techniques.

Within quantum cluster approaches the pseudogap phenomenon was first
studied by \textcite{maier:dca1} within the DCA/NCA formalism. In
contrast to $N_c=1$ where a Kondo-like quasiparticle peak emerges at
the chemical potential as the temperature is decreased (reminiscent of
the $D=\infty$ DMFT result \cite{georges:dmftrev}), a pseudogap
was found when non-local correlations were included in $N_c=4$
simulations.
\begin{figure}[htb]
  \centerline{
    \includegraphics*[width=3.0in,angle=0]{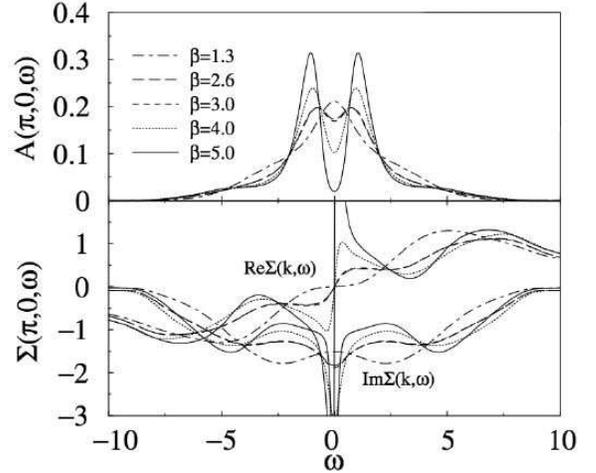}}
  \caption{Spectral function $A(\k,\omega)$ and the real $\Re
    e\Sigma(\k,\omega)$ and imaginary $\Im m \Sigma(\k,\omega)$ parts
    of the self-energy for various temperatures at $\k=(\pi,0)$ in the
    2D half-filled Hubbard model for $U=5.2t$ calculated with DCA/QMC
    for a 64-site cluster ($N_c=64$).
    Taken from \textcite{huscroft:dca1}.}
  \label{fig:res2dhmaf4} 
\end{figure} 
For larger cluster sizes, the emergence of the pseudogap in the DOS
was explored by DCA/QMC in \textcite{huscroft:dca1} and DCA/FLEX in
\textcite{karan:dca1}.  Figure~\ref{fig:res2dhmaf4} displays the
DCA/QMC results for the spectral function $A(\k,\omega)$ and the
self-energy $\Sigma(\k,\omega)$ at the Fermi wave-vector $\k=(\pi,0)$
in the paramagnetic state for a 64-site cluster ($N_c=64$) at various
temperatures. With decreasing temperature a pseudogap develops in
$A(\k,\omega)$ at the Fermi wave-vector $\k=(\pi,0)$.  Simultaneously
the slope of $\Re e \Sigma(\k,0)$ becomes positive at $\k=(\pi,0)$
signaling the appearance of two new solutions in the quasiparticle
equation $\omega-\epsilon_\k+\mu-\Re e\Sigma(\k,\omega)=0$.  In
addition to the strongly damped solution at $\omega=0$ which is also
present in the non-interacting system these two new quasiparticle
solutions appear on both sides of $\omega=0$. A consequence of the
antiferromagnetic order on short time- and length-scales, they can be
viewed as precursers of the doubling of the unit cell in the
antiferromagnetic state. The pseudogap is generated by the local
minimum in $\Im m\Sigma(\k,\omega)$ which signals the breakdown of
Fermi-liquid behavior.

\begin{figure}[htb]
  \centerline{
    \includegraphics*[width=2.5in,angle=0]{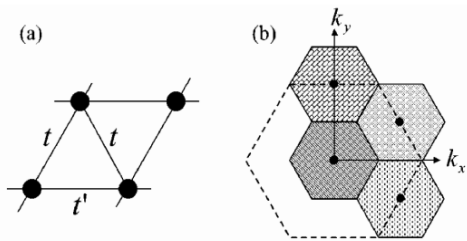}}
  \caption{Illustration of the triangular lattice with hopping amplitudes $t$ and $t'$ (a) and DCA coarse-graining cells (b) in the first Brillouin zone (dashed line) of the triangular lattice when $N_c=4$. Cluster $\K$-points are indicated by the dots. Taken from \textcite{imai:dca}.}
\label{fig:res2dhmaf5} 
\end{figure}

By studying the system on a triangular lattice, \textcite{imai:dca}
investigated the effects of frustration on the pseudogap in the
half-filled 2D Hubbard model using the DCA/NCA and DCA/FLEX
approaches. Figure \ref{fig:res2dhmaf5} schematically illustrates the
triangular lattice and the choice of cluster wave-vectors in the
corresponding Brillouin zone for $N_c=4$.

\begin{figure}[htb]
  \centerline{
    \includegraphics*[width=2.5in,angle=0]{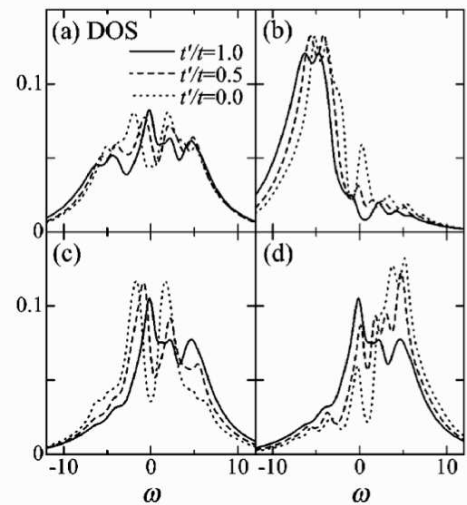}}
  \caption{The DOS (a) and coarse-grained single-particle spectral functions $\bar{A}(\K,\omega)$ in the 2D half-filled frustrated Hubbard model for $\K=(0,0)$ (b), $\K=(\pi,\pi/\sqrt{3})$ (c) and $\K=(0,2\pi/\sqrt{3})$ (d) for various values of the frustration $t'$ when $U=6t$ and $T=0.6t$. Taken from
    \textcite{imai:dca}.}
\label{fig:res2dhmaf6} 
\end{figure}

For $t'=0$ this setup corresponds to the unfrustrated system and the
effects of frustration can be systematically studied as $t'$ is
increased to its maximal value $t'=t$. Figure \ref{fig:res2dhmaf6}
reproduces the results for the DOS and coarse-grained spectra
$\bar{A}(\K,\omega)$ for different values of the frustration $t'$.
As the geometrical frustration increases from $t'=0$ to $t'=t$,
antiferromagnetic spin fluctuations are suppressed. Consequntly the
pseudogap in the unfrustrated system diminishes and a quasiparticle
peak develops at the chemical potential. The change in the DOS mainly
originates in the region in momentum space around
$\K=(\pi,\pi/\sqrt{3})$ where the Fermi surface is located. These
results are thus consistent with an antiferromagnetic spin fluctuation
driven pseudogap.

\subsubsection{Pseudogap at finite doping}
\label{subsubsec:pg}

The properties of the Hubbard model away from half-filling are of
great interest especially in the context of HTSC.  Contrary to
Fermi-liquid theory, low-energy spin-excitations in HTSC are
suppressed at low temperatures as evidenced by Knight-shift
experiments.  Concomitantly, the Fermi surface is gapped along certain
directions in the Brillouin zone as indicated in ARPES experiments.
This pseudogap phenomenon\footnote{For a review on the pseudogap
  phenomenon see \textcite{timusk:99}.} has proven a great challenge
for theories of strongly correlated systems.

DMFT has provided great insight into the evolution of spectra in doped
Mott-insulators. Exact results based on the self-consistent mapping
onto an Anderson impurity model show that the system is a Fermi-liquid
in the metallic state in the absence of symmetry breaking below a
coherence temperature reminiscent of the Kondo-temperature
\cite{georges:92}. Hence the spin-susceptibility becomes finite at low
temperatures in contrast to the experimental results in underdoped
cuprates.  Furthermore \textcite{hartmann:89b} showed that because the
self-energy is momentum independent, volume and shape of the Fermi
surface are identical to the non-interacting Fermi-surface. Thus DMFT
does not include the effects that lead to the emergence of a pseudogap
in the spin- and quasiparticle spectrum and cluster extensions are
necessary.

Within quantum cluster approaches the pseudogap phenomenon in the
doped 2D Hubbard model was studied with DCA/NCA in
\textcite{maier:dca1} and \textcite{maier:dca0}, with DCA/QMC in
\textcite{jarrell:dca2} and \textcite{maier:dca5}, with the two-site
composite operator method in
\textcite{phillips:cluster3,phillips:cluster2} and with the CPT in
\textcite{senechal:cluster3}.

\begin{figure}[htb]
  \centerline{
    \includegraphics*[width=2.5in,angle=0]{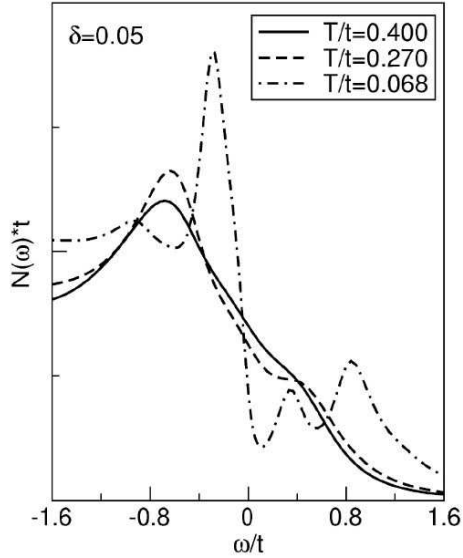}}
  \caption{DOS for various temperatures in the 2D Hubbard model at 5\%
    doping when $U=12t$ near the chemical potential ($\omega=0$)
    calculated with DCA/NCA for a 4-site cluster, $N_c=4$. Taken from
    \textcite{maier:dca0}.}
\label{fig:res2dhmpg1} 
\end{figure}

Figure \ref{fig:res2dhmpg1} shows the DCA/NCA result for the
low-energy DOS in the 2D Hubbard model at 5\% doping ($\delta=0.05$)
around the chemical potential ($\omega=0$) calculated on a 4-site
cluster ($N_c=4$). At high temperature $T>0.3t$ no pseudogap is seen
in the DOS. As the temperature is lowered the DOS distorts at the
chemical potential and a pseudogap emerges at $\omega=0$ when
$T\lesssim 0.3t$.

\begin{figure}[htb]
  \centerline{
    \includegraphics*[width=3.5in,angle=0]{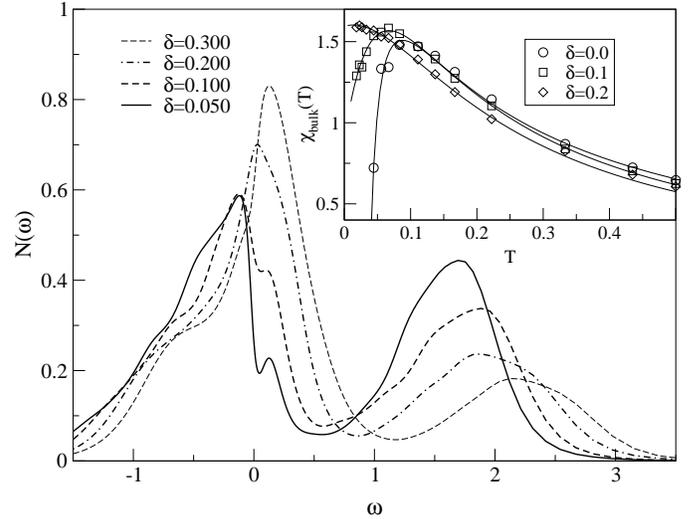}}
  \caption{DOS for various dopings $\delta$ in the 2D Hubbard model at $T=0.092t$ and $U=8t$ calculated with DCA/QMC for a 4-site cluster, $N_c=4$. Inset: Uniform spin-susceptibility as a function of temperature. Taken from \textcite{jarrell:dca2}.} 
\label{fig:res2dhmpg2} 
\end{figure}

These perturbative results\footnote{The NCA is perturbative in the
  coupling to the host, not in the interaction, as detailed in
  Sec.~\ref{subsubsec:nca}.} were confirmed in DCA/QMC simulations by
\textcite{jarrell:dca2} which we reproduce in
Fig.~\ref{fig:res2dhmpg2}. At low temperatures a pseudogap is observed
in the DOS at dopings $\delta \lesssim 0.2$.  This depression of
quasiparticle states at the chemical potential is accompanied by a
downturn of the uniform magnetic susceptibility shown in the inset.
For low to intermediate doping it develops a maximum defining a
crossover temperature $T^*$.  Below $T^*$ quasiparticle and low-energy
spin excitations are suppressed by non-local correlations similar to
what is observed in the experiment.
At the same time, the charge susceptibility (not shown) displays
qualitatively different behavior forming a strong low-energy peak at
low temperatures (see \onlinecite{maier:dca5,maier:dca6} and
Fig.~\ref{fig:res2dhmsc3}) which signals the emergence of coherent
charge excitations below $T^*$. The $N_c=4$ DCA results are thus
consistent with a spin-charge separated picture as in Anderson's RVB
theory. This is not surprising since, as we discussed in
Sec.~\ref{subsubsec:af}, fluctuations are enhanced in the $N_c=4$
cluster due to the ``too small'' coordination number. It is known that
small coordination numbers favor the spin-charge separated RVB state
\cite{anderson:rvb} over the N\'{e}el state. Indeed the RVB state was
shown to be the ground state of a 2$\times$2 Heisenberg model with
periodic boundary conditions with a large gap to the first excited
state \cite{dagotto:88}.

\begin{figure}[htb]
  \centerline{
    \includegraphics*[width=2.5in,angle=0]{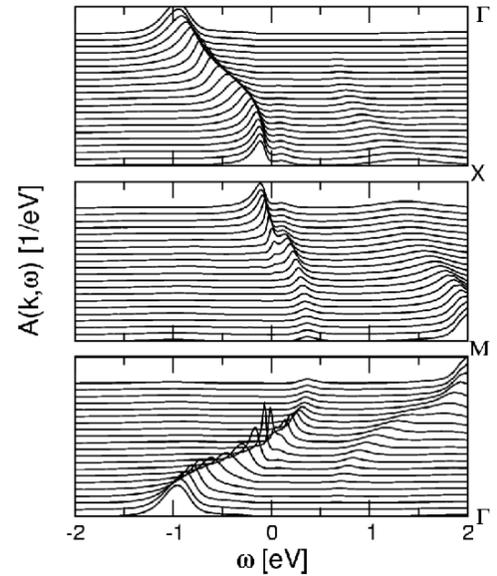}}
    \caption{The spectral function $A(\k,\omega)$ near the chemical potential in the 2D Hubbard model at $T=0.088t$ and $U=8t$ along high-symmetry directions in the first Brillouin-zone between $\Gamma=(0,0)$, $X=(\pi,0)$ and $M=(\pi,\pi)$ calculated with DCA/QMC for a 4-site cluster $N_c=4$.} 
\label{fig:res2dhmpg3} 
\end{figure} 

The corresponding 4-site DCA/QMC result for the momentum resolved
spectral function $A(\k,\omega)$ in the pseudogap regime is displayed
in Fig.~\ref{fig:res2dhmpg3} for energies near the chemical potential
between points of high symmetry $\Gamma=(0,0)$, ${\rm X}=(\pi,0)$ and
${\rm M}=(\pi,\pi)$ in the first Brillouin zone. The overall
dispersion of the band crossing the chemical potential $\omega=0$
follows that of the non-interacting system $\epsilon_\k$. 
While coherent quasiparticles exist along $\Gamma\rightarrow {\rm M}$,
the pseudogap is seen near ${\rm X}=(\pi,0)$ at the chemical potential
$\omega=0$. The anisotropy of the pseudogap is thus consistent with
that observed in ARPES measurements on underdoped hole-doped cuprates.

\begin{figure}[htb]
  \centerline{
    \includegraphics*[width=3.0in,angle=0]{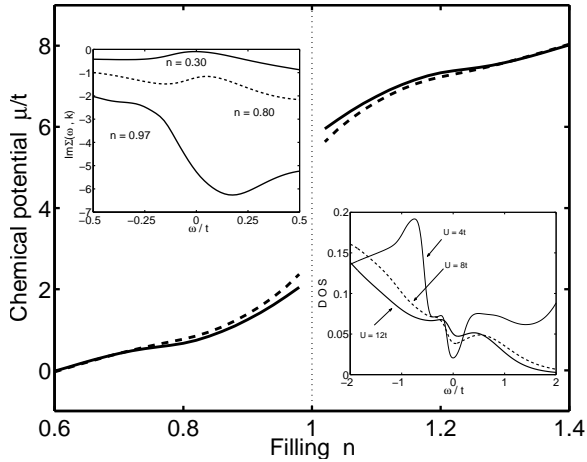}}
  \caption{The doping dependence of the chemical potential in the 2D Hubbard model calculated with the two-site composite operator method for $T=0.15t$ (dashed line) and $T=0.07t$ (solid line). Right inset: DOS for various values of $U$ at 5\% doping ($n=0.95$). Left inset: Imaginary part of the self-energy evaluated at a Fermi momentum (0.3,2.10) for $n=0.97$, (0.3,1.84) for $n=0.80$ and (0.3,1.06) for $n=0.3$. Taken from \textcite{phillips:cluster3}.} 
\label{fig:res2dhmpg4} 
\end{figure}

Qualitatively similar results for the emergence of the quasiparticle
spectrum in the doped 2D Hubbard model were obtained by
\textcite{phillips:cluster3} using the two-site composite operator
approach discussed in Sec.~\ref{subsec:alt} (see also
\onlinecite{phillips:cluster2}). Figure \ref{fig:res2dhmpg4}
illustrates their results for the doping dependence of the chemical
potential $\mu$, the imaginary part of the self-energy $\Im m \Sigma$
and the $U$ dependence of the low-energy DOS.
At half-filling the chemical potential has a discontinuity indicating
the absence of mid-gap states. In agreement with DCA/QMC
\cite{jarrell:dca2,jarrell:dca3} and DCA/NCA \cite{maier:dca0}
results, $|\Im m\Sigma|$ is large in the underdoped pseudogap regime
($n=0.97$) and acquires Fermi-liquid behavior at larger doping $n\leq
0.80$ indicated by the parabolic minimum at the chemical potential. As
illustrated in the inset, the depth of the pseudogap decreases as $U$
increases suggesting a pseudogap scale compatible with $t^2/U$.

\begin{figure}[htb]
  \centerline{
    \includegraphics*[width=3.0in,angle=0]{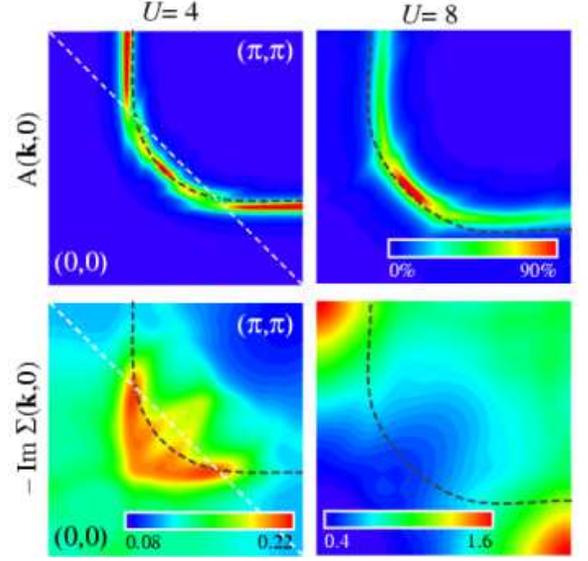}}
  \caption{Intensity plot of the spectral function $A(\k,0)$ (top) and imaginary part of the self-energy $\Im m\Sigma(\k,0)$ (bottom) of the 2D Hubbard model calculated with CPT on a $3\times 4$-site cluster. The left side shows the results in the first quadrant of the Brillouin zone for the 17\% electron-doped system at $U=4t$ and the right side for the  17\% hole-doped system at $U=8t$.  Taken from \textcite{senechal:cluster3}.} 
\label{fig:res2dhmpg5} 
\end{figure}

\textcite{senechal:cluster3} recently investigated the difference in
pseudogap behavior between electron and hole doped HTSC using CPT for
the 2D $t-t'-t''$ Hubbard model. As illustrated in the Fermi-surface
plots in Fig.~\ref{fig:res2dhmpg5}, their results at $U=8t$
demonstrate that the pseudogap in hole-doped systems (right side)
occurs near ${\rm X}=(\pi,0)$ at optimal doping consistent with the
results discussed above. In electron-doped systems (left side)
however, the pseudogap appears at the crossing points of the Fermi
surface with the antiferromagnetic Brillouin zone boundary at moderate
interaction $U=4t$. When $U$ is large however (not shown), the Fermi
surface only survives in the neighborhood of $(\pi,0)$ and $(0,\pi)$.
As shown in the lower panels of Fig.~\ref{fig:res2dhmpg5} the
pseudogap in both cases is generated by a large scattering rate $|\Im
m\Sigma (\k,0)|$ at the chemical potential.
A unified picture of the spectral properties of the electron- and
hole-doped cuprates thus emerges from these results if the interaction
strength $U$ is allowed to be doping dependent.  To reproduce the
experimental observations in optimally doped cuprates, large values of
$U$ seem necessary in hole-doped ($U\approx 8t$) systems, while
smaller values of $U$ describe the electron-doped systems ($U\lesssim
6t$) \cite{senechal:cluster3}.

\subsubsection{Superconductivity}

FSS QMC simulations for the doped 2D Hubbard model in the intermediate
coupling regime $U\sim W$ support the idea of a spin-fluctuation
driven interaction mediating $d$-wave superconductivity (for a review
see \onlinecite{scalapino:99}). However the fermion sign problem
limits these calculations to temperatures too high to study a possible
transition. These calculations are also restricted to relatively small
system sizes, making statements for the thermodynamic limit
problematic, and inhibiting studies of the low-energy physics.  These
shortcomings do not apply to embedded cluster theories which are built
for the thermodynamic limit. Cluster sizes larger than one are
necessary however, to describe a possible transition to a state with a
non-local ($d$-wave) order parameter as discussed in
Sec.~\ref{subsec:brokensym}.

It is well known from weak coupling FSS FLEX results \cite{bickers:89}
and phenomenological theories \cite{monthoux:91,scalapino:99} that
antiferromagnetic spin-fluctuations mediate pairing with $d$-wave
symmetry and cause a pseudogap in underdoped systems. In optimally
doped cuprates, these spin-fluctuations are known to be short-ranged,
extending over a few lattice spacings. Hence quantum cluster
approaches should provide an adequate methodology to study
superconductivity in these systems. Pairing in the 2D Hubbard model
was studied using DCA/NCA by \textcite{maier:dca2}, and with DCA/QMC
by \textcite{jarrell:dca2,jarrell:dca3} and \textcite{maier:dca6}. The
possible coexistence of superconductivity with antiferromagnetic order
was investigated by \textcite{kats:dca}.

\begin{figure}[htb]
  \centerline{
    \includegraphics*[width=3.5in,angle=0]{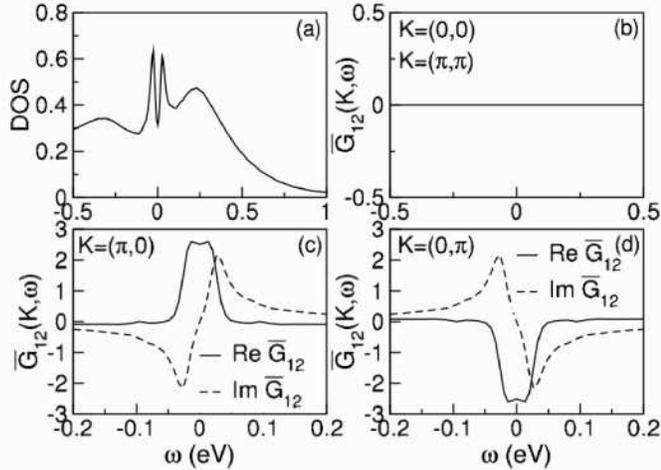}}
  \caption{DOS near the chemical potential (a) and coarse-grained anomalous Green function $\bar{G}_{12}(\K,\omega)\equiv \bar{F}(\K,\omega)$ for different DCA cells (b-c) in the superconducting state of the 2D Hubbard model at 19\% doping, $T=0.047t$, $U=12t$ calculated with DCA/NCA for a 4-site cluster, $N_c=4$. Taken from \textcite{maier:dca2}.} 
\label{fig:res2dhmsc1} 
\end{figure}

The results of 4-site ($N_c=4$) DCA simulations for the doped 2D
Hubbard model show an instability to a superconducting phase with a
$d_{x^2-y^2}$-wave order parameter at low enough temperatures. As a
typical example of this transition, Fig.~\ref{fig:res2dhmsc1} shows
the DCA/NCA result for the DOS and the coarse-grained anomalous Green
function $\Gbar_{12}(\K,\omega)\equiv{\bar F}(\K,\omega)$ defined in
Eq.~(\ref{eq:30}) at different cluster $\K$-points near the chemical
potential in the superconducting state.
$\Gbar_{12}(\K,\omega)$ vanishes at the cluster $\K=(0,0)$ and
$(\pi,\pi)$ but is finite at $(\pi,0)$ and $(0,\pi)$ with opposite
signs. Since the $\K$-dependence of the coarse-grained order parameter
${\bar \Delta}_\K$ is given by the $\K$-dependence of the
coarse-grained anomalous Green function (see Eq.~(\ref{eq:26})), this
result is consistent with a $d_{x^2-y^2}$-symmetry of the order
parameter.  The finite pair amplitude is also reflected in the DOS
depicted in the upper left part where the lower subband of the full
spectrum is shown. It displays the superconducting pseudogap at zero
frequency as expected for a $d$-wave order parameter.

\textcite{jarrell:dca2,jarrell:dca3} used DCA/QMC to search for many
different types of superconductivity, including $s$-, extended $s$-,
$p$- and $d$-wave, of both even and odd frequency. Of these, only the
odd-frequency $s$-wave and even-frequency $d$-wave pair-field
susceptibilities were strongly enhanced, and only the $d$-wave
susceptibility diverged.
\begin{figure}[htb]
  \centerline{
    \includegraphics*[width=3.5in,angle=0]{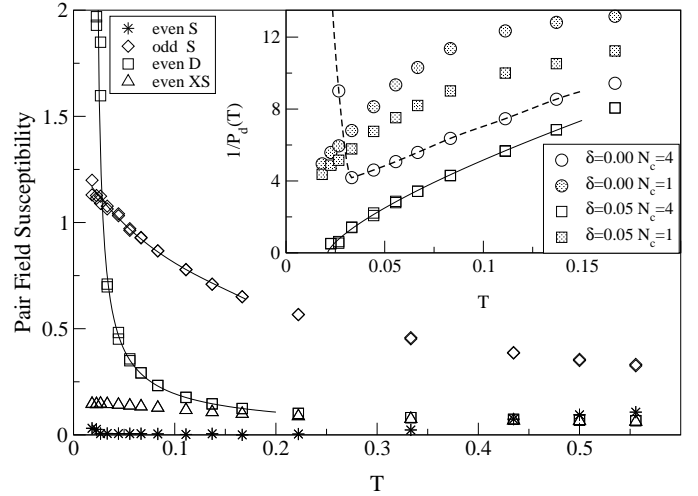}}
  \caption{Pair-field susceptibilities versus temperature in the even
    frequency $s$-wave, extended $s$-wave ($xs$), $d$-dave and
    odd-frequency $s$-wave channels in the 2D Hubbard model at 5\%
    doping, $U=8t$ calculated with DCA/QMC for a 4-site cluster,
    $N_c=4$. Inset: Inverse $d$-wave pair-field susceptibility versus
    temperature for different dopings and cluster sizes. The line is a
    fit to $b(T-T_c)^\gamma$ with $T_c=0.084t$ and $\gamma=0.72$.
    Taken from \textcite{jarrell:dca2}.}
\label{fig:res2dhmsc2} 
\end{figure} 
This is illustrated in Fig.~\ref{fig:res2dhmsc2} where the pair-field
susceptibilities are plotted versus temperature at 5\% doping. The
behavior of the inverse $d$-wave pair-field susceptibility as a
function of temperature for $N_c =1$ and 4 and $\delta=0$ and 0.5 is
shown in the inset. For $N_c = 1$ there is no tendency towards
pairing.  As detailed in Sec.~\ref{subsec:brokensym}, the DMFT is not
able to describe pairing with symmetries lower than the lattice
symmetry (i.e., $p$-, $d$-wave, etc.).  For $N_c=4$ and $\delta = 0$
the inverse susceptibility rises abruptly as the temperature is
lowered and the Mott gap opens in the DOS. The Mott gap becomes more
pronounced as $N_c$ increases (see Sec.~\ref{subsubsec:mi}), so that
for larger clusters the gap prevents superconductivity even for $U<W$.
If charge excitations are gapped, then pairing is suppressed. At
half-filling, for $U = 8t$ the gap is of order $U$, and thus much
larger than the magnetic exchange energy $J\sim 4t^2/U=0.5t$. Hence
the opening of the Mott gap suppresses any magnetically mediated
pairing. Away from half-filling the width of the pseudogap in the
charge excitation spectrum is much smaller, of the order of $J$ (see
Sec.~\ref{subsubsec:pg}), so that magnetically mediated pairing is
possible.

More insight in the nature of pairing was gained from further DCA/QMC
studies of the 2D Hubbard model \cite{maier:dca6}.
\begin{figure}[htb]
  \centerline{
    \includegraphics*[width=1.2in,angle=0]{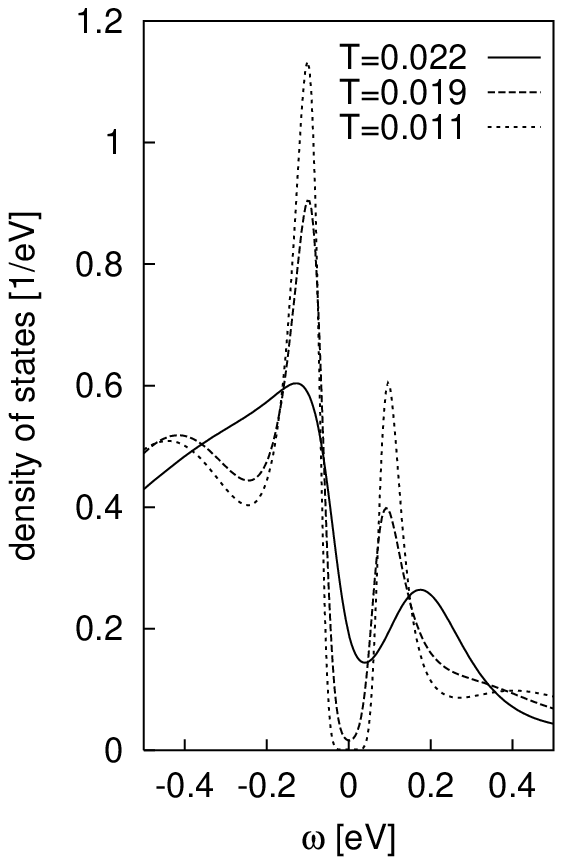}
    \includegraphics*[width=1.2in,angle=0]{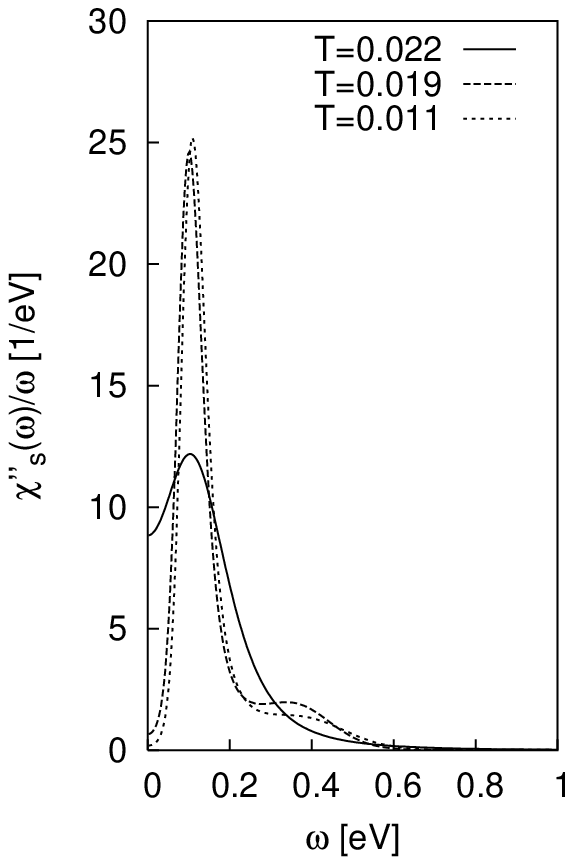}
    \includegraphics*[width=1.2in,angle=0]{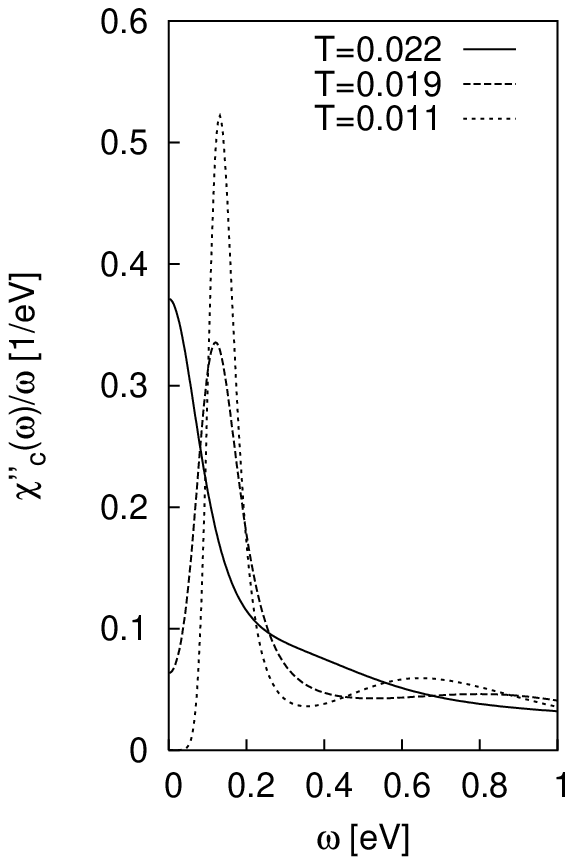}}
  \caption{DOS (left), local dynamic spin susceptibility (center) and
    local dynamic charge susceptibility (right) in the 2D Hubbard
    model at 5\% doping at different temperatures above and below the
    critical temperature $T_c=0.0218=0.087t$ calculated with DCA/QMC
    for a 4-site cluster, $N_c=4$. Taken from \textcite{maier:dca6}.}
  \label{fig:res2dhmsc3}
\end{figure} 
Figure \ref{fig:res2dhmsc3} shows the DCA/QMC result for the evolution
of the DOS, the local dynamic spin- and local dynamic
charge-susceptibility\footnote{Note that in the DCA, local quantities
  are identical in the lattice and on the cluster and thus are easily
  calculated.} as the temperature decreases below the critical
temperature $T_c$. As discussed in Sec.~\ref{subsubsec:pg}, the normal
state low-temperature DOS and spin-susceptibility display a pseudogap,
i.e. a depression of low-energy quasiparticle and spin excitations.
Both quantities evolve smoothly across the superconducting transition
with the pseudogap changing to a superconducting gap\footnote{Note
  that due to the finite resolution in momentum space, the DCA
  underestimates low-energy spectral weight in superconductors where
  the gap has nodes on the Fermi surface. As a result, a fully
  developed gap is found at low temperatures instead of a DOS that
  vanishes linearly in frequency as expected for a $d$-wave
  superconductor.} below $T_c$. However since the charge
susceptibility is peaked at zero frequency even slightly above $T_c$,
it changes abruptly upon pairing to show the same behavior as the
spin-susceptibility, including the superconducting gap at low
frequencies.  Remarkably, well below $T_c$ all quantities display
narrow peaks at $\omega\approx 0.1 {\rm eV}$, delimiting the
superconducting gap.  This result clearly indicates the formation of
quasiparticles below $T_c$. The absence of quasiparticles in the
normal state however undermines the very foundation of the BCS theory
of conventional superconductors where pairing is a result of a Fermi
surface instability that relies on the existence of quasiparticles in
a Fermi-liquid \cite{schrieffer:sc}.

\begin{figure}[htb]
  \centerline{
    \includegraphics*[width=3.7in,angle=0]{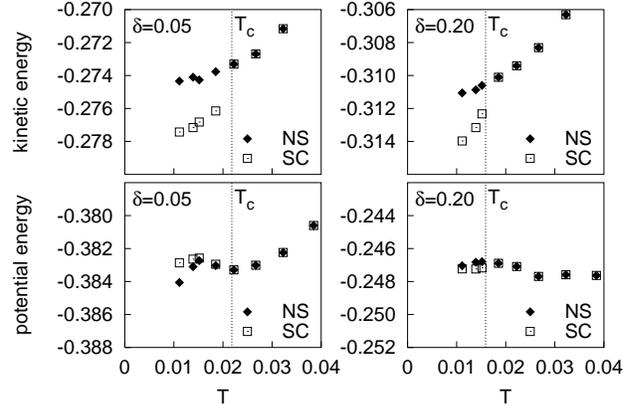}}
  \caption{Kinetic (top) and potential (bottom) energies versus temperature in the normal (NS) and superconducting (SC) states in the 2D Hubbard model at 5\% (left) and 20\% (right) doping for $U=8t$. $T_c$ is indicated by the vertical dotted lines. Taken from \textcite{maier:dca6}.}
\label{fig:res2dhmsc4} 
\end{figure}

DCA/QMC results for the condensation energy further establish the
unconventional character of superconductivity in the 2D Hubbard model
\cite{maier:dca6}.  Figure \ref{fig:res2dhmsc4} presents the kinetic
(top) and potential (bottom) energies, ${\rm Tr}(\t\G)$ and ${\rm
  Tr}(\bSigma\G)$ respectively, of the superconducting (SC) and normal
state (NS) solution as a function of temperature at low doping $\delta=0.05$ (left) and optimal doping $\delta=0.20$ (right).
For both doping levels, the kinetic energy in the superconducting
state is reduced compared to the normal state, while the potential
energies are almost identical. This result is in agreement with recent
optical experiments which show that the superconducting transition in
the cuprates is due to a lowering of the electronic kinetic energy
\cite{marel:kin}. It further supports the evidence that pairing in the
Hubbard model is fundamentally different from BCS pairing which occurs
through a reduction of the electronic potential energy accompanied by
a slight increase in the kinetic energy.

\begin{figure}[htb]
  \centerline{
    \includegraphics*[width=3.5in,angle=0]{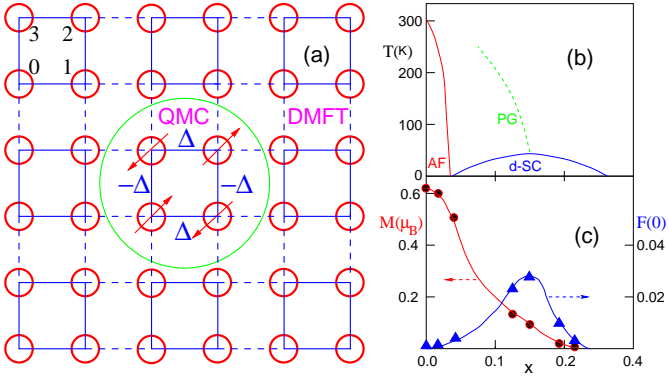}}
  \caption{(a) Schematic representation of an antiferromagnetic $d$-wave $2\times 2$ periodically repeated cluster. (b) Generic phase diagram of HTSC. (c) Magnetic (M) and $d$-wave superconducting (F) order parameters versus hole-doping in the 2D Hubbard model at $\beta t=15$, $t'=-0.15 t$, $U=4.8 t$ calculated with a 4-site cluster approach similar to the  DCA/QMC. Taken from \textcite{kats:dca}.}
\label{fig:res2dhmsc5} 
\end{figure}

The possibility of coexisting $d$-wave superconducting and
antiferromagnetic order in the 2D Hubbard model was investigated by
\textcite{kats:dca} using a 4-site cluster approach similar to the
DCA/QMC method (see Fig.~\ref{fig:res2dhmsc5}(a)).
In this approach, an $8\times 8$ matrix representation of the Green
function is required to account for both the antiferromagnetic order
parameter $\langle c^\dagger_{i\uparrow} c^{}_{j\downarrow} \rangle$
and the superconducting order parameter $\langle c_{i\downarrow}
c_{j\uparrow} \rangle$. Figure \ref{fig:res2dhmsc5}(c) reproduces the
results for the two order parameters as a function of doping at fixed
temperature in the weak coupling regime ($U=4.8t$). The authors find
that the antiferromagnetic order parameter coexists with the $d$-wave
superconducting order parameter over a wide range of doping.
Consistent with the DCA/QMC results, the antiferromagnetic order
parameter is maximal at zero doping where the superconducting order
parameter vanishes due to the opening of the gap.

\subsubsection{Phase diagram}
\label{subsubsec:pd}

\begin{figure}[htb]
  \centerline{
    \includegraphics*[width=3.5in,angle=0]{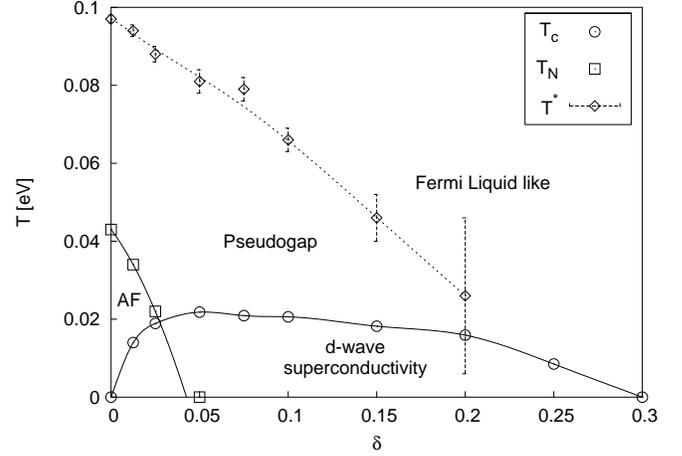}}
  \caption{Temperature-doping phase diagram of the 2D Hubbard model when $U=8t$ calculated with DCA/QMC for a 4-site cluster, $N_c=4$. The error bars on $T^*$ result from the difficulty in locating the maximum in the uniform spin susceptibility. Regions of antiferromagnetism, $d$-wave superconducting and pseudogap behavior are seen.}
\label{fig:res2dhmpd1} 
\end{figure}

The results reviewed in the preceding sections illustrate that quantum
cluster approaches applied to the 2D Hubbard model are able to capture
the complex behavior observed in HTSC. The qualitative agreement with
experiments is summarized in the $N_c=4$ DCA/QMC temperature-doping
($T$-$\delta$) phase diagram of the 2D Hubbard model in the
intermediate coupling regime $U=W$ shown in Fig.~\ref{fig:res2dhmpd1}.
The phase boundaries were determined by the instabilities of the
paramagnetic phase as indicated by the divergence of the corresponding
susceptibilities.  Therefore these results do not allow any
conclusions about a possible coexistence of the antiferromagnetic and
$d$-wave superconducting phases for $\delta<0.5$. The results by
\textcite{kats:dca} however suggest this coexistence at least for weak
coupling (see Fig.~\ref{fig:res2dhmsc5}(c)).

The pseudogap crossover temperature $T^*$ determined by the peak in
the uniform spin susceptibility (see Fig.~\ref{fig:res2dhmpg2}) serves
as a boundary separating the observed Fermi liquid and non-Fermi
liquid behavior. For $T<T^*$ the self-energy shows non-Fermi liquid
character for the parts on the Fermi surface near $\k=(\pi,0)$ (see
Fig.~\ref{fig:res2dhmpg4}).  Quasiparticle and low-energy spin
excitations are suppressed as indicated by the pseudogap in the DOS
and the spin-susceptibility (see Fig.~\ref{fig:res2dhmsc3}). At
$\delta \gtrsim 0.2$ Fermi-liquid behavior is recovered (see
Fig.~\ref{fig:res2dhmpg4}). At low temperatures, the systems is
antiferromagnetic near half-filling. The $d$-wave superconducting
phase at finite doping has its maximum transition temperature at
$\delta\approx 0.05$.

As indicated in Sec.~\ref{subsubsec:pg}, the $N_c=4$ DCA cluster
favors the spin-charge separated RVB state. The $N_c=4$ results may
thus be interpreted within the RVB picture \cite{anderson:rvb}: The
pairing of spins in singlets below the crossover temperature $T^*$
results in the suppression of low-energy spin excitations and
consequently in a pseudogap in the density of states.  Charge
excitations are quasi-free as indicated by the zero-frequency peak in
the charge susceptibility (see Fig.~\ref{fig:res2dhmsc3}).  Well below
the transition spin and charge degrees of freedom recombine, forming
electrons which pair.  Frustrated kinetic energy is recovered as
indicated by the reduction of the kinetic energy as the system goes
superconducting (see Fig.~\ref{fig:res2dhmsc4}).

Although the properties of HTSC are well described by the $N_c=4$
results, it is important to ask the question whether the phase diagram
and in particular the observed RVB nature of the results are stable
when the cluster size is increased. While the pseudogap temperature
$T^*$ may be expected to persist with increasing $N_c$, the transition
temperatures to the antiferromagnetic and superconducting state are
expected to fall to zero, as discussed in Sec.~\ref{subsec:ccw} (see
also Fig.~\ref{fig:res2dhmaf2}). This has been confirmed through
simulations of larger clusters \cite{tmaier:lcluster}.  A finite
inter-planar coupling between an infinite set of Hubbard planes may
then be used to stabilize the transitions at larger $N_c$, at least to
the antiferromagnetic state \cite{jarrell:dca3}.  The stability of the
RVB behavior and even the superconductivity is less clear and more
complex models may be necessary for them to persist in larger
clusters.

\subsubsection{Studies of related models}
\label{subsubsec:rm}

\paragraph{Stripes in the t-J-model.}

Several numerical studies indicate that there is a tendency for doped
holes to form stripes separated by antiferromagnetic domains in
strongly correlated systems (for a review see
\onlinecite{dagotto:rmp}). CPT is the quantum cluster method of choice
to study the large unit cells in the inhomogeneous stripe phase, which
complicate if not preclude the application of the numerically more
expensive embedded cluster techniques (CDMFT, DCA).  A thorough study
of stripes in the large $U$ limit of the Hubbard model, the t-J model,
was conducted by \textcite{zacher:cpt1,zacher:cpt2} within the CPT. To
implement the CPT for this problem, the authors divided the lattice
into alternating clusters of metallic stripes and antiferromagnetic
domains. The inter-cluster hopping linking these clusters was treated
perturbatively within the CPT. The enforced stripe pattern in this
implementation prohibits to explore the stability of stripes, but
allows to investigate the effects of the stripe pattern on the
single-particle excitations. In systems with less than 12\% doping the
technique was shown to reproduce salient ARPES features in selected
HTSC if a site-centered 3+1 stripe pattern, i.e. half-filled
antiferromagnetic three-leg ladders separated by doped one-leg chains,
was chosen \cite{zacher:cpt1}. At higher dopings the comparison with
ARPES indicates that the weight of bond-centered stripes with a 2+2
pattern increases in which excess holes proliferate out of the stripes
into the antiferromagnetic domain \cite{zacher:cpt2}.

\paragraph{Spectral properties of the three-band Hubbard model.}

In the context of HTSC, the single-band Hubbard model can be viewed as
a low-energy approximation of the more complex and more realistic
three-band Hubbard model \cite{emery:87,emery:88}. The three-band
Hubbard model takes into account the $p_x$ and $p_y$ oxygen orbitals
in addition to the Cu $d$ degrees of freedom in the superconducting
CuO$_2$ planes. The CPT study of its spectral properties by
\textcite{hanke:cluster} shows very good agreement with ARPES data on
HTSC at half-filling as well as in the doped system including a hole-like
Fermi surface at high doping which splits into an electron- and hole-like
branch when a bilayer hopping $t_\perp$ is included.

\paragraph{Cluster simulations of the periodic Anderson model.}

The periodic Anderson model (PAM) is widely considered to be a paradigm
for a variety of rare-earth and actinide compounds including the
heavy Fermion systems.  It is composed of a strongly hybridizing
band of $d$ electrons and a weakly hybridizing band of correlated
$f$-electrons described by the Hamiltonian
\begin{eqnarray}
H &=& -t\sum_{ij,\sigma}(d^\dagger_{i\sigma} d^{}_{j\sigma} + h.c.) 
       + \sum_{i} \epsilon_d ( n^d_{i\uparrow}+n^d_{i\downarrow} )
       \nonumber \\
  &+& \sum_{i} \epsilon_f  ( n^f_{i\uparrow}+n^f_{i\downarrow} ) 
       + U\sum_i n^f_{i\uparrow} n^f_{\downarrow}  
\end{eqnarray} 


DMFT simulations of the PAM reveal an antiferromagnetic insulating
phase at half filling of both the $f$ and $d$ bands.  The gap is set
by the Kondo coherence scale $T_0$ which is strongly enhanced compared
to the single-impurity model (SIAM) scale. When the $d$-band is doped
away from half filling while the $f$-band remains roughly half filled,
the system becomes more metallic, and the Kondo scale is strongly
suppressed compared to that of the SIAM.  In both the insulator and
metal, the temperature dependence of the impurity susceptibility and
many other low temperature quantities deviate strongly from that of
the SIAM \cite{niki:pam1}.

Non-local corrections were studied by \textcite{shimizu:pam} who used
the DCA together with the NCA as a cluster solver to study the
single-particle spectra.  He finds large deviations from the DMFT
result due to the effects of RKKY exchange.  At half filling, where
the RKKY exchange is strong and antiferromagnetic, he finds a large
gap of the order of the RKKY exchange energy.  When the filling of the
conduction band is small and the RKKY exchange is weaker and
ferromagnetic, the coherence peak is restored.








\section{Conclusions and perspectives}
\label{sec:conc}

In this review we tried to convey the message that quantum cluster
approaches provide powerful theoretical tools to study the rich
phenomenology in systems dominated by strong electronic interactions,
such as most notably transition metal oxides, heavy Fermion and
one-dimensional systems including superconducting and magnetic
compounds. Quantum cluster approaches are non-perturbative in nature,
their quality can be systematically improved by increasing the cluster
size, and they provide complementary information to finite size
simulations. By mapping the lattice problem to a finite size cluster
they describe short-ranged correlations within the cluster accurately
while approximating longer-ranged physics on the mean-field level. Of
the various attempts to add non-local corrections to local
approximations such as DMFT, we focused in this review on three
established quantum cluster approaches which we believe to play the
major role in the description of many-particle systems.

The cluster perturbation theory provides a very economic way to
calculate the single-particle dynamics by continuing the results of an
isolated finite size cluster to the thermodynamic limit.  For a
cluster consisting of a single site only this method is identical to
the Hubbard-I approximation, while it recovers the exact result in the
infinite size cluster limit. When combined with the self-energy
functional approach, the CPT can also be used to study instabilities
to symmetry broken phases.

Both embedded cluster techniques, the dynamical cluster approximation
and the cellular dynamical mean field theory are superior to the CPT
in that they map the lattice to an embedded cluster instead of the
CPT's isolated cluster.  This leads to a self-consistent theory with a
single-particle coupling between the cluster and the host. As a result
both DCA and CDMFT naturally allow for the study of phase-transitions
and they provide thermodynamically consistent results on the one- and
two-particle level.

CDMFT can be viewed as a direct generalization of DMFT to a cluster in
real space.  The mapping between lattice and cluster problems is
identical to that of the long established MCPA for disordered systems.
It leads to a cluster with open boundary conditions which violates
translational symmetries. In contrast the DCA cluster is defined in
cluster reciprocal space. Hence the DCA cluster has periodic boundary
conditions and therefore preserves the translational symmetries of the
lattice. This difference in boundary conditions translates directly to
different asymptotic behaviors for large linear cluster sizes $L_c$,
and the decision which method to use for a given problem strongly
depends on the quantities of interest: Local quantities, such as the
local density of states when defined on central cluster sites converge
faster in the CDMFT since they do not directly couple to the
mean-field. Due to the large mean-field coupling of the surface sites
however, the CDMFT converges slowly, with corrections of order ${\cal
  O}(1/L_c)$, for quantities extended over the cluster. The DCA
converges more quickly, as ${\cal O}(1/L_c^2)$, due to the periodic
boundary conditions on the cluster.

Quantum cluster approaches reduce the complexity of the infinite
lattice problem by mapping it to a cluster with fewer degrees of
freedom. The numerous methods employed to solve the DMFT equations are
in principle available to study the effective cluster model.  However
as the complexity of this task rapidly increases with cluster size,
potential cluster solvers are faced with severe size limitations.
As the most promising 
techniques we reviewed perturbative approaches including the
fluctuation-exchange approximation and the non-crossing approximation
as well as non-perturbative techniques including quantum Monte Carlo
and numerical renormalization group.

We discussed the application of quantum cluster approaches to a wide
range of problems in condensed matter physics. The information gained
from these studies has led to significant progress in the field of
strongly correlated electron systems. Even studies using small cluster
sizes opened up new insight in problems such as one-dimensional
systems, the Mott-Hubbard transition in two dimensions and
high-temperature superconductivity.

To gain conclusive evidence however, larger cluster size studies are
inevitable to verify the information obtained from small clusters.
This task is severely hampered by the rapidly increasing complexity
with cluster size. Future progress therefore is closely linked to
improvements in the efficiency and flexibility of the techniques used
to solve the effective cluster problem. Within quantum cluster
approaches we explore a coarse-graining approximation in $k$-space. To
further reduce the complexity, the same idea could be transfered to
the frequency domain. By coarse-graining the frequency dependence of
irreducible quantities, correlations on long time scales are neglected, 
while the short time scale behavior is described accurately. An important 
aspect in this context is again the causality question. First test results 
are encouraging; they show that coarse-graining in Matsubara space leads 
to acausalities, while coarse-graining the real frequency axis does not 
face this problem \cite{karan:dca1}.

Another route to defeat the cluster size problem is to develop hybrid
algorithms that treat different length-scales in the problem with
different accuracy. As a promising step in this direction,
\textcite{hague:03} have developed a hybrid technique which maps the
infinite lattice onto two embedded clusters of different size, thus
dividing the problem into three length-scales. Short-ranged
correlations described by the small cluster are treated accurately
within QMC, correlations of intermediate length-scale are treated
perturbatively in the large cluster using FLEX, and the long-ranged
physics beyond the size of the larger cluster is again approximated on
the mean-field level.

To improve comparisons with experiments and to achieve 
predictive capability, the inclusion of the specifics 
of the actual materials is required. Along the lines of the LDA+DMFT
approach, one can use  electronic structure calculations to parameterize 
the models studied by quantum cluster approaches.
First steps in this direction have been made by
\textcite{poteryaev:03}. A more integrated approach to the ab-initio
description of strongly correlated systems by combining the ideas of
density functional theory and quantum cluster approaches remains
an important and challenging task.


\section*{Acknowledgments} 
It is a pleasure to acknowledge useful discussions with 
H.~R.~Krishnamurthy, 
J.~Keller,
H.~Keiter, 
E.~M\"uller-Hartmann 
P.~van Dongen 
A.~Gonis, 
B.~Gyorffy,
M.~Mukherjee, 
A.~N.~Tavildar-Zadeh, 
G.~Baskaran,
M.~Vojta, 
R.~Bulla, 
A.~Lichtenstein, 
K.~Aryanpour
and 
G. Kotliar.  This work was
supported in part by NSF grants DMR-0312680, DMR-0113574, and DMR-0073308
and by the DFG through SFB 484.  We acknowledge supercomputer support by 
the Leibniz Computer Center, the Max-Planck Computer Center Garching 
under grant h0301, the Ohio Supercomputer Center, the Pittsburgh Supercomputer
Center (NSF) and the Center for Computational Sciences at the Oak Ridge 
National Laboratory. Part of this research was performed by TM as Eugene 
P.\  Wigner Fellow and staff member at the Oak Ridge National Laboratory, 
managed by UT-Battelle, LLC, for the U.S. Department of Energy under Contract
DE-AC05-00OR22725.

\bibliography{mybib1,mhh}

\begin{thebibliography}{182}
\expandafter\ifx\csname natexlab\endcsname\relax\def\natexlab#1{#1}\fi
\expandafter\ifx\csname bibnamefont\endcsname\relax
  \def\bibnamefont#1{#1}\fi
\expandafter\ifx\csname bibfnamefont\endcsname\relax
  \def\bibfnamefont#1{#1}\fi
\expandafter\ifx\csname citenamefont\endcsname\relax
  \def\citenamefont#1{#1}\fi
\expandafter\ifx\csname url\endcsname\relax
  \def\url#1{\texttt{#1}}\fi
\expandafter\ifx\csname urlprefix\endcsname\relax\def\urlprefix{URL }\fi
\providecommand{\bibinfo}[2]{#2}
\providecommand{\eprint}[2][]{\url{#2}}

\bibitem[{\citenamefont{Abrikosov} \emph{et~al.}(1963)\citenamefont{Abrikosov,
  Gorkov, and Dzyaloshinski}}]{AGD}
\bibinfo{author}{\bibnamefont{Abrikosov}, \bibfnamefont{A.}},
  \bibinfo{author}{\bibfnamefont{L.}~\bibnamefont{Gorkov}}, and
  \bibinfo{author}{\bibfnamefont{I.}~\bibnamefont{Dzyaloshinski}},
  \bibinfo{year}{1963}, \emph{\bibinfo{title}{Methods of Quantum Field Theory
  in Statistical Physics}} (\bibinfo{publisher}{Dover}).

\bibitem[{\citenamefont{Aichhorn} \emph{et~al.}(2003)\citenamefont{Aichhorn,
  Daghofer, Evertz, and v.d. Linden}}]{aichhorn:03}
\bibinfo{author}{\bibnamefont{Aichhorn}, \bibfnamefont{M.}},
  \bibinfo{author}{\bibfnamefont{M.}~\bibnamefont{Daghofer}},
  \bibinfo{author}{\bibfnamefont{H.}~\bibnamefont{Evertz}}, and
  \bibinfo{author}{\bibfnamefont{W.}~\bibnamefont{v.d. Linden}},
  \bibinfo{year}{2003}, \bibinfo{journal}{Phys. Rev. B}
  \textbf{\bibinfo{volume}{67}}, \bibinfo{pages}{161103}.

\bibitem[{\citenamefont{Anderson}(1961)}]{pwa:siam}
\bibinfo{author}{\bibnamefont{Anderson}, \bibfnamefont{P.}},
  \bibinfo{year}{1961}, \bibinfo{journal}{Phys. Rev.}
  \textbf{\bibinfo{volume}{124}}, \bibinfo{pages}{41}.

\bibitem[{\citenamefont{Anderson}(1997{\natexlab{a}})}]{anderson:1dhm}
\bibinfo{author}{\bibnamefont{Anderson}, \bibfnamefont{P.}},
  \bibinfo{year}{1997}{\natexlab{a}}, \bibinfo{journal}{Adv. Phys.}
  \textbf{\bibinfo{volume}{46}}, \bibinfo{pages}{3}.

\bibitem[{\citenamefont{Anderson}(1987)}]{anderson:rvb}
\bibinfo{author}{\bibnamefont{Anderson}, \bibfnamefont{P.~W.}},
  \bibinfo{year}{1987}, \bibinfo{journal}{Science}
  \textbf{\bibinfo{volume}{235}}, \bibinfo{pages}{1196}.

\bibitem[{\citenamefont{Anderson}(1997{\natexlab{b}})}]{anderson:htsc}
\bibinfo{author}{\bibnamefont{Anderson}, \bibfnamefont{P.~W.}},
  \bibinfo{year}{1997}{\natexlab{b}}, \emph{\bibinfo{title}{The Theory of
  Superconductivity in the High-$T_c$ Cuprates}} (\bibinfo{publisher}{Princeton
  University Press}, \bibinfo{address}{Princeton, NJ}).

\bibitem[{\citenamefont{Aryanpour} \emph{et~al.}(2002)\citenamefont{Aryanpour,
  Hettler, and Jarrell}}]{karan:dca3}
\bibinfo{author}{\bibnamefont{Aryanpour}, \bibfnamefont{K.}},
  \bibinfo{author}{\bibfnamefont{M.~H.} \bibnamefont{Hettler}}, and
  \bibinfo{author}{\bibfnamefont{M.}~\bibnamefont{Jarrell}},
  \bibinfo{year}{2002}, \bibinfo{journal}{Phys. Rev. B}
  \textbf{\bibinfo{volume}{65}}, \bibinfo{pages}{153102}.

\bibitem[{\citenamefont{Aryanpour}
  \emph{et~al.}(2003{\natexlab{a}})\citenamefont{Aryanpour, Hettler, and
  Jarrell}}]{karan:dca1}
\bibinfo{author}{\bibnamefont{Aryanpour}, \bibfnamefont{K.}},
  \bibinfo{author}{\bibfnamefont{M.~H.} \bibnamefont{Hettler}}, and
  \bibinfo{author}{\bibfnamefont{M.}~\bibnamefont{Jarrell}},
  \bibinfo{year}{2003}{\natexlab{a}}, \bibinfo{journal}{Phys. Rev. B}
  \textbf{\bibinfo{volume}{67}}, \bibinfo{pages}{085101}.

\bibitem[{\citenamefont{Aryanpour}
  \emph{et~al.}(2003{\natexlab{b}})\citenamefont{Aryanpour, Maier, and
  Jarrell}}]{karan:dca2}
\bibinfo{author}{\bibnamefont{Aryanpour}, \bibfnamefont{K.}},
  \bibinfo{author}{\bibfnamefont{T.~A.} \bibnamefont{Maier}}, and
  \bibinfo{author}{\bibfnamefont{M.}~\bibnamefont{Jarrell}},
  \bibinfo{year}{2003}{\natexlab{b}}, \bibinfo{journal}{preprint
  cond-mat/0301460} .

\bibitem[{\citenamefont{Baym}(1962)}]{baym}
\bibinfo{author}{\bibnamefont{Baym}, \bibfnamefont{G.}}, \bibinfo{year}{1962},
  \bibinfo{journal}{Phys. Rev.} \textbf{\bibinfo{volume}{127}},
  \bibinfo{pages}{1391}.

\bibitem[{\citenamefont{Baym and Kadanoff}(1961)}]{baym:61}
\bibinfo{author}{\bibnamefont{Baym}, \bibfnamefont{G.}}, and
  \bibinfo{author}{\bibfnamefont{L.}~\bibnamefont{Kadanoff}},
  \bibinfo{year}{1961}, \bibinfo{journal}{Phys. Rev.}
  \textbf{\bibinfo{volume}{124}}, \bibinfo{pages}{287}.

\bibitem[{\citenamefont{Becker and Keller}(1987)}]{keller:nca}
\bibinfo{author}{\bibnamefont{Becker}, \bibfnamefont{K.}}, and
  \bibinfo{author}{\bibfnamefont{J.}~\bibnamefont{Keller}},
  \bibinfo{year}{1987}, \bibinfo{journal}{Z.\ Phys.\ {\bf B}}
  \textbf{\bibinfo{volume}{36}}, \bibinfo{pages}{2036}.

\bibitem[{\citenamefont{Beeby and Edwards}(1962)}]{ATA:beeby}
\bibinfo{author}{\bibnamefont{Beeby}, \bibfnamefont{J.~L.}}, and
  \bibinfo{author}{\bibfnamefont{S.~F.} \bibnamefont{Edwards}},
  \bibinfo{year}{1962}, \bibinfo{journal}{Proc. Roy. Soc. (London)}
  \textbf{\bibinfo{volume}{A274}}, \bibinfo{pages}{395}.

\bibitem[{\citenamefont{Bethe}(1935)}]{bethe:cluster}
\bibinfo{author}{\bibnamefont{Bethe}, \bibfnamefont{H.~A.}},
  \bibinfo{year}{1935}, \bibinfo{journal}{Proc. R. Soc. London}
  \textbf{\bibinfo{volume}{A150}}, \bibinfo{pages}{552}.

\bibitem[{\citenamefont{Betts and Stewart}(1997)}]{betts:cubicclusters}
\bibinfo{author}{\bibnamefont{Betts}, \bibfnamefont{D.}}, and
  \bibinfo{author}{\bibfnamefont{G.}~\bibnamefont{Stewart}},
  \bibinfo{year}{1997}, \bibinfo{journal}{Can.\ J.\ Phys.}
  \textbf{\bibinfo{volume}{75}}, \bibinfo{pages}{47}.

\bibitem[{\citenamefont{Bickers}(1987)}]{Bickers:NCA}
\bibinfo{author}{\bibnamefont{Bickers}, \bibfnamefont{N.}},
  \bibinfo{year}{1987}, \bibinfo{journal}{Rev. Mod. Phys.}
  \textbf{\bibinfo{volume}{59}}, \bibinfo{pages}{845}.

\bibitem[{\citenamefont{Bickers} \emph{et~al.}(1987)\citenamefont{Bickers, Cox,
  and Wilkins}}]{BCW:NCA}
\bibinfo{author}{\bibnamefont{Bickers}, \bibfnamefont{N.}},
  \bibinfo{author}{\bibfnamefont{D.}~\bibnamefont{Cox}}, and
  \bibinfo{author}{\bibfnamefont{J.}~\bibnamefont{Wilkins}},
  \bibinfo{year}{1987}, \bibinfo{journal}{Phys. Rev. B}
  \textbf{\bibinfo{volume}{36}}, \bibinfo{pages}{2036}.

\bibitem[{\citenamefont{Bickers} \emph{et~al.}(1989)\citenamefont{Bickers,
  Scalapino, and White}}]{bickers:89}
\bibinfo{author}{\bibnamefont{Bickers}, \bibfnamefont{N.}},
  \bibinfo{author}{\bibfnamefont{D.}~\bibnamefont{Scalapino}}, and
  \bibinfo{author}{\bibfnamefont{S.}~\bibnamefont{White}},
  \bibinfo{year}{1989}, \bibinfo{journal}{Phys. Rev. Lett.}
  \textbf{\bibinfo{volume}{62}}, \bibinfo{pages}{961}.

\bibitem[{\citenamefont{Bickers and White}(1990)}]{flex:bickers2}
\bibinfo{author}{\bibnamefont{Bickers}, \bibfnamefont{N.}}, and
  \bibinfo{author}{\bibfnamefont{S.~R.} \bibnamefont{White}},
  \bibinfo{year}{1990}, \bibinfo{journal}{Phys. Rev. B}
  \textbf{\bibinfo{volume}{43}}, \bibinfo{pages}{8044}.

\bibitem[{\citenamefont{Biroli and Kotliar}(2002)}]{kotliar:cdmft2}
\bibinfo{author}{\bibnamefont{Biroli}, \bibfnamefont{G.}}, and
  \bibinfo{author}{\bibfnamefont{G.}~\bibnamefont{Kotliar}},
  \bibinfo{year}{2002}, \bibinfo{journal}{Phys. Rev. B}
  \textbf{\bibinfo{volume}{65}}, \bibinfo{pages}{155112}.

\bibitem[{\citenamefont{Biroli} \emph{et~al.}(2003)\citenamefont{Biroli,
  Parcollet, and Kotliar}}]{biroli:cdmft}
\bibinfo{author}{\bibnamefont{Biroli}, \bibfnamefont{G.}},
  \bibinfo{author}{\bibfnamefont{O.}~\bibnamefont{Parcollet}}, and
  \bibinfo{author}{\bibfnamefont{G.}~\bibnamefont{Kotliar}},
  \bibinfo{year}{2003}, \bibinfo{journal}{preprint con-mat/0307587} .

\bibitem[{\citenamefont{Bolech} \emph{et~al.}(2003)\citenamefont{Bolech,
  Kancharla, and Kotliar}}]{bolech:cluster}
\bibinfo{author}{\bibnamefont{Bolech}, \bibfnamefont{C.}},
  \bibinfo{author}{\bibfnamefont{S.}~\bibnamefont{Kancharla}}, and
  \bibinfo{author}{\bibfnamefont{G.}~\bibnamefont{Kotliar}},
  \bibinfo{year}{2003}, \bibinfo{journal}{Phys. Rev. B}
  \textbf{\bibinfo{volume}{67}}, \bibinfo{pages}{075110}.

\bibitem[{\citenamefont{Brandt and Mielsch}(1989)}]{brandt:fkm3}
\bibinfo{author}{\bibnamefont{Brandt}, \bibfnamefont{U.}}, and
  \bibinfo{author}{\bibfnamefont{U.}~\bibnamefont{Mielsch}},
  \bibinfo{year}{1989}, \bibinfo{journal}{Z. Phys. B}
  \textbf{\bibinfo{volume}{75}}, \bibinfo{pages}{365}.

\bibitem[{\citenamefont{Brandt and Schmidt}(1986)}]{brandt:fkm1}
\bibinfo{author}{\bibnamefont{Brandt}, \bibfnamefont{U.}}, and
  \bibinfo{author}{\bibfnamefont{R.}~\bibnamefont{Schmidt}},
  \bibinfo{year}{1986}, \bibinfo{journal}{Z. Phys. B}
  \textbf{\bibinfo{volume}{63}}, \bibinfo{pages}{45}.

\bibitem[{\citenamefont{Brandt and Schmidt}(1987)}]{brandt:fkm2}
\bibinfo{author}{\bibnamefont{Brandt}, \bibfnamefont{U.}}, and
  \bibinfo{author}{\bibfnamefont{R.}~\bibnamefont{Schmidt}},
  \bibinfo{year}{1987}, \bibinfo{journal}{Z. Phys. B}
  \textbf{\bibinfo{volume}{67}}, \bibinfo{pages}{43}.

\bibitem[{\citenamefont{Bulla}(1999)}]{dmftnrg:bulla1}
\bibinfo{author}{\bibnamefont{Bulla}, \bibfnamefont{R.}}, \bibinfo{year}{1999},
  \bibinfo{journal}{Phys. Rev. Lett.} \textbf{\bibinfo{volume}{83}},
  \bibinfo{pages}{136}.

\bibitem[{\citenamefont{Bulla} \emph{et~al.}(2001)\citenamefont{Bulla, Costi,
  and Vollhardt}}]{dmftnrg:bulla2}
\bibinfo{author}{\bibnamefont{Bulla}, \bibfnamefont{R.}},
  \bibinfo{author}{\bibfnamefont{T.}~\bibnamefont{Costi}}, and
  \bibinfo{author}{\bibfnamefont{D.}~\bibnamefont{Vollhardt}},
  \bibinfo{year}{2001}, \bibinfo{journal}{Phys. Rev. B}
  \textbf{\bibinfo{volume}{64}}, \bibinfo{pages}{045103}.

\bibitem[{\citenamefont{Bulla} \emph{et~al.}(1998)\citenamefont{Bulla, Hewson,
  and Pruschke}}]{nrg:bulla1}
\bibinfo{author}{\bibnamefont{Bulla}, \bibfnamefont{R.}},
  \bibinfo{author}{\bibfnamefont{A.}~\bibnamefont{Hewson}}, and
  \bibinfo{author}{\bibfnamefont{T.}~\bibnamefont{Pruschke}},
  \bibinfo{year}{1998}, \bibinfo{journal}{J. Phys.: Condens. Matter}
  \textbf{\bibinfo{volume}{10}}, \bibinfo{pages}{8365}.

\bibitem[{\citenamefont{Caffarel and Krauth}(1994)}]{caffarel}
\bibinfo{author}{\bibnamefont{Caffarel}, \bibfnamefont{M.}}, and
  \bibinfo{author}{\bibfnamefont{W.}~\bibnamefont{Krauth}},
  \bibinfo{year}{1994}, \bibinfo{journal}{Phys. Rev. Lett.}
  \textbf{\bibinfo{volume}{72}}, \bibinfo{pages}{1545}.

\bibitem[{\citenamefont{Cox and Grewe}(1987)}]{CoxGrewe}
\bibinfo{author}{\bibnamefont{Cox}, \bibfnamefont{D.}}, and
  \bibinfo{author}{\bibfnamefont{N.}~\bibnamefont{Grewe}},
  \bibinfo{year}{1987}, \bibinfo{journal}{Z.\ Phys.\ {\bf B}}
  \textbf{\bibinfo{volume}{71}}, \bibinfo{pages}{321}.

\bibitem[{\citenamefont{Creffield} \emph{et~al.}(1995)\citenamefont{Creffield,
  Klepfish, Pike, and Sarkar}}]{creffield:95}
\bibinfo{author}{\bibnamefont{Creffield}, \bibfnamefont{C.}},
  \bibinfo{author}{\bibfnamefont{E.}~\bibnamefont{Klepfish}},
  \bibinfo{author}{\bibfnamefont{E.}~\bibnamefont{Pike}}, and
  \bibinfo{author}{\bibfnamefont{S.}~\bibnamefont{Sarkar}},
  \bibinfo{year}{1995}, \bibinfo{journal}{Phys. Rev. Lett.}
  \textbf{\bibinfo{volume}{75}}, \bibinfo{pages}{517}.

\bibitem[{\citenamefont{Dagotto}(1994)}]{dagotto:rmp}
\bibinfo{author}{\bibnamefont{Dagotto}, \bibfnamefont{E.}},
  \bibinfo{year}{1994}, \bibinfo{journal}{Rev. Mod. Phys}
  \textbf{\bibinfo{volume}{66}}, \bibinfo{pages}{763}.

\bibitem[{\citenamefont{Dagotto and Moreo}(1988)}]{dagotto:88}
\bibinfo{author}{\bibnamefont{Dagotto}, \bibfnamefont{E.}}, and
  \bibinfo{author}{\bibfnamefont{A.}~\bibnamefont{Moreo}},
  \bibinfo{year}{1988}, \bibinfo{journal}{Phys. Rev. B}
  \textbf{\bibinfo{volume}{38}}, \bibinfo{pages}{5078}.

\bibitem[{\citenamefont{Dahnken} \emph{et~al.}(2003)\citenamefont{Dahnken,
  Aichhorn, Hanke, Arrigoni, and Potthoff}}]{potthoff:cluster3}
\bibinfo{author}{\bibnamefont{Dahnken}, \bibfnamefont{C.}},
  \bibinfo{author}{\bibfnamefont{M.}~\bibnamefont{Aichhorn}},
  \bibinfo{author}{\bibfnamefont{W.}~\bibnamefont{Hanke}},
  \bibinfo{author}{\bibfnamefont{E.}~\bibnamefont{Arrigoni}}, and
  \bibinfo{author}{\bibfnamefont{M.}~\bibnamefont{Potthoff}},
  \bibinfo{year}{2003}, \bibinfo{journal}{preprint cond-mat/0309407} .

\bibitem[{\citenamefont{Dahnken} \emph{et~al.}(2002)\citenamefont{Dahnken,
  Arrigoni, and Hanke}}]{hanke:cluster}
\bibinfo{author}{\bibnamefont{Dahnken}, \bibfnamefont{C.}},
  \bibinfo{author}{\bibfnamefont{E.}~\bibnamefont{Arrigoni}}, and
  \bibinfo{author}{\bibfnamefont{W.}~\bibnamefont{Hanke}},
  \bibinfo{year}{2002}, \bibinfo{journal}{J. Low Temp. Phys.}
  \textbf{\bibinfo{volume}{126}}, \bibinfo{pages}{949}.

\bibitem[{\citenamefont{Deisz} \emph{et~al.}(1996)\citenamefont{Deisz, Hess,
  and Serene}}]{deisz:96}
\bibinfo{author}{\bibnamefont{Deisz}, \bibfnamefont{J.}},
  \bibinfo{author}{\bibfnamefont{D.}~\bibnamefont{Hess}}, and
  \bibinfo{author}{\bibfnamefont{J.}~\bibnamefont{Serene}},
  \bibinfo{year}{1996}, \bibinfo{journal}{Phys. Rev. Lett.}
  \textbf{\bibinfo{volume}{76}}, \bibinfo{pages}{1312}.

\bibitem[{\citenamefont{Deisz} \emph{et~al.}(2003)\citenamefont{Deisz, Hess,
  and Serene}}]{flex:dhs}
\bibinfo{author}{\bibnamefont{Deisz}, \bibfnamefont{J.}},
  \bibinfo{author}{\bibfnamefont{D.}~\bibnamefont{Hess}}, and
  \bibinfo{author}{\bibfnamefont{J.}~\bibnamefont{Serene}},
  \bibinfo{year}{2003}, \emph{\bibinfo{title}{Recent Progress in many Body
  Theories}}, volume~\bibinfo{volume}{4} (\bibinfo{publisher}{Plenum, New
  York}).

\bibitem[{\citenamefont{Ducastelle}(1974)}]{duc:mcpa}
\bibinfo{author}{\bibnamefont{Ducastelle}, \bibfnamefont{F.}},
  \bibinfo{year}{1974}, \bibinfo{journal}{J. Phys C}
  \textbf{\bibinfo{volume}{7}}, \bibinfo{pages}{1795}.

\bibitem[{\citenamefont{Emery}(1987)}]{emery:87}
\bibinfo{author}{\bibnamefont{Emery}, \bibfnamefont{V.}}, \bibinfo{year}{1987},
  \bibinfo{journal}{Phys. Rev. Lett.} \textbf{\bibinfo{volume}{58}},
  \bibinfo{pages}{2794}.

\bibitem[{\citenamefont{Emery and Reiter}(1988)}]{emery:88}
\bibinfo{author}{\bibnamefont{Emery}, \bibfnamefont{V.}}, and
  \bibinfo{author}{\bibfnamefont{G.}~\bibnamefont{Reiter}},
  \bibinfo{year}{1988}, \bibinfo{journal}{Phys. Rev. B}
  \textbf{\bibinfo{volume}{38}}, \bibinfo{pages}{4547}.

\bibitem[{\citenamefont{Fetter and Walecka}(1971)}]{QTMPS:fetter}
\bibinfo{author}{\bibnamefont{Fetter}, \bibfnamefont{A.}}, and
  \bibinfo{author}{\bibfnamefont{J.}~\bibnamefont{Walecka}},
  \bibinfo{year}{1971}, \emph{\bibinfo{title}{Quantum Theory of Many-Particle
  Systems}}, International Series in Pure and Applied Physics
  (\bibinfo{publisher}{McGraw-Hill}, \bibinfo{address}{New York}).

\bibitem[{\citenamefont{Fischer}(1997)}]{Fischer:NCA}
\bibinfo{author}{\bibnamefont{Fischer}, \bibfnamefont{K.}},
  \bibinfo{year}{1997}, \bibinfo{journal}{Phys. Rev. B}
  \textbf{\bibinfo{volume}{55}}, \bibinfo{pages}{13575}.

\bibitem[{\citenamefont{Fisher and Barber}(1972)}]{fisher:72}
\bibinfo{author}{\bibnamefont{Fisher}, \bibfnamefont{M.}}, and
  \bibinfo{author}{\bibfnamefont{M.}~\bibnamefont{Barber}},
  \bibinfo{year}{1972}, \bibinfo{journal}{Phys. Rev. Lett.}
  \textbf{\bibinfo{volume}{28}}, \bibinfo{pages}{1516}.

\bibitem[{\citenamefont{Georges and Kotliar}(1992)}]{georges:92}
\bibinfo{author}{\bibnamefont{Georges}, \bibfnamefont{A.}}, and
  \bibinfo{author}{\bibfnamefont{G.}~\bibnamefont{Kotliar}},
  \bibinfo{year}{1992}, \bibinfo{journal}{Phys. Rev. B}
  \textbf{\bibinfo{volume}{45}}, \bibinfo{pages}{6479}.

\bibitem[{\citenamefont{Georges} \emph{et~al.}(1996)\citenamefont{Georges,
  Kotliar, Krauth, and Rozenberg}}]{georges:dmftrev}
\bibinfo{author}{\bibnamefont{Georges}, \bibfnamefont{A.}},
  \bibinfo{author}{\bibfnamefont{G.}~\bibnamefont{Kotliar}},
  \bibinfo{author}{\bibfnamefont{W.}~\bibnamefont{Krauth}}, and
  \bibinfo{author}{\bibfnamefont{M.}~\bibnamefont{Rozenberg}},
  \bibinfo{year}{1996}, \bibinfo{journal}{Rev. Mod. Phys.}
  \textbf{\bibinfo{volume}{68}}, \bibinfo{pages}{13}.

\bibitem[{\citenamefont{Goldenfeld}(1992)}]{tmatrix:goldenfeld}
\bibinfo{author}{\bibnamefont{Goldenfeld}, \bibfnamefont{N.}},
  \bibinfo{year}{1992}, \emph{\bibinfo{title}{Lectures on Phase Transitions and
  the Renormalization Group}}, Frontiers in Physics
  (\bibinfo{publisher}{Addison Wesley}, \bibinfo{address}{Reading,
  Massachusetts}).

\bibitem[{\citenamefont{Gonis}(1992)}]{gonis:gf}
\bibinfo{author}{\bibnamefont{Gonis}, \bibfnamefont{A.}}, \bibinfo{year}{1992},
  \emph{\bibinfo{title}{Green Functions for Ordered and Disordered Systems}},
  Studies in Mathematical Physics (\bibinfo{publisher}{North-Holland},
  \bibinfo{address}{Amsterdam}).

\bibitem[{\citenamefont{Grewe}(1983)}]{Grewe:NCA}
\bibinfo{author}{\bibnamefont{Grewe}, \bibfnamefont{N.}}, \bibinfo{year}{1983},
  \bibinfo{journal}{Z.\ Phys.\ {\bf B}} \textbf{\bibinfo{volume}{53}},
  \bibinfo{pages}{271}.

\bibitem[{\citenamefont{Grewe}(1987)}]{Grewe:LNCA}
\bibinfo{author}{\bibnamefont{Grewe}, \bibfnamefont{N.}}, \bibinfo{year}{1987},
  \bibinfo{journal}{Z.\ Phys.\ {\bf B}} \textbf{\bibinfo{volume}{67}},
  \bibinfo{pages}{323}.

\bibitem[{\citenamefont{Grewe} \emph{et~al.}(1988)\citenamefont{Grewe, Keiter,
  and Pruschke}}]{Grewe:LNCA2}
\bibinfo{author}{\bibnamefont{Grewe}, \bibfnamefont{N.}},
  \bibinfo{author}{\bibfnamefont{H.}~\bibnamefont{Keiter}}, and
  \bibinfo{author}{\bibfnamefont{T.}~\bibnamefont{Pruschke}},
  \bibinfo{year}{1988}, \bibinfo{journal}{Z.\ Phys.\ {\bf B}}
  \textbf{\bibinfo{volume}{71}}, \bibinfo{pages}{75}.

\bibitem[{\citenamefont{Gros and Valenti}(1994)}]{gross:cluster}
\bibinfo{author}{\bibnamefont{Gros}, \bibfnamefont{C.}}, and
  \bibinfo{author}{\bibfnamefont{R.}~\bibnamefont{Valenti}},
  \bibinfo{year}{1994}, \bibinfo{journal}{Annalen der Phys.}
  \textbf{\bibinfo{volume}{3}}, \bibinfo{pages}{460}.

\bibitem[{\citenamefont{Hague}(2003)}]{hague:03b}
\bibinfo{author}{\bibnamefont{Hague}, \bibfnamefont{J.}}, \bibinfo{year}{2003},
  \bibinfo{journal}{J. Phys.: Cond. Mat.} \textbf{\bibinfo{volume}{15}},
  \bibinfo{pages}{2535}.

\bibitem[{\citenamefont{Hague} \emph{et~al.}(2003)\citenamefont{Hague, Jarrell,
  and Schulthess}}]{hague:03}
\bibinfo{author}{\bibnamefont{Hague}, \bibfnamefont{J.}},
  \bibinfo{author}{\bibfnamefont{M.}~\bibnamefont{Jarrell}}, and
  \bibinfo{author}{\bibfnamefont{T.}~\bibnamefont{Schulthess}},
  \bibinfo{year}{2003}, \bibinfo{journal}{preprint cond-mat/0312155} .

\bibitem[{\citenamefont{Haydock} \emph{et~al.}(1975)\citenamefont{Haydock,
  Heine, and Kelly}}]{haydock}
\bibinfo{author}{\bibnamefont{Haydock}, \bibfnamefont{R.}},
  \bibinfo{author}{\bibfnamefont{V.}~\bibnamefont{Heine}}, and
  \bibinfo{author}{\bibfnamefont{M.~J.} \bibnamefont{Kelly}},
  \bibinfo{year}{1975}, \bibinfo{journal}{J. Phys. C}
  \textbf{\bibinfo{volume}{8}}, \bibinfo{pages}{2591}.

\bibitem[{\citenamefont{Heindl} \emph{et~al.}(2000)\citenamefont{Heindl,
  Pruschke, and Keller}}]{Heindl:NCA}
\bibinfo{author}{\bibnamefont{Heindl}, \bibfnamefont{W.}},
  \bibinfo{author}{\bibfnamefont{T.}~\bibnamefont{Pruschke}}, and
  \bibinfo{author}{\bibfnamefont{J.}~\bibnamefont{Keller}},
  \bibinfo{year}{2000}, \bibinfo{journal}{J. Phys.: Condens. Matter}
  \textbf{\bibinfo{volume}{12}}, \bibinfo{pages}{2245}.

\bibitem[{\citenamefont{Hettler} \emph{et~al.}(2000)\citenamefont{Hettler,
  Mukherjee, Jarrell, and Krishnamurthy}}]{hettler:dca2}
\bibinfo{author}{\bibnamefont{Hettler}, \bibfnamefont{M.~H.}},
  \bibinfo{author}{\bibfnamefont{M.}~\bibnamefont{Mukherjee}},
  \bibinfo{author}{\bibfnamefont{M.}~\bibnamefont{Jarrell}}, and
  \bibinfo{author}{\bibfnamefont{H.~R.} \bibnamefont{Krishnamurthy}},
  \bibinfo{year}{2000}, \bibinfo{journal}{Phys. Rev. B}
  \textbf{\bibinfo{volume}{61}}, \bibinfo{pages}{12739}.

\bibitem[{\citenamefont{Hettler} \emph{et~al.}(1998)\citenamefont{Hettler,
  Tahvildar-Zadeh, Jarrell, Pruschke, and Krishnamurthy}}]{hettler:dca1}
\bibinfo{author}{\bibnamefont{Hettler}, \bibfnamefont{M.~H.}},
  \bibinfo{author}{\bibfnamefont{A.~N.} \bibnamefont{Tahvildar-Zadeh}},
  \bibinfo{author}{\bibfnamefont{M.}~\bibnamefont{Jarrell}},
  \bibinfo{author}{\bibfnamefont{T.}~\bibnamefont{Pruschke}}, and
  \bibinfo{author}{\bibfnamefont{H.~R.} \bibnamefont{Krishnamurthy}},
  \bibinfo{year}{1998}, \bibinfo{journal}{Phys. Rev. B}
  \textbf{\bibinfo{volume}{58}}, \bibinfo{pages}{R7475}.

\bibitem[{\citenamefont{Hewson}(1993)}]{ah:siam}
\bibinfo{author}{\bibnamefont{Hewson}, \bibfnamefont{A.}},
  \bibinfo{year}{1993}, \emph{\bibinfo{title}{The Kondo Problem to Heavy
  Fermions}}, Cambridge Studies in Magnetism (\bibinfo{publisher}{Cambridge
  UNiversity Press}, \bibinfo{address}{Cambridge}).

\bibitem[{\citenamefont{Hirsch}(1983)}]{hirsch:HHS}
\bibinfo{author}{\bibnamefont{Hirsch}, \bibfnamefont{J.}},
  \bibinfo{year}{1983}, \bibinfo{journal}{Phys.\ Rev.\ B}
  \textbf{\bibinfo{volume}{28}}, \bibinfo{pages}{4059}.

\bibitem[{\citenamefont{Hirsch and Fye}(1986)}]{hirsch:QMC}
\bibinfo{author}{\bibnamefont{Hirsch}, \bibfnamefont{J.}}, and
  \bibinfo{author}{\bibfnamefont{R.}~\bibnamefont{Fye}}, \bibinfo{year}{1986},
  \bibinfo{journal}{Phys.\ Rev.\ Lett.} \textbf{\bibinfo{volume}{56}},
  \bibinfo{pages}{2521}.

\bibitem[{\citenamefont{Hohenadler}
  \emph{et~al.}(2003)\citenamefont{Hohenadler, Aichhorn, and v.d.
  Linden}}]{hohenadler:03}
\bibinfo{author}{\bibnamefont{Hohenadler}, \bibfnamefont{M.}},
  \bibinfo{author}{\bibfnamefont{M.}~\bibnamefont{Aichhorn}}, and
  \bibinfo{author}{\bibfnamefont{W.}~\bibnamefont{v.d. Linden}},
  \bibinfo{year}{2003}, \bibinfo{journal}{Phys. Rev. B}
  \textbf{\bibinfo{volume}{68}}, \bibinfo{pages}{184304}.

\bibitem[{\citenamefont{Hubbard}(1963)}]{hubbard:63}
\bibinfo{author}{\bibnamefont{Hubbard}, \bibfnamefont{J.}},
  \bibinfo{year}{1963}, \bibinfo{journal}{Proc. Royal. Soc. London}
  \textbf{\bibinfo{volume}{276}}, \bibinfo{pages}{238}.

\bibitem[{\citenamefont{Huscroft} \emph{et~al.}(2000)\citenamefont{Huscroft,
  Gass, and Jarrell}}]{huscroft:thermo}
\bibinfo{author}{\bibnamefont{Huscroft}, \bibfnamefont{C.}},
  \bibinfo{author}{\bibfnamefont{R.}~\bibnamefont{Gass}}, and
  \bibinfo{author}{\bibfnamefont{M.}~\bibnamefont{Jarrell}},
  \bibinfo{year}{2000}, \bibinfo{journal}{Phys. Rev. B}
  \textbf{\bibinfo{volume}{61}}, \bibinfo{pages}{9300}.

\bibitem[{\citenamefont{Huscroft} \emph{et~al.}(2001)\citenamefont{Huscroft,
  Jarrell, Maier, Moukouri, and Tahvildarzadeh}}]{huscroft:dca1}
\bibinfo{author}{\bibnamefont{Huscroft}, \bibfnamefont{C.}},
  \bibinfo{author}{\bibfnamefont{M.}~\bibnamefont{Jarrell}},
  \bibinfo{author}{\bibfnamefont{T.}~\bibnamefont{Maier}},
  \bibinfo{author}{\bibfnamefont{S.}~\bibnamefont{Moukouri}}, and
  \bibinfo{author}{\bibfnamefont{A.}~\bibnamefont{Tahvildarzadeh}},
  \bibinfo{year}{2001}, \bibinfo{journal}{Phys. Rev. Lett.}
  \textbf{\bibinfo{volume}{86}}, \bibinfo{pages}{139}.

\bibitem[{\citenamefont{Imai and Kawakami}(2002)}]{imai:dca}
\bibinfo{author}{\bibnamefont{Imai}, \bibfnamefont{Y.}}, and
  \bibinfo{author}{\bibfnamefont{N.}~\bibnamefont{Kawakami}},
  \bibinfo{year}{2002}, \bibinfo{journal}{Phys. Rev. B}
  \textbf{\bibinfo{volume}{65}}, \bibinfo{pages}{233103}.

\bibitem[{\citenamefont{Itzykson and Drouffe}(1989)}]{itzykson:89}
\bibinfo{author}{\bibnamefont{Itzykson}, \bibfnamefont{C.}}, and
  \bibinfo{author}{\bibfnamefont{J.-M.} \bibnamefont{Drouffe}},
  \bibinfo{year}{1989}, \emph{\bibinfo{title}{Statistical Field Theory}}
  (\bibinfo{publisher}{Cambridge University Press}).

\bibitem[{\citenamefont{{Jakli\v c} and {Prelov\v sek}}(2000)}]{prelovsek:00}
\bibinfo{author}{\bibnamefont{{Jakli\v c}}, \bibfnamefont{J.}}, and
  \bibinfo{author}{\bibfnamefont{P.}~\bibnamefont{{Prelov\v sek}}},
  \bibinfo{year}{2000}, \bibinfo{journal}{Adv. Phys.}
  \textbf{\bibinfo{volume}{49}}, \bibinfo{pages}{1}.

\bibitem[{\citenamefont{Jarrell}(1992)}]{jarrell:92}
\bibinfo{author}{\bibnamefont{Jarrell}, \bibfnamefont{M.}},
  \bibinfo{year}{1992}, \bibinfo{journal}{Phys. Rev. Lett.}
  \textbf{\bibinfo{volume}{69}}, \bibinfo{pages}{168}.

\bibitem[{\citenamefont{Jarrell} \emph{et~al.}(1993)\citenamefont{Jarrell,
  Akhlaghpour, and Pruschke}}]{jarrell:qmc}
\bibinfo{author}{\bibnamefont{Jarrell}, \bibfnamefont{M.}},
  \bibinfo{author}{\bibfnamefont{H.}~\bibnamefont{Akhlaghpour}}, and
  \bibinfo{author}{\bibfnamefont{T.}~\bibnamefont{Pruschke}},
  \bibinfo{year}{1993}, \emph{\bibinfo{title}{Quantum Monte Carlo Methods in
  Condensed Matter Physics}} (\bibinfo{publisher}{Worl Scientific},
  \bibinfo{address}{Singapore}), chapter \bibinfo{chapter}{Quantum Monte Carlo
  in the Infinite Dimensional Limit}, pp. \bibinfo{pages}{221--234}.

\bibitem[{\citenamefont{Jarrell and Gubernatis}(1996)}]{jarrell:MEM}
\bibinfo{author}{\bibnamefont{Jarrell}, \bibfnamefont{M.}}, and
  \bibinfo{author}{\bibfnamefont{J.}~\bibnamefont{Gubernatis}},
  \bibinfo{year}{1996}, \bibinfo{journal}{Physics Reports}
  \textbf{\bibinfo{volume}{269}}, \bibinfo{pages}{133}.

\bibitem[{\citenamefont{Jarrell and Johnson}(2004)}]{jarrell:ddca}
\bibinfo{author}{\bibnamefont{Jarrell}, \bibfnamefont{M.}}, and
  \bibinfo{author}{\bibfnamefont{D.}~\bibnamefont{Johnson}},
  \bibinfo{year}{2004}, \bibinfo{journal}{in preparation} .

\bibitem[{\citenamefont{Jarrell and Krishnamurthy}(2001)}]{jarrell:dca1}
\bibinfo{author}{\bibnamefont{Jarrell}, \bibfnamefont{M.}}, and
  \bibinfo{author}{\bibfnamefont{H.~R.} \bibnamefont{Krishnamurthy}},
  \bibinfo{year}{2001}, \bibinfo{journal}{Phys. Rev. B}
  \textbf{\bibinfo{volume}{63}}, \bibinfo{pages}{125102}.

\bibitem[{\citenamefont{Jarrell}
  \emph{et~al.}(2001{\natexlab{a}})\citenamefont{Jarrell, Maier, Hettler, and
  Tahvildarzadeh}}]{jarrell:dca2}
\bibinfo{author}{\bibnamefont{Jarrell}, \bibfnamefont{M.}},
  \bibinfo{author}{\bibfnamefont{T.}~\bibnamefont{Maier}},
  \bibinfo{author}{\bibfnamefont{M.~H.} \bibnamefont{Hettler}}, and
  \bibinfo{author}{\bibfnamefont{A.~N.} \bibnamefont{Tahvildarzadeh}},
  \bibinfo{year}{2001}{\natexlab{a}}, \bibinfo{journal}{Europhys. Lett.}
  \textbf{\bibinfo{volume}{56}}, \bibinfo{pages}{563}.

\bibitem[{\citenamefont{Jarrell}
  \emph{et~al.}(2001{\natexlab{b}})\citenamefont{Jarrell, Maier, Huscroft, and
  Moukouri}}]{jarrell:dca3}
\bibinfo{author}{\bibnamefont{Jarrell}, \bibfnamefont{M.}},
  \bibinfo{author}{\bibfnamefont{T.}~\bibnamefont{Maier}},
  \bibinfo{author}{\bibfnamefont{C.}~\bibnamefont{Huscroft}}, and
  \bibinfo{author}{\bibfnamefont{S.}~\bibnamefont{Moukouri}},
  \bibinfo{year}{2001}{\natexlab{b}}, \bibinfo{journal}{Phys. Rev. B}
  \textbf{\bibinfo{volume}{64}}, \bibinfo{pages}{195130}.

\bibitem[{\citenamefont{Jarrell} \emph{et~al.}(2004)\citenamefont{Jarrell,
  Maier, and Schulthess}}]{tmaier:lcluster}
\bibinfo{author}{\bibnamefont{Jarrell}, \bibfnamefont{M.}},
  \bibinfo{author}{\bibfnamefont{T.}~\bibnamefont{Maier}}, and
  \bibinfo{author}{\bibfnamefont{T.}~\bibnamefont{Schulthess}},
  \bibinfo{year}{2004}, \bibinfo{journal}{in preparation} .

\bibitem[{\citenamefont{Jarrell and Pruschke}(1993)}]{jarrell:93}
\bibinfo{author}{\bibnamefont{Jarrell}, \bibfnamefont{M.}}, and
  \bibinfo{author}{\bibfnamefont{T.}~\bibnamefont{Pruschke}},
  \bibinfo{year}{1993}, \bibinfo{journal}{Z. Phys. B}
  \textbf{\bibinfo{volume}{90}}, \bibinfo{pages}{187}.

\bibitem[{\citenamefont{Jones and Varma}(1987)}]{jones:tiam}
\bibinfo{author}{\bibnamefont{Jones}, \bibfnamefont{B.}}, and
  \bibinfo{author}{\bibfnamefont{C.}~\bibnamefont{Varma}},
  \bibinfo{year}{1987}, \bibinfo{journal}{Phys. Rev. Lett.}
  \textbf{\bibinfo{volume}{58}}, \bibinfo{pages}{843}.

\bibitem[{\citenamefont{Jones} \emph{et~al.}(1988)\citenamefont{Jones, Varma,
  and Wilkins}}]{jones:tiam2}
\bibinfo{author}{\bibnamefont{Jones}, \bibfnamefont{B.}},
  \bibinfo{author}{\bibfnamefont{C.}~\bibnamefont{Varma}}, and
  \bibinfo{author}{\bibfnamefont{J.}~\bibnamefont{Wilkins}},
  \bibinfo{year}{1988}, \bibinfo{journal}{Phys. Rev. Lett.}
  \textbf{\bibinfo{volume}{61}}, \bibinfo{pages}{125}.

\bibitem[{\citenamefont{Kakehashi}(2002)}]{Kakehashi:DCPA}
\bibinfo{author}{\bibnamefont{Kakehashi}, \bibfnamefont{Y.}},
  \bibinfo{year}{2002}, \bibinfo{journal}{Phys. Rev. B}
  \textbf{\bibinfo{volume}{66}}, \bibinfo{pages}{104428}.

\bibitem[{\citenamefont{Kampf and Schrieffer}(1990)}]{kampf:prec}
\bibinfo{author}{\bibnamefont{Kampf}, \bibfnamefont{A.}}, and
  \bibinfo{author}{\bibfnamefont{J.}~\bibnamefont{Schrieffer}},
  \bibinfo{year}{1990}, \bibinfo{journal}{Phys. Rev. B}
  \textbf{\bibinfo{volume}{42}}, \bibinfo{pages}{7967}.

\bibitem[{\citenamefont{Keiter and Czycholl}(1983)}]{Keiter:NCA3}
\bibinfo{author}{\bibnamefont{Keiter}, \bibfnamefont{H.}}, and
  \bibinfo{author}{\bibfnamefont{C.}~\bibnamefont{Czycholl}},
  \bibinfo{year}{1983}, \bibinfo{journal}{J. Magn. and Magn. Mat.}
  \textbf{\bibinfo{volume}{31--34}}, \bibinfo{pages}{477}.

\bibitem[{\citenamefont{Keiter and Kimball}(1970)}]{Keiter:NCA1}
\bibinfo{author}{\bibnamefont{Keiter}, \bibfnamefont{H.}}, and
  \bibinfo{author}{\bibfnamefont{J.}~\bibnamefont{Kimball}},
  \bibinfo{year}{1970}, \bibinfo{journal}{Phys. Rev. Lett.}
  \textbf{\bibinfo{volume}{25}}, \bibinfo{pages}{672}.

\bibitem[{\citenamefont{Keiter and Kimball}(1971)}]{Keiter:NCA2}
\bibinfo{author}{\bibnamefont{Keiter}, \bibfnamefont{H.}}, and
  \bibinfo{author}{\bibfnamefont{J.}~\bibnamefont{Kimball}},
  \bibinfo{year}{1971}, \bibinfo{journal}{Int. J. Magn.}
  \textbf{\bibinfo{volume}{1}}, \bibinfo{pages}{233}.

\bibitem[{\citenamefont{Kikuchi}(1951)}]{Kikuchi:BCising}
\bibinfo{author}{\bibnamefont{Kikuchi}, \bibfnamefont{R.}},
  \bibinfo{year}{1951}, \bibinfo{journal}{Phys. Rev.}
  \textbf{\bibinfo{volume}{81}}, \bibinfo{pages}{988}.

\bibitem[{\citenamefont{Kim} \emph{et~al.}(1990)\citenamefont{Kim, Kuramoto,
  and Kasoya}}]{Kim:NCA}
\bibinfo{author}{\bibnamefont{Kim}, \bibfnamefont{C.}},
  \bibinfo{author}{\bibfnamefont{Y.}~\bibnamefont{Kuramoto}}, and
  \bibinfo{author}{\bibfnamefont{T.}~\bibnamefont{Kasoya}},
  \bibinfo{year}{1990}, \bibinfo{journal}{J. Phys. Soc. Jpn.}
  \textbf{\bibinfo{volume}{59}}, \bibinfo{pages}{2414}.

\bibitem[{\citenamefont{Kotliar} \emph{et~al.}(2001)\citenamefont{Kotliar,
  Savrasov, Pallson, and Biroli}}]{kotliar:cdmft}
\bibinfo{author}{\bibnamefont{Kotliar}, \bibfnamefont{G.}},
  \bibinfo{author}{\bibfnamefont{S.}~\bibnamefont{Savrasov}},
  \bibinfo{author}{\bibfnamefont{G.}~\bibnamefont{Pallson}}, and
  \bibinfo{author}{\bibfnamefont{G.}~\bibnamefont{Biroli}},
  \bibinfo{year}{2001}, \bibinfo{journal}{Phys. Rev. Lett.}
  \textbf{\bibinfo{volume}{87}}, \bibinfo{pages}{186401}.

\bibitem[{\citenamefont{Krishnamurthy}
  \emph{et~al.}(1980{\natexlab{a}})\citenamefont{Krishnamurthy, Wilkins, and
  Wilson}}]{nrg:krish1}
\bibinfo{author}{\bibnamefont{Krishnamurthy}, \bibfnamefont{H.}},
  \bibinfo{author}{\bibfnamefont{J.}~\bibnamefont{Wilkins}}, and
  \bibinfo{author}{\bibfnamefont{K.}~\bibnamefont{Wilson}},
  \bibinfo{year}{1980}{\natexlab{a}}, \bibinfo{journal}{Phys. Rev. B}
  \textbf{\bibinfo{volume}{21}}, \bibinfo{pages}{1003}.

\bibitem[{\citenamefont{Krishnamurthy}
  \emph{et~al.}(1980{\natexlab{b}})\citenamefont{Krishnamurthy, Wilkins, and
  Wilson}}]{nrg:krish2}
\bibinfo{author}{\bibnamefont{Krishnamurthy}, \bibfnamefont{H.}},
  \bibinfo{author}{\bibfnamefont{J.}~\bibnamefont{Wilkins}}, and
  \bibinfo{author}{\bibfnamefont{K.}~\bibnamefont{Wilson}},
  \bibinfo{year}{1980}{\natexlab{b}}, \bibinfo{journal}{Phys. Rev. B}
  \textbf{\bibinfo{volume}{21}}, \bibinfo{pages}{1044}.

\bibitem[{\citenamefont{Kuramoto}(1983)}]{Kuramoto:NCA}
\bibinfo{author}{\bibnamefont{Kuramoto}, \bibfnamefont{Y.}},
  \bibinfo{year}{1983}, \bibinfo{journal}{Z. Phys.} \textbf{\bibinfo{volume}{B
  53}}, \bibinfo{pages}{37}.

\bibitem[{\citenamefont{Kuramoto}(1985)}]{Kuramoto:NCA2}
\bibinfo{author}{\bibnamefont{Kuramoto}, \bibfnamefont{Y.}},
  \bibinfo{year}{1985}, \bibinfo{journal}{Springer Series in Sol. State} ,
  \bibinfo{pages}{152}.

\bibitem[{\citenamefont{Laad and van~den Bossche}(2000)}]{laad:cluster}
\bibinfo{author}{\bibnamefont{Laad}, \bibfnamefont{M.~S.}}, and
  \bibinfo{author}{\bibfnamefont{M.}~\bibnamefont{van~den Bossche}},
  \bibinfo{year}{2000}, \bibinfo{journal}{J. Phys.: Condens. Matter}
  \textbf{\bibinfo{volume}{12}}, \bibinfo{pages}{2209}.

\bibitem[{\citenamefont{Landau}(1976)}]{landau:76}
\bibinfo{author}{\bibnamefont{Landau}, \bibfnamefont{D.}},
  \bibinfo{year}{1976}, \bibinfo{journal}{Phys. Rev. B}
  \textbf{\bibinfo{volume}{13}}, \bibinfo{pages}{2997}.

\bibitem[{\citenamefont{Lichtenstein and Katsnelson}(2000)}]{kats:dca}
\bibinfo{author}{\bibnamefont{Lichtenstein}, \bibfnamefont{A.~I.}}, and
  \bibinfo{author}{\bibfnamefont{M.~I.} \bibnamefont{Katsnelson}},
  \bibinfo{year}{2000}, \bibinfo{journal}{Phys. Rev. B}
  \textbf{\bibinfo{volume}{62}}, \bibinfo{pages}{R9283}.

\bibitem[{\citenamefont{Lieb and Wu}(1968)}]{lieb:1dhm}
\bibinfo{author}{\bibnamefont{Lieb}, \bibfnamefont{E.}}, and
  \bibinfo{author}{\bibfnamefont{F.}~\bibnamefont{Wu}}, \bibinfo{year}{1968},
  \bibinfo{journal}{Phys. Rev. Lett.} \textbf{\bibinfo{volume}{20}},
  \bibinfo{pages}{1445}.

\bibitem[{\citenamefont{Lombardo} \emph{et~al.}(1996)\citenamefont{Lombardo,
  Avignon, Schmalian, and Bennemann}}]{Schmalian:NCA2}
\bibinfo{author}{\bibnamefont{Lombardo}, \bibfnamefont{P.}},
  \bibinfo{author}{\bibfnamefont{M.}~\bibnamefont{Avignon}},
  \bibinfo{author}{\bibfnamefont{J.}~\bibnamefont{Schmalian}}, and
  \bibinfo{author}{\bibfnamefont{K.}~\bibnamefont{Bennemann}},
  \bibinfo{year}{1996}, \bibinfo{journal}{Phys. Rev. B}
  \textbf{\bibinfo{volume}{54}}, \bibinfo{pages}{5317}.

\bibitem[{\citenamefont{Maier}(2001)}]{maier:dca0}
\bibinfo{author}{\bibnamefont{Maier}, \bibfnamefont{T.}}, \bibinfo{year}{2001},
  \emph{\bibinfo{title}{Nonlocal Dynamical Correlations in Strongly Interacting
  Fermion Systems}} (\bibinfo{publisher}{Logos Verlag Berlin},
  \bibinfo{address}{Berlin, Germany}).

\bibitem[{\citenamefont{Maier}(2003)}]{maier:03}
\bibinfo{author}{\bibnamefont{Maier}, \bibfnamefont{T.}}, \bibinfo{year}{2003},
  \bibinfo{journal}{preprint cond-mat/0312447} .

\bibitem[{\citenamefont{Maier and Jarrell}(2002)}]{maier:dca3}
\bibinfo{author}{\bibnamefont{Maier}, \bibfnamefont{T.}}, and
  \bibinfo{author}{\bibfnamefont{M.}~\bibnamefont{Jarrell}},
  \bibinfo{year}{2002}, \bibinfo{journal}{Phys. Rev. B}
  \textbf{\bibinfo{volume}{65}}, \bibinfo{pages}{041104}.

\bibitem[{\citenamefont{Maier} \emph{et~al.}(2004)\citenamefont{Maier, Jarrell,
  Macridin, and Slezak}}]{maier:dca6}
\bibinfo{author}{\bibnamefont{Maier}, \bibfnamefont{T.}},
  \bibinfo{author}{\bibfnamefont{M.}~\bibnamefont{Jarrell}},
  \bibinfo{author}{\bibfnamefont{A.}~\bibnamefont{Macridin}}, and
  \bibinfo{author}{\bibfnamefont{C.}~\bibnamefont{Slezak}},
  \bibinfo{year}{2004}, \bibinfo{journal}{Phys. Rev. Lett.}
  \textbf{\bibinfo{volume}{92}}, \bibinfo{pages}{027005}.

\bibitem[{\citenamefont{Maier}
  \emph{et~al.}(2002{\natexlab{a}})\citenamefont{Maier, Jarrell, Macridin, and
  Zhang}}]{maier:dca5}
\bibinfo{author}{\bibnamefont{Maier}, \bibfnamefont{T.}},
  \bibinfo{author}{\bibfnamefont{M.}~\bibnamefont{Jarrell}},
  \bibinfo{author}{\bibfnamefont{A.}~\bibnamefont{Macridin}}, and
  \bibinfo{author}{\bibfnamefont{F.-C.} \bibnamefont{Zhang}},
  \bibinfo{year}{2002}{\natexlab{a}}, \bibinfo{journal}{preprint
  cond-mat/0208419} .

\bibitem[{\citenamefont{Maier}
  \emph{et~al.}(2000{\natexlab{a}})\citenamefont{Maier, Jarrell, Pruschke, and
  Keller}}]{maier:dca2}
\bibinfo{author}{\bibnamefont{Maier}, \bibfnamefont{T.}},
  \bibinfo{author}{\bibfnamefont{M.}~\bibnamefont{Jarrell}},
  \bibinfo{author}{\bibfnamefont{T.}~\bibnamefont{Pruschke}}, and
  \bibinfo{author}{\bibfnamefont{J.}~\bibnamefont{Keller}},
  \bibinfo{year}{2000}{\natexlab{a}}, \bibinfo{journal}{Phys. Rev. Lett.}
  \textbf{\bibinfo{volume}{85}}, \bibinfo{pages}{1524}.

\bibitem[{\citenamefont{Maier}
  \emph{et~al.}(2000{\natexlab{b}})\citenamefont{Maier, Jarrell, Pruschke, and
  Keller}}]{maier:dca1}
\bibinfo{author}{\bibnamefont{Maier}, \bibfnamefont{T.}},
  \bibinfo{author}{\bibfnamefont{M.}~\bibnamefont{Jarrell}},
  \bibinfo{author}{\bibfnamefont{T.}~\bibnamefont{Pruschke}}, and
  \bibinfo{author}{\bibfnamefont{J.}~\bibnamefont{Keller}},
  \bibinfo{year}{2000}{\natexlab{b}}, \bibinfo{journal}{Eur. Phys. J B}
  \textbf{\bibinfo{volume}{13}}, \bibinfo{pages}{613}.

\bibitem[{\citenamefont{Maier}
  \emph{et~al.}(1999{\natexlab{a}})\citenamefont{Maier, Z\"olfl, Pruschke, and
  Keller}}]{Maier:NCA1}
\bibinfo{author}{\bibnamefont{Maier}, \bibfnamefont{T.}},
  \bibinfo{author}{\bibfnamefont{M.}~\bibnamefont{Z\"olfl}},
  \bibinfo{author}{\bibfnamefont{T.}~\bibnamefont{Pruschke}}, and
  \bibinfo{author}{\bibfnamefont{J.}~\bibnamefont{Keller}},
  \bibinfo{year}{1999}{\natexlab{a}}, \bibinfo{journal}{Eur. Phys. J. B}
  \textbf{\bibinfo{volume}{7}}, \bibinfo{pages}{377}.

\bibitem[{\citenamefont{Maier}
  \emph{et~al.}(1999{\natexlab{b}})\citenamefont{Maier, Z\"olfl, Pruschke, and
  Keller}}]{Maier:NCA2}
\bibinfo{author}{\bibnamefont{Maier}, \bibfnamefont{T.}},
  \bibinfo{author}{\bibfnamefont{M.}~\bibnamefont{Z\"olfl}},
  \bibinfo{author}{\bibfnamefont{T.}~\bibnamefont{Pruschke}}, and
  \bibinfo{author}{\bibfnamefont{J.}~\bibnamefont{Keller}},
  \bibinfo{year}{1999}{\natexlab{b}}, \bibinfo{journal}{Physica B}
  \textbf{\bibinfo{volume}{259}}, \bibinfo{pages}{747}.

\bibitem[{\citenamefont{Maier}
  \emph{et~al.}(2002{\natexlab{b}})\citenamefont{Maier, Gonzalez, Jarrell, and
  Schulthess}}]{maier:dca4}
\bibinfo{author}{\bibnamefont{Maier}, \bibfnamefont{T.~A.}},
  \bibinfo{author}{\bibfnamefont{O.}~\bibnamefont{Gonzalez}},
  \bibinfo{author}{\bibfnamefont{M.}~\bibnamefont{Jarrell}}, and
  \bibinfo{author}{\bibfnamefont{T.}~\bibnamefont{Schulthess}},
  \bibinfo{year}{2002}{\natexlab{b}}, \emph{\bibinfo{title}{Two Quantum Cluster
  Approximations}}, Springer Proceedings in Physics.

\bibitem[{\citenamefont{Matsumoto and Mancini}(1997)}]{mancini:cluster}
\bibinfo{author}{\bibnamefont{Matsumoto}, \bibfnamefont{H.}}, and
  \bibinfo{author}{\bibfnamefont{F.}~\bibnamefont{Mancini}},
  \bibinfo{year}{1997}, \bibinfo{journal}{Phys. Rev. B}
  \textbf{\bibinfo{volume}{55}}, \bibinfo{pages}{2095}.

\bibitem[{\citenamefont{Metzner and Vollhardt}(1989)}]{metzner:dmft}
\bibinfo{author}{\bibnamefont{Metzner}, \bibfnamefont{W.}}, and
  \bibinfo{author}{\bibfnamefont{D.}~\bibnamefont{Vollhardt}},
  \bibinfo{year}{1989}, \bibinfo{journal}{Phys. Rev. Lett.}
  \textbf{\bibinfo{volume}{62}}, \bibinfo{pages}{324}.

\bibitem[{\citenamefont{Minh-Tien}(1998)}]{tran:cluster1}
\bibinfo{author}{\bibnamefont{Minh-Tien}, \bibfnamefont{T.}},
  \bibinfo{year}{1998}, \bibinfo{journal}{Phys. Rev. B}
  \textbf{\bibinfo{volume}{58}}, \bibinfo{pages}{R15965}.

\bibitem[{\citenamefont{Minh-Tien}(1999{\natexlab{a}})}]{tran:cluster2}
\bibinfo{author}{\bibnamefont{Minh-Tien}, \bibfnamefont{T.}},
  \bibinfo{year}{1999}{\natexlab{a}}, \bibinfo{journal}{Europhys. Lett.}
  \textbf{\bibinfo{volume}{47}}, \bibinfo{pages}{582}.

\bibitem[{\citenamefont{Minh-Tien}(1999{\natexlab{b}})}]{tran:cluster3}
\bibinfo{author}{\bibnamefont{Minh-Tien}, \bibfnamefont{T.}},
  \bibinfo{year}{1999}{\natexlab{b}}, \bibinfo{journal}{Phys. Rev. B}
  \textbf{\bibinfo{volume}{60}}, \bibinfo{pages}{16371}.

\bibitem[{\citenamefont{Minh-Tien}(2001)}]{tran:cluster4}
\bibinfo{author}{\bibnamefont{Minh-Tien}, \bibfnamefont{T.}},
  \bibinfo{year}{2001}, \bibinfo{journal}{Phys. Rev. B}
  \textbf{\bibinfo{volume}{63}}, \bibinfo{pages}{165117}.

\bibitem[{\citenamefont{Molegraaf} \emph{et~al.}(2002)\citenamefont{Molegraaf,
  Presura, van~der Marel, Kes, and Li}}]{marel:kin}
\bibinfo{author}{\bibnamefont{Molegraaf}, \bibfnamefont{H.~J.~A.}},
  \bibinfo{author}{\bibfnamefont{C.}~\bibnamefont{Presura}},
  \bibinfo{author}{\bibfnamefont{D.}~\bibnamefont{van~der Marel}},
  \bibinfo{author}{\bibfnamefont{P.~H.} \bibnamefont{Kes}}, and
  \bibinfo{author}{\bibfnamefont{M.}~\bibnamefont{Li}}, \bibinfo{year}{2002},
  \bibinfo{journal}{Science} \textbf{\bibinfo{volume}{295}},
  \bibinfo{pages}{2239}.

\bibitem[{\citenamefont{Monthoux} \emph{et~al.}(1991)\citenamefont{Monthoux,
  Balatsky, and Pines}}]{monthoux:91}
\bibinfo{author}{\bibnamefont{Monthoux}, \bibfnamefont{P.}},
  \bibinfo{author}{\bibfnamefont{A.}~\bibnamefont{Balatsky}}, and
  \bibinfo{author}{\bibfnamefont{D.}~\bibnamefont{Pines}},
  \bibinfo{year}{1991}, \bibinfo{journal}{Phys. Rev. Lett.}
  \textbf{\bibinfo{volume}{67}}, \bibinfo{pages}{3448}.

\bibitem[{\citenamefont{Moukouri} \emph{et~al.}(1999)\citenamefont{Moukouri,
  Allen, Lemay, Kyong, Poulin, Vilk, and Tremblay}}]{moukouri:99}
\bibinfo{author}{\bibnamefont{Moukouri}, \bibfnamefont{S.}},
  \bibinfo{author}{\bibfnamefont{S.}~\bibnamefont{Allen}},
  \bibinfo{author}{\bibfnamefont{F.}~\bibnamefont{Lemay}},
  \bibinfo{author}{\bibfnamefont{B.}~\bibnamefont{Kyong}},
  \bibinfo{author}{\bibfnamefont{D.}~\bibnamefont{Poulin}},
  \bibinfo{author}{\bibfnamefont{Y.}~\bibnamefont{Vilk}}, and
  \bibinfo{author}{\bibfnamefont{A.-M.} \bibnamefont{Tremblay}},
  \bibinfo{year}{1999}, \bibinfo{journal}{Phys. Rev. B}
  \textbf{\bibinfo{volume}{61}}, \bibinfo{pages}{7887}.

\bibitem[{\citenamefont{Moukouri and Jarrell}(2001)}]{moukouri:dca1}
\bibinfo{author}{\bibnamefont{Moukouri}, \bibfnamefont{S.}}, and
  \bibinfo{author}{\bibfnamefont{M.}~\bibnamefont{Jarrell}},
  \bibinfo{year}{2001}, \bibinfo{journal}{Phys. Rev. Lett.}
  \textbf{\bibinfo{volume}{87}}, \bibinfo{pages}{167010}.

\bibitem[{\citenamefont{M\"uller-Hartmann}(1984)}]{MuHa:NCA}
\bibinfo{author}{\bibnamefont{M\"uller-Hartmann}, \bibfnamefont{E.}},
  \bibinfo{year}{1984}, \bibinfo{journal}{Z.\ Phys.\ {\bf B}}
  \textbf{\bibinfo{volume}{57}}, \bibinfo{pages}{293}.

\bibitem[{\citenamefont{M\"uller-Hartmann}(1989{\natexlab{a}})}]{hartmann:89b}
\bibinfo{author}{\bibnamefont{M\"uller-Hartmann}, \bibfnamefont{E.}},
  \bibinfo{year}{1989}{\natexlab{a}}, \bibinfo{journal}{Z.\ Phys.\ {\bf B}}
  \textbf{\bibinfo{volume}{76}}, \bibinfo{pages}{211}.

\bibitem[{\citenamefont{M\"uller-Hartmann}(1989{\natexlab{b}})}]{hartmann:dmft}
\bibinfo{author}{\bibnamefont{M\"uller-Hartmann}, \bibfnamefont{E.}},
  \bibinfo{year}{1989}{\natexlab{b}}, \bibinfo{journal}{Z.\ Phys.\ {\bf B}}
  \textbf{\bibinfo{volume}{74}}, \bibinfo{pages}{507}.

\bibitem[{\citenamefont{Nordheim}(1931{\natexlab{a}})}]{VCA:nordheim1}
\bibinfo{author}{\bibnamefont{Nordheim}, \bibfnamefont{J.}},
  \bibinfo{year}{1931}{\natexlab{a}}, \bibinfo{journal}{Ann. Physik}
  \textbf{\bibinfo{volume}{9}}, \bibinfo{pages}{607}.

\bibitem[{\citenamefont{Nordheim}(1931{\natexlab{b}})}]{VCA:nordheim2}
\bibinfo{author}{\bibnamefont{Nordheim}, \bibfnamefont{J.}},
  \bibinfo{year}{1931}{\natexlab{b}}, \bibinfo{journal}{Ann. Physik}
  \textbf{\bibinfo{volume}{9}}, \bibinfo{pages}{641}.

\bibitem[{\citenamefont{Okamoto} \emph{et~al.}(2003)\citenamefont{Okamoto,
  Millis, Monien, and Fuhrmann}}]{okamoto:cluster}
\bibinfo{author}{\bibnamefont{Okamoto}, \bibfnamefont{S.}},
  \bibinfo{author}{\bibfnamefont{A.}~\bibnamefont{Millis}},
  \bibinfo{author}{\bibfnamefont{H.}~\bibnamefont{Monien}}, and
  \bibinfo{author}{\bibfnamefont{A.}~\bibnamefont{Fuhrmann}},
  \bibinfo{year}{2003}, \bibinfo{journal}{Phys. Rev. B}
  \textbf{\bibinfo{volume}{68}}, \bibinfo{pages}{195121}.

\bibitem[{\citenamefont{Oliveira and Oliveira}(1994)}]{NRG:oliveira}
\bibinfo{author}{\bibnamefont{Oliveira}, \bibfnamefont{W.}}, and
  \bibinfo{author}{\bibfnamefont{L.}~\bibnamefont{Oliveira}},
  \bibinfo{year}{1994}, \bibinfo{journal}{Phys. Rev. B}
  \textbf{\bibinfo{volume}{49}}, \bibinfo{pages}{11986}.

\bibitem[{\citenamefont{Pairault} \emph{et~al.}(1998)\citenamefont{Pairault,
  S\'en\'echal, and Tremblay}}]{pairault:strc2}
\bibinfo{author}{\bibnamefont{Pairault}, \bibfnamefont{S.}},
  \bibinfo{author}{\bibfnamefont{D.}~\bibnamefont{S\'en\'echal}}, and
  \bibinfo{author}{\bibfnamefont{A.-M.} \bibnamefont{Tremblay}},
  \bibinfo{year}{1998}, \bibinfo{journal}{Phys. Rev. Lett.}
  \textbf{\bibinfo{volume}{80}}, \bibinfo{pages}{5389}.

\bibitem[{\citenamefont{Pairault} \emph{et~al.}(2000)\citenamefont{Pairault,
  S\'en\'echal, and Tremblay}}]{pairault:strc}
\bibinfo{author}{\bibnamefont{Pairault}, \bibfnamefont{S.}},
  \bibinfo{author}{\bibfnamefont{D.}~\bibnamefont{S\'en\'echal}}, and
  \bibinfo{author}{\bibfnamefont{A.-M.} \bibnamefont{Tremblay}},
  \bibinfo{year}{2000}, \bibinfo{journal}{Eur. Phys. J. B}
  \textbf{\bibinfo{volume}{16}}, \bibinfo{pages}{85}.

\bibitem[{\citenamefont{Parcollet} \emph{et~al.}(2003)\citenamefont{Parcollet,
  Biroli, and Kotliar}}]{parcollet:cdmft}
\bibinfo{author}{\bibnamefont{Parcollet}, \bibfnamefont{O.}},
  \bibinfo{author}{\bibfnamefont{G.}~\bibnamefont{Biroli}}, and
  \bibinfo{author}{\bibfnamefont{G.}~\bibnamefont{Kotliar}},
  \bibinfo{year}{2003}, \bibinfo{journal}{preprint cond-mat/0308577} .

\bibitem[{\citenamefont{Parmenter}(1955)}]{VCA:parmenter}
\bibinfo{author}{\bibnamefont{Parmenter}, \bibfnamefont{R.~H.}},
  \bibinfo{year}{1955}, \bibinfo{journal}{Phys. Rev.}
  \textbf{\bibinfo{volume}{97}}, \bibinfo{pages}{587}.

\bibitem[{\citenamefont{Paula} \emph{et~al.}(1999)\citenamefont{Paula, Silva,
  and Oliveira}}]{NRG:paula}
\bibinfo{author}{\bibnamefont{Paula}, \bibfnamefont{C.}},
  \bibinfo{author}{\bibfnamefont{M.}~\bibnamefont{Silva}}, and
  \bibinfo{author}{\bibfnamefont{L.}~\bibnamefont{Oliveira}},
  \bibinfo{year}{1999}, \bibinfo{journal}{Phys. Rev. B}
  \textbf{\bibinfo{volume}{59}}, \bibinfo{pages}{85}.

\bibitem[{\citenamefont{Poteryaev} \emph{et~al.}(2003)\citenamefont{Poteryaev,
  Lichtenstein, and Kotliar}}]{poteryaev:03}
\bibinfo{author}{\bibnamefont{Poteryaev}, \bibfnamefont{A.}},
  \bibinfo{author}{\bibfnamefont{A.}~\bibnamefont{Lichtenstein}}, and
  \bibinfo{author}{\bibfnamefont{G.}~\bibnamefont{Kotliar}},
  \bibinfo{year}{2003}, \bibinfo{journal}{preprint cond-mat/0311319} .

\bibitem[{\citenamefont{Potthoff}(2003{\natexlab{a}})}]{potthoff:cluster4}
\bibinfo{author}{\bibnamefont{Potthoff}, \bibfnamefont{M.}},
  \bibinfo{year}{2003}{\natexlab{a}}, \bibinfo{journal}{preprint
  cond-mat/0309407} .

\bibitem[{\citenamefont{Potthoff}(2003{\natexlab{b}})}]{potthoff:cluster2}
\bibinfo{author}{\bibnamefont{Potthoff}, \bibfnamefont{M.}},
  \bibinfo{year}{2003}{\natexlab{b}}, \bibinfo{journal}{Eur. Phys. J. B}
  \textbf{\bibinfo{volume}{32}}, \bibinfo{pages}{429}.

\bibitem[{\citenamefont{Potthoff} \emph{et~al.}(2003)\citenamefont{Potthoff,
  Aichhorn, and Dahnken}}]{potthoff:cluster1}
\bibinfo{author}{\bibnamefont{Potthoff}, \bibfnamefont{M.}},
  \bibinfo{author}{\bibfnamefont{M.}~\bibnamefont{Aichhorn}}, and
  \bibinfo{author}{\bibfnamefont{C.}~\bibnamefont{Dahnken}},
  \bibinfo{year}{2003}, \bibinfo{journal}{Phys. Rev. Lett.}
  \textbf{\bibinfo{volume}{91}}, \bibinfo{pages}{206402}.

\bibitem[{\citenamefont{Pruschke} \emph{et~al.}(2000)\citenamefont{Pruschke,
  Bulla, and Jarrell}}]{dmftnrg:pruschke1}
\bibinfo{author}{\bibnamefont{Pruschke}, \bibfnamefont{T.}},
  \bibinfo{author}{\bibfnamefont{R.}~\bibnamefont{Bulla}}, and
  \bibinfo{author}{\bibfnamefont{M.}~\bibnamefont{Jarrell}},
  \bibinfo{year}{2000}, \bibinfo{journal}{Phys. Rev. B}
  \textbf{\bibinfo{volume}{61}}, \bibinfo{pages}{12799}.

\bibitem[{\citenamefont{Pruschke}
  \emph{et~al.}(1993{\natexlab{a}})\citenamefont{Pruschke, Cox, and
  Jarrell}}]{Pruschke:NCA1}
\bibinfo{author}{\bibnamefont{Pruschke}, \bibfnamefont{T.}},
  \bibinfo{author}{\bibfnamefont{D.}~\bibnamefont{Cox}}, and
  \bibinfo{author}{\bibfnamefont{M.}~\bibnamefont{Jarrell}},
  \bibinfo{year}{1993}{\natexlab{a}}, \bibinfo{journal}{Europhys. Lett.}
  \textbf{\bibinfo{volume}{21}}, \bibinfo{pages}{593}.

\bibitem[{\citenamefont{Pruschke}
  \emph{et~al.}(1993{\natexlab{b}})\citenamefont{Pruschke, Cox, and
  Jarrell}}]{Pruschke:NCA2}
\bibinfo{author}{\bibnamefont{Pruschke}, \bibfnamefont{T.}},
  \bibinfo{author}{\bibfnamefont{D.}~\bibnamefont{Cox}}, and
  \bibinfo{author}{\bibfnamefont{M.}~\bibnamefont{Jarrell}},
  \bibinfo{year}{1993}{\natexlab{b}}, \bibinfo{journal}{Phys. Rev. B}
  \textbf{\bibinfo{volume}{47}}, \bibinfo{pages}{3553}.

\bibitem[{\citenamefont{Pruschke and Grewe}(1989)}]{Pruschke:NCA}
\bibinfo{author}{\bibnamefont{Pruschke}, \bibfnamefont{T.}}, and
  \bibinfo{author}{\bibfnamefont{N.}~\bibnamefont{Grewe}},
  \bibinfo{year}{1989}, \bibinfo{journal}{Z.\ Phys.\ {\bf B}}
  \textbf{\bibinfo{volume}{74}}, \bibinfo{pages}{439}.

\bibitem[{\citenamefont{Pruschke} \emph{et~al.}(1995)\citenamefont{Pruschke,
  Jarrell, and Freericks}}]{pruschke:dmftrev}
\bibinfo{author}{\bibnamefont{Pruschke}, \bibfnamefont{T.}},
  \bibinfo{author}{\bibfnamefont{M.}~\bibnamefont{Jarrell}}, and
  \bibinfo{author}{\bibfnamefont{J.}~\bibnamefont{Freericks}},
  \bibinfo{year}{1995}, \bibinfo{journal}{Adv. in Phys.}
  \textbf{\bibinfo{volume}{44}}, \bibinfo{pages}{187}.

\bibitem[{\citenamefont{Pruschke and Zitzler}(2003)}]{dmftnrg:zitzler3}
\bibinfo{author}{\bibnamefont{Pruschke}, \bibfnamefont{T.}}, and
  \bibinfo{author}{\bibfnamefont{R.}~\bibnamefont{Zitzler}},
  \bibinfo{year}{2003}, \bibinfo{journal}{J. Phys. Chem.}
  \textbf{\bibinfo{volume}{15}}, \bibinfo{pages}{7867}.

\bibitem[{\citenamefont{Roth}(1969)}]{roth:sda}
\bibinfo{author}{\bibnamefont{Roth}, \bibfnamefont{L.~M.}},
  \bibinfo{year}{1969}, \bibinfo{journal}{Phys. Rev.}
  \textbf{\bibinfo{volume}{184}}, \bibinfo{pages}{461}.

\bibitem[{\citenamefont{Sakai and Kuramoto}(1994)}]{dmftnrg:sakai}
\bibinfo{author}{\bibnamefont{Sakai}, \bibfnamefont{O.}}, and
  \bibinfo{author}{\bibfnamefont{Y.}~\bibnamefont{Kuramoto}},
  \bibinfo{year}{1994}, \bibinfo{journal}{Solid State Commun.}
  \textbf{\bibinfo{volume}{89}}, \bibinfo{pages}{307}.

\bibitem[{\citenamefont{Sakai and Shimizu}(1992)}]{sakai:tiam2}
\bibinfo{author}{\bibnamefont{Sakai}, \bibfnamefont{O.}}, and
  \bibinfo{author}{\bibfnamefont{Y.}~\bibnamefont{Shimizu}},
  \bibinfo{year}{1992}, \bibinfo{journal}{J. Phys. Soc. Jpn.}
  \textbf{\bibinfo{volume}{61}}, \bibinfo{pages}{2333}.

\bibitem[{\citenamefont{Sakai} \emph{et~al.}(1989)\citenamefont{Sakai, Shimizu,
  and Kasuya}}]{NRG:sakai}
\bibinfo{author}{\bibnamefont{Sakai}, \bibfnamefont{O.}},
  \bibinfo{author}{\bibfnamefont{Y.}~\bibnamefont{Shimizu}}, and
  \bibinfo{author}{\bibfnamefont{T.}~\bibnamefont{Kasuya}},
  \bibinfo{year}{1989}, \bibinfo{journal}{J. Phys. Soc. Jpn.}
  \textbf{\bibinfo{volume}{59}}, \bibinfo{pages}{3666}.

\bibitem[{\citenamefont{Sakai} \emph{et~al.}(1990)\citenamefont{Sakai, Shimizu,
  and Kasuya}}]{sakai:tiam}
\bibinfo{author}{\bibnamefont{Sakai}, \bibfnamefont{O.}},
  \bibinfo{author}{\bibfnamefont{Y.}~\bibnamefont{Shimizu}}, and
  \bibinfo{author}{\bibfnamefont{T.}~\bibnamefont{Kasuya}},
  \bibinfo{year}{1990}, \bibinfo{journal}{Solid State Comm.}
  \textbf{\bibinfo{volume}{75}}, \bibinfo{pages}{81}.

\bibitem[{\citenamefont{Scalapino}(1999)}]{scalapino:99}
\bibinfo{author}{\bibnamefont{Scalapino}, \bibfnamefont{D.}},
  \bibinfo{year}{1999}, \bibinfo{journal}{J. Low Temp. Phys.}
  \textbf{\bibinfo{volume}{117}}(\bibinfo{number}{3-4}), \bibinfo{pages}{179}.

\bibitem[{\citenamefont{Schiller and Ingersent}(1995)}]{schiller:cluster}
\bibinfo{author}{\bibnamefont{Schiller}, \bibfnamefont{A.}}, and
  \bibinfo{author}{\bibfnamefont{K.}~\bibnamefont{Ingersent}},
  \bibinfo{year}{1995}, \bibinfo{journal}{Phys. Rev. Lett.}
  \textbf{\bibinfo{volume}{75}}, \bibinfo{pages}{113}.

\bibitem[{\citenamefont{Schmalian} \emph{et~al.}(1996)\citenamefont{Schmalian,
  Lombardo, Avignon, and Bennemann}}]{Schmalian:NCA1}
\bibinfo{author}{\bibnamefont{Schmalian}, \bibfnamefont{J.}},
  \bibinfo{author}{\bibfnamefont{P.}~\bibnamefont{Lombardo}},
  \bibinfo{author}{\bibfnamefont{M.}~\bibnamefont{Avignon}}, and
  \bibinfo{author}{\bibfnamefont{K.}~\bibnamefont{Bennemann}},
  \bibinfo{year}{1996}, \bibinfo{journal}{Physica B}
  \textbf{\bibinfo{volume}{222--224}}, \bibinfo{pages}{602}.

\bibitem[{\citenamefont{Schoen}(1969)}]{VCA:schoen}
\bibinfo{author}{\bibnamefont{Schoen}, \bibfnamefont{J.~M.}},
  \bibinfo{year}{1969}, \bibinfo{journal}{Phys. Rev.}
  \textbf{\bibinfo{volume}{184}}, \bibinfo{pages}{858}.

\bibitem[{\citenamefont{Schrieffer}(1993)}]{schrieffer:sc}
\bibinfo{author}{\bibnamefont{Schrieffer}, \bibfnamefont{J.}},
  \bibinfo{year}{1993}, \emph{\bibinfo{title}{Theory of Superconductivity}}
  (\bibinfo{publisher}{Addison Wesley}, \bibinfo{address}{Reading, MA}).

\bibitem[{\citenamefont{Schwartz} \emph{et~al.}(1971)\citenamefont{Schwartz,
  Brouers, Vedyayev, and Ehrenreich}}]{ATA:ehrenreich}
\bibinfo{author}{\bibnamefont{Schwartz}, \bibfnamefont{L.}},
  \bibinfo{author}{\bibfnamefont{F.}~\bibnamefont{Brouers}},
  \bibinfo{author}{\bibfnamefont{A.~V.} \bibnamefont{Vedyayev}}, and
  \bibinfo{author}{\bibfnamefont{H.}~\bibnamefont{Ehrenreich}},
  \bibinfo{year}{1971}, \bibinfo{journal}{Phys. Rev. B}
  \textbf{\bibinfo{volume}{4}}, \bibinfo{pages}{3383}.

\bibitem[{\citenamefont{S\'en\'echal}
  \emph{et~al.}(2000)\citenamefont{S\'en\'echal, Perez, and
  Pioro-Ladri\'ere}}]{senechal:cluster2}
\bibinfo{author}{\bibnamefont{S\'en\'echal}, \bibfnamefont{D.}},
  \bibinfo{author}{\bibfnamefont{D.}~\bibnamefont{Perez}}, and
  \bibinfo{author}{\bibfnamefont{M.}~\bibnamefont{Pioro-Ladri\'ere}},
  \bibinfo{year}{2000}, \bibinfo{journal}{Phys. Rev. Lett.}
  \textbf{\bibinfo{volume}{84}}, \bibinfo{pages}{522}.

\bibitem[{\citenamefont{S\'en\'echal}
  \emph{et~al.}(2002)\citenamefont{S\'en\'echal, Perez, and
  Plouffe}}]{senechal:cluster}
\bibinfo{author}{\bibnamefont{S\'en\'echal}, \bibfnamefont{D.}},
  \bibinfo{author}{\bibfnamefont{D.}~\bibnamefont{Perez}}, and
  \bibinfo{author}{\bibfnamefont{D.}~\bibnamefont{Plouffe}},
  \bibinfo{year}{2002}, \bibinfo{journal}{Phys. Rev. B}
  \textbf{\bibinfo{volume}{66}}, \bibinfo{pages}{075129}.

\bibitem[{\citenamefont{S\'en\'echal and Tremblay}(2003)}]{senechal:cluster3}
\bibinfo{author}{\bibnamefont{S\'en\'echal}, \bibfnamefont{D.}}, and
  \bibinfo{author}{\bibfnamefont{A.-M.} \bibnamefont{Tremblay}},
  \bibinfo{year}{2003}, \bibinfo{journal}{preprint cond-mat/0308625} .

\bibitem[{\citenamefont{Shiba}(1971)}]{shiba:RCPA}
\bibinfo{author}{\bibnamefont{Shiba}, \bibfnamefont{H.}}, \bibinfo{year}{1971},
  \bibinfo{journal}{Prog.\ Theo.\ Phys.} \textbf{\bibinfo{volume}{46}},
  \bibinfo{pages}{77}.

\bibitem[{\citenamefont{Shimizu}(2002)}]{shimizu:pam}
\bibinfo{author}{\bibnamefont{Shimizu}, \bibfnamefont{Y.}},
  \bibinfo{year}{2002}, \bibinfo{journal}{Journal of the Physical Society of
  Japan} \textbf{\bibinfo{volume}{71}}, \bibinfo{pages}{1166}.

\bibitem[{\citenamefont{Shimizu and Sakai}(1995)}]{dmftnrg:shimizu}
\bibinfo{author}{\bibnamefont{Shimizu}, \bibfnamefont{Y.}}, and
  \bibinfo{author}{\bibfnamefont{O.}~\bibnamefont{Sakai}},
  \bibinfo{year}{1995}, \emph{\bibinfo{title}{Computational Physics as a New
  Frontier in Condensed Matter Research}} (\bibinfo{publisher}{The Physical
  Society of Japan}), p.~\bibinfo{pages}{42}.

\bibitem[{\citenamefont{Si} \emph{et~al.}(1994)\citenamefont{Si, Rozenberg,
  Kotliar, and Ruckenstein}}]{si:ed}
\bibinfo{author}{\bibnamefont{Si}, \bibfnamefont{Q.}},
  \bibinfo{author}{\bibfnamefont{M.~J.} \bibnamefont{Rozenberg}},
  \bibinfo{author}{\bibfnamefont{G.}~\bibnamefont{Kotliar}}, and
  \bibinfo{author}{\bibfnamefont{A.~E.} \bibnamefont{Ruckenstein}},
  \bibinfo{year}{1994}, \bibinfo{journal}{Phys. Rev. Lett.}
  \textbf{\bibinfo{volume}{72}}, \bibinfo{pages}{2761}.

\bibitem[{\citenamefont{Smith and Si}(2000)}]{edmft:si}
\bibinfo{author}{\bibnamefont{Smith}, \bibfnamefont{J.}}, and
  \bibinfo{author}{\bibfnamefont{Q.}~\bibnamefont{Si}}, \bibinfo{year}{2000},
  \bibinfo{journal}{Phys. Rev. B} \textbf{\bibinfo{volume}{61}},
  \bibinfo{pages}{5184}.

\bibitem[{\citenamefont{Soven}(1967)}]{soven:CPA}
\bibinfo{author}{\bibnamefont{Soven}, \bibfnamefont{P.}}, \bibinfo{year}{1967},
  \bibinfo{journal}{Phys. Rev.} \textbf{\bibinfo{volume}{156}},
  \bibinfo{pages}{809}.

\bibitem[{\citenamefont{Stanescu and Phillips}(2001)}]{phillips:cluster1}
\bibinfo{author}{\bibnamefont{Stanescu}, \bibfnamefont{T.~D.}}, and
  \bibinfo{author}{\bibfnamefont{P.}~\bibnamefont{Phillips}},
  \bibinfo{year}{2001}, \bibinfo{journal}{Phys. Rev. B}
  \textbf{\bibinfo{volume}{64}}, \bibinfo{pages}{235117}.

\bibitem[{\citenamefont{Stanescu and
  Phillips}(2003{\natexlab{a}})}]{phillips:cluster2}
\bibinfo{author}{\bibnamefont{Stanescu}, \bibfnamefont{T.~D.}}, and
  \bibinfo{author}{\bibfnamefont{P.}~\bibnamefont{Phillips}},
  \bibinfo{year}{2003}{\natexlab{a}}, \bibinfo{journal}{preprint
  cond-mat/0301254} .

\bibitem[{\citenamefont{Stanescu and
  Phillips}(2003{\natexlab{b}})}]{phillips:cluster3}
\bibinfo{author}{\bibnamefont{Stanescu}, \bibfnamefont{T.~D.}}, and
  \bibinfo{author}{\bibfnamefont{P.}~\bibnamefont{Phillips}},
  \bibinfo{year}{2003}{\natexlab{b}}, \bibinfo{journal}{Phys. Rev. Lett.}
  \textbf{\bibinfo{volume}{91}}, \bibinfo{pages}{017002}.

\bibitem[{\citenamefont{Sumi}(1974)}]{Sumi:DCPA}
\bibinfo{author}{\bibnamefont{Sumi}, \bibfnamefont{H.}}, \bibinfo{year}{1974},
  \bibinfo{journal}{J. Phys. Soc. Jpn.} \textbf{\bibinfo{volume}{36}},
  \bibinfo{pages}{770}.

\bibitem[{\citenamefont{Suzuki}(1986)}]{Suzuki:CAM}
\bibinfo{author}{\bibnamefont{Suzuki}, \bibfnamefont{M.}},
  \bibinfo{year}{1986}, \bibinfo{journal}{J.\ Phys.\ Soc.\ Jpn}
  \textbf{\bibinfo{volume}{55}}, \bibinfo{pages}{4205}.

\bibitem[{\citenamefont{Tahvildar-Zadeh}
  \emph{et~al.}(1997)\citenamefont{Tahvildar-Zadeh, Jarrell, and
  Freericks}}]{niki:pam1}
\bibinfo{author}{\bibnamefont{Tahvildar-Zadeh}, \bibfnamefont{A.}},
  \bibinfo{author}{\bibfnamefont{M.}~\bibnamefont{Jarrell}}, and
  \bibinfo{author}{\bibfnamefont{J.}~\bibnamefont{Freericks}},
  \bibinfo{year}{1997}, \bibinfo{journal}{Phys. Rev. B}
  \textbf{\bibinfo{volume}{55}}, \bibinfo{pages}{R3332}.

\bibitem[{\citenamefont{Tanh-Hai and Minh-Tien}(2001)}]{tran:cluster5}
\bibinfo{author}{\bibnamefont{Tanh-Hai}, \bibfnamefont{D.}}, and
  \bibinfo{author}{\bibfnamefont{T.}~\bibnamefont{Minh-Tien}},
  \bibinfo{year}{2001}, \bibinfo{journal}{J. Phys.: Condens. Matter}
  \textbf{\bibinfo{volume}{13}}, \bibinfo{pages}{5625}.

\bibitem[{\citenamefont{Taylor}(1967)}]{taylor:CPA}
\bibinfo{author}{\bibnamefont{Taylor}, \bibfnamefont{D.}},
  \bibinfo{year}{1967}, \bibinfo{journal}{Phys. Rev.}
  \textbf{\bibinfo{volume}{156}}, \bibinfo{pages}{1017}.

\bibitem[{\citenamefont{Timusk and Statt}(1999)}]{timusk:99}
\bibinfo{author}{\bibnamefont{Timusk}, \bibfnamefont{T.}}, and
  \bibinfo{author}{\bibfnamefont{B.}~\bibnamefont{Statt}},
  \bibinfo{year}{1999}, \bibinfo{journal}{Rep. Prog. Phys.}
  \textbf{\bibinfo{volume}{62}}, \bibinfo{pages}{61}.

\bibitem[{\citenamefont{Tsukada}(1969)}]{tsukada:MCPA}
\bibinfo{author}{\bibnamefont{Tsukada}, \bibfnamefont{M.}},
  \bibinfo{year}{1969}, \bibinfo{journal}{J.\ Phs.\ Soc.\ Jpn}
  \textbf{\bibinfo{volume}{26}}, \bibinfo{pages}{684}.

\bibitem[{\citenamefont{Vekic and White}(1993)}]{vekic:93}
\bibinfo{author}{\bibnamefont{Vekic}, \bibfnamefont{M.}}, and
  \bibinfo{author}{\bibfnamefont{S.}~\bibnamefont{White}},
  \bibinfo{year}{1993}, \bibinfo{journal}{Phys. Rev. B}
  \textbf{\bibinfo{volume}{47}}, \bibinfo{pages}{1160}.

\bibitem[{\citenamefont{Vidberg and Serene}(1977)}]{vidberg}
\bibinfo{author}{\bibnamefont{Vidberg}, \bibfnamefont{H.~J.}}, and
  \bibinfo{author}{\bibfnamefont{J.~W.} \bibnamefont{Serene}},
  \bibinfo{year}{1977}, \bibinfo{journal}{J. Low. Temp. Phys.}
  \textbf{\bibinfo{volume}{19}}, \bibinfo{pages}{179}.

\bibitem[{\citenamefont{Voit}(1994)}]{voit:1d}
\bibinfo{author}{\bibnamefont{Voit}, \bibfnamefont{J.}}, \bibinfo{year}{1994},
  \bibinfo{journal}{Rep. Prog. Phys.} \textbf{\bibinfo{volume}{57}},
  \bibinfo{pages}{977}.

\bibitem[{\citenamefont{de~Vries}
  \emph{et~al.}(1993{\natexlab{a}})\citenamefont{de~Vries, Michelsen, , and
  De~Raedt}}]{deVries:fkm2}
\bibinfo{author}{\bibnamefont{de~Vries}, \bibfnamefont{P.}},
  \bibinfo{author}{\bibfnamefont{K.}~\bibnamefont{Michelsen}}, , and
  \bibinfo{author}{\bibfnamefont{H.}~\bibnamefont{De~Raedt}},
  \bibinfo{year}{1993}{\natexlab{a}}, \bibinfo{journal}{Z. Phys. B}
  \textbf{\bibinfo{volume}{92}}, \bibinfo{pages}{353}.

\bibitem[{\citenamefont{de~Vries}
  \emph{et~al.}(1993{\natexlab{b}})\citenamefont{de~Vries, Michelsen, and
  De~Raedt}}]{deVries:fkm1}
\bibinfo{author}{\bibnamefont{de~Vries}, \bibfnamefont{P.}},
  \bibinfo{author}{\bibfnamefont{K.}~\bibnamefont{Michelsen}}, and
  \bibinfo{author}{\bibfnamefont{H.}~\bibnamefont{De~Raedt}},
  \bibinfo{year}{1993}{\natexlab{b}}, \bibinfo{journal}{Phys. Rev. Lett.}
  \textbf{\bibinfo{volume}{70}}, \bibinfo{pages}{2463}.

\bibitem[{\citenamefont{de~Vries} \emph{et~al.}(1994)\citenamefont{de~Vries,
  Michelsen, and De~Raedt}}]{deVries:fkm3}
\bibinfo{author}{\bibnamefont{de~Vries}, \bibfnamefont{P.}},
  \bibinfo{author}{\bibfnamefont{K.}~\bibnamefont{Michelsen}}, and
  \bibinfo{author}{\bibfnamefont{H.}~\bibnamefont{De~Raedt}},
  \bibinfo{year}{1994}, \bibinfo{journal}{Z. Phys. B}
  \textbf{\bibinfo{volume}{95}}, \bibinfo{pages}{475}.

\bibitem[{\citenamefont{Weiss}(1907)}]{weiss:1907}
\bibinfo{author}{\bibnamefont{Weiss}, \bibfnamefont{P.}}, \bibinfo{year}{1907},
  \bibinfo{journal}{J. Phys. Radium} \textbf{\bibinfo{volume}{6}},
  \bibinfo{pages}{661}.

\bibitem[{\citenamefont{Wilson}(1975)}]{nrg:wilson}
\bibinfo{author}{\bibnamefont{Wilson}, \bibfnamefont{K.}},
  \bibinfo{year}{1975}, \bibinfo{journal}{Rev. Mod. Phys.}
  \textbf{\bibinfo{volume}{47}}, \bibinfo{pages}{773}.

\bibitem[{\citenamefont{Zacher}
  \emph{et~al.}(2002{\natexlab{a}})\citenamefont{Zacher, Eder, Arrigoni, and
  Hanke}}]{zacher:cpt2}
\bibinfo{author}{\bibnamefont{Zacher}, \bibfnamefont{M.~G.}},
  \bibinfo{author}{\bibfnamefont{R.}~\bibnamefont{Eder}},
  \bibinfo{author}{\bibfnamefont{E.}~\bibnamefont{Arrigoni}}, and
  \bibinfo{author}{\bibfnamefont{W.}~\bibnamefont{Hanke}},
  \bibinfo{year}{2002}{\natexlab{a}}, \bibinfo{journal}{Phys. Rev. B}
  \textbf{\bibinfo{volume}{65}}, \bibinfo{pages}{045109}.

\bibitem[{\citenamefont{Zacher}
  \emph{et~al.}(2002{\natexlab{b}})\citenamefont{Zacher, Eder, Arrigoni, and
  Hanke}}]{zacher:cpt1}
\bibinfo{author}{\bibnamefont{Zacher}, \bibfnamefont{M.~G.}},
  \bibinfo{author}{\bibfnamefont{R.}~\bibnamefont{Eder}},
  \bibinfo{author}{\bibfnamefont{E.}~\bibnamefont{Arrigoni}}, and
  \bibinfo{author}{\bibfnamefont{W.}~\bibnamefont{Hanke}},
  \bibinfo{year}{2002}{\natexlab{b}}, \bibinfo{journal}{Phys. Rev. Lett.}
  \textbf{\bibinfo{volume}{85}}, \bibinfo{pages}{2585}.

\bibitem[{\citenamefont{Zhang and Rice}(1988)}]{zhang:88}
\bibinfo{author}{\bibnamefont{Zhang}, \bibfnamefont{F.}}, and
  \bibinfo{author}{\bibfnamefont{T.}~\bibnamefont{Rice}}, \bibinfo{year}{1988},
  \bibinfo{journal}{Phys. Rev. B} \textbf{\bibinfo{volume}{37}},
  \bibinfo{pages}{3759}.

\bibitem[{\citenamefont{Zhang and Rice}(1990)}]{zhang:90}
\bibinfo{author}{\bibnamefont{Zhang}, \bibfnamefont{F.}}, and
  \bibinfo{author}{\bibfnamefont{T.}~\bibnamefont{Rice}}, \bibinfo{year}{1990},
  \bibinfo{journal}{Phys. Rev. B} \textbf{\bibinfo{volume}{41}},
  \bibinfo{pages}{7243}.

\bibitem[{\citenamefont{Zitzler} \emph{et~al.}(2002)\citenamefont{Zitzler,
  Pruschke, and Bulla}}]{dmftnrg:zitzler1}
\bibinfo{author}{\bibnamefont{Zitzler}, \bibfnamefont{R.}},
  \bibinfo{author}{\bibfnamefont{T.}~\bibnamefont{Pruschke}}, and
  \bibinfo{author}{\bibfnamefont{R.}~\bibnamefont{Bulla}},
  \bibinfo{year}{2002}, \bibinfo{journal}{Eur. Phys. J. B}
  \textbf{\bibinfo{volume}{27}}, \bibinfo{pages}{473}.

\bibitem[{\citenamefont{Zitzler} \emph{et~al.}(2003)\citenamefont{Zitzler,
  Tong, Pruschke, and Bulla}}]{dmftnrg:zitzler2}
\bibinfo{author}{\bibnamefont{Zitzler}, \bibfnamefont{R.}},
  \bibinfo{author}{\bibfnamefont{N.}~\bibnamefont{Tong}},
  \bibinfo{author}{\bibfnamefont{T.}~\bibnamefont{Pruschke}}, and
  \bibinfo{author}{\bibfnamefont{R.}~\bibnamefont{Bulla}},
  \bibinfo{year}{2003}, \bibinfo{journal}{preprint cond-mat/0308202} .

\bibitem[{\citenamefont{Z\"olfl} \emph{et~al.}(2000)\citenamefont{Z\"olfl,
  Maier, Pruschke, and Keller}}]{Zoelfl:NCA}
\bibinfo{author}{\bibnamefont{Z\"olfl}, \bibfnamefont{M.}},
  \bibinfo{author}{\bibfnamefont{T.}~\bibnamefont{Maier}},
  \bibinfo{author}{\bibfnamefont{T.}~\bibnamefont{Pruschke}}, and
  \bibinfo{author}{\bibfnamefont{J.}~\bibnamefont{Keller}},
  \bibinfo{year}{2000}, \bibinfo{journal}{Eur. Phys. J. B}
  \textbf{\bibinfo{volume}{13}}, \bibinfo{pages}{47}.

\end{thebibliography}

\end{document}